\documentclass{article}      

\usepackage{lineno,hyperref}
\modulolinenumbers[5]

\usepackage{geometry}
\usepackage{graphicx}
\usepackage{booktabs} 
\usepackage{lscape} 
\usepackage{tabularht}  
\usepackage{tabularx}  
\usepackage[thinlines]{easytable}
\usepackage{amsmath}   
\usepackage{multirow}  
\usepackage{longtable} 

\usepackage{float}
\usepackage[caption = false]{subfig}

\usepackage{epstopdf} 

\usepackage[export]{adjustbox}

\usepackage{outlines}
\usepackage{enumitem}
\setenumerate[1]{label=\Roman*.}
\setenumerate[2]{label=\Alph*.}
\setenumerate[3]{label=\roman*.}
\setenumerate[4]{label=\alph*.}

\usepackage{appendix}

\usepackage{longtable}

\DeclareUnicodeCharacter{2009}{FIXME}

\title{Interdependence of Growth, Structure, Size and Resource Consumption During an Economic Growth Cycle}

\author{
  Carey W. King*\\
  \texttt{Energy Institute, The University of Texas at Austin}\\
  \texttt{careyking@mail.utexas.edu}
}
\date{\today}

\begin{document}             

\maketitle                   

\abstract{All economies require physical resource consumption to grow and maintain their structure. The modern economy is additionally characterized by private debt.  The Human and Resources with MONEY (HARMONEY) economic growth model links these features using a stock and flow consistent framework in physical and monetary units.  Via an updated version, we explore the interdependence of growth and three major structural metrics of an economy.  First, we show that relative decoupling of gross domestic product (GDP) from resource consumption is an expected pattern that occurs because of physical limits to growth, not a response to avoid physical limits.  While an increase in resource efficiency of operating capital does increase the level of relative decoupling, so does a change in pricing from one based on full costs to one based only on marginal costs that neglects depreciation and interest payments leading to higher debt ratios.  Second, if assuming full labor bargaining power for wages, when a previously-growing economy reaches peak resource extraction and GDP, wages remain high but profits and debt decline to zero.  By removing bargaining power, profits can remain positive at the expense of declining wages.  Third, the distribution of intermediate transactions within the input-output table of the model follows the same temporal pattern as in the post-World War II U.S. economy.  These results indicate that the HARMONEY framework enables realistic investigation of interdependent structural change and trade-offs between economic distribution, size, and resources consumption.}\\
\noindent \textbf{Keywords:} energy, resources, structural change, macroeconomics, post-Keynesian, systems, long-term growth

\section{Introduction}

Scientists and economists often seek to understand the linkages among natural resources consumption and the cost of resources in tandem with the growth rate, size, and structure of complex systems.  These systems can be biological organisms, ecosystems, and national or global economies.   

The size of organisms and ecosystems is measured by their mass, volume, and population.  Their structure is measured by the flow of nutrients and energy in food webs and internal distributions systems (i.e., circulatory systems) as well as social networks, such as within eusocial insect colonies. 

The size of economies is measured by many metrics such as gross domestic product (or net output), gross output, population, the quantity of physical capital (in money and physical units), and others.  The structure of an economy can be measured by functions and metrics that summarize the distribution of stocks (e.g., capital) and flows (e.g., income, power) among people and jobs, companies, economic sectors \cite{Hidalgo2007,Hidalgo2009,King2016}, or other categories through which money and natural resources flow.  

The purpose of this paper is discuss the interdependence of growth, size, and structure of an economy using outputs from an updated version of the Human and Resources with MONEY (HARMONEY) model of King (2020) \cite{King2020}.  This new version is HARMONEY v1.1.  Because of its structure, the HARMONEY model helps ``narrow the differences'' between economic and ecological viewpoints, which as the late Martin Weitzmann suggested \cite{Nordhaus1992}, provides value by creating enhanced understanding of economic dynamics.  That is to say, because the model simultaneously tracks physical and monetary stocks and flows, by including physical resources and constraints along with macroeconomic accounting and debt, HARMONEY speaks the language of \textit{both} economists \textit{and} physical and natural scientists.


All models are simplified representations of real-world processes, yet stylized models are still useful for providing insight into real-world data.  A model that can accurately represent the dynamic interdependence between growth, size, and structure has more explanatory power than those that cannot.   While HARMONEY v1.1 is not calibrated to a real-world economy, it has critical features and structural assumptions that make it applicable and valuable for comparing its trends to long-term trends in real-world data.  

Macroecological and biological growth literature \cite{Sibly2014,Brown2004} has accumulated an extensive number of studies seeking to explain the universality and robustness of the finding that adult individual organisms \cite{West2001,West2005,Banavar2010,West2017,Ballesteros2018} and groups of organisms, such as eusocial insect colonies \cite{Hou2010,Shik2010,Waters2010,Shik2014,Fewell2016,Waters2017}, follow sublinear scaling, or allometry, relating their metabolism to mass.  That is to say, once an organism obtains a mature structure, basal metabolism (B) increases more slowly than mass (M), as $B \propto M^b$, where $b<1$.  Before reaching its mature structure, an organism can exhibit superlinear scaling ($b>1$) when basal metabolism increases faster than mass, such as in fish \cite{Clarke1999,Mueller2011} and trees \cite{Mori2010}.  Also, when increasing in size from very small single-celled organisms (bacteria) to larger eukaryotic single-celled organisms (protists) to multi-celled organisms, DeLong \textit{et al.} (2010) indicate the scaling of metabolism to mass transitions from superlinear to linear to sublinear, respectively \cite{DeLong2010a}.  Hatton \textit{et al.} (2019) suggest that linear scaling more accurately relates metabolism to mass across all eukaryotes (neglecting bacteria), and that it is an organism's growth (mass/time) that scales sublinearly, near $b=3/4$, with mass \cite{Hatton2019}.  The implication is that the growth rate (\%/time) declines with size.  

This brings us to contemporary questions regarding the rate of economic growth as the global economy continues to increase in size.  Must the global economy necessarily slow its growth rate as it increases in size?  If so, are the reasons similar to those of biological organisms and ecosystems?  

Neoclassical growth theory is of no help in answering these questions.  It posits declining returns to growth if ``technological change'' ceases as capital and labor increase.  In mathematical terms, assuming the usual Cobb-Douglas aggregate production function $Y=A(t) K^\alpha L^{1-\alpha}$, with $0 < \alpha < 1$ and $A(t)$ as total factor productivity, if $dA(t)/dt = 0$, then output $Y$ grows more slowly than capital ($K$) or labor ($L$).  Critics of neoclassical growth theory indicate that the choice of the exponents, or output elasticities, of the factors of production is arbitrary and based on macroeconomic accounting assumptions rather than fundamental features of the economy \cite{Shaikh1974,Felipe2003,Felipe2006}.  If one includes primary energy consumption (PEC) or useful work (U) as a third factor of production, one can accurately represent historical GDP \cite{Ayres2005b,Ayres2008,Kuemmel2011}.  However, one must also abandon the so-called cost-share theorem assumption of neoclassical theory that ``... says that the economic weight of a production factor, which is called the \textit{output elasticity} of that factor, should always be equal to the factor's share in total factor cost.'' \cite{Kuemmel2014}  Both Ayres and Warr \cite{Ayres2005b,Ayres2008} and K\"{u}mmel and Lindenberger \cite{Kuemmel2011,Kuemmel2014} have shown that the output elasticity of primary energy or useful work is near 0.4-0.7, an order of magnitude larger than its cost share.  




Unlike neoclassical growth theory, the post-Keynesian and biophysical structure of the HARMONEY model does enable it to test whether the economy has similar energy-GDP scaling as biological systems and for the same reasons.  Several authors have indicated the similar scaling relation of primary energy consumption to country and global GDP as exists for metabolism and mass for biological organisms \cite{Odum1971,Odum1997,Odum2007,GeorgescuRoegen1971,GeorgescuRoegen1975,Hagens2020,King2021}.  Jarvis and King (in review) summarize how global primary energy consumption (PEC) and GDP scale approximately linearly from 1900-1970, and since scale sublinearly at $PEC \propto GDP^{\frac{2}{3}}$ \cite{Jarvis2020}, and Giraud and Kahraman (2014) confirm a similar finding \cite{Giraud2014}.  Brown \textit{et al.} (2011) \cite{Brown2011} indicate how post-1970 trends show PEC of countries scales with their GDP nearly as $PEC \propto GDP^{\frac{3}{4}}$, the same as basal metabolism scales with mass in mammals.   As stated in \cite{Brown2014}: ``Regardless of whether the approximately 3/4 power scaling is due to a deep causal relationship or an amazing coincidence, both relationships reflect similar underlying causes -- the energy cost of maintaining the structure and function of a large, complex system.''  

A contribution of this paper is to address this ``underlying cause'' in the context of energy-size scaling by using an economic growth model that, among other things, explicitly considers the ``energy cost of maintaining the structure and function'' of an economy as a complex system.   This paper does not address the exact scaling (i.e., value of $b$) between energy consumption and GDP, but it explains why we expect a transition from superlinear or linear scaling to sublinear scaling, just as observed in biological systems.  

Thus, this paper also contributes to the discussion of \textit{decoupling} of GDP from PEC via increases in energy efficiency. Sublinear scaling in the economy, often referred to as a state of declining energy intensity (= PEC/GDP), is often seen as a consequence of increasing energy efficiency.  The issue of decoupling is important because economy-wide rebound effects might erode more than half the reductions in engineering energy efficiency  investments \cite{Brockway2021}. Futher, because the mechanisms of the rebound effect are largely overlooked by integrated assessment models and global energy models that guide policy \cite{King2021,Brockway2021,Keysser2021}, policymakers and energy efficiency advocates are unaware that their efforts to reduce carbon emissions by increasing device efficiency are not nearly as effective as they assume.  That is to say, proponents of energy efficiency measures claim that declining energy intensity is caused by specific actions to increase energy conversion efficiency in machines and distribution networks, and that this reduces energy consumption since less energy is needed for a unit of work or GDP (taking the notion from Warr and Ayres that useful work scales linearly with GDP \cite{Ayres2009,Warr2010,Keen2019}).  However, animals, including each of our own bodies as \textit{Homo sapiens}, exhibit sublinear scaling without any conscious action or choice to do so.  

If we do not make decisions to reduce the ``energy intensity'' of our own bodies, then how we can we be so sure that a declining economy-wide energy intensity is a consequence of our collective conscious actions?  One appropriate way to address this question is via a complex system growth model with enough of features to realistically address the question.  Such is the HARMONEY model of this paper.

%
%
%
%
%
%
%
%
%
%

\section{Methods}

Here we summarize the formulation of HARMONEY v1.1 model by discussing features both similar to and different from v1.0.  We also summarize the information theoretic metrics, as in King (2016), used to characterize the structure of the modeled HARMONEY economy during simulated growth \cite{King2016}.  Using these metrics we check if and how the features of the theoretical HARMONEY model are consistent with trends from U.S. data as in King (2016).  

\subsection{Description of HARMONEY Model (same features as v1.0)}

HARMONEY is an economic growth model that is stock and flow consistent in both money and physical variables \cite{King2020}.  Conceptually it combines the Goodwin-Keen model \cite{Keen1995,Keen2013} (that adds private debt to the Goodwin business cycle model \cite{Goodwin1967}), to the Lotka-Volterra framework of the Human and Nature Dynamics (HANDY) model \cite{Motesharrei2014} of a population that survives by extracting a single regenerative (e.g., forest) natural resource.  To these base frameworks, HARMONEY separates economic production into two industrial sectors: resource extraction and capital goods production.  The goods production sector makes capital for both sectors, and the extraction sector extracts resources required to operate capital in each sector.  Each sector has its own capital ($K_i$), labor ($L_i$), price of output ($P_i$), debt as loans ($D_i$) from a private bank, and physical stock of inventory.

%

\subsubsection{Production and Natural Resource Extraction}


The rate of change of natural resource in the environment, $y$, is equal to resource regeneration minus gross extraction (Equation \ref{eq:nature}), where extraction is a Leontief production function of capital, $K_e$, and labor, $L_e$, as in Equation \ref{eq:gross_output_extraction}.  Any of labor, extraction capital, or resource consumption (to operate capital) can constrain output such that the equality in Equation \ref{eq:gross_output_extraction} holds and capacity utilization, $CU_e$, adjusts as needed.  Extraction technology is described by parameter $\delta_y$ that is the rate resources extracted at full capacity utilization.  For the regenerative natural resource, regeneration is a function of the maximum size of the resource, $\lambda_{y}$, the resource regeneration rate, $\gamma$, and the available stock of resource in the environment, $y$. The maximum regeneration rate occurs when $y = \lambda_{y} / 2$.

\begin{eqnarray}\label{eq:nature}
\dot{y} & = & \textmd{regeneration - extraction} \\ \nonumber
\dot{y} & = & \gamma y(\lambda_y - y) - \delta_{y} y K_e CU_e
\end{eqnarray}

\begin{equation}\label{eq:gross_output_extraction}
X_e =  \delta_y y K_e CU_e = L_e a_e
\end{equation}


The gross output of capital goods, $X_g$, from the goods sector is as in Equation \ref{eq:gross_output_goods}.  As with extraction, it is a Leontief production function of capital, $K_g$, and labor, $L_g$, where $\nu_g$ is constant capital:output ratio, and $a_g$ is a constant sector-specific labor productivity.  As with the extraction sector, given capital and labor, capacity utilization, $CU_g$, adjusts to ensure equality in Equation \ref{eq:gross_output_goods}.

\begin{equation}\label{eq:gross_output_goods}
X_g =  \frac{K_g CU_g}{\nu_g} = L_g a_g
\end{equation}

\subsubsection{Intermediate Demands}\label{sec:IntermediateDemands}

The 2-sector model has four technical coefficients for its Leontief input-output matrix, \textbf{A}, in Equation \ref{eq:Amatrix} and Table \ref{tab:Amatrix}.  The technical coefficients are the same as in HARMONEY v1.0 of King (2020).  We assume the technical coefficients $a_{ge}$ and $a_{gg}$ are constant.  The coefficient $a_{ee}$ indicates the amount of resource consumption required to extract a unit of resources where $\eta_e$ characterizes the level of resources consumption required to operate a unit of capital at full capacity utilization.  Coefficient $a_{eg}$ has component $a^o_{eg}$ to account for resource consumption to operate goods capital and component $a^I_{eg}$ to account for resources that become physically embodied in capital. The factor $a^I_{eg}$ is analogous to the finding in biology that a constant amount of energy is required per unit of animal mass including offspring \cite{Brown2018}, where offspring are the analog of new capital investment in the economy.

\begin{equation}\label{eq:Amatrix}
\textbf{A} = 
\begin{bmatrix}
a_{gg} & a_{ge} \\
a_{eg} & a_{ee}
\end{bmatrix}
= 
\begin{bmatrix}
\frac{x_{gg}}{X_g} & \frac{x_{ge}}{X_e} \\
\frac{x_{eg}}{X_g} & \frac{x_{ee}}{X_e} 
\end{bmatrix}
\end{equation}

\begin{table}[h]
\centering
\caption{Equations for elements of technical requirements matrix, \textbf{A}.}
\begin{tabular}{llc}
\hline
$a_{ee}$ & resources to operate extraction capital & $\frac{\eta_{e}K_{e}CU_{e}}{\delta_y y K_{e} CU_{e}} = \frac{\eta_{e}}{\delta_y y}$ \\
$a_{eg}$ & resources to operate goods capital and invest & $a_{eg} = a^o_{eg} + a^I_{eg}$ \\
$a^o_{eg}$ & resources to operate goods capital & $\frac{\eta_{g}K_{g}CU_{g}}{(K_g CU_g / \nu_g)} = \eta_{g} \nu_g$ \\
$a^I_{eg}$ & resources that become physical goods (incl. investment) & $\frac{y_{X_g}K_{g}CU_{g}}{(K_g CU_g / \nu_g)} = y_{X_g} \nu_g$ \\
$a_{ge}$ & goods input for extraction & constant \\
$a_{gg}$ & goods input to produce goods & constant \\
\hline
\end{tabular}\label{tab:Amatrix}
\end{table}

\subsubsection{Inventories and Capacity Utilization}
Equations \ref{eq:wealth_dot} and \ref{eq:goods_dot} show the rate of change of the physical quantity of inventory for goods, $g$, and wealth, $w_H$, respectively.  As in King (2020), the term ``wealth'' for the physical inventory of extracted resources maintains the nomenclature of the HANDY model \cite{Motesharrei2014}.  The change in physical inventory for each sector is the difference between the reference ($IC_{ref,i}$) and current inventory coverage ($IC_{i}$) multiplied by the targeted physical consumption of each sector output.  If the inventory coverage is higher than the set reference, then inventory decreases, and vice versa.  In essence, the inventories scale up with demand.  

\begin{eqnarray}\label{eq:wealth_dot}
\dot{w}_H & = & (\textmd{reference inventory coverage - inventory coverage})(\textmd{targeted consumption of resources}) \nonumber \\ 
\dot{w}_H & = & (IC_{ref,e} - IC_e)(C_e/P_e + a_{eg}X_g + a_{ee}X_e) 
\end{eqnarray}

\begin{eqnarray}\label{eq:goods_dot}
\dot{g} & = & (\textmd{reference inventory coverage - inventory coverage})(\textmd{targeted consumption of goods}) \nonumber \\ 
\dot{g} & = & (IC_{ref,g} - IC_g)((C_g+I_g+I_e)/P_g + a_{ge}X_e + a_{gg}X_g) 
\end{eqnarray}\\

Wealth and goods inventories can rise and fall with business cycles.   We model capital capacity utilization ($CU_i$) as a function of perceived inventory coverage, following Sterman (2000) \cite{Sterman2000}.  Perceived inventory coverage for sector $i$, $IC_{i,perceived}$, is defined as the quantity of physical inventory divided by a time lag, $\tau$, and the targeted consumption for that sector output.  The higher the total consumption for a given output, the larger the inventory stock needed to buffer consumption over the period of the time lag.  See Supplemental Section \ref{app:LaggedStates} for inventory equations describing inventory coverage and capacity utilization.  Section \ref{sec:Method_Thresholds} summarizes how we determine capacity utilization under resource and participation rate constraints (also see Supplemental Figure S.2 of King (2020)).

\subsubsection{Monetary Net Output and Consumption}
The sectoral monetary gross output is equal to intermediate sales plus total net output, $Y$, of the economy (See Supplementary Table \ref{tab:IOmatrix}).  Since we specify gross extraction and intermediate sales, we solve for monetary net output vector, $Y$, as in Equation \ref{eq:Y_netoutput_matrix}.  In Equation \ref{eq:Y_netoutput_matrix}, $X$ is a vector of sectoral gross output, $\hat{P}$ is a diagonal matrix with sectoral prices on the diagonal, and $\mathbf{1}$ is the identity matrix.  
 
\begin{equation}\label{eq:Y_netoutput_matrix}
Y = \hat{P}X - \hat{P} \textbf{A} X = \hat{P}(\mathbf{1} - \textbf{A})X
\end{equation}

The value of inventory in sector $i$, $\textmd{INV}_i$, is equal to the current unit production cost times the physical quantity of inventory (Equations \ref{eq:value_inventory_goods} and \ref{eq:value_inventory_extract}).  The change in the value of inventory, $\Delta\textmd{INV}$, is the current value of inventory minus the value from the previous time period. We model the value of inventory using lagged equations (see Section \ref{sec:Lagged}). 

\begin{equation}\label{eq:value_inventory_goods}
\textmd{INV}_g =  c_g g
\end{equation}
\begin{equation}\label{eq:value_inventory_extract}
\textmd{INV}_e =  c_e w_H
\end{equation}

The present model is of a closed economy (no imports or exports) with no government.  Then by convention, net monetary output is equal to final consumption plus investment plus change in value of inventories.   We assume household consumption, $C_i$, is fully accommodating and is the residual left from subtracting investment and change in value of inventories from net output (Equations \ref{eq:Cg} and \ref{eq:Cext}).  Since we assume only the goods sector produces investment goods, there is no investment goods output from the extraction sector, and extraction sector net output is equal to sector consumption minus change in the value of inventory. 

\begin{equation}\label{eq:Cg}
C_{g} =  Y_{g} - P_g (I^g_{g} + I^e_g) - \Delta\textmd{INV}_g
\end{equation}
\begin{equation}\label{eq:Cext}
C_{e} =  Y_{e} - \Delta\textmd{INV}_e
\end{equation}

\subsubsection{Population}

Population, $N$, changes via constant birth rate, $\beta_N$, and death rate, $\alpha_N$, as a function of per capita physical consumption of extracted resources, $\frac{C_e}{N P_e}$, where $P_e$ is the unit price of extracted resources (Equation \ref{eq:population}). This death rate function, ${\alpha_N\left(\frac{C_e}{N P_e}\right)}$ decreases from a maximum value at zero resource consumption to a minimum positive death rate at some specified per capita resource consumption (see Supplemental Equation \ref{eq:death_rate}). 

\begin{equation}\label{eq:population}
\dot{N} = \beta_N N - \alpha_N\left(\frac{C_e}{N P_e}\right) N
\end{equation}

\subsubsection{Debt}

Debt for each sector, $D_i$, increases when total monetary investment for that sector, $I_i$, exceeds depreciation and net profits, $\Pi_i$.

\begin{equation}\label{eq:Ddot}
\dot{D}_{i} = I_i - P_g \delta K_i - \Pi_{i}
\end{equation}

\subsubsection{Net Power Accounting}\label{sec:NetPowerAccounting}
Net power accounting using metrics of net external power ratio (NEPR) of the extraction sector and net power ratio at the economy-wide level (NPR) are the same as in King (2020) and described in Supplemental Section \ref{app:NetPowerAccounting}.

\subsubsection{Lagged Equations for Simulation}\label{sec:Lagged}
In the real world, data are only available for decision making after some amount of time.  For example, firms and governments know profits and net output from the previous year, but they generally don't know those values for last month, yesterday, or the previous hour.  To make certain variables available within the simulation code, we model their values from the ``previous time period'' as a first order lag (see Supplemental Section \ref{app:LaggedStates} and Equation \ref{eq:lag1}) \cite{Sterman2000}.  For each sector $i$, we update the following variables using a first order lag: capacity utilization ($CU_i$), perceived inventory coverage ($IC_{i,perceived}$), profit ($\Pi_i$), value added ($V_i$), and value of inventory ($\textmd{INV}_i$).  For further reference see the Appendix \ref{app:EquationsList} that lists the core differential equations of the model.

\subsubsection{Biophysical Constraining Thresholds}\label{sec:Method_Thresholds}
Resource extraction is allocated among the operation of each type of capital, household consumption, and resource embodied in investment (Figure 1 of King (2020)).  It is possible that the extraction rate of resources is insufficient to fully satisfy minimum levels of household consumption and operational inputs for capital along with the desired level of investment.  To account for output constraints based on labor or resource flows, the model dynamics operate within one of eight possible modes based on three binary threshold criteria (i.e., $2^3 = 8$). 

The first threshold criterion is the maximum participation rate ($\lambda_{N\textmd{,max}}$). If there is not enough labor to operate capital at full utilization, then labor is at maximum participation rate and capacity utilization decreases to ensure the equality of Equations \ref{eq:gross_output_extraction} and \ref{eq:gross_output_goods}.

The second threshold criterion is a minimum household consumption of resources per person ($\rho_e > 0$).  If per capita household resource consumption ($ = C_e/(P_e N)$) would otherwise be less than $\rho_e$, we set $C_e/(P_e N)=\rho_e$ and reduce physical investment to match total resource consumption to extraction.  A reduction in investment reduces gross output of goods which in turn reduces total resource consumption since resources are embodied in physical capital.  

The third and final threshold criterion is a minimum household consumption of goods per person ($\rho_g \geq 0$) that we set to zero.  If this threshold is met, gross investment is reduced to reduce resource consumption. If gross investment declines to zero, intermediate demands account for all goods consumption.

\subsection{Description of HARMONEY Model (Differences in this v.1.1 from v1.0)}
This section describes HARMONEY v1.1 differences as compared to HARMONEY 1.0 in King (2020).  These changes generally make the model more robust to parameter changes.

\subsubsection{Wages and Labor}

The participation rate, $\lambda_N$ (employment), is the labor of both sectors divided by population, $N$.  We specify a maximum participation rate, $\lambda_{N\textmd{,max}} \leq 1$, (equal to 80\% in this paper), to represent that some fraction of the population is too young, old, or otherwise unable to work. 

Following Keen (2013) we model the rate of changes of wages ($w$) as a function of participation rate and inflation \cite{Keen2013}:

\begin{equation}\label{eq:wages}
\frac{\dot{w}}{w} = \phi(\lambda_N) + w_1 i + w_2 \frac{1}{\lambda_N}\frac{d \lambda_N}{dt} 
\end{equation}

\noindent where $\lambda_N$ is participation rate, $\phi(\lambda_N)$ is a short-run Phillips curve (see Supplemental Section \ref{app:WageFunction}), $0 \leq w_1 \leq 1$ weights how much inflation affects the nominal wage, inflation, $i$, is calculated as the consumption-weighted average change in prices (Equation \ref{eq:inflation}), and $w_2$ weights how much the rate of change of employment affects nominal wage.  The difference from King (2020) is the addition of the second and third terms in Equation \ref{eq:wages}.   When $w_1=1$, the participation rate can come to equilibrium to its nominal value $\lambda_{N,o}$ as defined in the Phillips curve.

\begin{equation}\label{eq:inflation}
i = \frac{C_g}{C_g+C_e}\frac{\dot{P_g}}{P_g} + \frac{C_e}{C_g+C_e}\frac{\dot{P_e}}{P_e}
\end{equation}

\subsubsection{Investment and Capital Accumulation}\label{sec:InvestmentCapitalDebt}

We model investment the same as in King (2020) but add an option to include what we term Ponzi investment.  Ponzi investment increases debt but does not contribute to new physical capital.   It is unrealistic to think that firms will continue to invest in physical capital if that physical capital continues to accumulate but operate at declining capacity utilization.  However, speculative and Ponzi-style investment does occur, as described by Hyman Minsky and by Keen \cite{Minsky1977,Keen2013}. 

%


Total monetary investment in each sector is as in King (2020) and shown in Equation \ref{eq:investment} where $\kappa_{0,i}$ and  $\kappa_{1,i}$ describe investment as multipliers on depreciation and net profit share, $\Pi_i$, respectively. The Ponzi fraction of investment is $fP_{I_i}$ ($0 \leq fP_{I_i} \leq 1 $).  The non-Ponzi fraction, $(1 - fP_{I_i})$, of this total sectoral investment is allocated to new capital formation (Equation \ref{eq:investment_newK}).  The Ponzi fraction of monetary investment (Equation \ref{eq:investment_Ponzi}) does not increase the existing capital stock, but only increases debt since all monetary investment, $I_i$, increases debt (Equation \ref{eq:Ddot}).   Our modeling of Ponzi investment is inspired by, but different from, that defined in Keen (2009) and Grasselli and Costa Lima (2012) who model Ponzi investment as debt that increases as a function of the real GDP growth rate \cite{Keen2009,Grasselli2012}.  We model the Ponzi fraction of investment as a function of capacity utilization as in Equation \ref{eq:PonziFraction} where $CU_{i,ref}$ is the reference, or target, capacity utilization which we set at 85\%.  The parameter $a_{Ponzi,i} \geq 0$ governs the magnitude of Ponzi investment, with larger values shifting more investment from physical capital to Ponzi.

\begin{equation}\label{eq:investment}
I_{i} = max\{0, \kappa_{0,i}P_g \delta K_i + \kappa_{1,i} \Pi_i\}
\end{equation}

\begin{equation}\label{eq:investment_newK}
I_{newK,i} = (1 - fP_{I_i}) I_{i}
\end{equation}

\begin{equation}\label{eq:investment_Ponzi}
I_{Ponzi,i} = fP_{I_i} I_{i}
\end{equation}

\begin{equation}\label{eq:PonziFraction}
fP_{I_i} = min \left\lbrace 1, max \left\lbrace 0, a_{Ponzi,i} \left( \frac{CU_{i,ref} - CU_i}{CU_{i,ref}} \right) \right\rbrace \right\rbrace
\end{equation}

$I^g_e$ and $I^g_g$ represent physical investment in $K_e$ and $K_g$, respectively, where the superscript $g$ indicates capital has units of goods.  Physical investment in new capital for each sector is thus equal to the non-Ponzi monetary investment divided by the price of goods, or $I^g_i = \frac{(1 - fP_{I_i})I_i}{P_g}$.  We use the perpetual inventory method for capital accumulation as physical investment minus physical depreciation occurring at constant rate, $\delta$, for each sector (Equation \ref{eq:Kdot}). 

\begin{equation}\label{eq:Kdot}
\dot{K}_{i} = I^g_i - \delta K_{i} = \frac{(1 - fP_{I_i})I_i}{P_g} - \delta K_{i}
\end{equation}

\subsubsection{Prices and Costs}
Similarly to King (2020) we calculate prices, $P_i$, based on a constant markup, $\mu_i \geq 0$, multiplied by the cost of production, or $P_i = (1+\mu_i)c_i$.  Thus, prices change based upon the difference between the marked-up cost and price as in Equations \ref{eq:dP_over_P} and \ref{eq:Price_i}, and these equations are equivalent to those in King (2020). However, unlike King (2020), we no longer solve for all sector prices simultaneously (using a matrix inversion), but use Equation \ref{eq:Price_i} to solve for the change in price for each sector.  Equations \ref{eq:cost_goods} and \ref{eq:cost_extract} define the \textit{full cost} of production. In the results we explore differences that arise from the assumption that producers set prices on the full cost versus only the \textit{marginal costs} that neglect the cost of both interest payments ($r_L D_i$) and depreciation ($P_g\delta K_i$).

\begin{equation}\label{eq:dP_over_P}
\frac{\dot{P}_{i}}{P_i} = \left(\frac{1}{\tau_{P_i}}\right)\left((1 + \mu_i)(c_i/P_i) - 1 \right)
\end{equation}

\begin{equation}\label{eq:Price_i}
\dot{P}_{i} = \left(\frac{1}{\tau_{P_i}}\right) \left((1 + \mu_i) c_i - P_i \right)
\end{equation}

\begin{equation}\label{eq:cost_goods}
c_g = P_g a_{gg} + P_e a_{eg} + (wL_g + r_L D_g + P_g \delta K_g)/X_g
\end{equation}

\begin{equation}\label{eq:cost_extract}
c_e = P_e a_{ee} + P_g a_{ge} + (wL_e + r_L D_e + P_g \delta K_e)/X_e
\end{equation}

\subsection{Information Theory and Self Organization}\label{sec:InformationTheory_SelfOrganize}

\subsubsection{Information Theory to Assess Economy Structure}\label{sec:InformationTheory_Ulanowicz}

Over the course of a few decades Robert Ulanowicz developed the use of information theoretic metrics to quantify the structure of food webs \cite{Ulanowicz2009a,Ulanowicz2009}.  King (2016) applied those methods to the U.S. economy \cite{King2016}.  We use these mathematics to quantify the internal structure of the HARMONEY model economy.  We are interested in structure because in network science, ecology, and economics, system structures that distribute flows more evenly are sometimes considered more resilient and complex.  By internal structure, we refer to the proportional distribution of economic transactions within the model's 2$\times$2 intermediate transactions matrix, \textbf{X} (Equation \ref{eq:Xmatrix}).  By discussing structural metrics of information theory along with measures of size and growth (population, debt, resource extraction rate, net output, etc.) we enable a more holistic description of economic evolution.  

In Results and Supplemental material we discuss three information theory metrics: information entropy ($H$), conditional entropy ($\Psi$), and mutual constraint ($X_{MC}$).\footnote{Here we use $X_{MC}$ to represent mutual constraint, instead of $X$ as in much of the literature because this paper already uses $X$ to represent physical gross output of each sector in our model. Mutual constraint is mathematically equal to the terms mutual information and average mutual information \cite{Shannon1962} that are used in the field of information theory.}  Equations \ref{eq:H}--\ref{eq:psi} show the mathematics for these metrics, and we summarize them here but refer the reader to King (2016) for full details.  We use economic rather than network terminology where a network node is a sector, and network flow is the transaction between sectors.  A monetary purchase (flow) within input-output (I-O) matrix \textbf{X} from sector $j$ to sector $i$ is represented as $\textbf{X}_{ij}$ (Equation \ref{eq:Xmatrix}).  The `dot' subscript on \textbf{X} in Equations \ref{eq:H}--\ref{eq:psi} indicates the sum of items over that dimension. For example, ~$\textbf{X}_{.j}$ is the sum of all purchases by sector $j$, and ~$\textbf{X}_{i.}$ is the sum of all sales by sector $i$.  Also, ~$\textbf{X}_{..}$ is the total system throughput (TST), or the sum of all transactions in the I-O table \textbf{X} (see Equation \ref{eq:TST}).

\begin{equation}\label{eq:Xmatrix}
\textbf{X} = 
\begin{bmatrix}
x_{gg} & x_{ge} \\
x_{eg} & x_{ee} 
\end{bmatrix}
\end{equation}

The economy information entropy, or indeterminacy, H is defined in Equation \ref{eq:H}, and is equal to the sum of mutual constraint and conditional entropy (Equation \ref{eq:H2}).  The economy mutual constraint, $X_{MC}$ (Equation \ref{eq:X}), measures the degree to which an economy efficiently distributes flows among its sectors or its average degrees of constraint.  The conditional entropy, $\Psi$ (Equation \ref{eq:psi}), is a measure of the average degrees of freedom of the economic network for all transactions $X_{ij}$, or the remaining choice of flow pathways for transactions going from sector $i$ to sector $j$.  Ulanowicz interprets $X_{MC}$ as what is known about the network and $\Psi$ as what is not known, but what is possible in terms of flows moving through the network \cite{Ulanowicz2009}. 

\begin{equation}\label{eq:H}
H = -\sum_{i,j}\frac{\textbf{X}_{ij}}{\textbf{X}_{..}}log_2\left(\frac{\textbf{X}_{ij}}{\textbf{X}_{..}}\right)
\end{equation}

\begin{equation}\label{eq:X}
X_{MC} = \sum_{i,j}\frac{\textbf{X}_{ij}}{\textbf{X}_{..}}log_2\left(\frac{\textbf{X}_{ij}\textbf{X}_{..}}{\textbf{X}_{i.}\textbf{X}_{.j}}\right)
\end{equation}

\begin{equation}\label{eq:psi}
\Psi = -\sum_{i,j}\frac{\textbf{X}_{ij}}{\textbf{X}_{..}}log_2\left(\frac{\textbf{X}_{ij}^2}{\textbf{X}_{i.}\textbf{X}_{.j}}\right)
\end{equation}

\begin{equation}\label{eq:TST}
\textbf{X}_{..} = \sum_{i,j}\textbf{X}_{ij}
\end{equation}

\begin{equation}\label{eq:H2}
H = X_{MC} + \Psi
\end{equation}

\begin{figure}
\begin{center}
\includegraphics[width=.4\columnwidth]{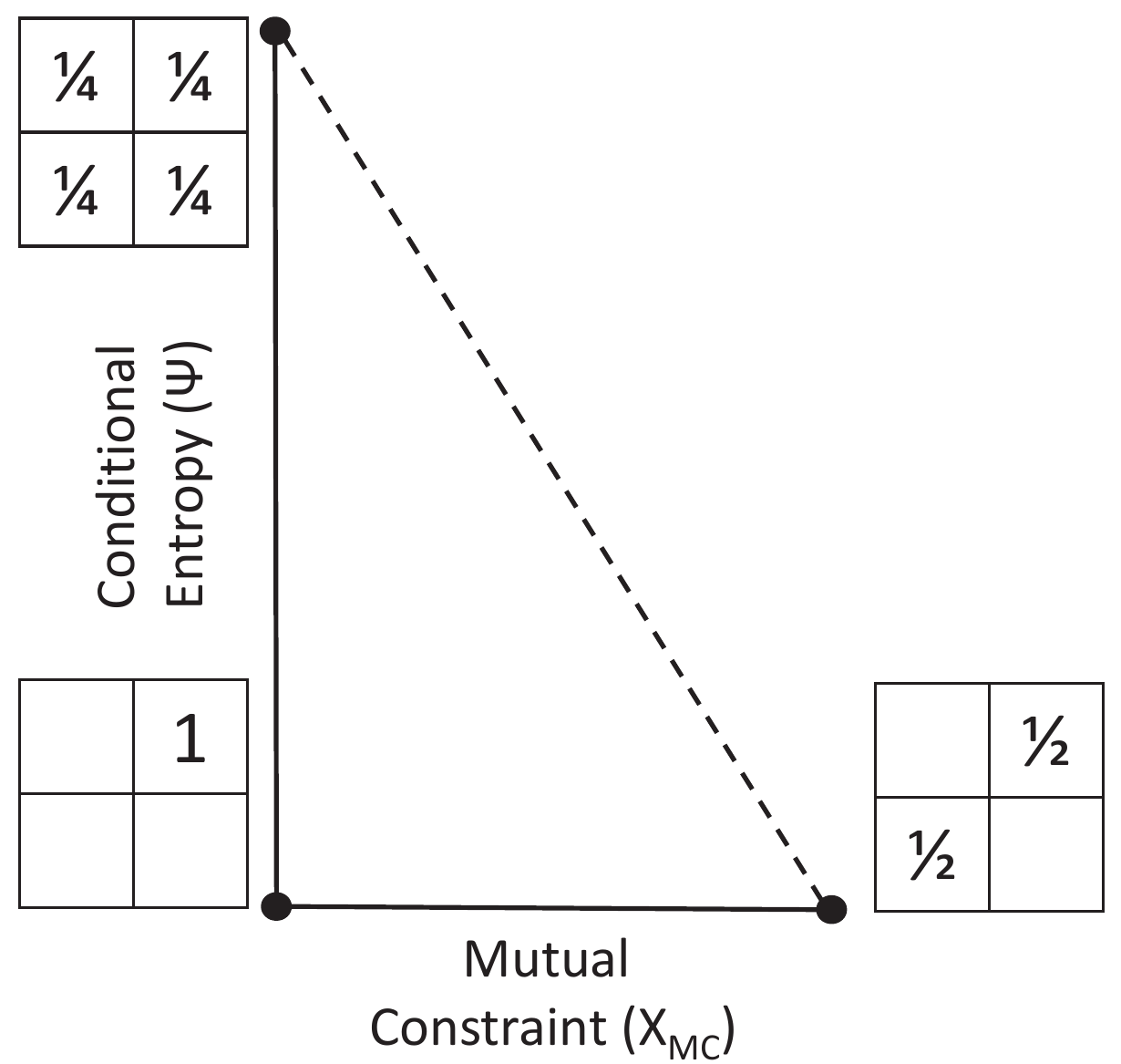}
\caption{The conditional entropy versus mutual constraint phase space is used to interpret the proportion of all intermediate transactions that occur within any sector-to-sector transaction and in total for each sector. This figure indicates the 2-sector model's 2$\times$2 I-O table values, as fractions of the total, at the extreme points of the phase space.}\label{fig:InformationTheory}
\end{center}
\end{figure}

Figure \ref{fig:InformationTheory} helps interpret the metrics.  The calculations of $X_{MC}$ and $\Psi$ are restricted to the triangular area, or phase space, encompassed by the solid and dashed lines.  
The maximum number for each metric increases with the number of nodes, $n$, of the network ($H_{max}=\Psi_{max}=log_2(n^2)$, $X_{MC,max}=log_2(n)$).  
The higher the conditional entropy, the more equal is each intersectoral transaction. At maximum conditional entropy (also maximum information entropy and zero mutual constraint) all intersectoral transactions are equal (upper boundary point in Figure \ref{fig:InformationTheory}). At maximum mutual constraint each sector transacts with only one other sector, and each of these single transactions are equal (lower-right boundary point in Figure \ref{fig:InformationTheory}).  At zero conditional entropy and mutual constraint, there is only one intersectoral transaction (lower-left boundary point in Figure \ref{fig:InformationTheory}).  \textit{Ceteris paribus}, an economy with higher information entropy is more resilient to changing conditions and has a more diverse economy because many sectors contribute a significant share of economic transactions. However, in general, physical constraints in the economy prevent achieving a state of maximum conditional entropy of a monetary I-O matrix (e.g., the ``petroleum refining'' sector inherently purchases more from the ``oil and gas extraction'' sector than the other way around).

\subsection{Endogenous and Exogenous Variables}\label{sec:VariablesTypes}

Table \ref{tab:EndogenousExogenousVariables} lists the endogenous and exogenous variables included in the model.  

\begin{table}
\centering
\caption{A list of endogenous and exogenous variables within the model.}
\begin{tabular}{cc}
\hline
Endogenous Variables & Exogenous Variables  \\
\hline
population via death rates ($\alpha_N$) & birth rates ($\beta_N$) \\
prices ($P_e, P_g$) & natural resource definition  \\
wages ($w$) & wage function parameters (Phillips curve, $w_1, w_2$)  \\
debt of firms ($D_e, D_g$) & interest rates ($r_L, r_M$) \\
capacity utilization ($CU_e, CU_g$) & sector-specific labor productivity ($a_e, a_g$) \\
inventories & goods sector productivity (capital:output ratio, $\nu_g$) \\
economic output ($Y_e, Y_g$) & resource requirements to make capital ($y_{X_g}$) \\
physical output ($X_e, X_g$) & resource requirement to operate capital ($\eta_e, \eta_g$) \\
investment ($I_e, I_g$) & time constants ($\tau$) \\
capital ($K_e, K_g$) &  \\
labor ($L_e, L_g$) &  \\
household consumption ($C_e, C_g$) &  \\
net power ratios (e.g., NEPR) &  \\
\hline
\end{tabular}\label{tab:EndogenousExogenousVariables}
\end{table}

\newpage
\section{Results}\label{sec:Results}

\subsection{Scenario Definitions}

We run several scenarios to explore the influence of changing four major factors and assumptions (Table \ref{tab:ScenariosDefinitions}).  The first is the assumption whether prices are based on full cost (FC) or marginal cost (MC) pricing.  The second decreases the resource consumption to operate capital, $\eta_i$, to observe the effects of increasing efficiency.  Starting at time $T=0$, we decrease $\eta_i$ as a 3rd order function of investment (into physical capital) to approximate that improvements in capital stock occur via investing in new capital (i.e., learning by doing).  See Figures \ref{fig:World_and_A11A00F11F00_GrowthPEU_vs_GrowthGDP}(e) and (f) and Supplemental Section \ref{app:DecreasingEta} for description of $\eta_i$ as a function of investment.

The third concept defining the scenarios relates to wages.  We linearly decrease $w_1$ and $w_2$ from 1 to 0, during the time span indicated in Table \ref{tab:ScenariosDefinitions}, to simulate the loss of labor ``bargaining power.''  Finally, the fourth scenario factor is whether to include Ponzi investing (at all times) by changing $a_{Ponzi,i}$ from 0 (no Ponzi investing) to 3 (with Ponzi investing).

\begin{table}[h]
\centering
\caption[]{Definitions of simulated scenarios.  The shorthand format to describe scenarios is FC-XYZ and MC-XYZ such that FC-000 and MC-000 represent ``baseline'' scenarios to which changes are made where X, Y, and Z switch to 1 as follows.  FC=full cost pricing, MC=marginal cost pricing, X=0: $\eta_i$ remain constant, X=1: $\eta_i$ decrease (higher efficiency scenarios), Y=0: labor retains full bargaining power ($w_1=w_2=1$), Y=1: labor bargaining power reduces to zero ($w_1=w_2=0$), Z=0: no Ponzi investing, and Z=1: includes Ponzi investing.}\label{tab:ScenariosDefinitions}
\begin{tabular}{ccccc}
\hline
Scenario & $\eta_i$ & Pricing: Full or  & $(w_1$, $w_2)$  & Ponzi factor, \\
 &  & Marginal Costs &  & $a_{Ponzi,i}$ \\
 & Table \ref{tab:Amatrix} & Eq. \ref{eq:cost_extract}, \ref{eq:cost_goods} & Eq. \ref{eq:wages} & Eq. \ref{eq:PonziFraction}   \\
\hline
FC-000 & constant & Full & (1,1) & 0 \\
MC-000 & constant & Marginal & (1,1) & 0  \\
FC-010 & constant & Full & $(1,1) \rightarrow (0.0,0.0)@t=60$ to $t=160$ & 0 \\
MC-010 & constant & Marginal & $(1,1) \rightarrow (0,0)@t=100$ to $t=200$ & 0 \\
FC-011 & constant & Full & $(1,1) \rightarrow (0.0,0.0)@t=60$ to $t=160$ & 3 \\
MC-011 & constant & Marginal & $(1,1) \rightarrow (0,0)@t=100$ to $t=200$ & 3 \\
\hline
FC-100 & decreasing  & Full & (1,1) & 0 \\
MC-100 & decreasing & Marginal & (1,1) & 0 \\
FC-110 & decreasing & Full &  $(1,1) \rightarrow (0.0,0.0) @t=60$ to $t=160$ & 0 \\
MC-110 & decreasing & Marginal &  $(1,1) \rightarrow (0.0,0.0) @ t=100$ to $t=200$ & 0 \\
FC-111 & decreasing & Full &  $(1,1) \rightarrow (0.0,0.0) @ t=60$ to $t=160$ & 3 \\
MC-111 & decreasing & Marginal &  $(1,1) \rightarrow (0.0,0.0) @ t=100$ to $t=200$ & 3 \\
\hline
\hline
\end{tabular}
\end{table}

\subsection{High Level Results: Full and Marginal Cost Assumptions}
We first discuss some high level takeaways, and highlight differences from the first HARMONEY paper in King (2020).  Figure \ref{fig:1p8_A11A00F11F00} compares scenarios with and without gains in capital operating efficiency for both full and marginal cost assumptions.  Scenarios FC-000 and MC-000 assume no change in resource consumption per unit of capital output ($\eta_i$) whereas scenarios FC-100 and MC-100 assume an increase in machine efficiency as a decrease in $\eta_i$.   Each scenario begins from a steady state condition (e.g., constant level of stocks).  The equilibrium conditions are defined as the steady state values achieved by simulating each cost scenario at a relatively low value of the depletion parameter, $\delta_y=0.0072$, such that very little resource is exploited (less than 5\% of maximum level).  They start from these initial conditions (including zero debt and profit share) such that population and capital increase by gradually increasing $\delta_y$ to a value of $\delta_y=0.009$ to enable a larger quantity of the natural resource base to be profitably extracted (see Figure \ref{fig:World_and_A11A00F11F00_GrowthPEU_vs_GrowthGDP}).  We increase $\delta_y$ via a 3rd order time delay, reaching its halfway mark at $T=40$ and 99\% of its maximum value at $T=84$.  This assumption of increasing $\delta_y$ mimics an improvement in technological capability for a known resource base, and drives an initial decrease in technical coefficient $a_{ee}$.  Similar results and investigations of growth can be achieved in other ways, such as by increasing the maximum size of the natural resource, $\lambda_{y,max}$, that mimics finding more resources at constant extraction technology (see Section \ref{app:delta_y_vs_lambda_y} and Figure \ref{fig:delta_y_vs_lambda_y} for a brief comparison), but we do not further explore these other assumptions to induce growth.

\begin{figure}
\begin{center}
\subfloat[]{\includegraphics[width=0.33\columnwidth]{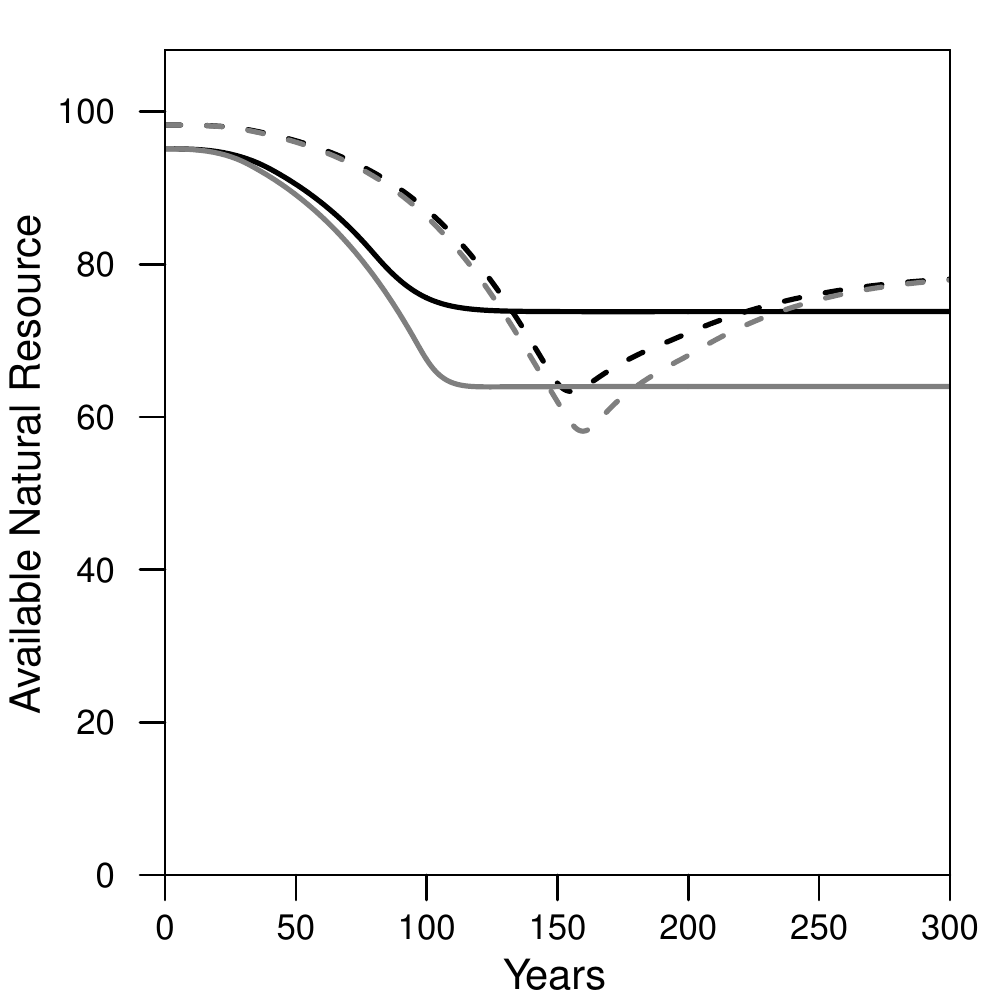}}
\subfloat[]{\includegraphics[width=0.33\columnwidth]{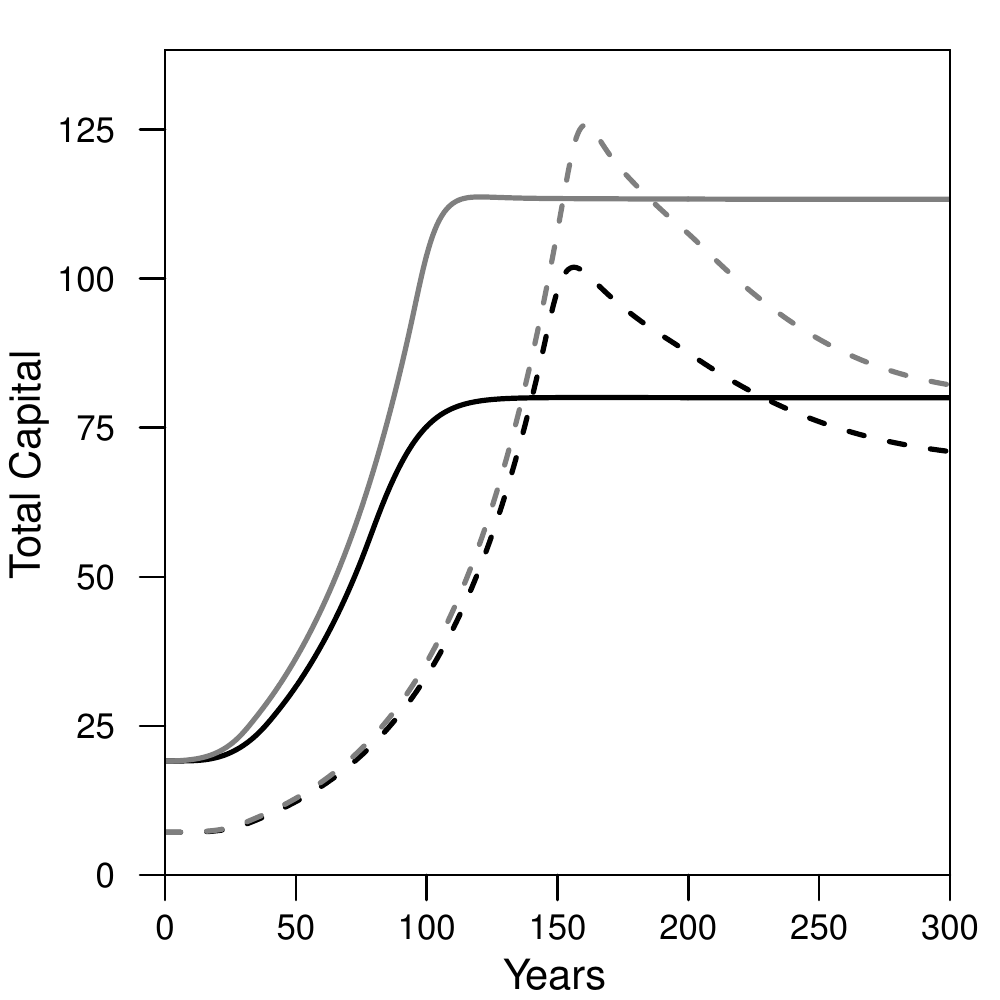}}
\subfloat[]{\includegraphics[width=0.33\columnwidth]{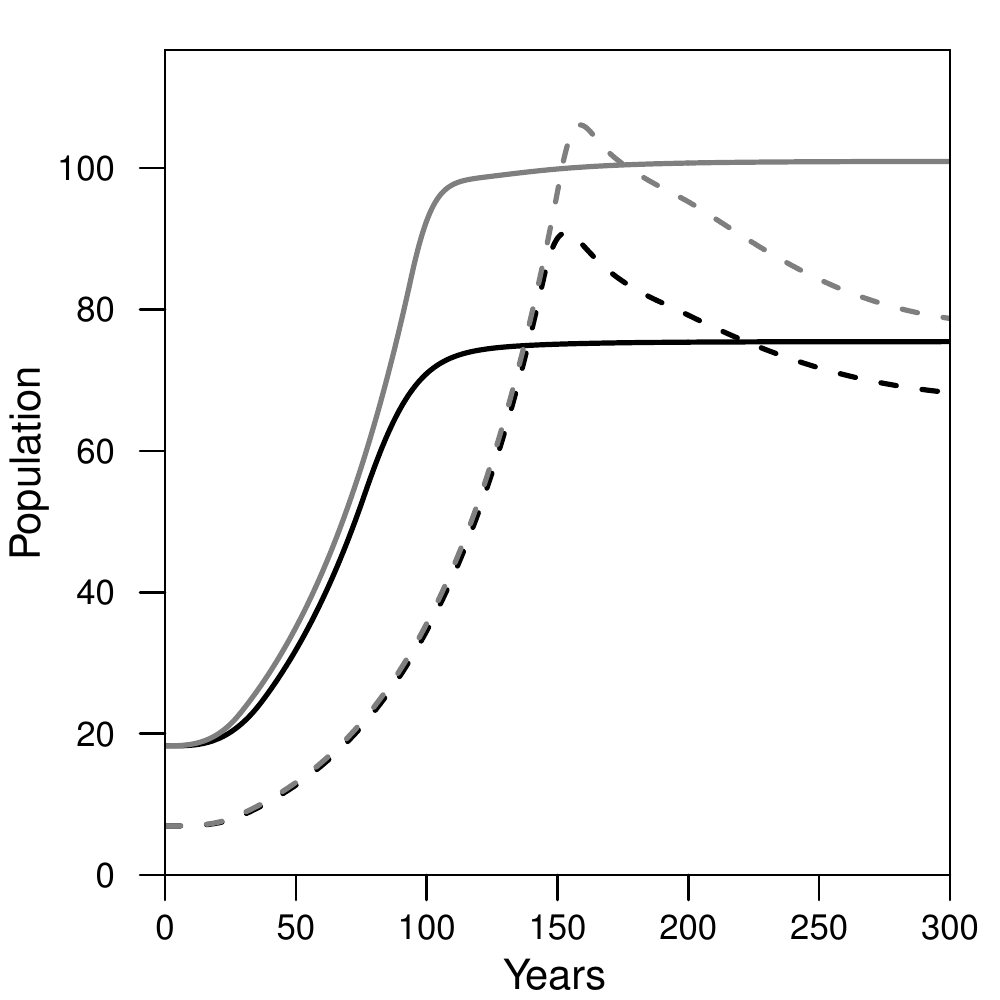}}\\
\subfloat[]{\includegraphics[width=0.33\columnwidth]{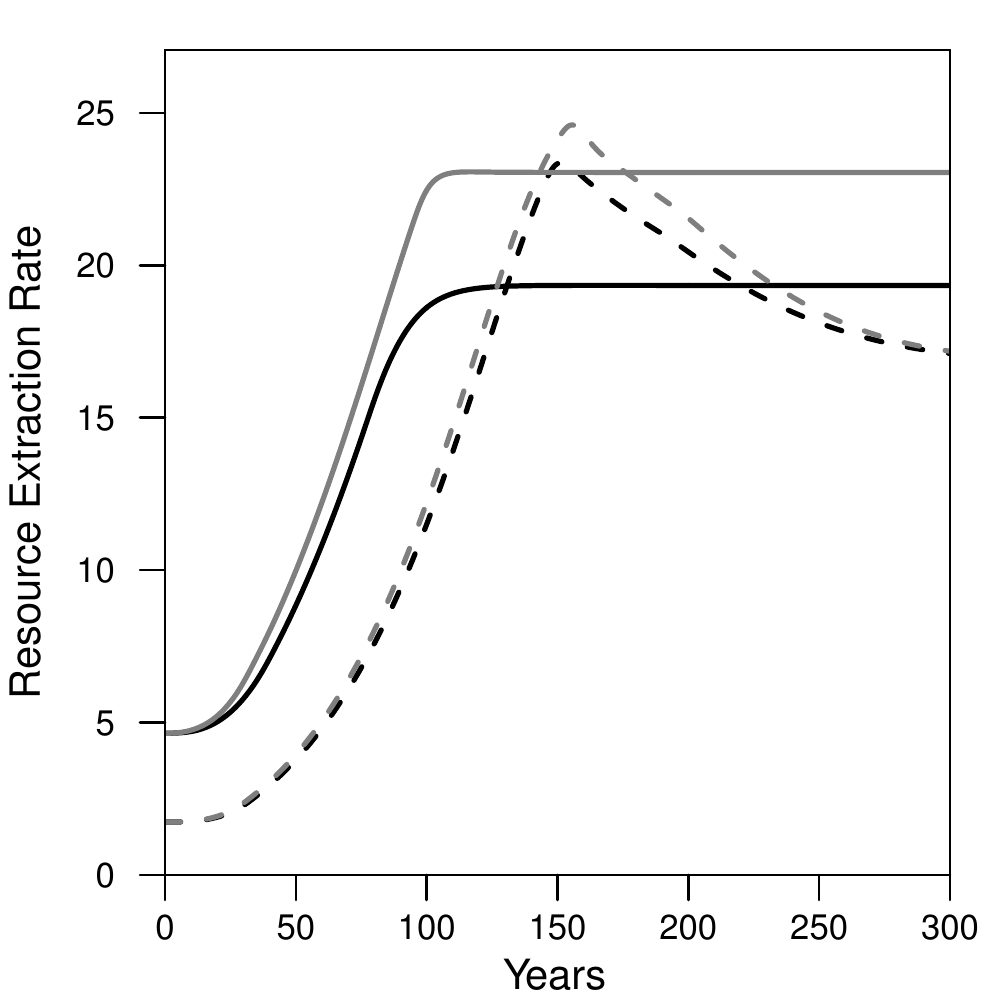}}
\subfloat[]{\includegraphics[width=0.33\columnwidth]{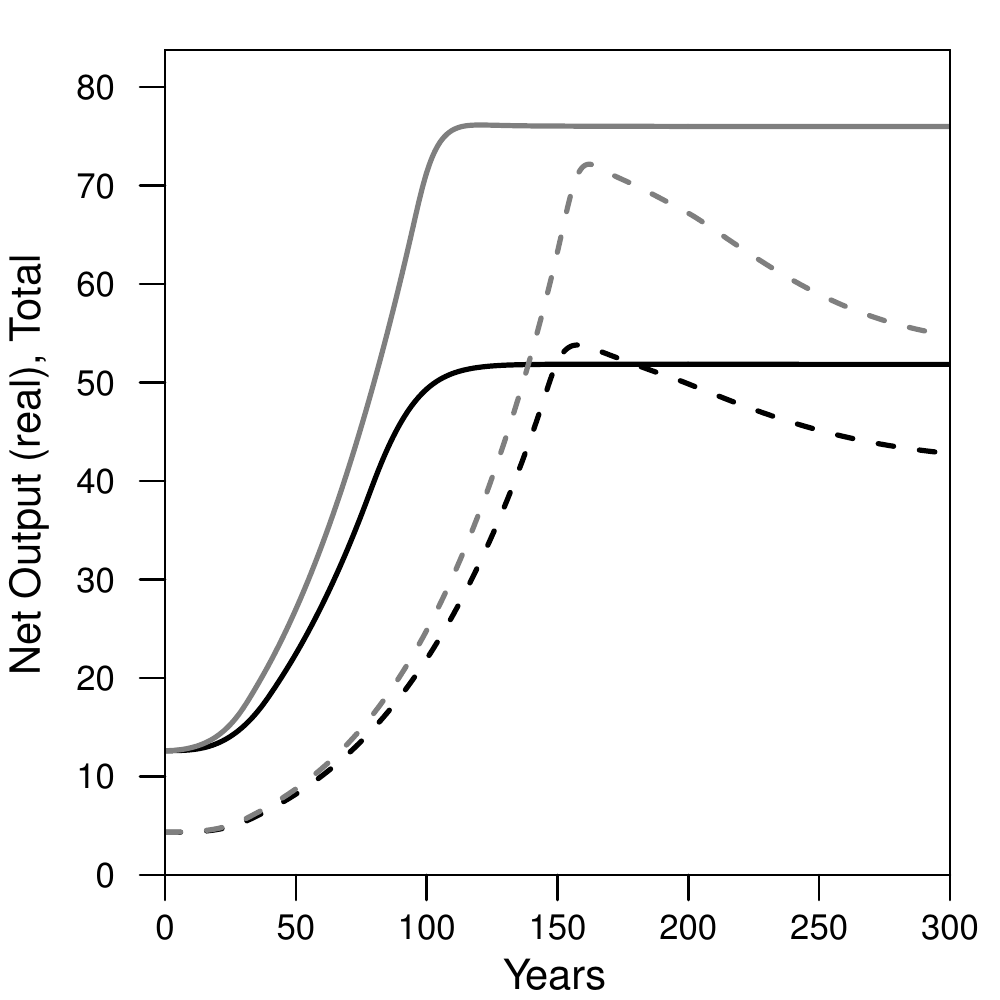}}
\subfloat[]{\includegraphics[width=0.33\columnwidth]{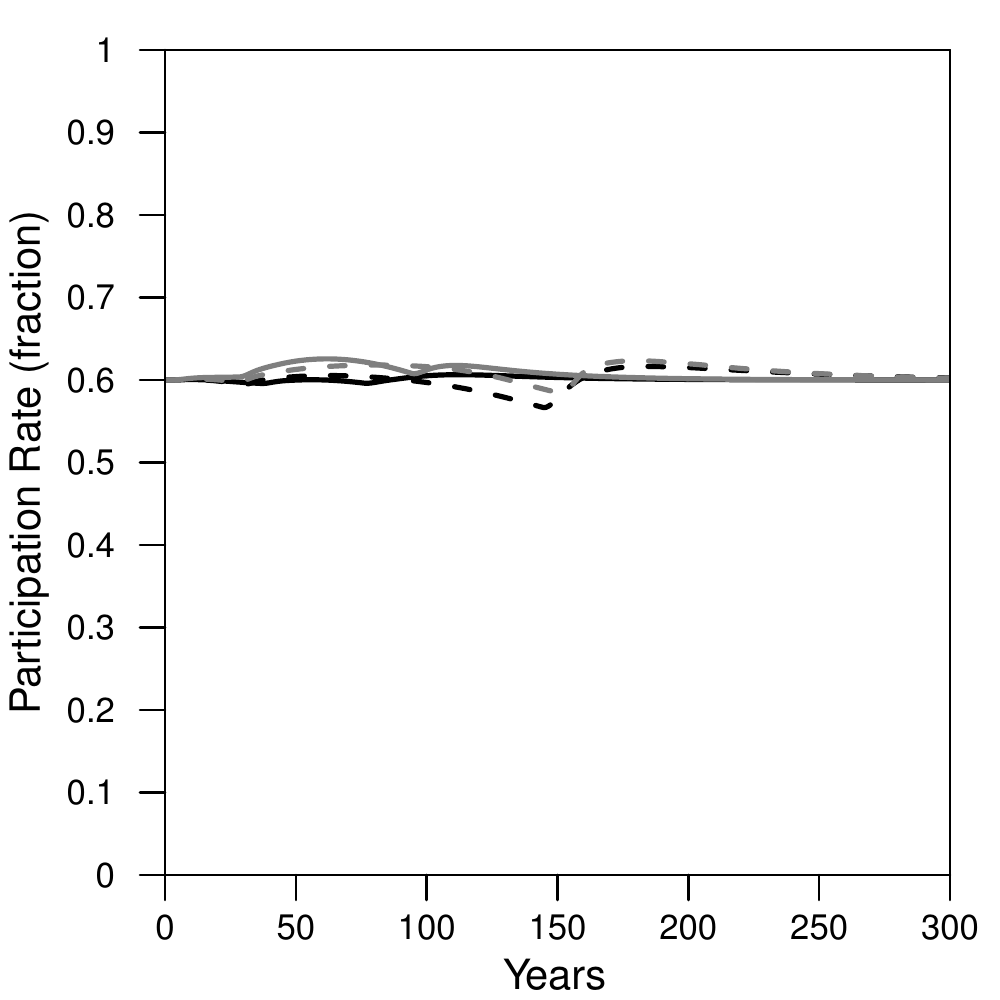}}\\
\subfloat[]{\includegraphics[width=0.33\columnwidth]{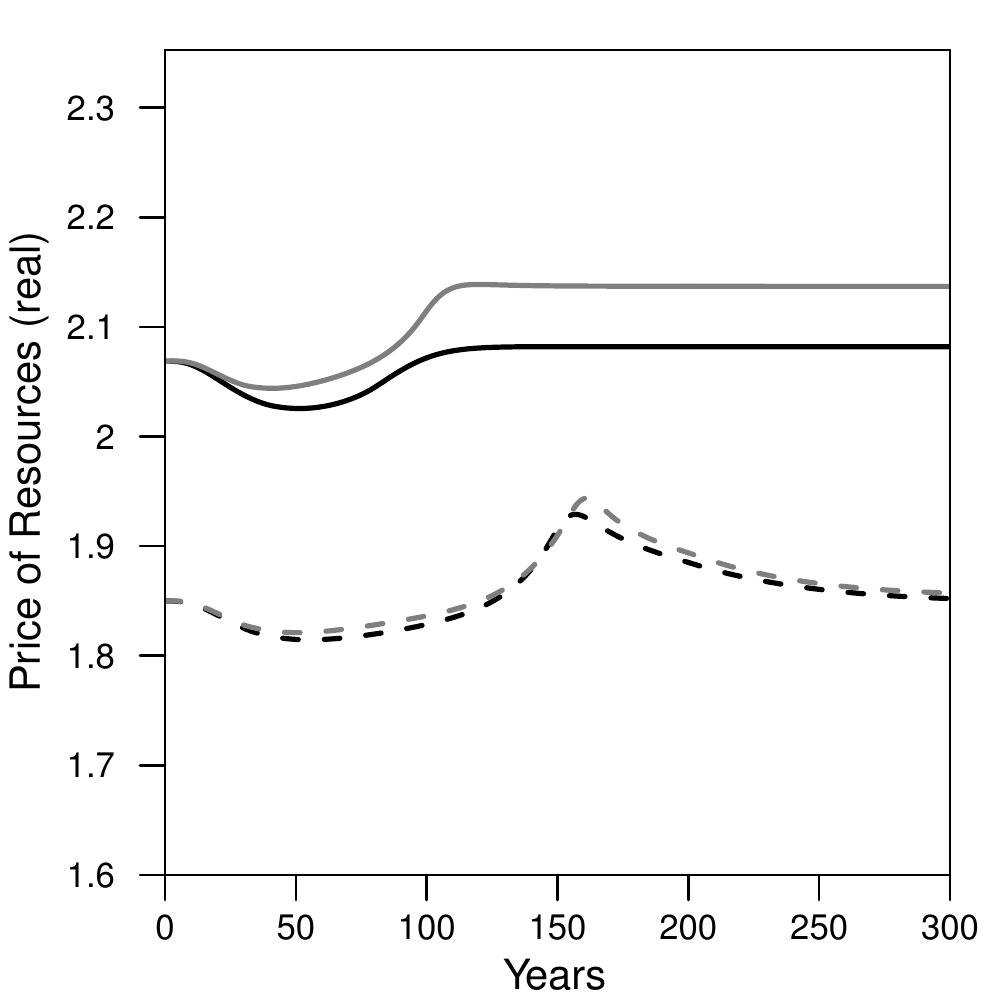}}
\subfloat[]{\includegraphics[width=0.33\columnwidth]{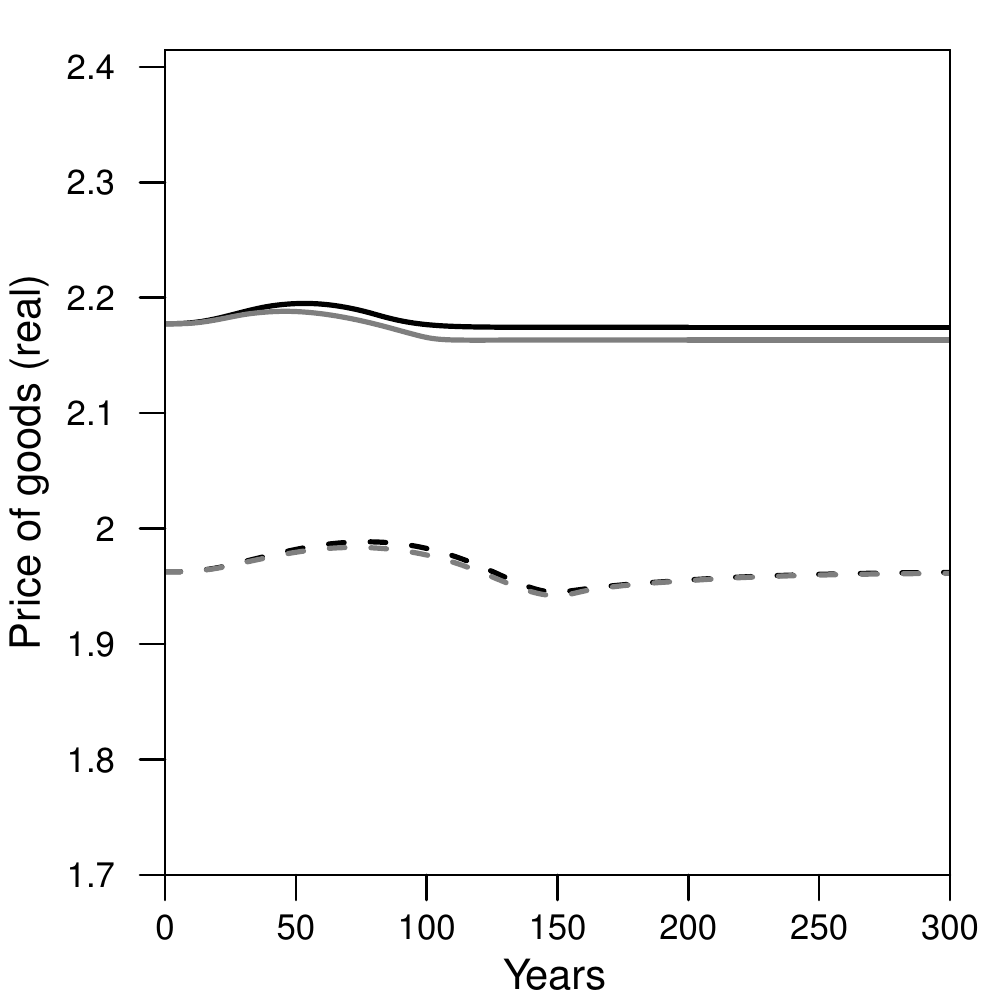}}
\subfloat[]{\includegraphics[width=0.33\columnwidth]{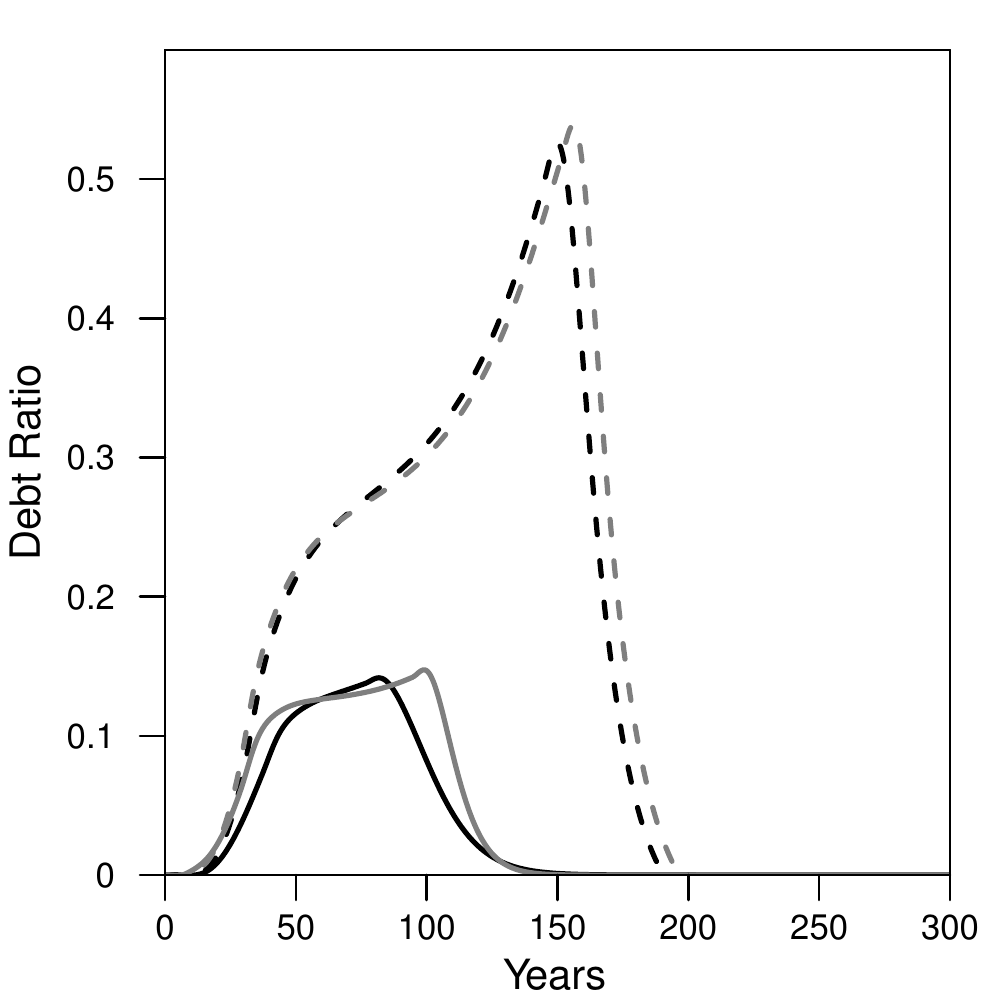}}\\
\caption{Scenarios FC-000, FC-100, MC-000, and MC-100 (black, gray, black dashed, gray dashed). (a) available resources (in the environment), (b)  total capital, (c) population, (d) resource extraction rate, (e) total real net output, (f) participation rate, (g) real price of extracted resources, (h) real price of goods, (i) debt ratio, and (continued) \dots}\label{fig:1p8_A11A00F11F00}
\end{center}
\end{figure}

\begin{figure}\ContinuedFloat
\begin{center}
\subfloat[]{\includegraphics[width=0.33\columnwidth]{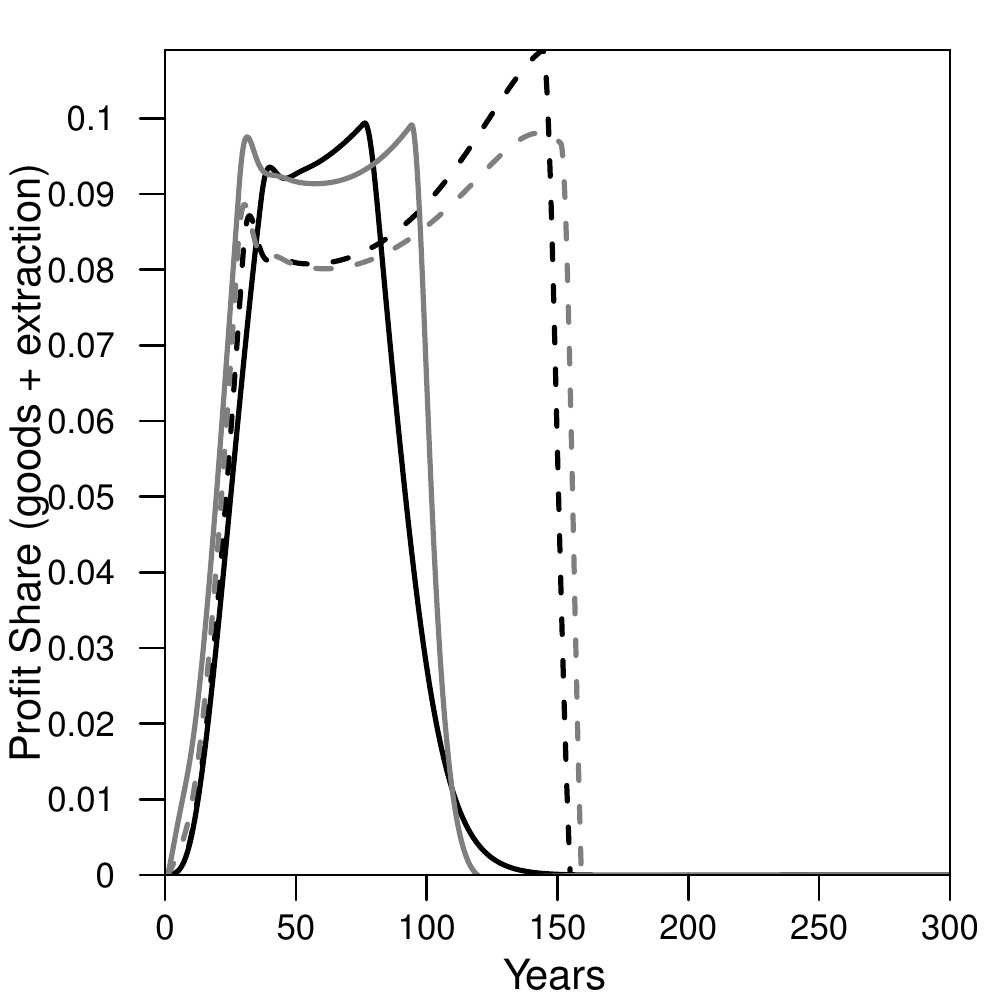}}
\subfloat[]{\includegraphics[width=0.33\columnwidth]{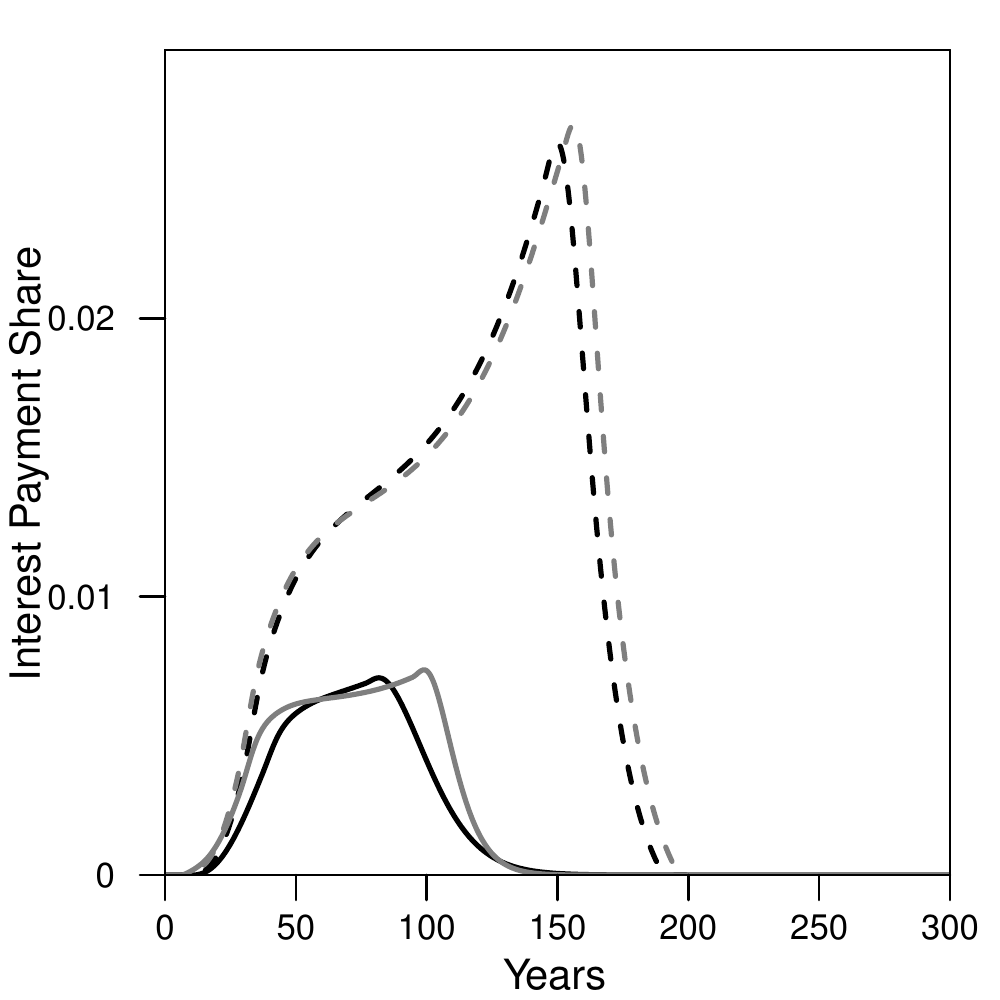}}
\subfloat[]{\includegraphics[width=0.33\columnwidth]{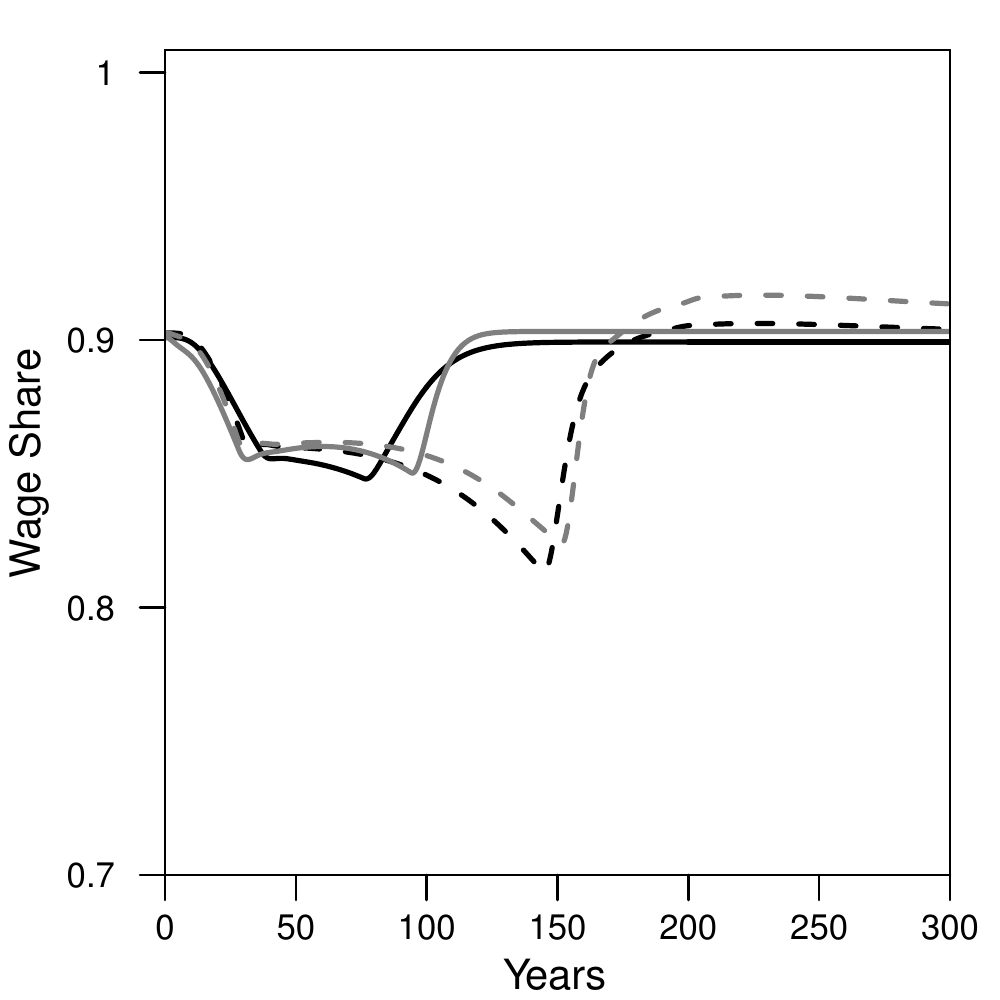}}\\
\subfloat[]{\includegraphics[width=0.33\columnwidth]{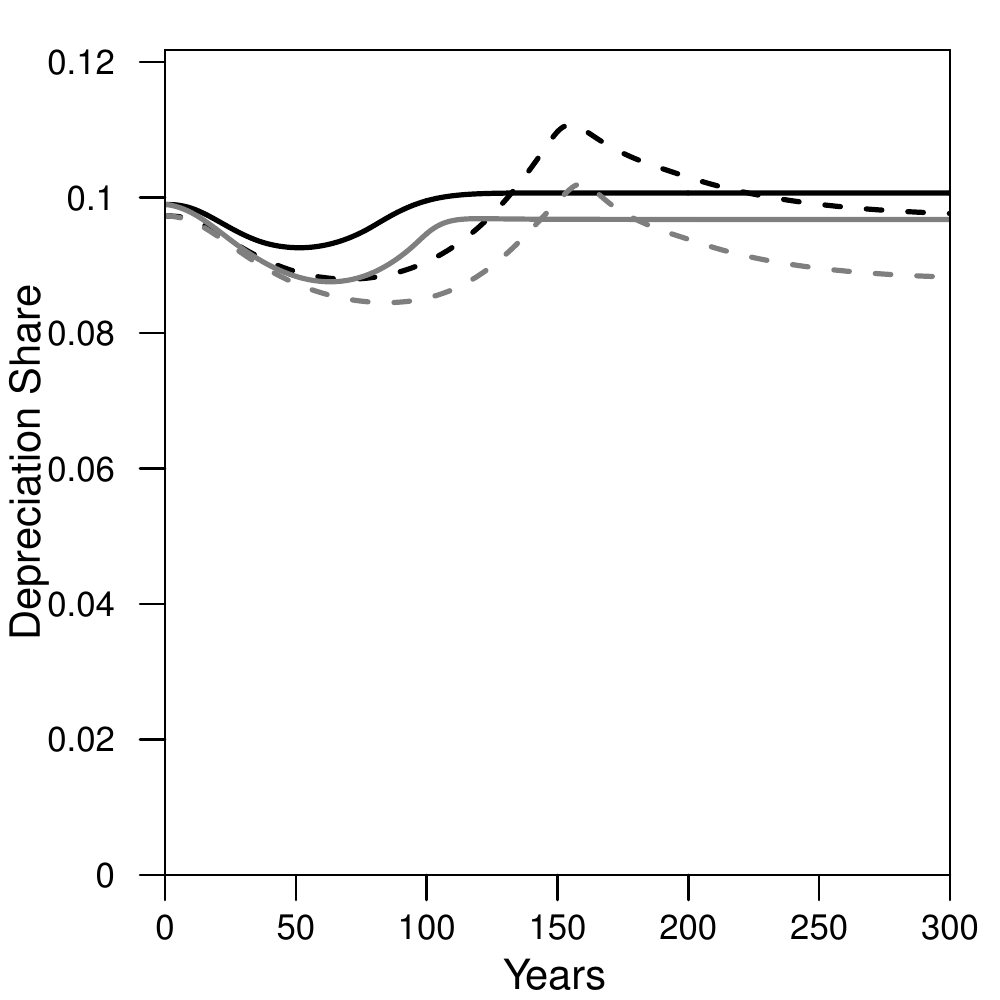}}
\subfloat[]{\includegraphics[width=0.33\columnwidth]{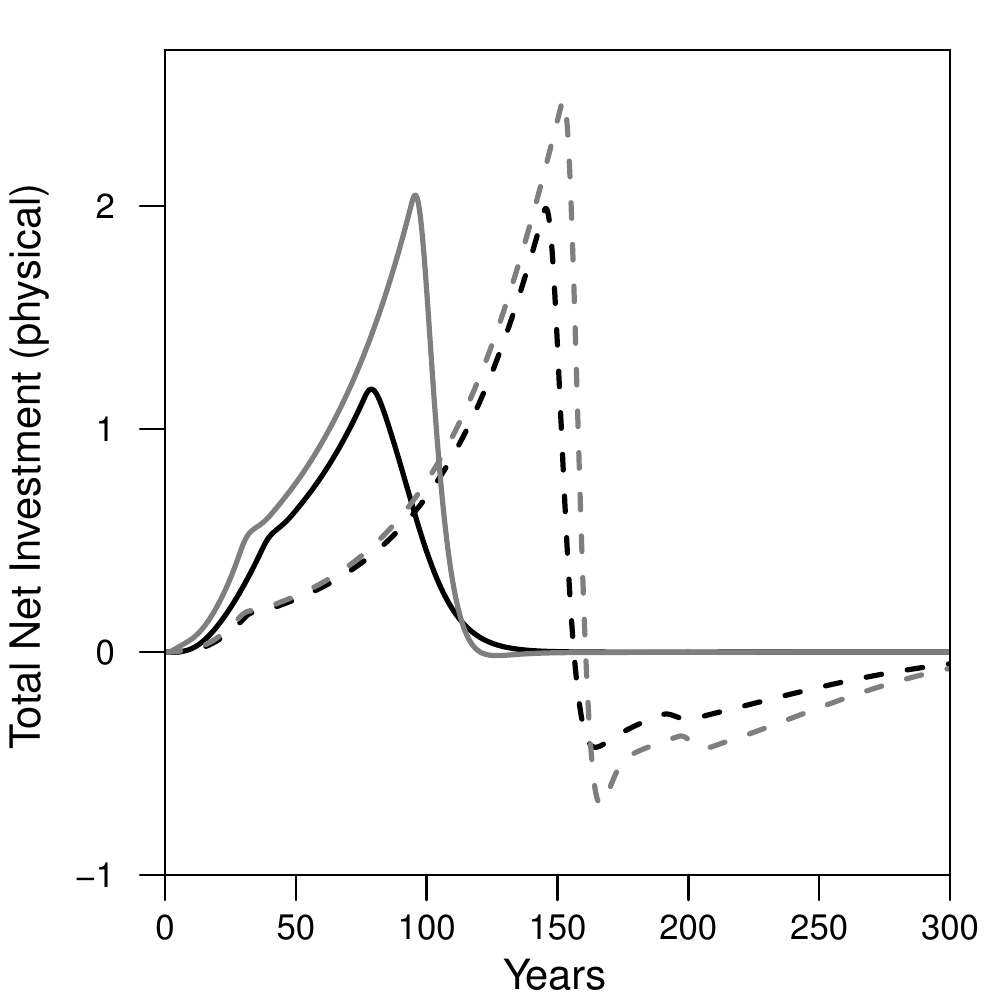}}
\subfloat[]{\includegraphics[width=0.33\columnwidth]{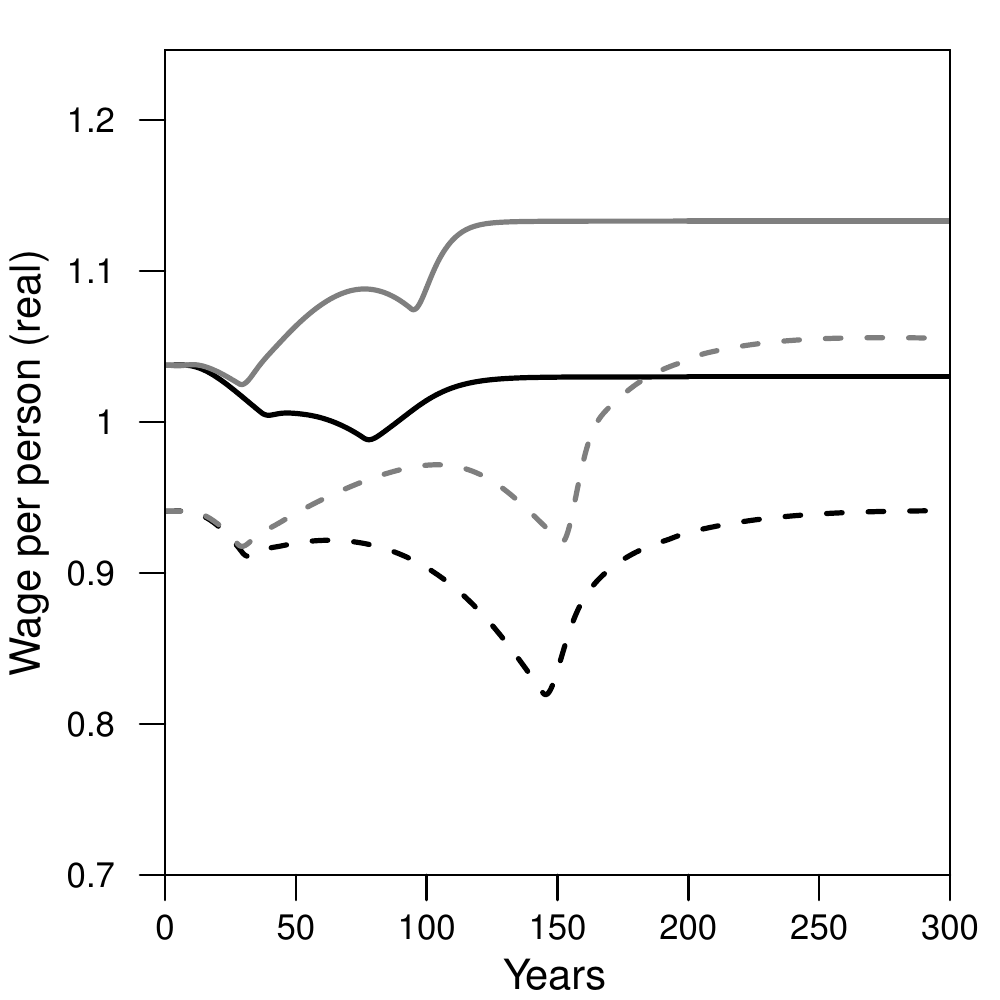}}\\
\subfloat[]{\includegraphics[width=0.33\columnwidth]{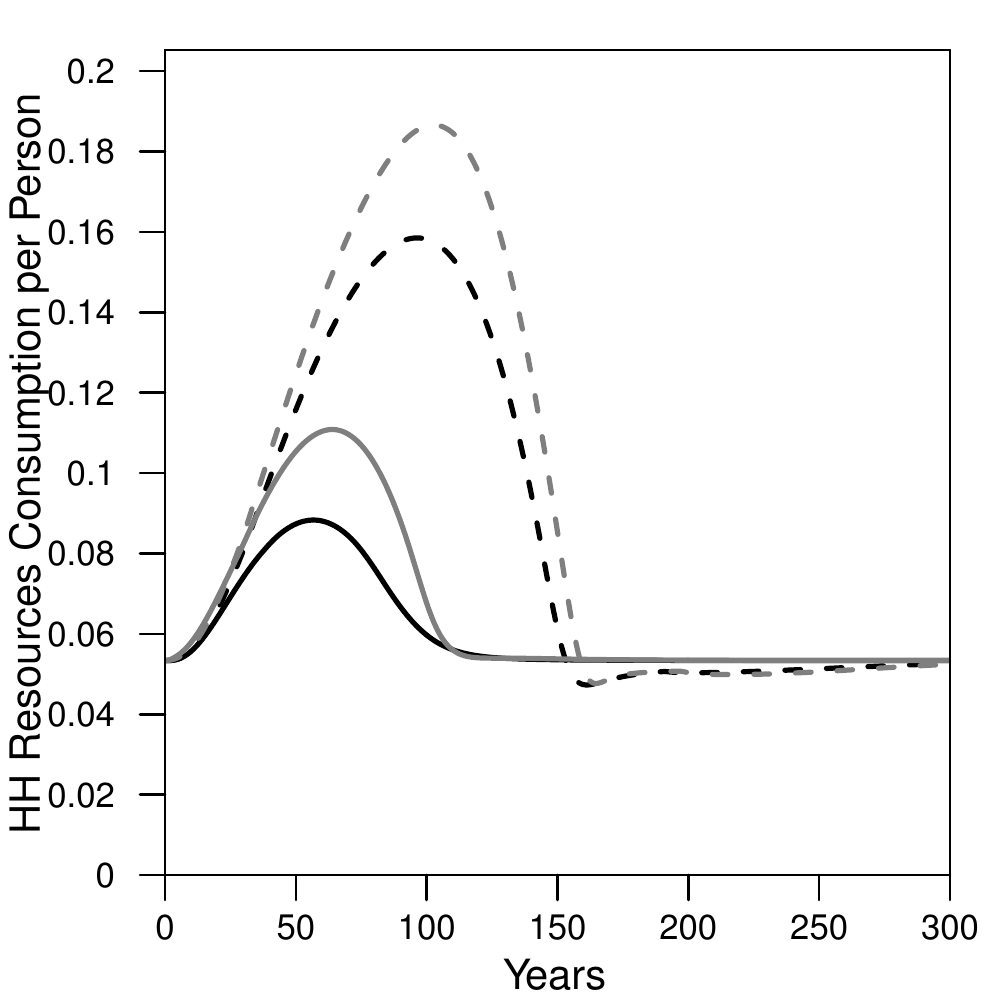}}
\subfloat[]{\includegraphics[width=0.33\columnwidth]{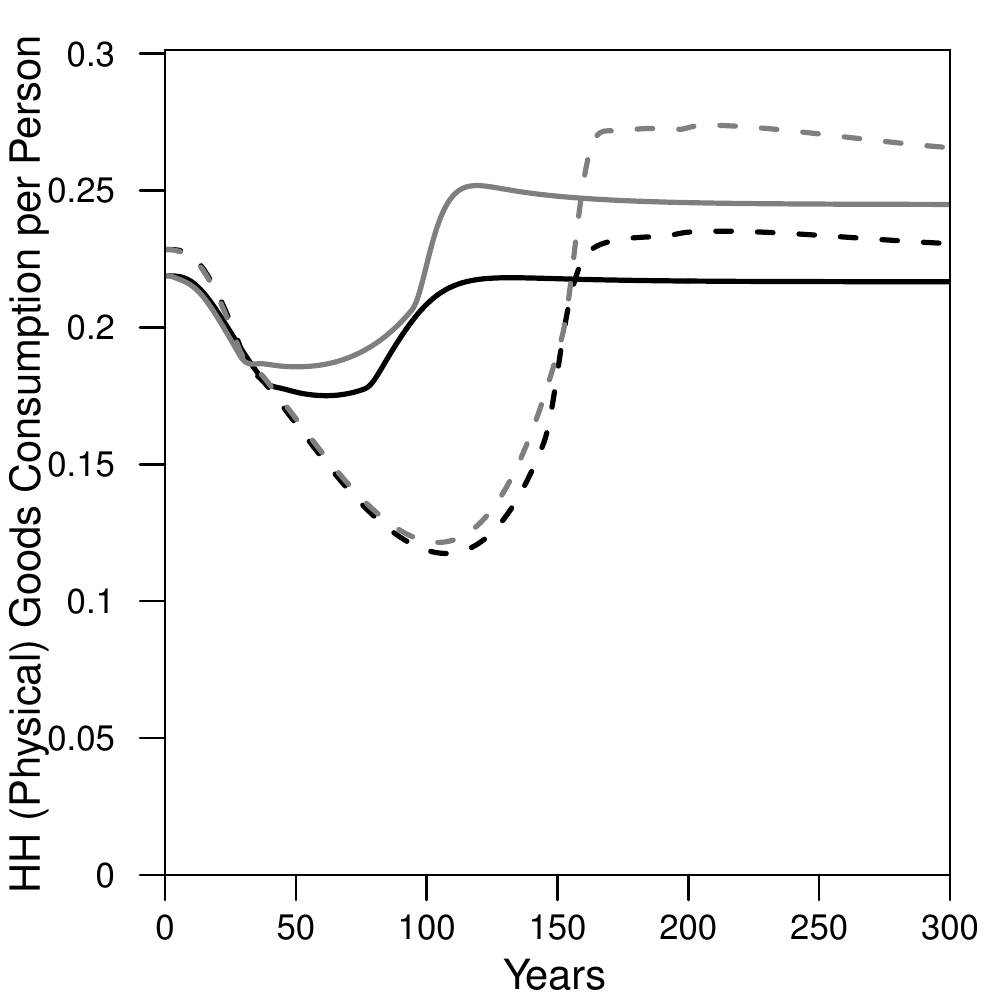}}
\caption{(continued) (j) profit share, (k) interest share, (l) wage share, (m) depreciation share, (n) physical net investment in new capital, (o) real wage per person, (p) household consumption of (physical) resources per person, (q) household consumption of (physical) goods per person, (continued) \dots} 
\end{center}
\end{figure}

\begin{figure}\ContinuedFloat
\begin{center}
\subfloat[]{\includegraphics[width=0.33\columnwidth]{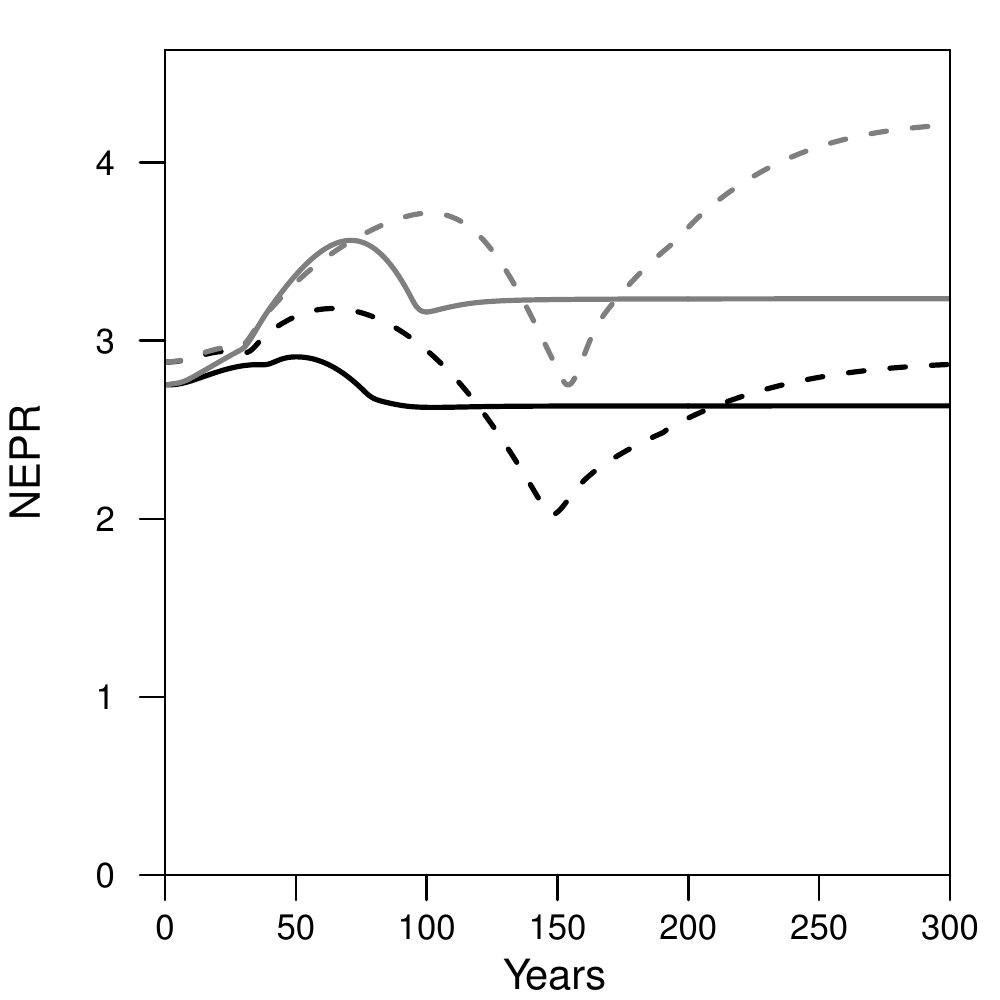}}
\subfloat[]{\includegraphics[width=0.33\columnwidth]{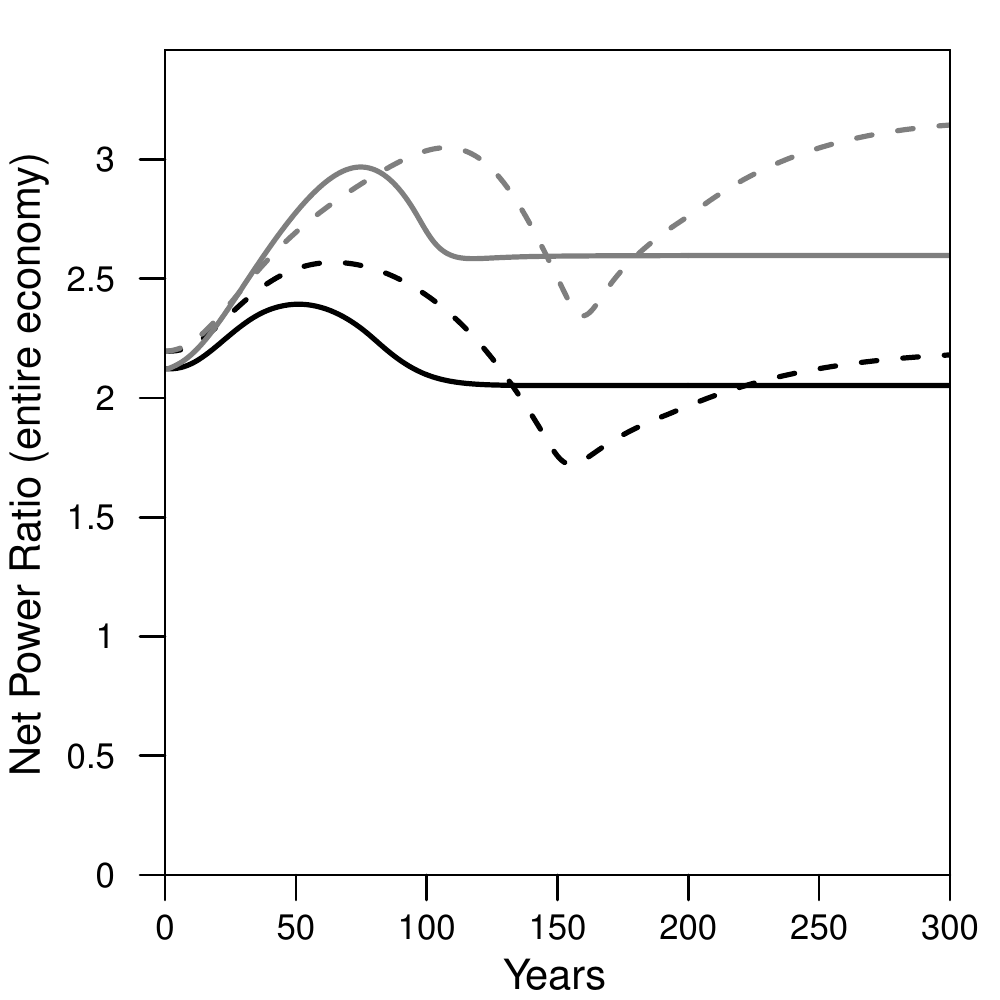}}
\subfloat[]{\includegraphics[width=0.33\columnwidth]{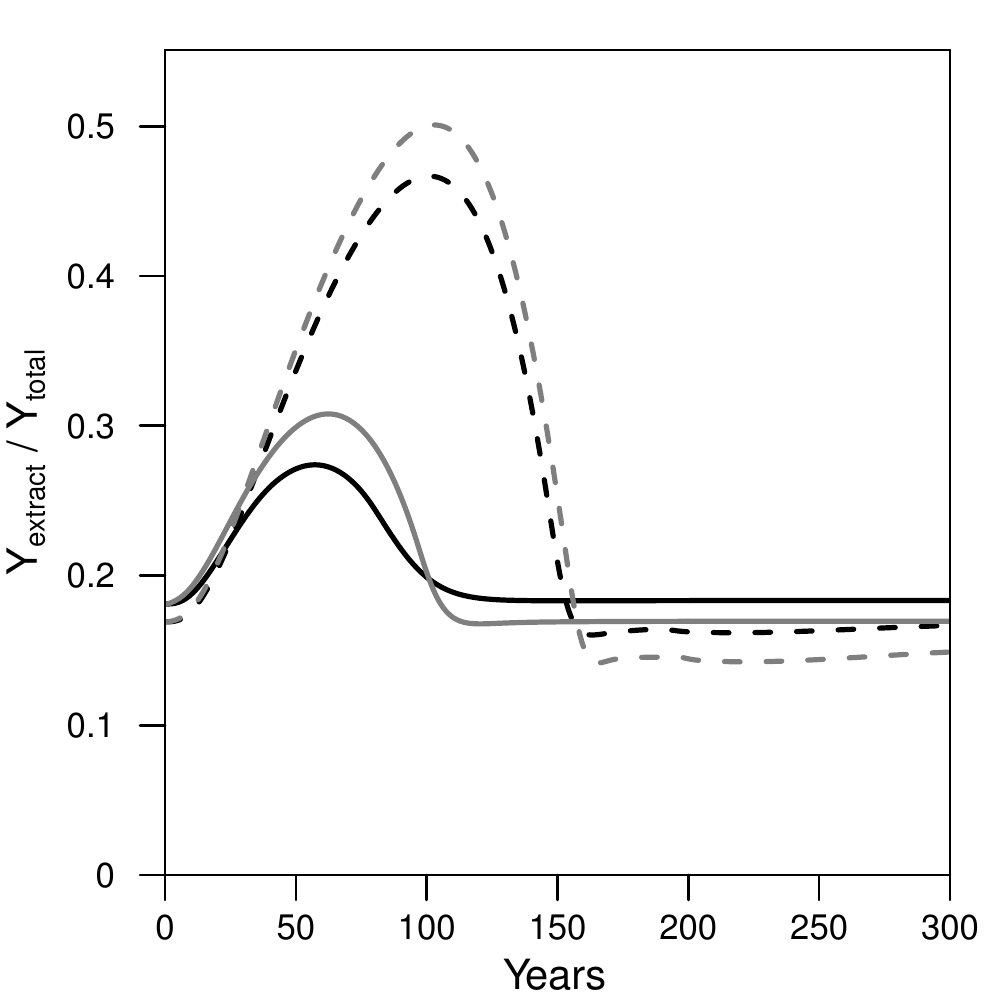}}\\
\subfloat[]{\includegraphics[width=0.33\columnwidth]{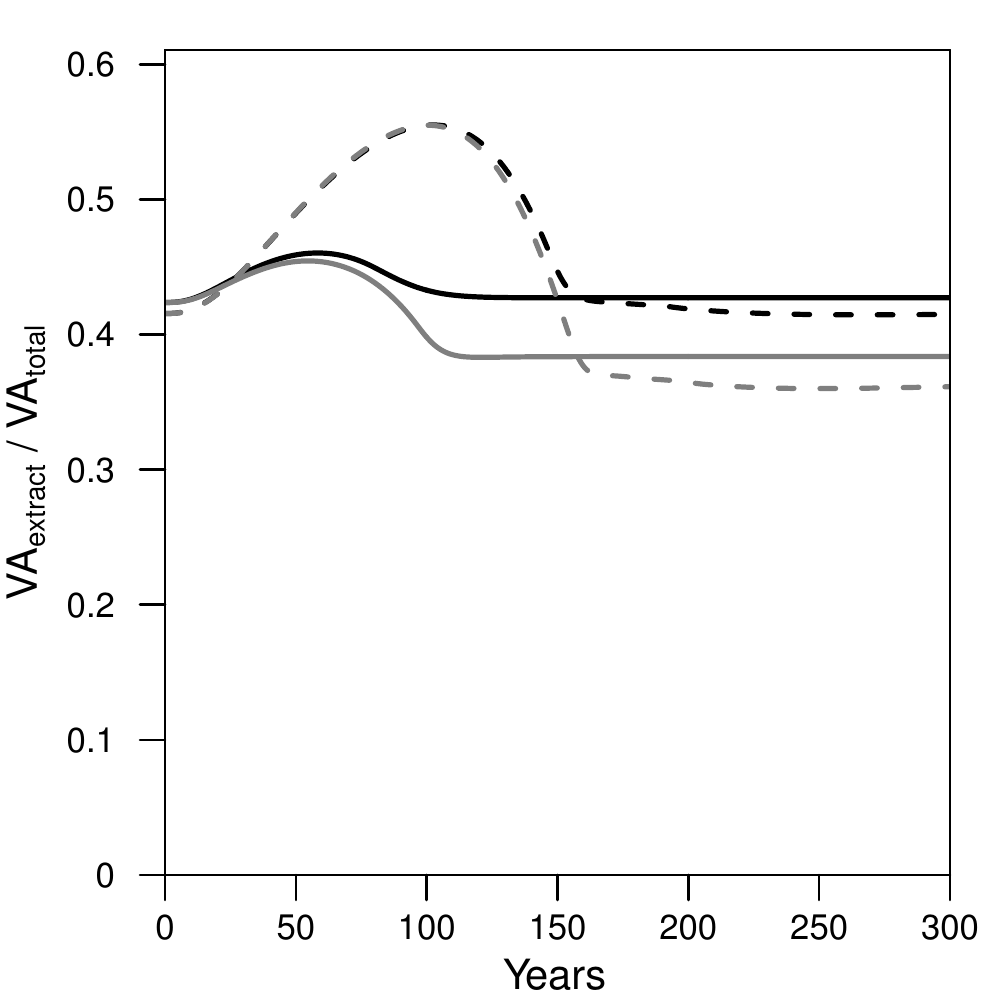}}
\subfloat[]{\includegraphics[width=0.33\columnwidth]{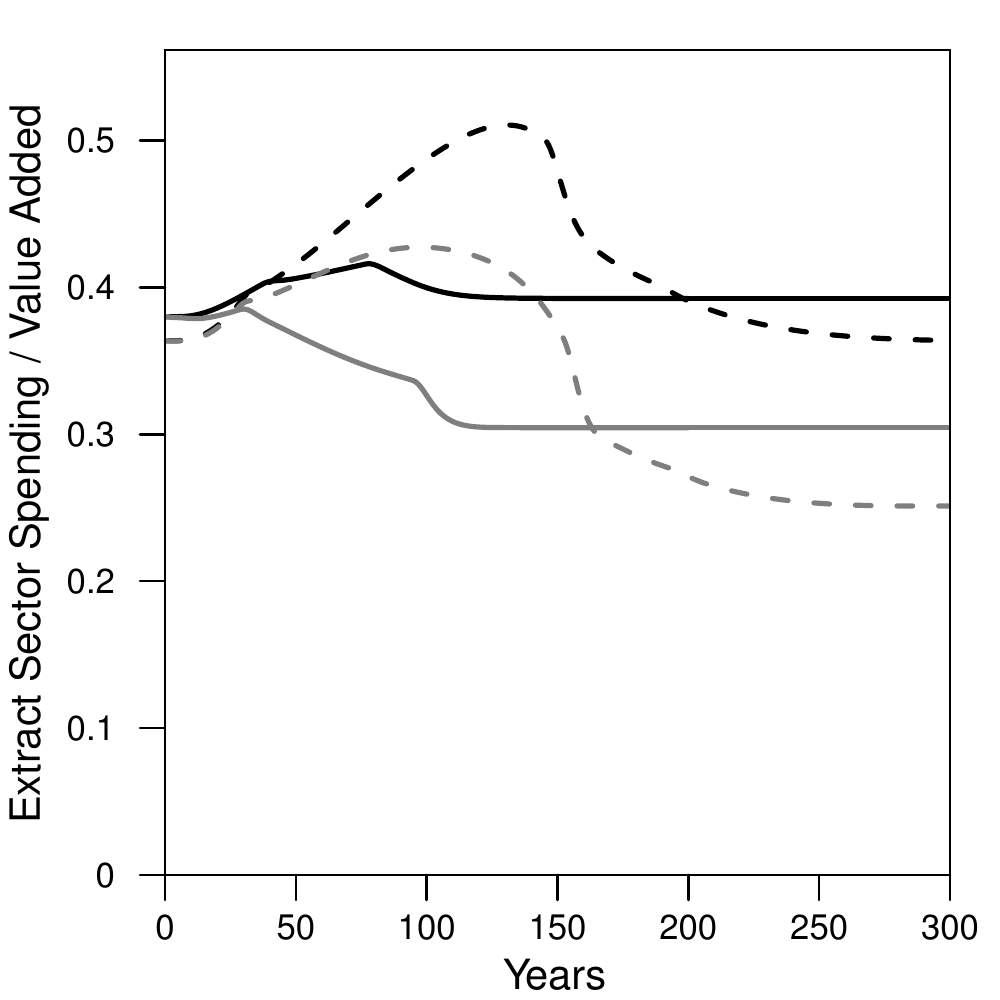}}
\subfloat[]{\includegraphics[width=0.33\columnwidth]{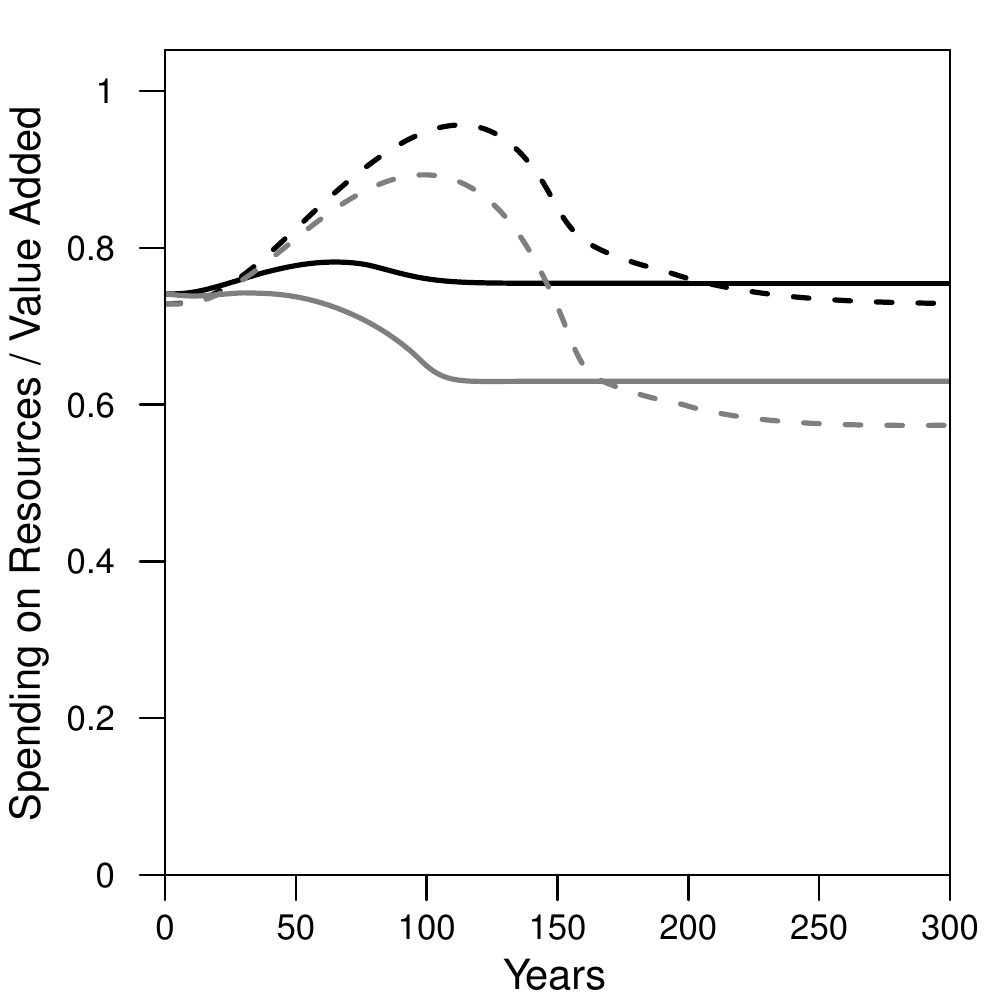}}\\
\subfloat[]{\includegraphics[width=0.33\columnwidth]{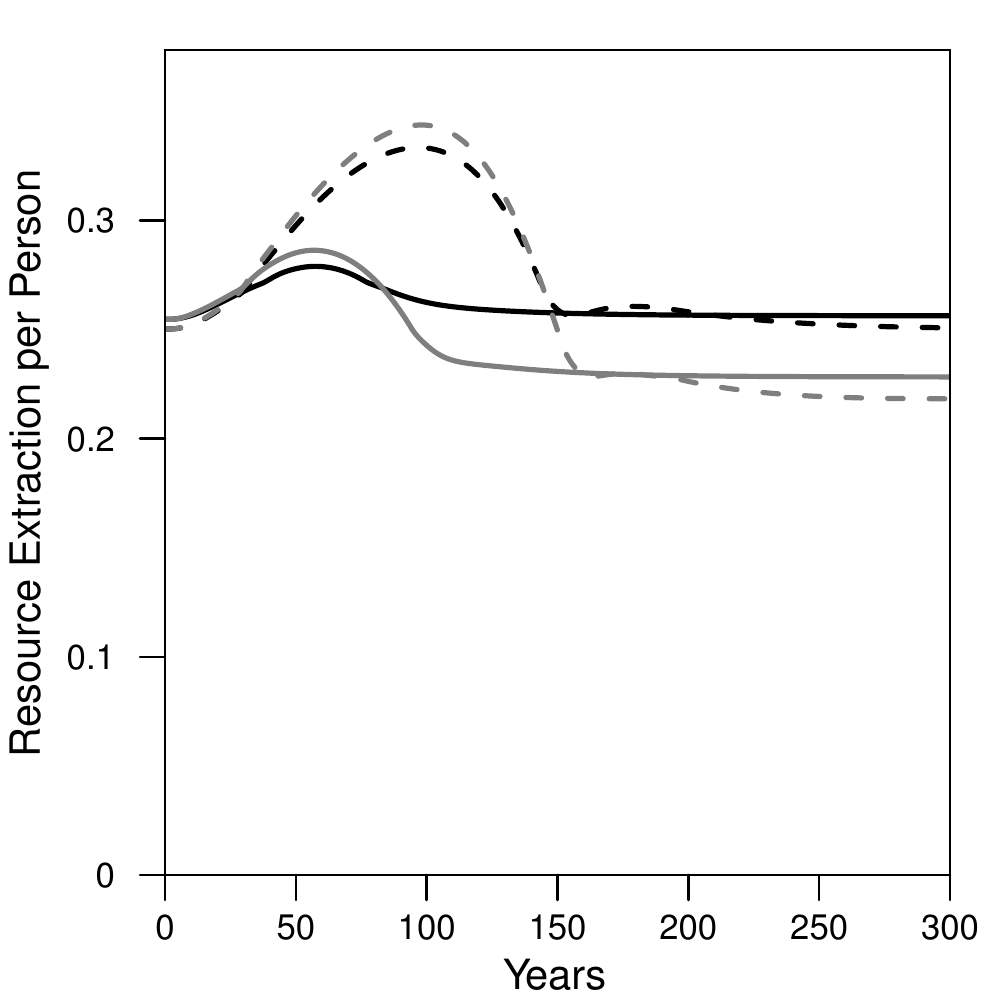}}
\subfloat[]{\includegraphics[width=0.33\columnwidth]{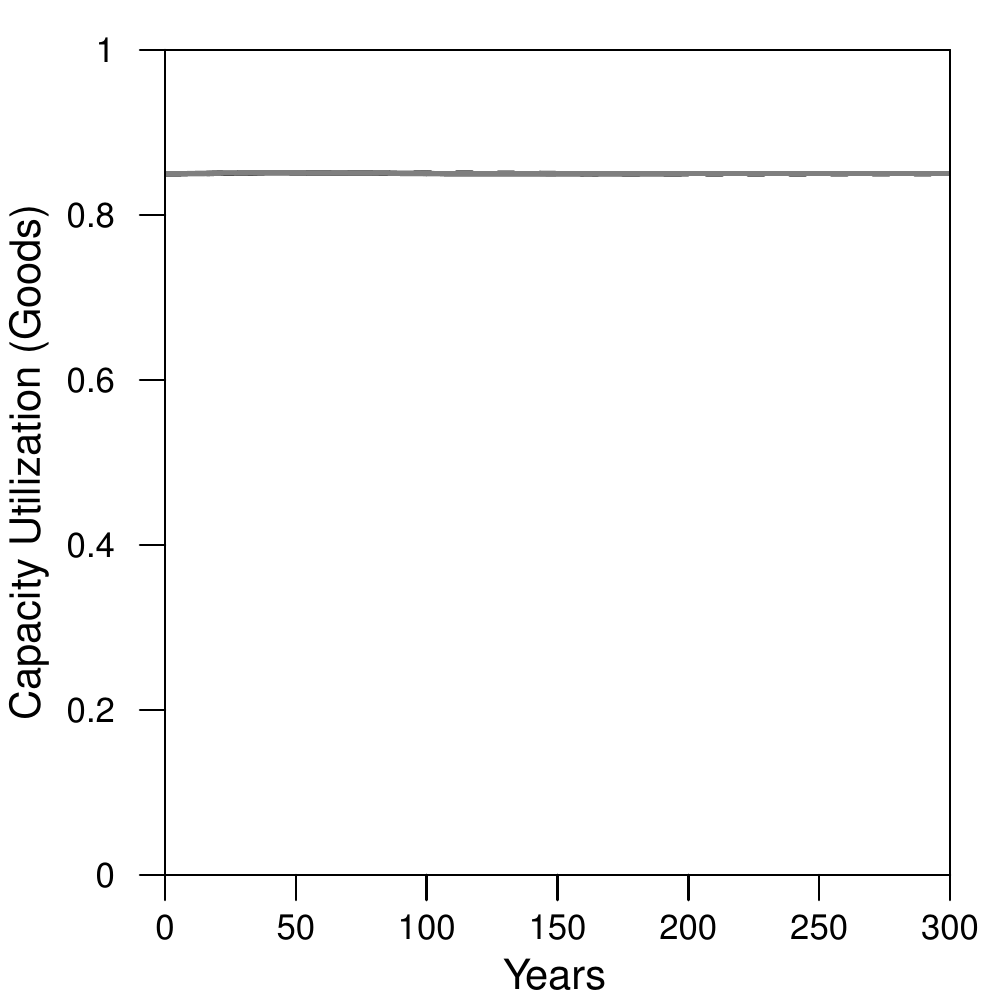}}
\subfloat[]{\includegraphics[width=0.33\columnwidth]{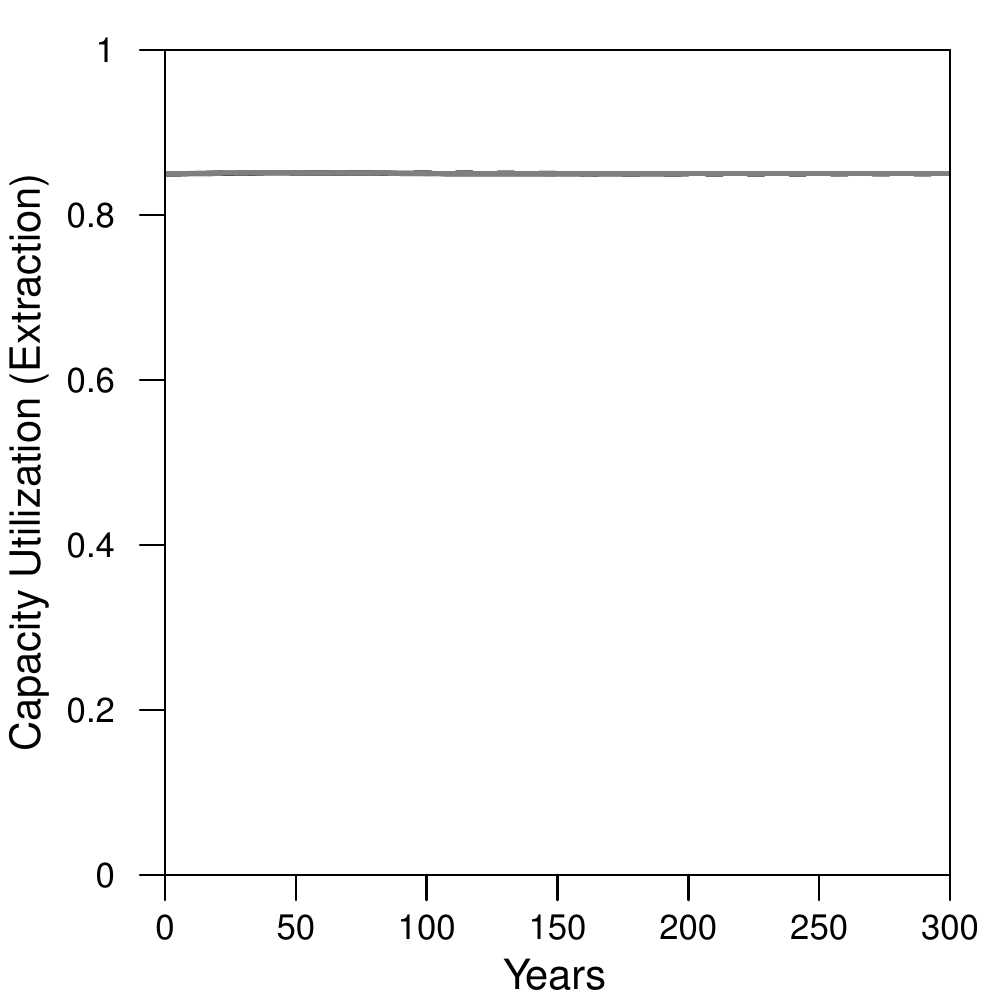}}
\caption{(continued) (r) net external power ratio (extraction sector, NEPR), (s) net power ratio (entire economy, NPR), (t) fraction of net output from extraction sector, (u) fraction of value added in extraction sector, (v) extraction sector spending per total value added (= per total net output), (w) spending on resources per total value added (= per total net output), (x) total resources extraction per person, (y) capacity utilization of goods capital, and (z) capacity utilization of extraction capital.} 
\end{center}
\end{figure}

%
%

The main differences from results in King (2020) \cite{King2020} stem from changes to the calculation of wages and price, investigation into the effects of increasing resource consumption efficiency in capital, changes in the definition of cost that informs pricing, and the consideration of several structural comparisons to global and U.S. data. 

\subsubsection{Full and Marginal Cost Pricing}
Scenario FC-000 with full cost pricing (comparable to King (2020)) and full bargaining power (not comparable to King (2020)) reaches an equilibrium steady state at the nominal participation rate (60\% in Figure \ref{fig:1p8_A11A00F11F00}(f)) with zero debt and profits (Figures \ref{fig:1p8_A11A00F11F00}(i) and (j)).   This steady state outcome was not attainable in HARMONEY v1.0.  

From the beginning, the economy grows, accumulating capital, population, and debt while depleting the natural resource.  During initial rapid growth profits and debt increase. They then level off before declining rapidly when peak resource extraction (Figure \ref{fig:1p8_A11A00F11F00}(d)) inhibits further growth at which time all resources are needed to maintain existing capital and population, thus driving net capital investment to zero (Figure \ref{fig:1p8_A11A00F11F00}(n)).  During the entire simulation, there is no ``overinvestment'' such that capacity utilization remains at the target level of 85\% (Figure \ref{fig:1p8_A11A00F11F00}(y) and (z)).

With full bargaining power, after net investment declines to zero, wages and wage share increase at the expense of profits.  Thus, HARMONEY puts the battle of labor versus capital into the context of resource consumption, and we expand on this later.  The depreciation share of value added does not appreciably change.

The net power ratios ((Figure \ref{fig:1p8_A11A00F11F00}(r) and (s)) first increase, mostly driven by the exogenous assumption of an increase in $\delta_y$ that drives down $a_{ee}$, but then decrease once $\delta_y$ reaches its final value and resource depletion starts to increase costs, and thus prices (Figure \ref{fig:1p8_A11A00F11F00}(g) and (h)), due to the need to consume more resource to extract the next unit of resource.  This rise and fall of NEPR generally mimics the expected trend that growth is characterized by increasing or high net energy resources.  For example, the NEPR of U.S. oil and gas (normally termed EROI = ``energy return on energy invested'') increased from 16 in 1919 to 24 in 1954 before declining, with volatility, to 11 in 2007 \cite{Guilford2011}.  However, growth can continue with declining NEPR, for a while.  

Population (Figure \ref{fig:1p8_A11A00F11F00}(c)) levels off as per capita household resource consumption declines enough (Figure \ref{fig:1p8_A11A00F11F00}(p)) to increase death rates to equal birth rates.

\subsubsection{Marginal Cost Pricing (Differences from Full Cost)}

There are a few notable differences between the marginal and full cost pricing results, and we expand on this in later sections.  First, marginal cost pricing enables an overshoot of steady state values for population and capital.  While marginal cost pricing reaches higher peak levels of capital, population, net output, debt, resource extraction rates, and resource depletion, the steady state values are lower than in the full cost assumption.  That is to say, with all other parameters the same, the marginal cost pricing reaches higher peaks in growth, but they are only temporarily higher than if assuming full cost pricing.  Another intuitive finding is that because costs are lower when assuming marginal cost pricing, the marginal cost prices are lower than full cost prices (Figure \ref{fig:1p8_A11A00F11F00}(g) and (h)).

\subsection{Explanation of Patterns in Growth Rates: Extraction and GDP}

\begin{figure}
\begin{center}
\subfloat[]{\includegraphics[width=0.33\columnwidth]{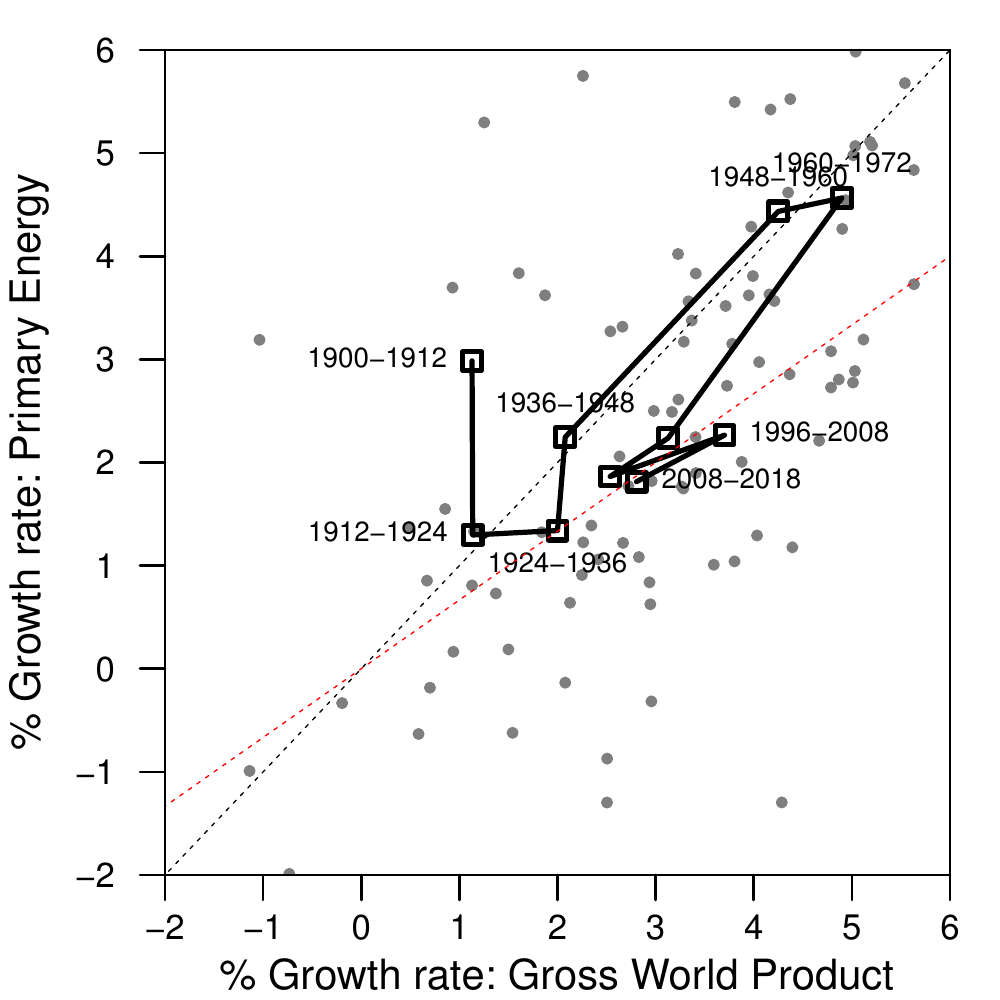}}
\subfloat[]{\includegraphics[width=0.33\columnwidth]{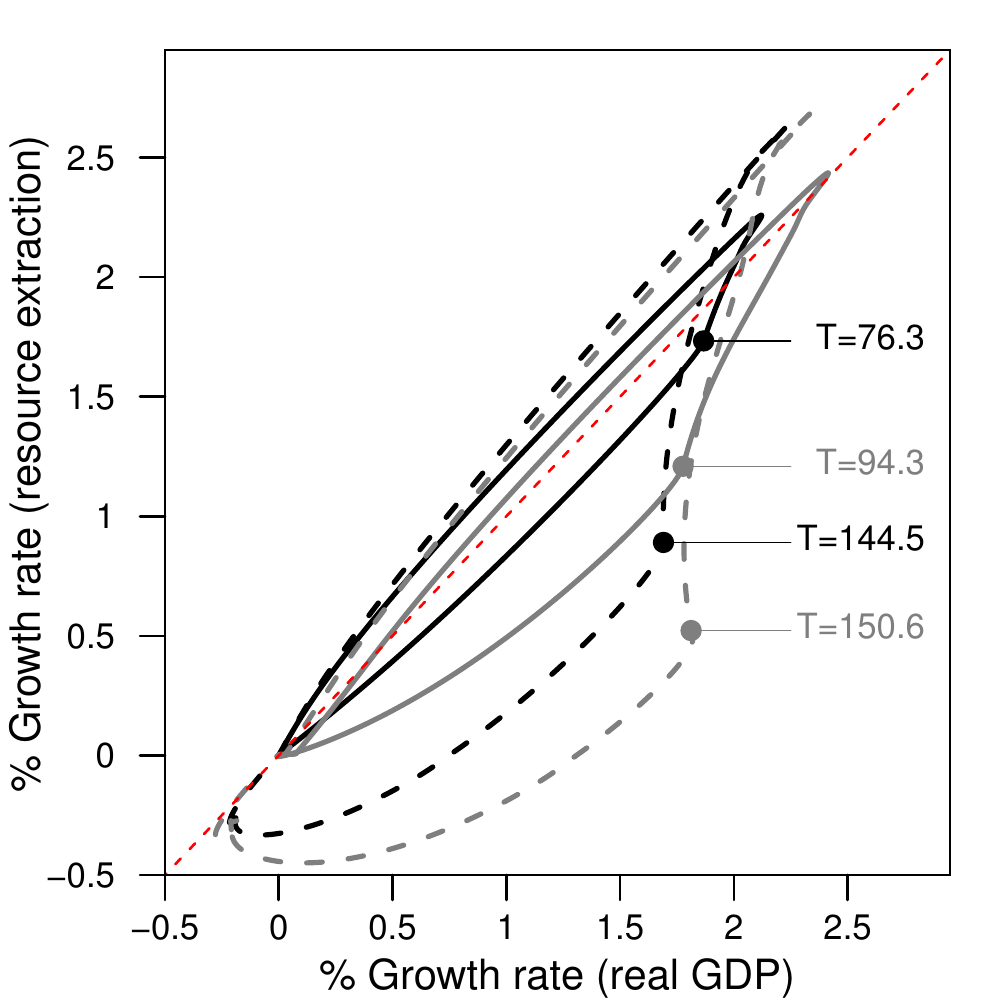}}
\subfloat[]{\includegraphics[width=0.33\columnwidth]{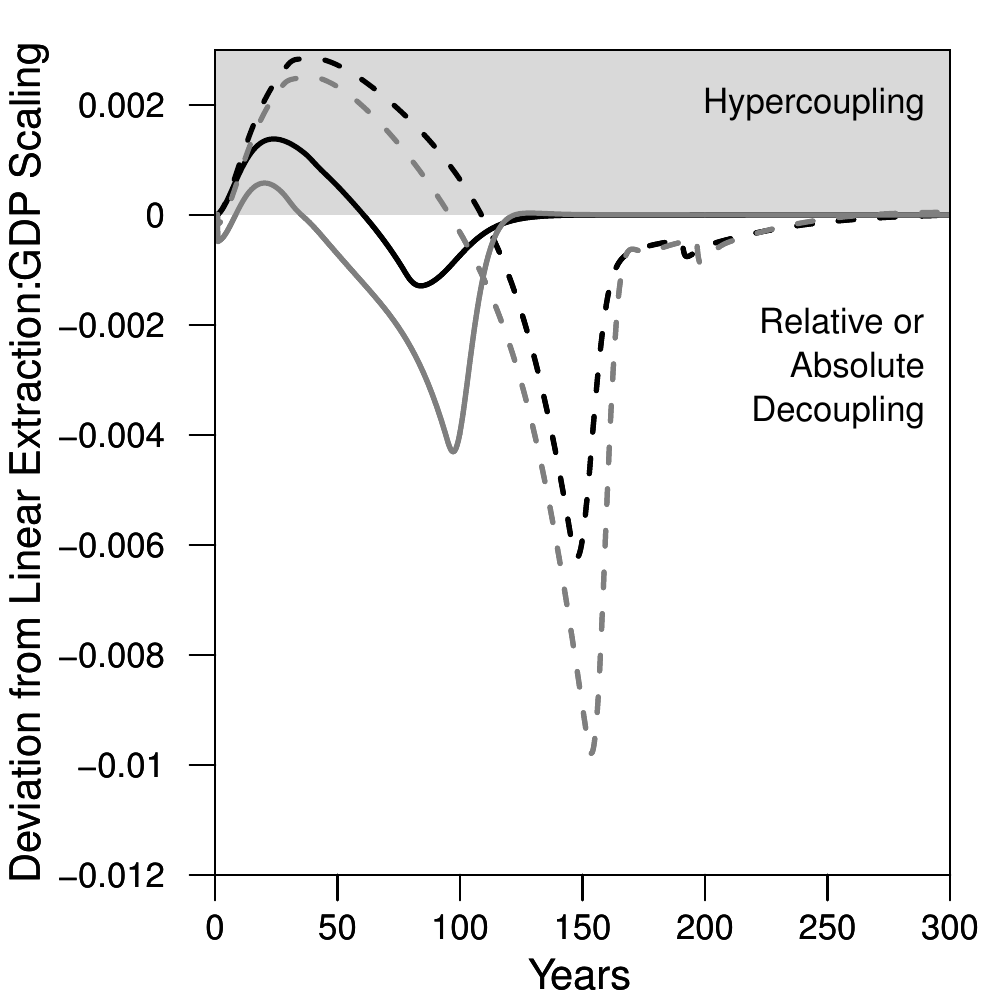}}\\
\subfloat[]{\includegraphics[width=0.33\columnwidth]{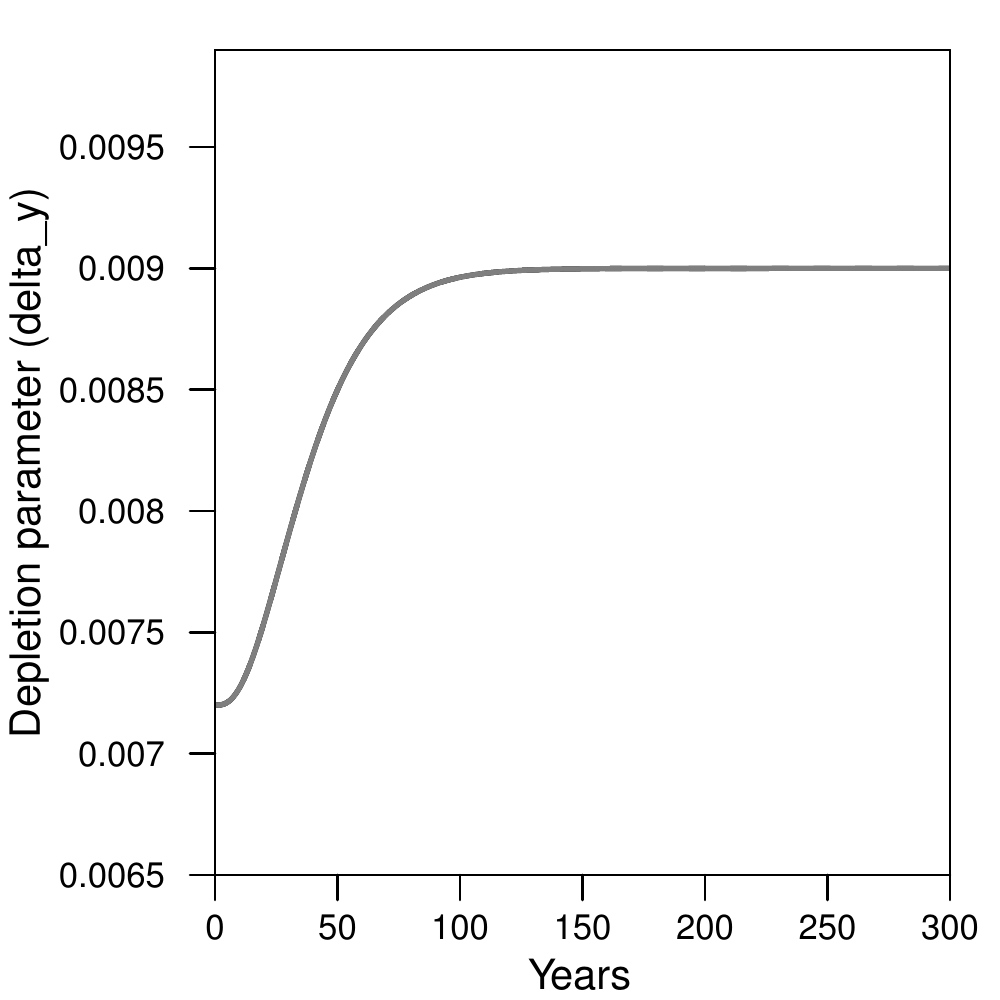}}
\subfloat[]{\includegraphics[width=0.33\columnwidth]{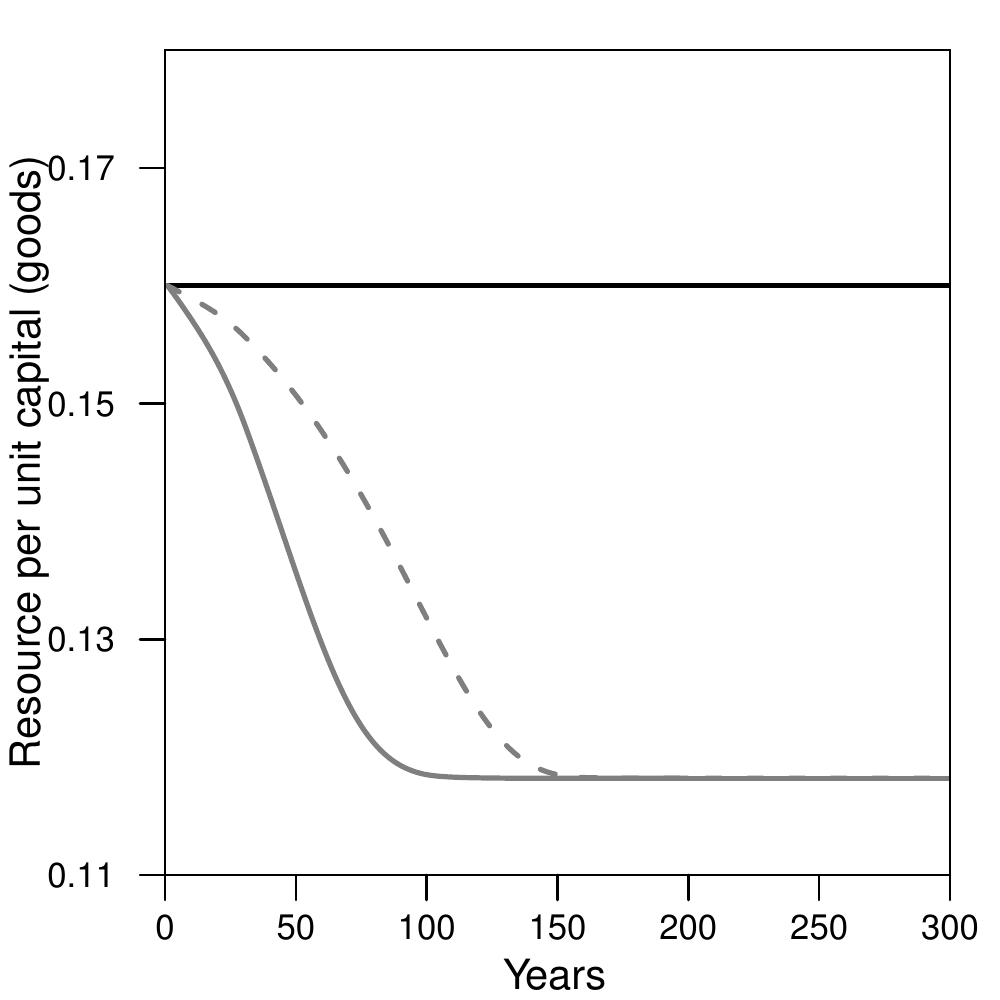}}
\subfloat[]{\includegraphics[width=0.33\columnwidth]{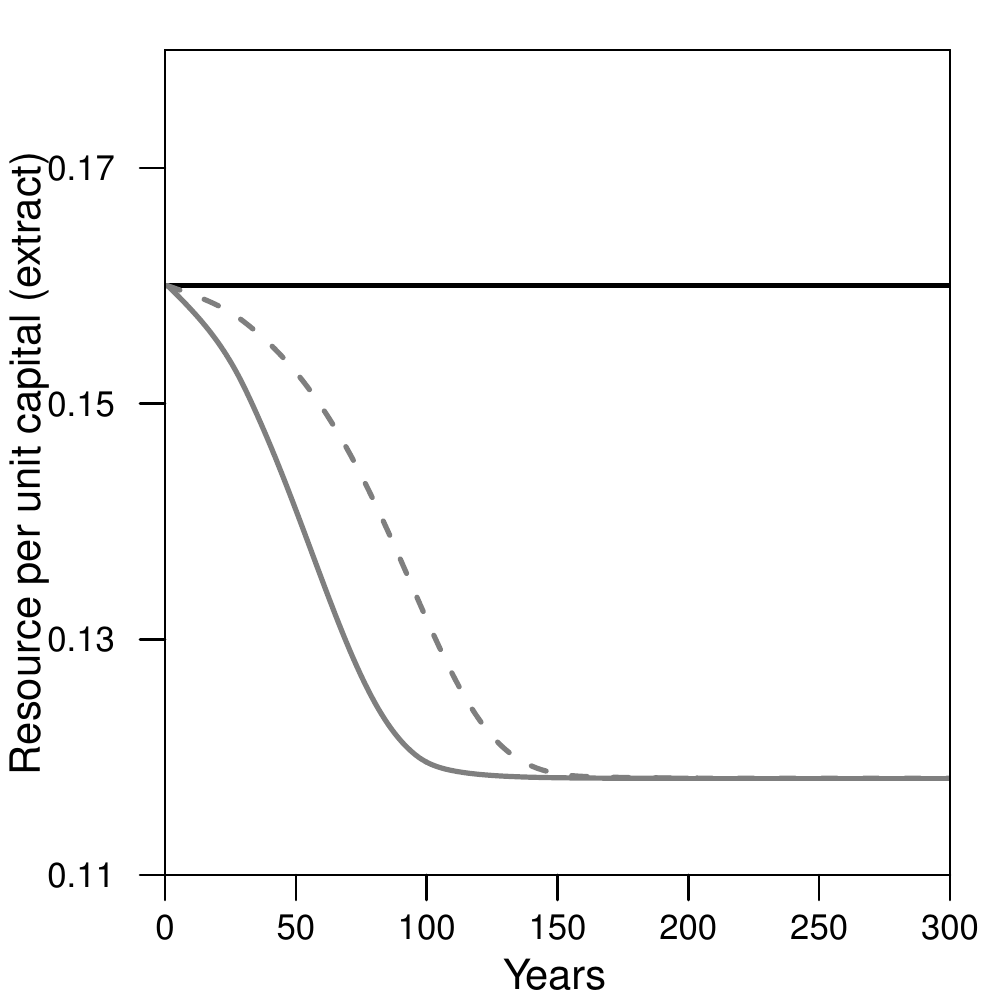}}\\
\subfloat[]{\includegraphics[width=0.33\columnwidth]{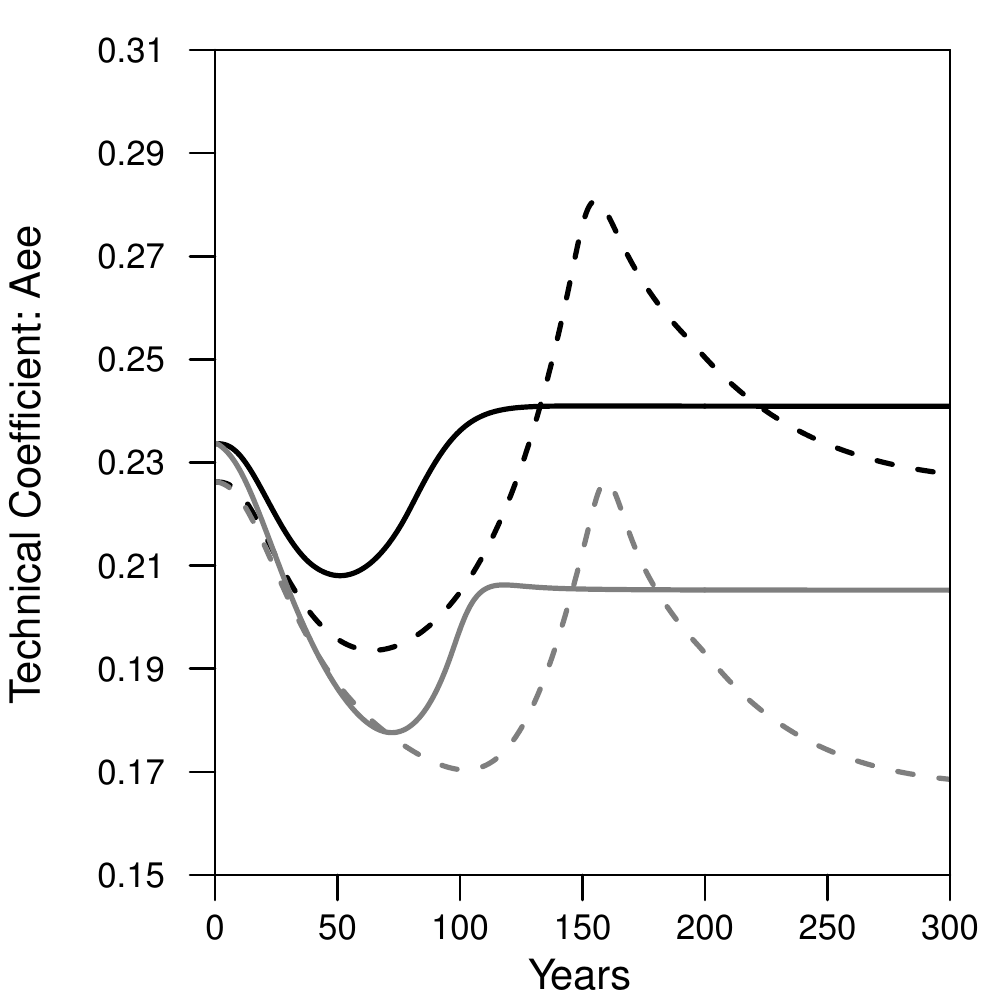}}
\caption{(a) Global data (1900-2018) indicating annual growth rates of primary energy consumption versus annual growth rates of gross world product. Squares indicate average growth rates for groups of 12-years. (b) The parallel figure of (a) for the HARMONEY scenarios showing the growth rates of natural resource extraction versus growth rates of real GDP (FC-000: black solid, MC-000: black dashed, FC-100: gray solid, and MC-100: gray dashed).  (c) The distance of the curves in subfigure (b) from the 1:1 slope.  Positive values indicate hypercoupling (above 1:1 line). Negative values indicate decoupling (below 1:1 line). (d) The exogenous change ($t=0$ to ~$t=100$) in depletion parameter, $\delta_y$ is the same for all simulations. (e) and (f) Scenarios FC-100 and MC-100 decrease the resource consumption requirements for goods and extraction capital, $\eta_g$ and $\eta_e$ respectively, from 0.16 to 0.12 as a 3rd-order delay function of the rate of investment in each respective type of capital. (g) Technical coefficient $a_{ee}$, resource consumption of extraction sector per unit of extraction.}\label{fig:World_and_A11A00F11F00_GrowthPEU_vs_GrowthGDP}
\end{center}
\end{figure}

Figure \ref{fig:World_and_A11A00F11F00_GrowthPEU_vs_GrowthGDP} compares global and model trends for growth rates in consumption of primary energy (global data) and natural resources (model) versus real GDP growth rates (data as in \cite{Jarvis2020}).  The global data from 1900 to 2018 show a general clockwise trend.  From 1900 to the early 1970s the trajectory moves from lower-left to upper-right along a near 1:1 ratio in growth rates. After the 1960s and early 1970s the data move to a region with a 2:3 ratio in growth rates, centered near a 2\% and 3\% growth rate in primary energy and gross world product (GWP), respectively.  The HARMONEY model simulation results follow the same clockwise trajectory as seen in the global data in moving from an increasing or near constant energy intensity to a declining energy intensity.  Thus, the HARMONEY 1.1 model endogenously recreates an important observed pattern in the real economy.  

These HARMONEY model results inform the discussion of both relative and absolute decoupling.  Here, relative decoupling refers as the situation in which growth of both GDP and energy (or resources) is positive, but GDP growth is higher. Absolute decoupling refers to positive growth in GDP but zero or declining growth of energy (or resource) consumption.   The model patterns in Figure \ref{fig:World_and_A11A00F11F00_GrowthPEU_vs_GrowthGDP}(b) exhibit time periods of both relative and absolute decoupling. We measure the level of coupling or decoupling in Figure \ref{fig:World_and_A11A00F11F00_GrowthPEU_vs_GrowthGDP}(c) as the distance of the growth rate trajectories in Figure \ref{fig:World_and_A11A00F11F00_GrowthPEU_vs_GrowthGDP}(b) from the 1:1 slope line that would represent that resource extraction growth rates equal those of real GDP.  Positive values indicate hyper coupling (or increasing resource intensity above the 1:1 line) and negative values indicate decoupling (or decreasing resource intensity below the 1:1 line).  

We now discuss three concepts to explain the relationship between growth in resource extraction and GDP: biophysical constraints, resource consumption efficiency, and the level of private debt as influenced by the definition of cost of production.

First, there are biophysical constraints.  Indefinite exponential growth (at a constant or increasing rate) from a finite resource (as a stock or flow) is not possible.  HARMONEY 1.1 explicitly assumes a resource with finite limits of extraction, but early stages of growth do occur at an increasing exponential growth rate.  The maximum growth rates (upper right extents in Figure \ref{fig:World_and_A11A00F11F00_GrowthPEU_vs_GrowthGDP}(b))  occur at times of $T=$40, 33, 32, and 32 for the FC-000, MC-000, FC-100, and MC-100 scenarios, respectively.  Biophysical constraints govern the growth rates of resource extraction, but the direct linkage to economic growth, as measured by GDP (or net output), is exhibited by the fact that after a peak in growth rates, both the historical data and the model results move ``down and to the left,'' meaning that both growth rates decrease. This contrasts with an alternative trajectory of, for example ``down and to the right,'' as if GDP could continue to increase at higher rates while resource extraction increased at decreasing growth rates.  At the peak growth rates, the economy has effectively grown to such a sufficient size relative to the size of its environment that it can no longer maintain existing capital and population while growing at an increasingly fast rate.  This reasoning regarding the comparison of data in Figures \ref{fig:World_and_A11A00F11F00_GrowthPEU_vs_GrowthGDP}(a) and (b) implies that the decreasing GWP growth rate can also be interpreted as the global economy becoming ``large'' relative to its environment from which it extracts energy.  

Each curve of Figure \ref{fig:World_and_A11A00F11F00_GrowthPEU_vs_GrowthGDP}(b) exhibits a ``knee'' in the sublinear scaling regime where the trajectory changes from downward (near vertical) to a near 45 degree angle down and to the left.  For each scenario, this ``knee'' occurs when the sum of firm profits and interest payments (as a share of GDP) is at its maximum (at $T=$ 76, 145, 94, and 151 for scenarios FC-000, MC-000, FC-100, and MC-100, respectively). 
Also, very soon after this time, total physical net investment peaks ($T=$ 78, 145, 95, and 152, respectively, Figure \ref{fig:1p8_A11A00F11F00}(n)) and debt ratios peak ($T=$ 82, 150, 99, and 156, respectively, Figure \ref{fig:1p8_A11A00F11F00}(i)) before the rapid repayment of debt and population death rates increase from their minimum value (death rates increase from the minimum value when household per person resource consumption is below $s=0.08$ in Figure \ref{fig:1p8_A11A00F11F00}(p)).  Importantly, these events occur nearly simultaneously but before resource extraction actually peaks (i.e., growth rate of extraction remains positive at these peaks).  Nonetheless, there is no longer enough physical flow of natural resources to invest in new capital at the previous rates while maintaining the existing levels of population and capital.  Something must give, and eventually everything does in the process of reaching a steady state without overshoot (full cost scenarios) or with overshoot (marginal cost scenarios).  Importantly, the assumption that wages fully adjust to price inflation ($w_2=1$) means that as profit and interest shares drop to zero, wages (Figure \ref{fig:1p8_A11A00F11F00}(o)) and the wage share (Figure \ref{fig:1p8_A11A00F11F00}(l)) rise.


Second, we now discuss resource consumption efficiency as a driver of decoupling.  A common interpretation of relative decoupling is that the economy becomes more efficient in its consumption of energy in producing the goods and services of which GDP is composed (e.g., UN Sustainable Development Goal 7: Affordable and Clean Energy \cite{UN2020_SDG}, climate mitigation \cite{IPCC2014_wg3}).   Because studies have shown that useful work scales proportionally to GDP, then increasing efficiency of converting primary fuels to useful work could also explain relative decoupling \cite{Warr2010,Jarvis2020}.   However, scenarios FC-000 and MC-000 (black solid and dashed lines) assume no exogenous increase in sector-specific labor productivity ($a_g=a_e=$constant), resource efficiency in operating capital ($\eta_g=\eta_e=$constant), or resource efficiency as embodied in new capital ($y_{X_g}=$constant), yet the results still show that such an economy can reside, for some time period, in the relative decoupling regime.  For the marginal cost scenario, there is a short time period in the absolute decoupling regime (more explanation later).  The explanation does not reside in the fact that all scenarios in Figures \ref{fig:1p8_A11A00F11F00} and \ref{fig:World_and_A11A00F11F00_GrowthPEU_vs_GrowthGDP} assume an initial increase in the resource depletion factor, $\delta_y$, that represents an exogenous increase in the amount of resources extracted per unit of extraction capital.  The scenarios show no relative decoupling during initial growth when $\delta_y$ increases, and they reside in the regime of relative decoupling well after $\delta_y$ reaches a constant level.

The technical coefficient $a_{ee} = \frac{\eta_e}{\delta_y y}$ (see Figure \ref{fig:World_and_A11A00F11F00_GrowthPEU_vs_GrowthGDP}(g)) is another parameter with which to interpret efficiency: a higher value represents a less efficient resource sector.  It initially decreases due to our exogenous initial increase in technological parameter $\delta_y$. However, it eventually increases as the resource, $y$, depletes and does so during the time when the economy appears most relatively decoupled (Figure \ref{fig:World_and_A11A00F11F00_GrowthPEU_vs_GrowthGDP}(c)). This is again more evidence that resource efficiency is not the full explanation of apparent relative decoupling.  Note that because of their definition as a function of $a_{ee}$, the net energy ratios, NEPR and NPR (Figures \ref{fig:1p8_A11A00F11F00}(r) and (s)), tend to have opposite trends as $a_{ee}$.

The model does provide justification, however, for the view that increasing energy efficiency can move the economy to a more decoupled state compared to an economy with no energy efficiency.  Scenarios FC-100 and MC-100 (gray lines) are identical to FC-000 and MC-000 (black lines), respectively, except that they assume a decrease in $\eta_g$ and $\eta_e$ (see Figure \ref{fig:World_and_A11A00F11F00_GrowthPEU_vs_GrowthGDP}(e) and (f)).  The increase in engineering resource consumption efficiency of capital (i.e., fuel consumption of machines) indeed causes a shift further into the relative decoupled zone compared to no increase.  

However, given the assumption that higher profits translate to increased investment, increasing energy efficiency clearly enables increased natural resource depletion (Figure \ref{fig:1p8_A11A00F11F00}(a)) while enabling the economy to achieve higher resource extraction rates (Figure \ref{fig:1p8_A11A00F11F00}(d)), population (Figure \ref{fig:1p8_A11A00F11F00}(c)), capital accumulation (Figure \ref{fig:1p8_A11A00F11F00}(b)), and net output (Figure \ref{fig:1p8_A11A00F11F00}(e)).  In other words, the HARMONEY model supports the Jevons Paradox, or rebound effect, in that higher fuel efficiency in operating capital \textit{increases} overall economic size and consumption \cite{Jevons1866}.  This theoretical finding is consistent with studies supporting the evidence for a strong rebound effect \cite{Brockway2021}, as well as the general observation that over the course of industrialization to date, the human economy has indeed invented more efficient processes while at the same time consumed more energy decade to decade.

Finally, we discuss decoupling in the context of the definition of cost that sets prices.  The maximum level of decoupling occurs at almost the same times ($T=$ 84, 148, 97, and 154, for FC-000, MC-000, FC-100, and MC-100 scenarios, respectively) as the peaks in debt ratios.   Compared to full cost price scenario FC-000, marginal cost price scenario MC-000 reaches 4.8 times higher decoupling and 3.7 times higher peak debt ratio.  For the efficiency scenarios, compared to full cost price scenario FC-100, marginal cost price scenario MC-100 reaches 2.3 times higher decoupling and 3.7 times higher peak debt ratio.   

In short, the more decoupled scenarios are those that assume marginal costs of production inform prices.  They also lead to higher debt ratios than the full cost pricing scenarios.  In addition, marginal cost pricing scenarios reach states of ``absolute decoupling'', whereas the full cost scenarios do not.  It is important to note this apparent absolute decoupling occurs transiently for a very short time from about $T=149$ to $T=158$ for the MC-000 scenario and $T=156$ to $T=162$ for the MC-100 scenario.  After these time intervals the economy again approaches full coupling as resource extraction and GDP decline, along with declines in both capital and population.

We conclude that while ``decoupling'' does increase with energy efficiency of machines, there are other factors that are at least as relevant, it not more so. Decoupling (as it appears) can come from producers using marginal cost pricing if they lack ability to, or choose not to, pass depreciation costs and interest payments into prices.  The exclusion of depreciation in cost, much more so than excluding interest payments, has the majority affect on apparent increases in decoupling.  Further, the HARMONEY model supports the conclusion that a decoupled state is a natural expected stage during a growth cycle that follows the inability to increase growth rates. That is to say, not only does a stage of relative decoupling occur during periods with no perceivable increase in device energy efficiency, this stage is evidence \textit{for} limits to (increasingly fast) growth, not evidence against limits to growth.

\begin{figure}
\begin{center}
\subfloat[]{\includegraphics[width=0.33\columnwidth]{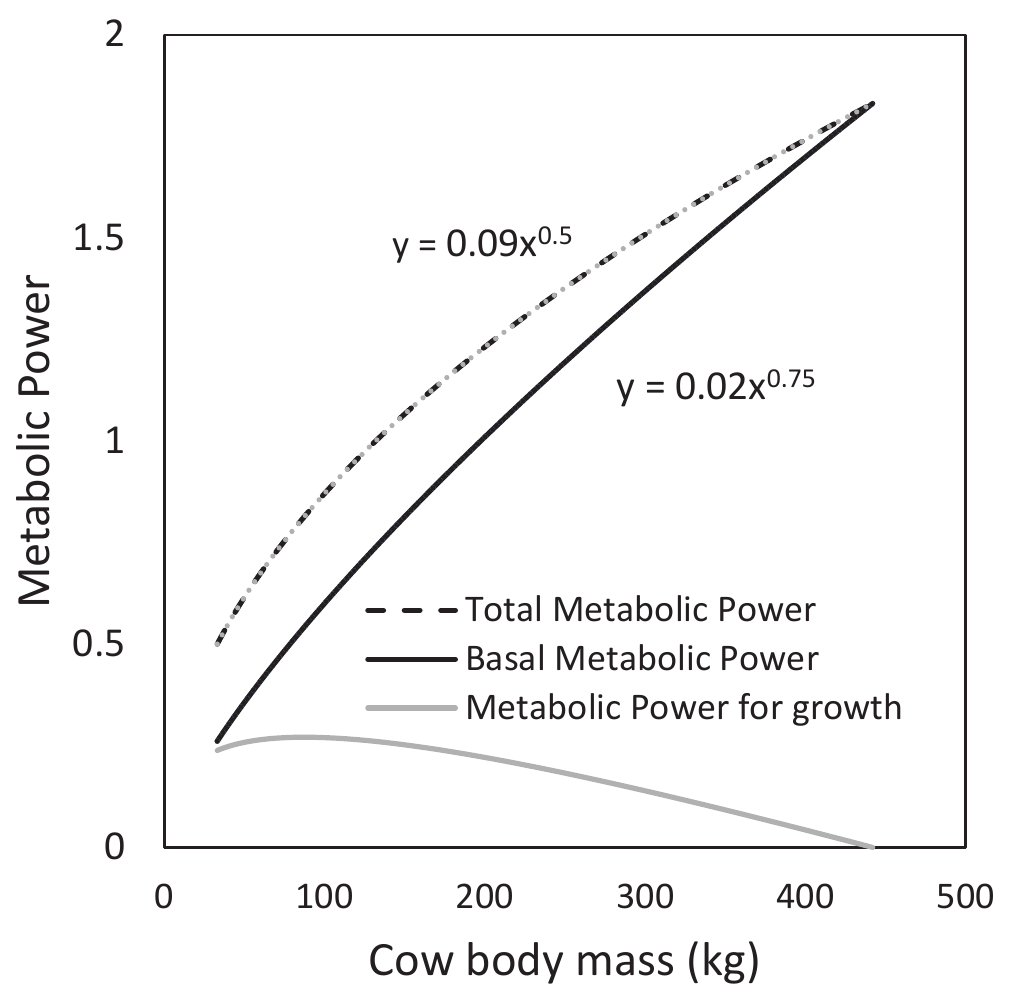}}
\subfloat[]{\includegraphics[width=0.33\columnwidth]{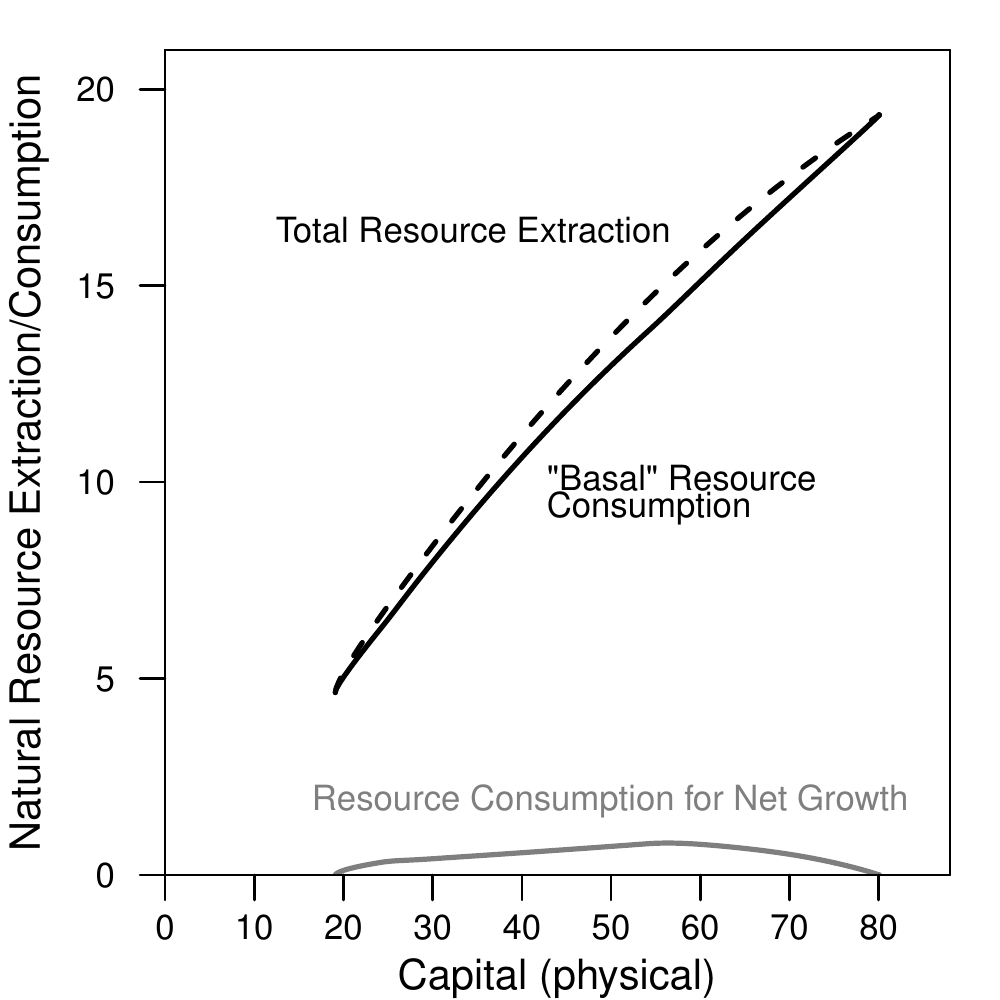}}\\
\subfloat[]{\includegraphics[width=0.33\columnwidth]{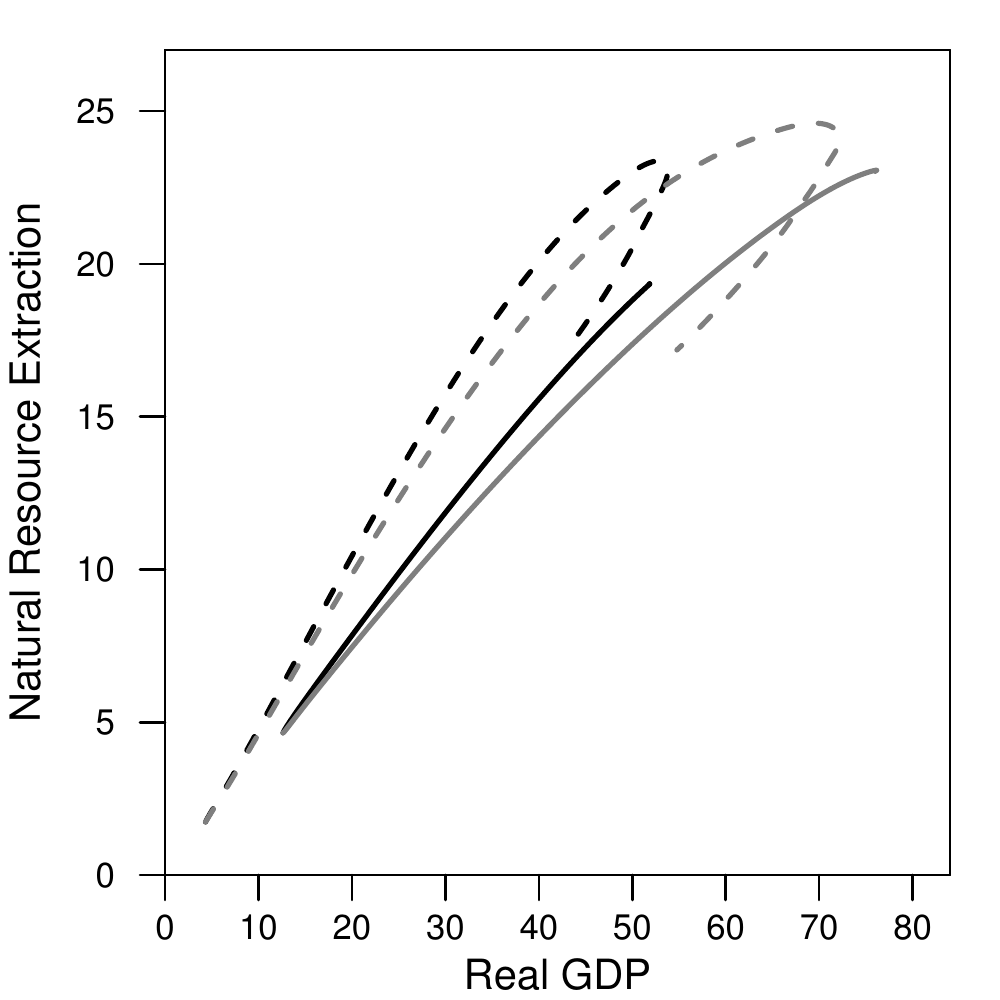}}
\subfloat[]{\includegraphics[width=0.33\columnwidth]{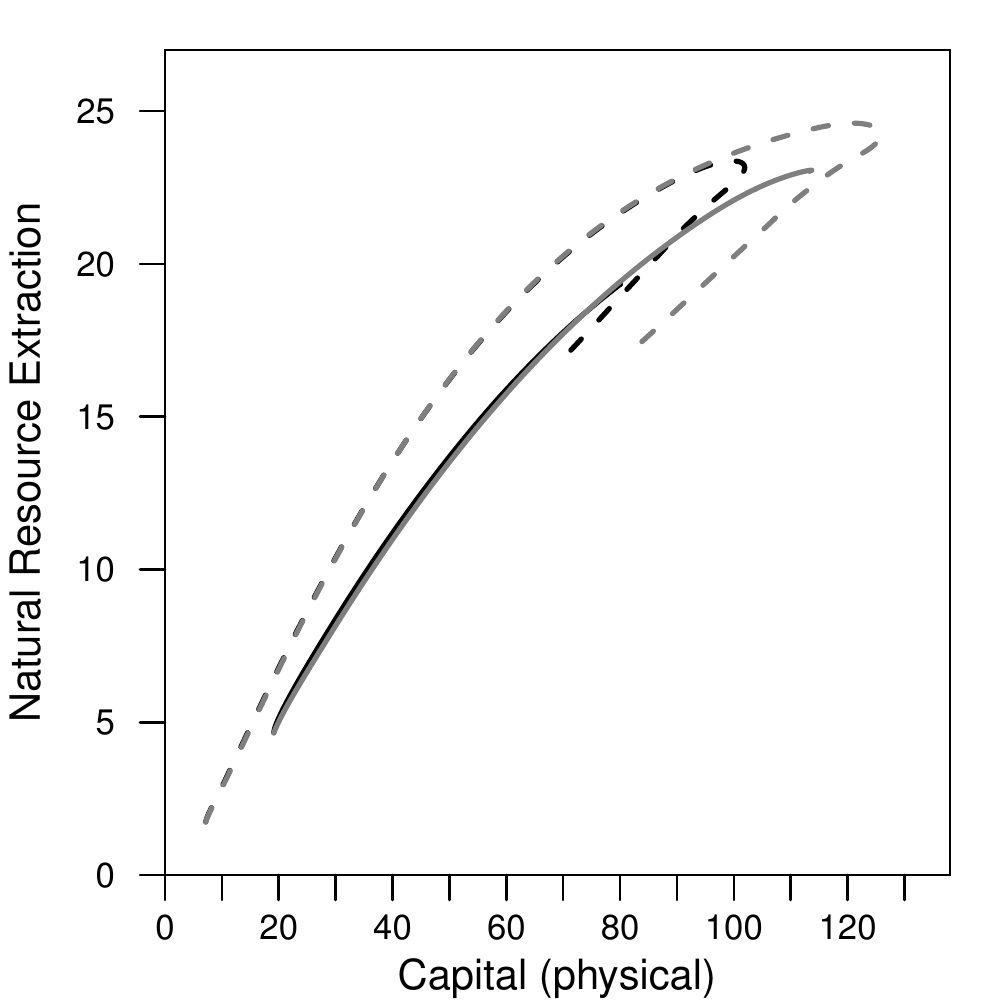}}\\
\caption{Comparison of (a) biological ontogenetic metabolic scaling (relation of energy consumption to mass), including power for growth, from \cite{West2001} with (b) a parallel plot from HARMONEY full cost pricing result with no change in efficiency.  HARMONEY plots of the same scenarios as in Figure \ref{fig:1p8_A11A00F11F00} of natural resource extraction versus (c) real GDP and (d) versus total physical capital ($K_e+K_g$). (c) and (d) show data for full cost scenarios FC-000 (black solid) and FC-100 (black dashed) with marginal cost scenarios MC-000(gray solid) and MC-100 (gray dashed).}\label{fig:Biological_and_Economic_ScalingLawComparison_A11A00F11F00}
\end{center}
\end{figure}

Before leaving the discussion of the relationship between energy consumption and economic growth, we make an explicit comparison to biological growth.  
Figure \ref{fig:Biological_and_Economic_ScalingLawComparison_A11A00F11F00} compares a typical metabolism versus mass trend (for a cow starting from birth per \cite{West2001}) to corresponding results from the HARMONEY model as total resource extraction versus GDP and capital.  Figure \ref{fig:Biological_and_Economic_ScalingLawComparison_A11A00F11F00}(a) displays curves for total metabolic power and basal metabolic power, scaling with mass to the 0.5 and 0.75 power, respectively.  The difference between these is the metabolic power allocated to growth of new mass.  One important point is that most organisms grow only to a certain size, with the trajectory of Figure \ref{fig:Biological_and_Economic_ScalingLawComparison_A11A00F11F00}(a) moving up and to the right, eventually stopping at some point.  The full cost HARMONEY scenarios show a similar trajectory. 


In Figures \ref{fig:Biological_and_Economic_ScalingLawComparison_A11A00F11F00}(c) and (d) the full cost pricing scenarios grow until a point at which they stop and the economy remains at a steady state value of resource extraction, GDP, and capital.  In contrast the marginal cost pricing scenarios move up and to the right before looping downward and to the left, eventually also resting at non-zero steady state values of extraction, GDP, and capital.   More clearly than in Figure \ref{fig:World_and_A11A00F11F00_GrowthPEU_vs_GrowthGDP}, Figure \ref{fig:Biological_and_Economic_ScalingLawComparison_A11A00F11F00} shows how the modeled marginal cost economy exists in an apparent state of absolute decoupling for only a very short time.  It is as if biological organisms, such as mammals, are forced to adhere to ``full cost pricing'' of their allocation of energy to mass because of a lack of cultural or economic choice among their cells.  In our human economy, however, we have the option to define rules based upon ``marginal cost'' accounting that appear to, but do not, thwart the necessity of resource consumption for growth and maintenance.  

%
%
%
%
%
%

\subsection{Impacts on Loss of Wage Bargaining Power and Ponzi Investing}\label{sec:WageBargaining_Ponzi}
The results in Figure \ref{fig:1p8_A11A00F11F00} assume that wages are fully indexed to inflation by assuming the ``wage bargaining'' factors as $w_1=1$ and $w_2=1$ in Equation \ref{eq:wages}.  We test how a reduction in these wage bargaining factors affects the real wage and wage share, as many posit a loss of bargaining power as a major explanation for stagnant U.S. wages since the 1970s \cite{Bivens2015}.  We choose the timing for lowering the wage bargaining factors to mimic what occurred in the U.S.  The data indicate that, starting in the early 1970s once the U.S. reached peak per capita energy consumption, real wages stopped increasing and wage share began to decline  \cite{King2020,King2021,Bivens2015}.  Thus, we begin reducing  $w_1=1$ and $w_2=1$ once the simulation reaches peak per capita resource extraction (as indicated in Table \ref{tab:ScenariosDefinitions}).  Figure \ref{fig:1p8_A11A11gA11iA00A00gA00i} (dashed lines) shows the impact of gradually reducing $w_1$ and $w_2$ to zero (see Supplemental Figures \ref{fig:1p8_A11A11gA11iF11F11gF11i} and \ref{fig:1p8_A00A00gA00iF00F00gF00i} for the full suite of Table \ref{tab:ScenariosDefinitions} scenarios exploring bargaining power reductions). 

In all cases, the removal of bargaining power leads to participation rate increasing to the maximum level of 80\% (Figure \ref{fig:1p8_A11A11gA11iA00A00gA00i}(f)).  When considering full cost pricing without increasing capital operating resource efficiency (scenario FC-010), the loss of bargaining power allocates a higher share of value added to profits (Figure \ref{fig:1p8_A11A11gA11iA00A00gA00i}(e)) and therefore increases investment(Figure \ref{fig:1p8_A11A11gA11iA00A00gA00i}(g)).  Capital then accumulates to a much higher level (Figure \ref{fig:1p8_A11A11gA11iA00A00gA00i}(c)) while operating at declining capacity utilization (Figure \ref{fig:1p8_A11A11gA11iA00A00gA00i}(k) and (l)).  Both real wage Figure \ref{fig:1p8_A11A11gA11iA00A00gA00i}((a)) and wage share (Figure \ref{fig:1p8_A11A11gA11iA00A00gA00i}(b)) decrease to zero as depreciation (Figures \ref{fig:1p8_A11A11gA11iF11F11gF11i}(m)) accounts for the dominant share of value added.  

The result from loss of bargaining power is quite different when using the marginal cost pricing assumption (MC-010, gray dashed line).  While real wage and wage share no longer rise as occurs when full bargaining power remains, they also do not decline but stay approximately level once bargaining power is removed (Figure \ref{fig:1p8_A11A11gA11iA00A00gA00i}(a) and (b)).  While investment increases with respect to maintaining full labor bargaining power, the inability to pass through depreciation costs limits net investment to similar levels as with full bargaining power (Figure \ref{fig:1p8_A11A11gA11iA00A00gA00i}(g)).

\subsubsection{Ponzi Investing with Loss of Bargaining Power}
As defined, Ponzi investing diverts private investment away from investment in physical capital.  If full bargaining power remains, capacity utilization remains at its target level, and Ponzi investment is zero (not shown).  

If wage bargaining power is removed, Ponzi investing (Figure \ref{fig:1p8_A11A11gA11iA00A00gA00i}(h)) significantly increases private debt ratios (Figure \ref{fig:1p8_A11A11gA11iA00A00gA00i}(i)).  For full cost pricing (scenario FC-011), Ponzi investing translates to a slower decline in wages (Figure \ref{fig:1p8_A11A11gA11iA00A00gA00i}(a)) compared with all investment going to physical capital. Thus, perhaps unintuitively from a wage perspective, if bargaining power is removed, Ponzi investing appears good for labor.  In such a case, Ponzi investing also appears good for firms since profit share is also higher than without Ponzi investing (Figure \ref{fig:1p8_A11A11gA11iA00A00gA00i}(e)).  However, the full cost pricing assumption still drives capacity utilization to low levels (near 65\%) that might not remain in a real-world economy (Figure \ref{fig:1p8_A11A11gA11iA00A00gA00i}(k) and (l)).  
 
If using marginal cost pricing with Ponzi investing (scenario MC-011), there is no meaningful change in wages and wage share as they remain constant at their lowest values.  Capacity utilization declines to near 75\% compared to the targeted 85\% (Figure \ref{fig:1p8_A11A11gA11iA00A00gA00i}(k) and (l)).  Private debt ratio continues to rise but to much lower levels than under the full cost pricing assumption (Figure \ref{fig:1p8_A11A11gA11iA00A00gA00i}(i)). 
  
\begin{figure}
\begin{center}
\subfloat[]{\includegraphics[width=0.33\columnwidth]{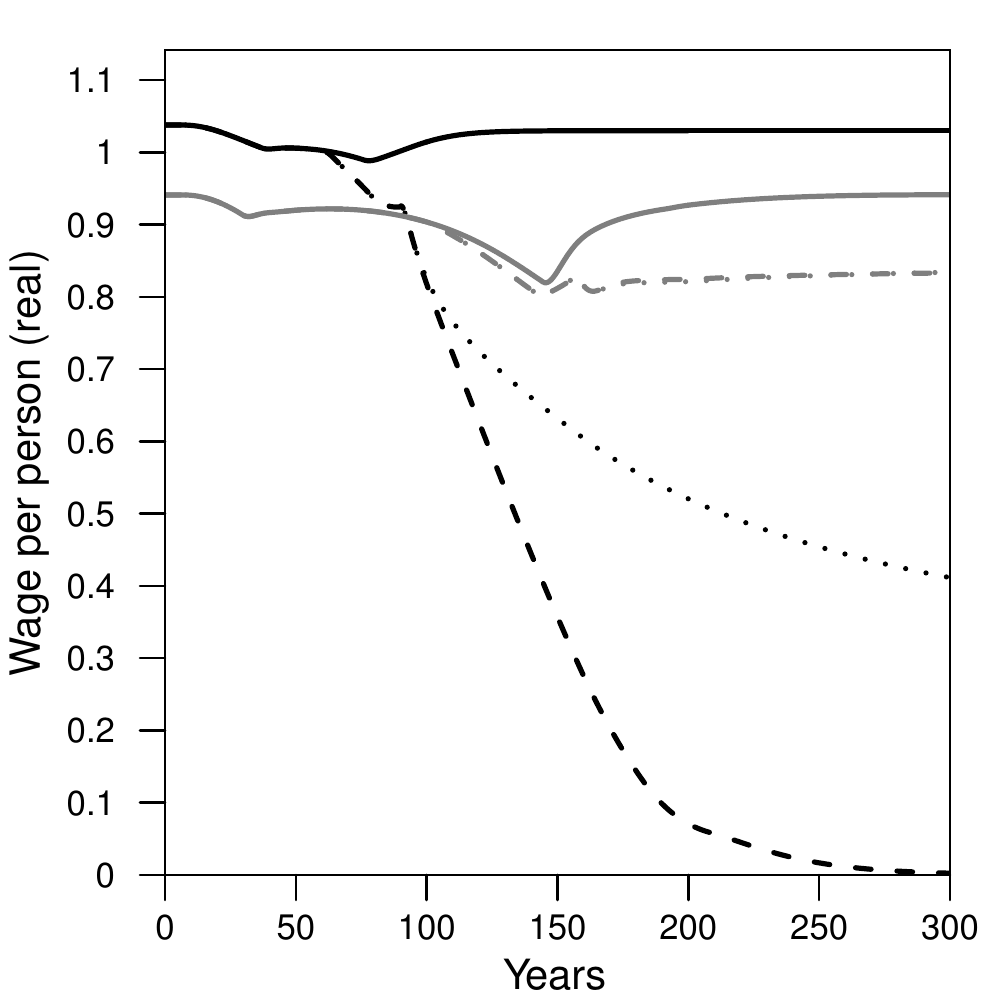}}
\subfloat[]{\includegraphics[width=0.33\columnwidth]{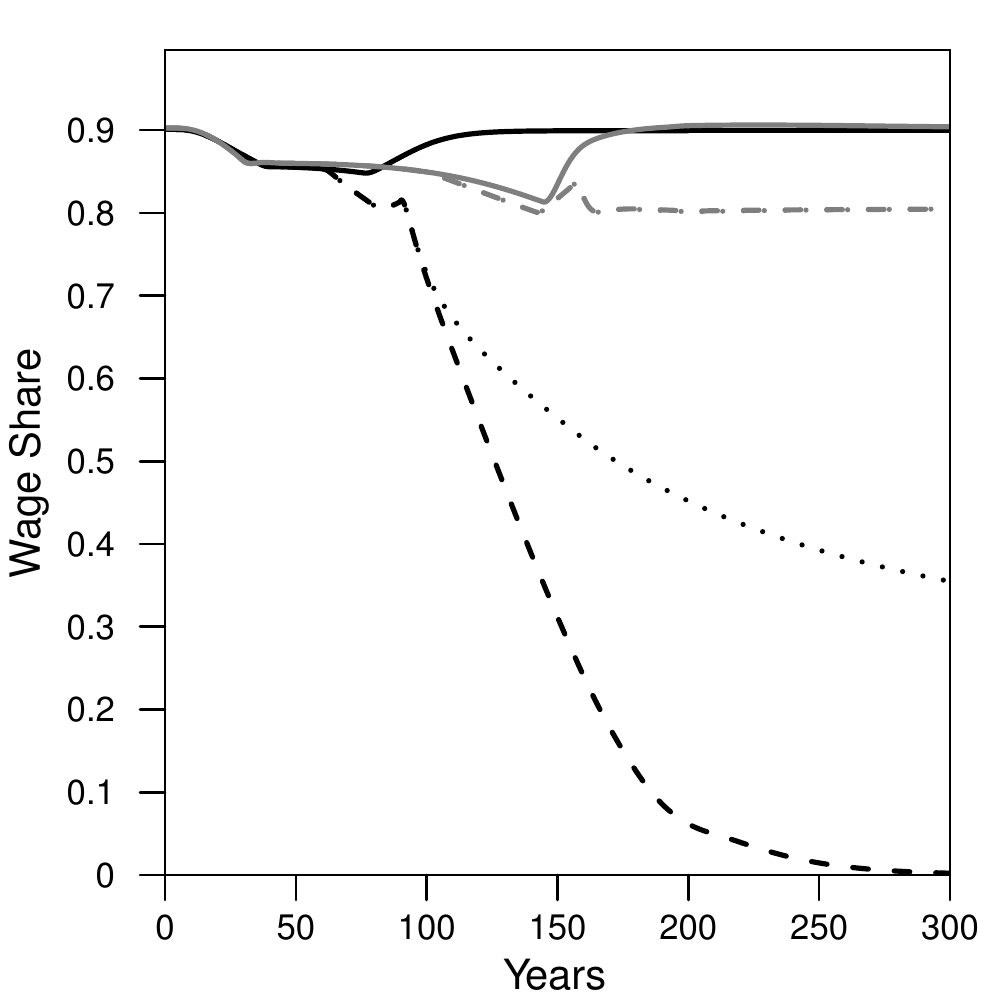}}
\subfloat[]{\includegraphics[width=0.33\columnwidth]{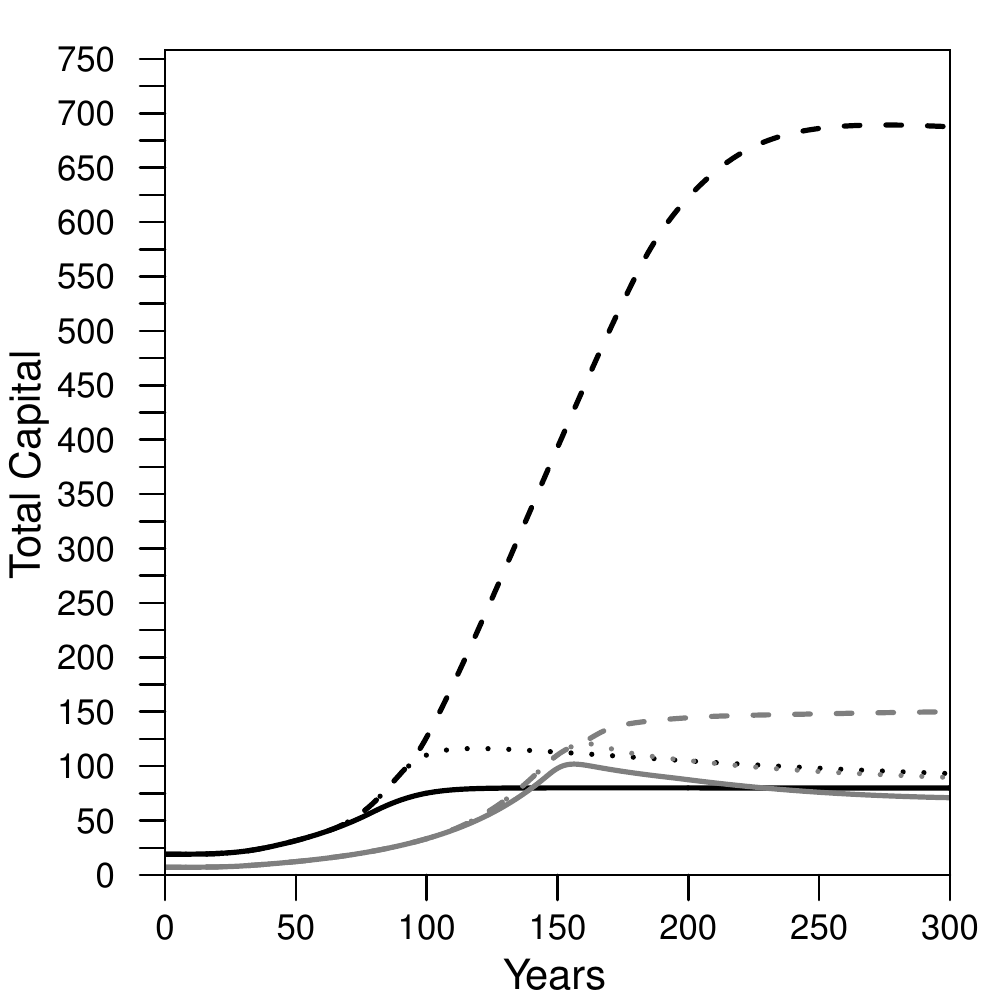}}\\
\subfloat[]{\includegraphics[width=0.33\columnwidth]{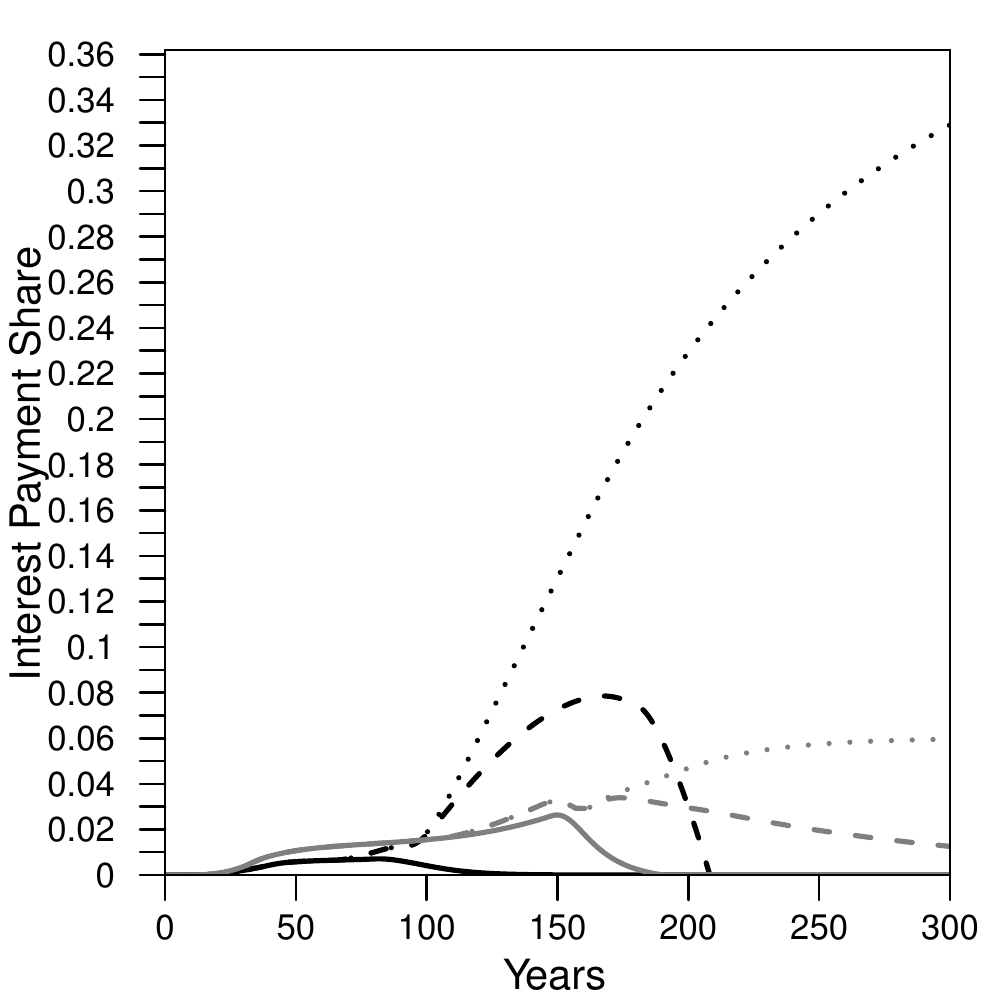}}
\subfloat[]{\includegraphics[width=0.33\columnwidth]{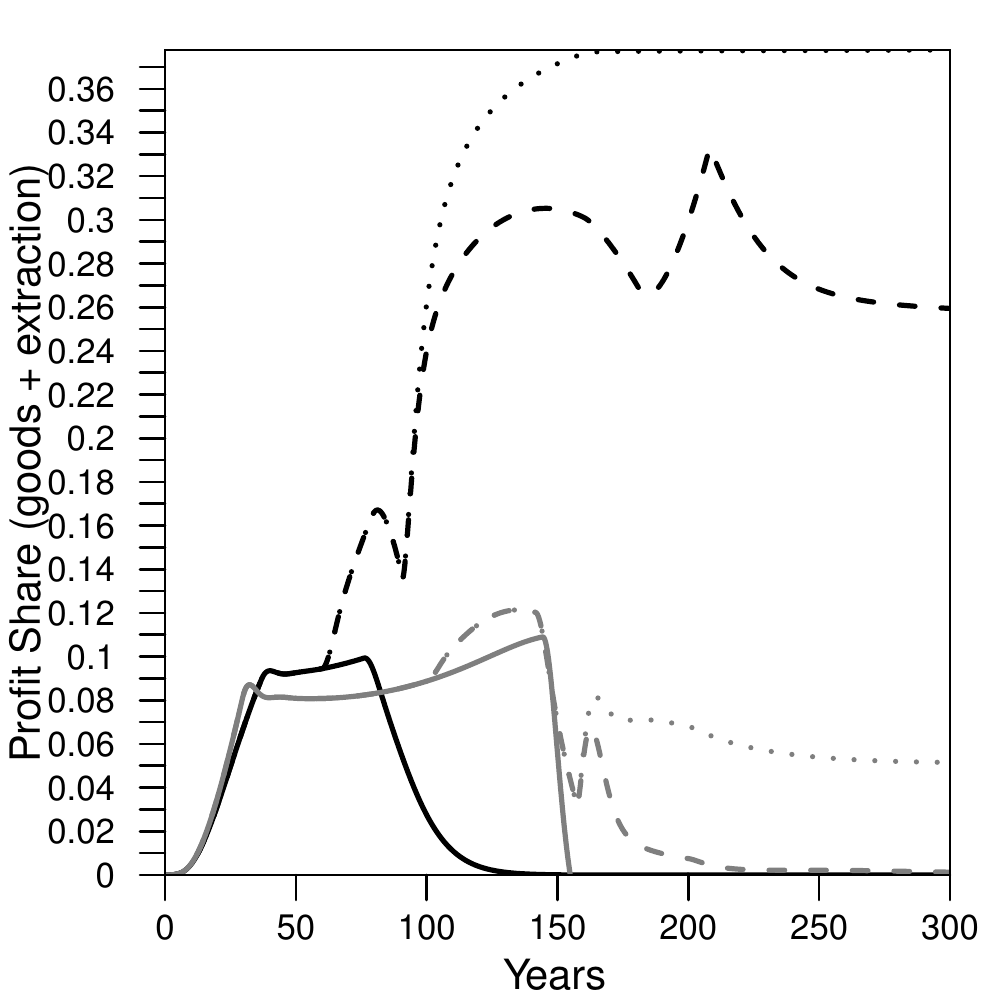}}
\subfloat[]{\includegraphics[width=0.33\columnwidth]{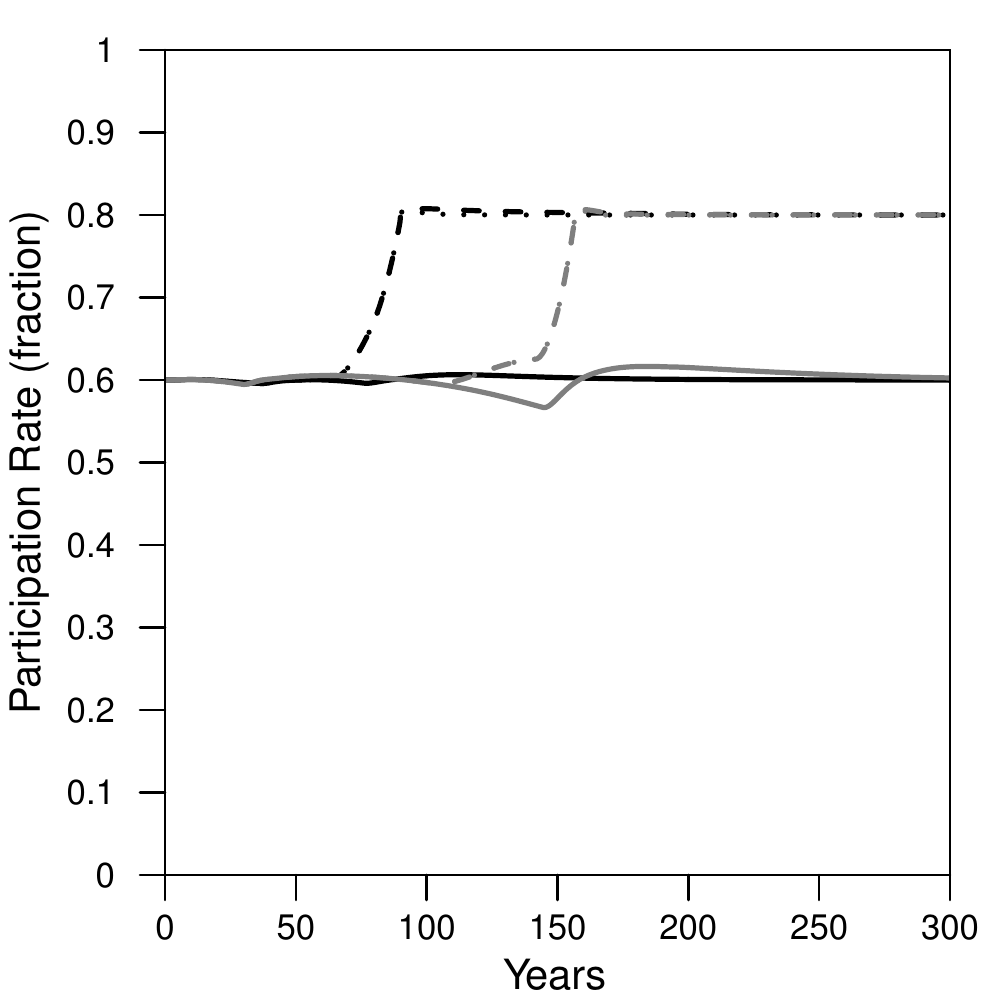}}\\
\subfloat[]{\includegraphics[width=0.33\columnwidth]{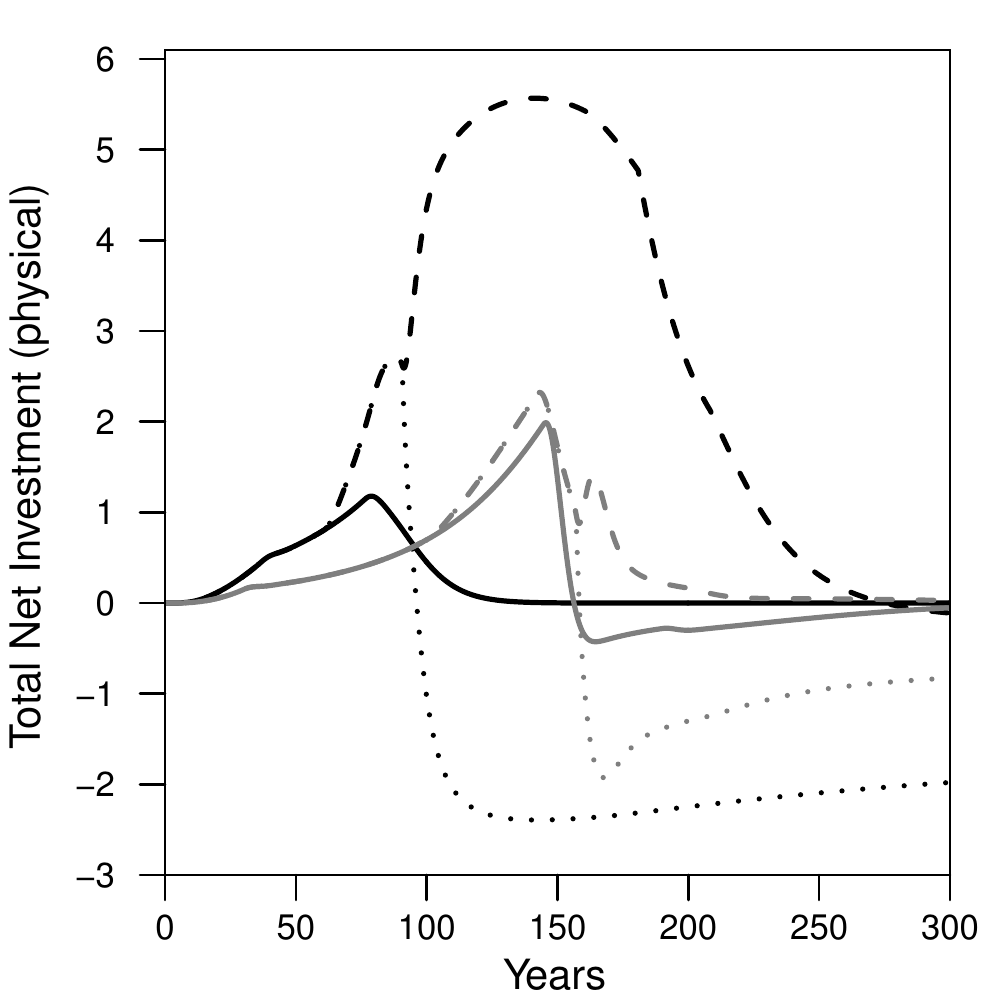}}
\subfloat[]{\includegraphics[width=0.33\columnwidth]{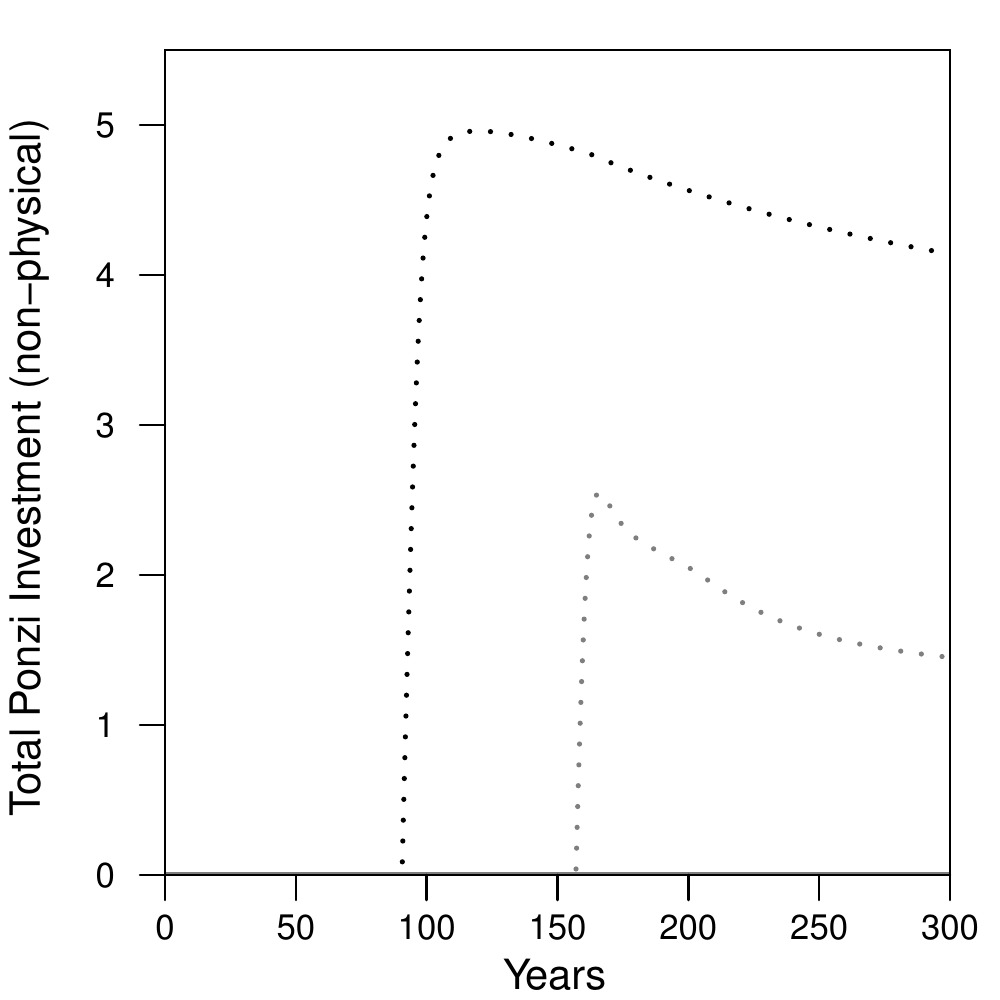}}
\subfloat[]{\includegraphics[width=0.33\columnwidth]{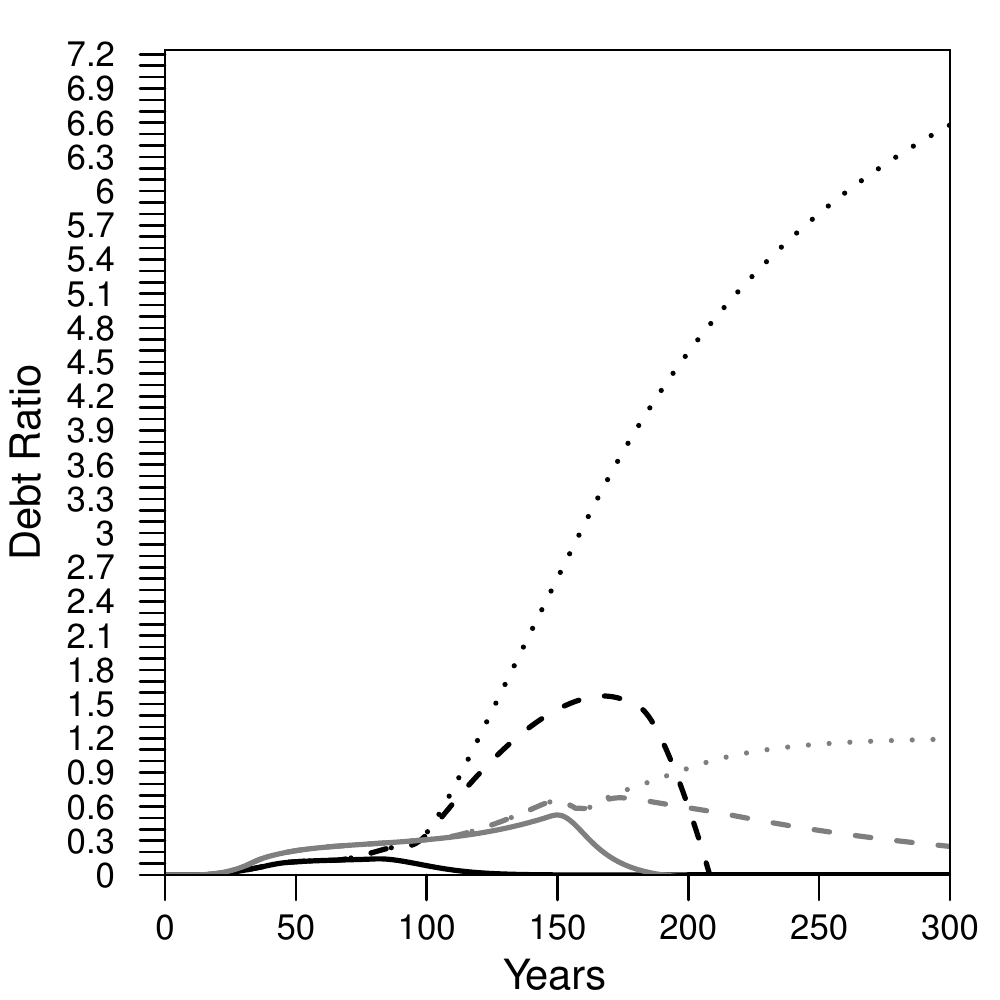}}\\
\caption{Scenarios FC-000, FC-010, FC-011, MC-000, MC-010, and MC-011 (black, black dashed, black dotted, gray, gray dashed, gray dotted) (a) wage per person (real), (b)  wage share, (c) total capital, (d) interest share, (e) profit share, (f) participation rate, (g) total net investment (physical capital), (h) Ponzi investment, (i) debt ratio, (continued) \dots}\label{fig:1p8_A11A11gA11iA00A00gA00i}
\end{center}
\end{figure}

\begin{figure}\ContinuedFloat
\begin{center}
\subfloat[]{\includegraphics[width=0.33\columnwidth]{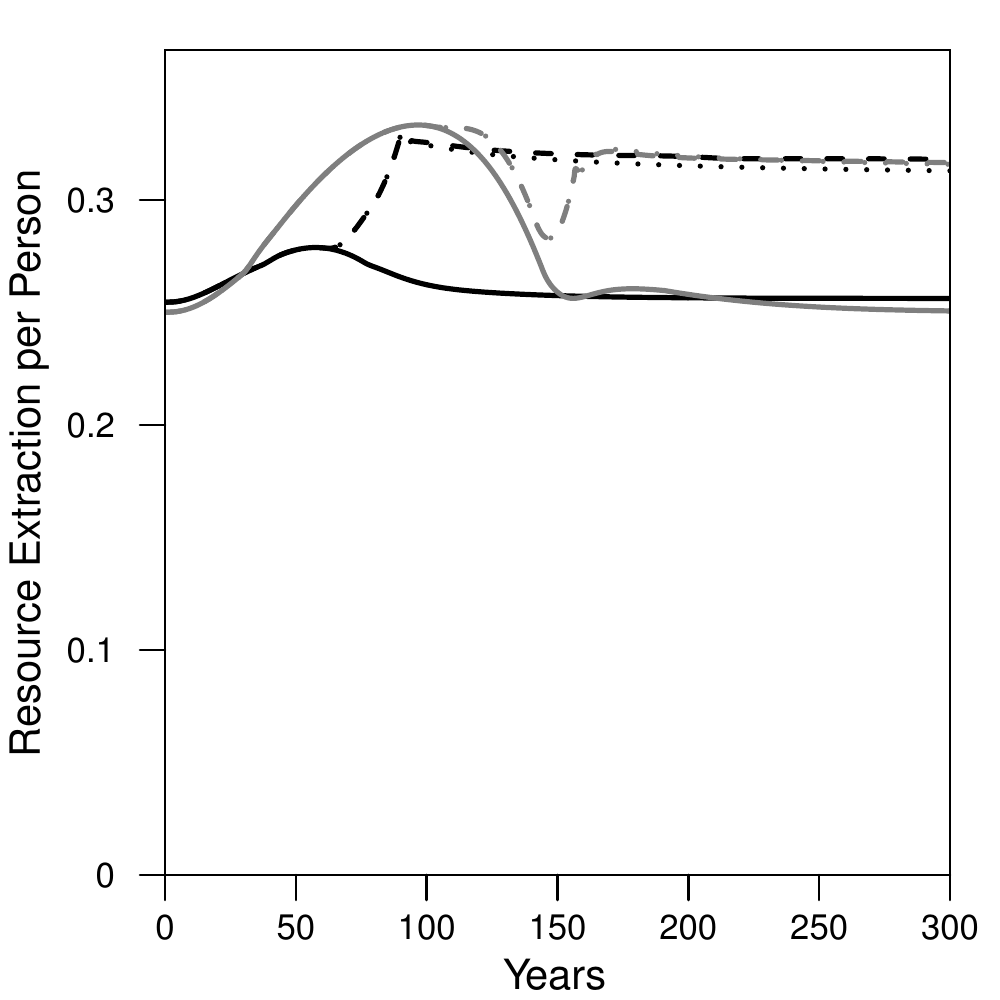}}
\subfloat[]{\includegraphics[width=0.33\columnwidth]{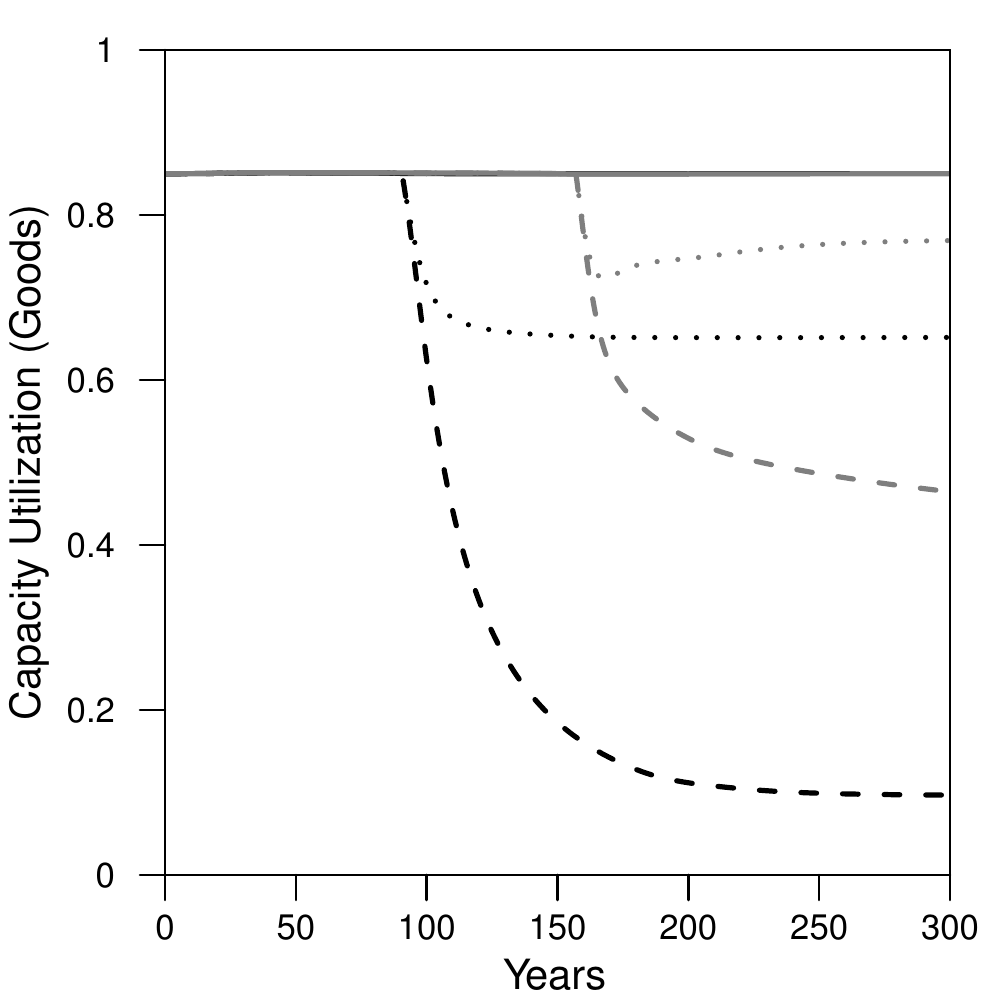}}
\subfloat[]{\includegraphics[width=0.33\columnwidth]{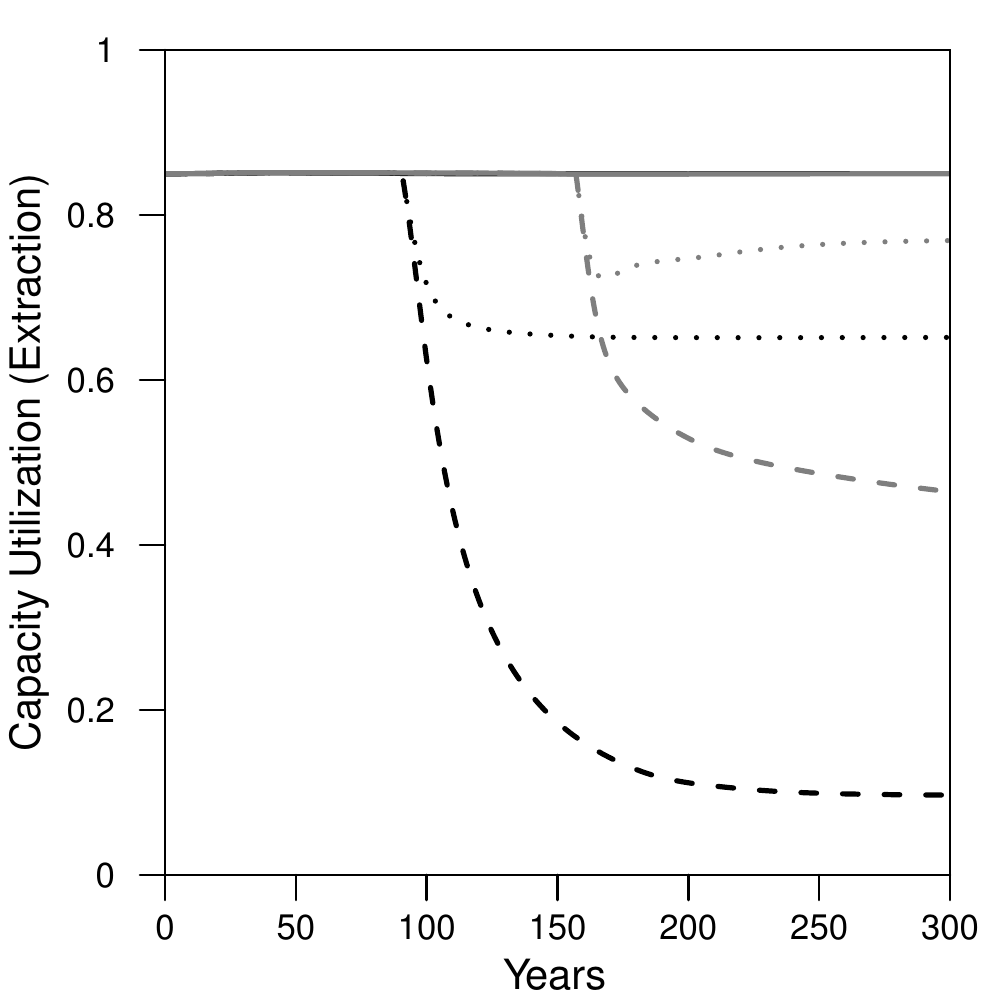}}
\caption{(continued) (j) total resources extraction per person, (k) capacity utilization of goods capital, and (l) capacity utilization of extraction capital.} 
\end{center}
\end{figure}

\subsection{U.S. Structural Trend Comparison}

Here we compare structural trends of the HARMONEY results to those of the U.S. since World War II. 

\subsubsection{Capacity Utilization}

\begin{figure}[h]
\begin{center}
\includegraphics[width=.4\columnwidth]{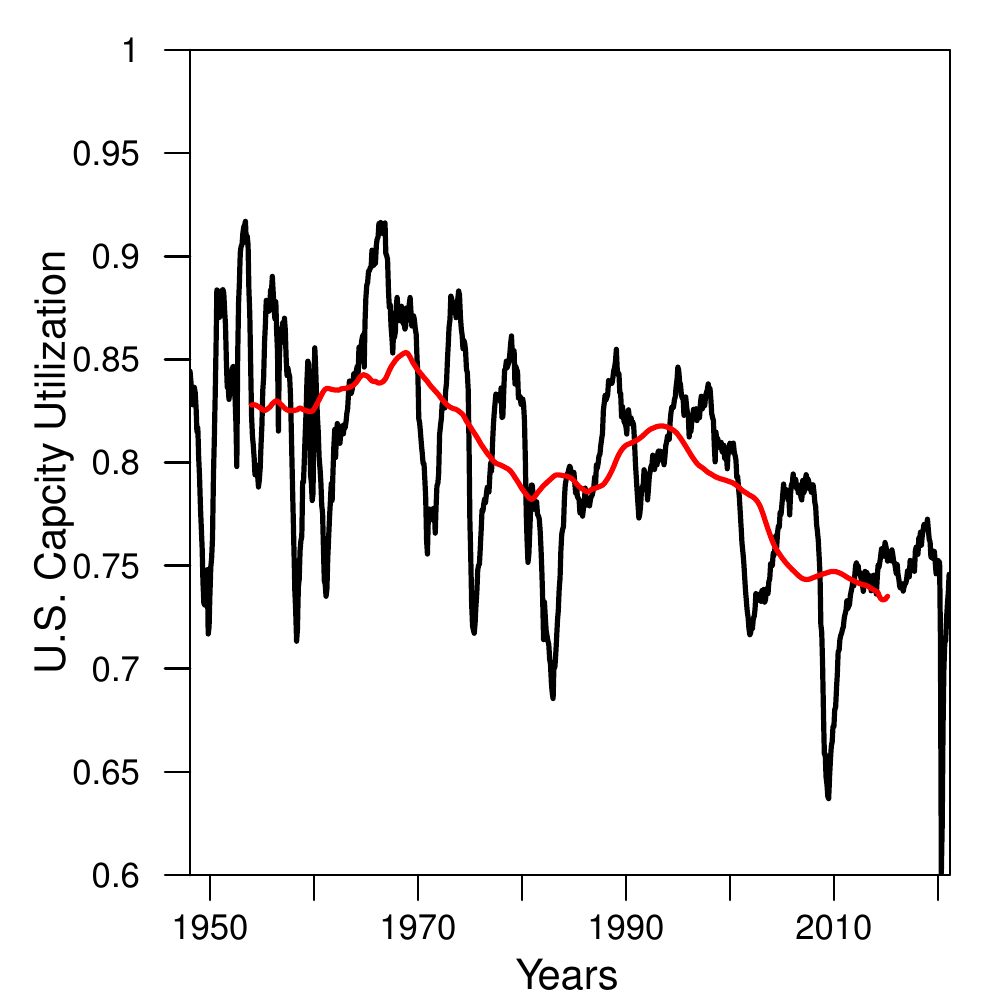}
\caption{Monthly capacity utilization of U.S. manufacturing (SIC) (black: monthly data; red: 144 month = 12 year rolling average).  Data from Federal Reserve Bank of St. Louis data set CUMFNS.}\label{fig:CapacityUtilization_US_FRED}
\end{center}
\end{figure}

Figure \ref{fig:CapacityUtilization_US_FRED} shows the capacity utilization of U.S. manufacturing from 1948--2020.  It was highest during the post-World War II decades through the early 1970s, with a 12-year running average between 0.83--0.85. U.S. capacity utilization declined after the 1970s with the running average remaining below 0.75 since 2005.

HARMONEY goods sector (i.e., manufacturing) capacity utilization shows an interesting parallel with that of U.S. manufacturing (see Figure \ref{fig:1p8_A11A11gA11iA00A00gA00i}(k)).  The HARMONEY scenarios with the loss of wage bargaining power also show a decline in capacity utilization.  Thus, independent of concerns about the offshoring of U.S. manufacturing jobs, since HARMONEY assumes a closed economy (i.e., no imports or exports), the model results provide justification for interdependence among overinvestment in capital (leading to lower capacity utilization), loss of wage bargaining power, and a slow-down (or stagnation) in per capita energy consumption.  

\subsubsection{Labor Productivity and Real Wages}
\begin{figure}
\begin{center}
\subfloat[Full Cost, Gross Labor Productivity]{\includegraphics[width=0.4\columnwidth]{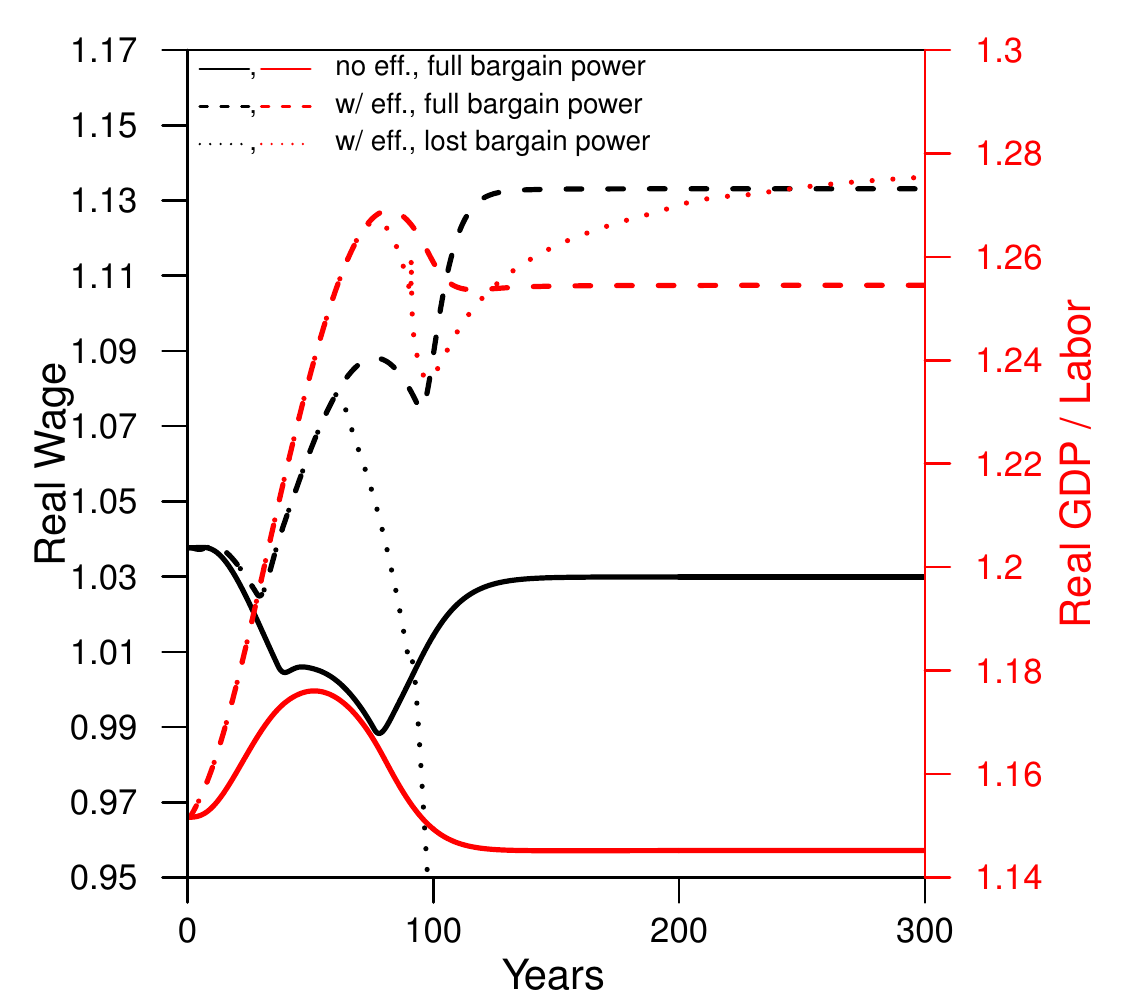}}
\subfloat[Marginal Cost, Gross Labor Productivity]{\includegraphics[width=0.4\columnwidth]{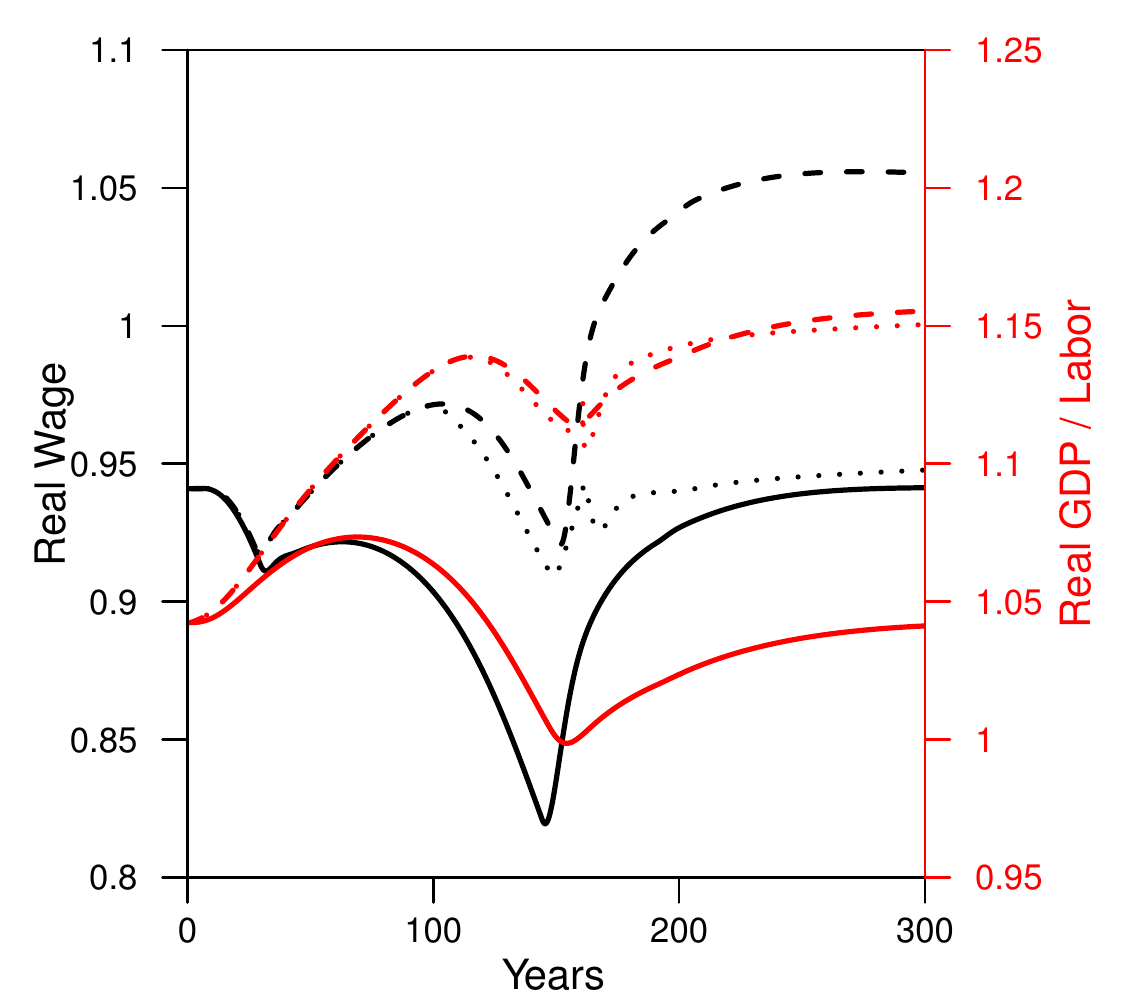}}\\
\subfloat[Full Cost, Net Labor Productivity]{\includegraphics[width=0.4\columnwidth]{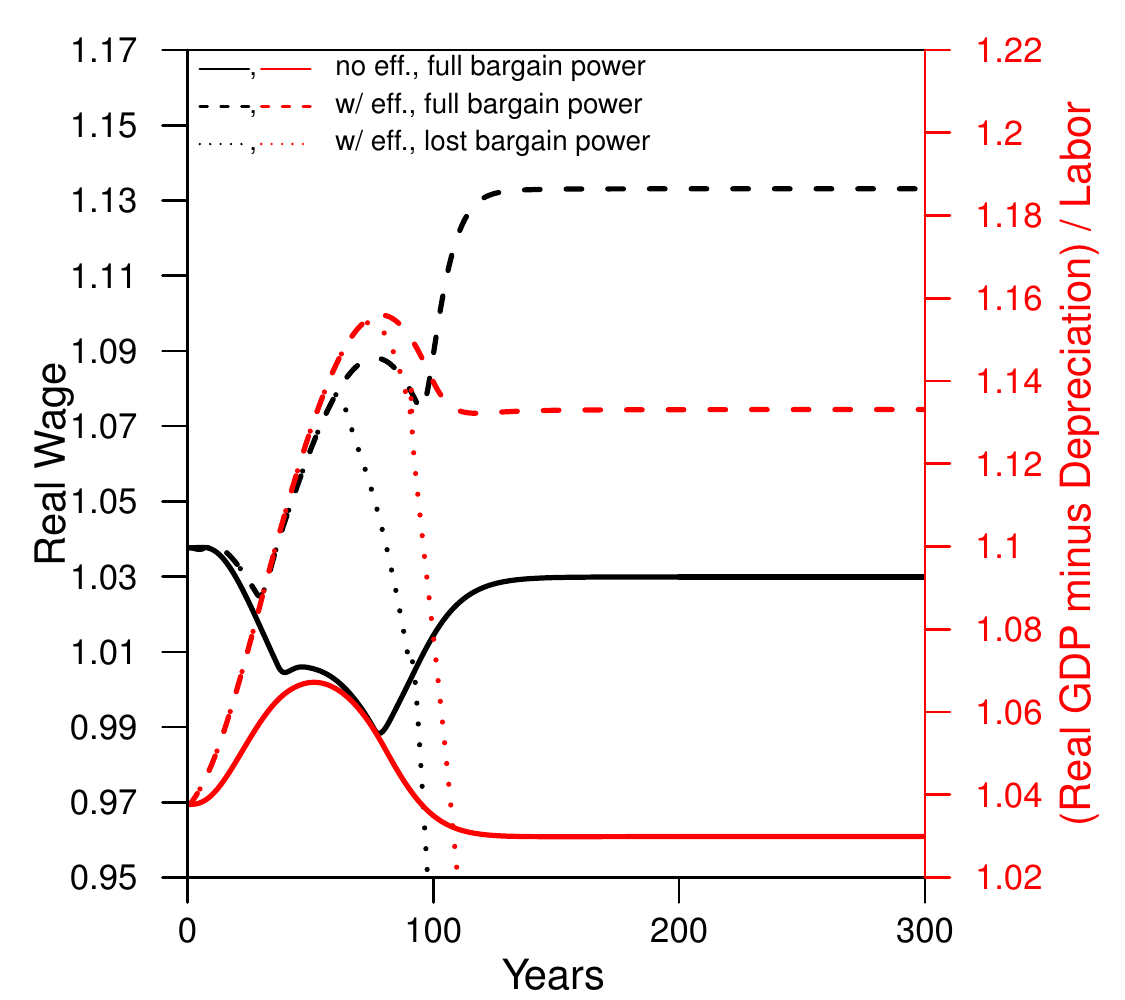}}
\subfloat[Marginal Cost, Net Labor Productivity]{\includegraphics[width=0.4\columnwidth]{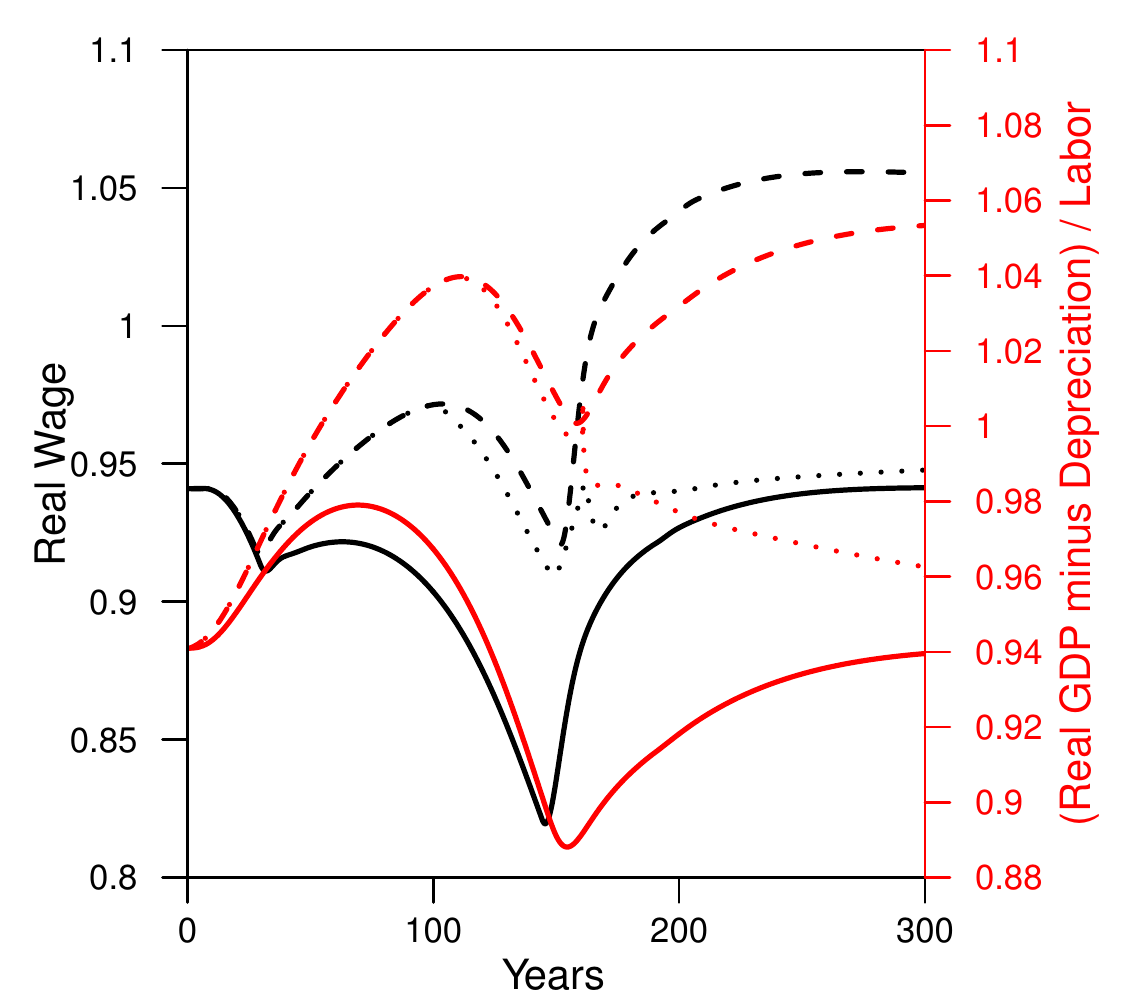}}
\caption{Real wages (left axis, black) plotted with total economy labor productivity (right axis, red) for the full cost (FC-000, FC-100, and FC-110 in (a) and (c)) and marginal cost (MC-000, MC-100, and MC-110 in (b) and (d)) pricing scenarios.  Subfigures (a) and (b) show ``full'' labor productivity, and subfigures (c) and (d) show net labor productivity that subtracts depreciation from GDP.  Per Table \ref{tab:ScenariosDefinitions} --- Solid lines: no efficiency increase, labor has full bargaining power (FC-000 and MC-000). Dashed lines: with efficiency increase, labor has full bargaining power (FC-100 and MC-100). Dotted lines: with efficiency increase, labor bargaining power declines starting at peak resource extraction per person (FC-110 and MC-110).}\label{fig:Wage_LaborProductivity_comparison}
\end{center}
\end{figure}

The effect of resource consumption on economy-wide labor productivity and wages is evident in Figure \ref{fig:Wage_LaborProductivity_comparison} which is meant to compare to the trend noted in Bishel and Mivens (2015) \cite{Bivens2015}.  They note that U.S. hourly compensation rises with net productivity (= growth of output of goods and services minus depreciation per hour worked) from 1948 to 1973, but afterwards hourly compensation is relatively constant while net productivity continues to increase.  In Figures \ref{fig:Wage_LaborProductivity_comparison}(a) and (b), gross labor productivity is real GDP divided by total labor, and net labor productivity in Figures \ref{fig:Wage_LaborProductivity_comparison}(c) and (d) subtracts depreciation from GDP.  

The trends are roughly the same between gross and net productivity, except for the effect when labor bargaining power is removed.  Aside from the initial 20-30 years of the simulation, real wages increase with increased productivity and decrease with decreasing productivity.  Also, both increase significantly more when assuming an increasing efficiency of resource consumption to operate capital.  This matches exactly with the U.S. data that show the rise in U.S. real wages from 1945 to the early 1970s coincided with the U.S.'s largest increase in conversion efficiency of primary energy to useful work from 6\% to 11\% \cite{Warr2010}.  From 1900--1940 and 1970--2000, this conversion efficiency was nearly stagnant.

The loss of bargaining power (starting at peak resource extraction per person per Table \ref{tab:ScenariosDefinitions}) does cause a divergence in trend between wages and productivity.  When removing labor bargaining power, gross labor productivity slightly increases with full cost pricing (Figure \ref{fig:Wage_LaborProductivity_comparison}(a)), and it follows a similar trend as with bargaining power for the marginal cost pricing scenario (Figure \ref{fig:Wage_LaborProductivity_comparison}(b)).  On the other hand, with the loss of bargaining power, net labor productivity decreases dramatically with full cost pricing (Figure \ref{fig:Wage_LaborProductivity_comparison}(c)) but only asymptotically declines with marginal cost pricing (Figure \ref{fig:Wage_LaborProductivity_comparison}(d)).  In short, when removing wage bargaining power, gross and net labor productivity move in opposite directions.  

None of the scenarios translates accurately to the U.S. trend from Bishel and Mivens (2015) \cite{Bivens2015}, as the real world U.S. situation is more complicated (e.g., imports and exports, offshore investment).  The HARMONEY results do, however, point to the need to consider resource consumption in tandem with policies and labor laws.

\subsubsection{Internal Structure of U.S. and HARMONEY I-O Tables}\label{sec:Results_InformationTheory}

\begin{figure}
\begin{center}
\subfloat[]{\includegraphics[width=.40\textwidth]{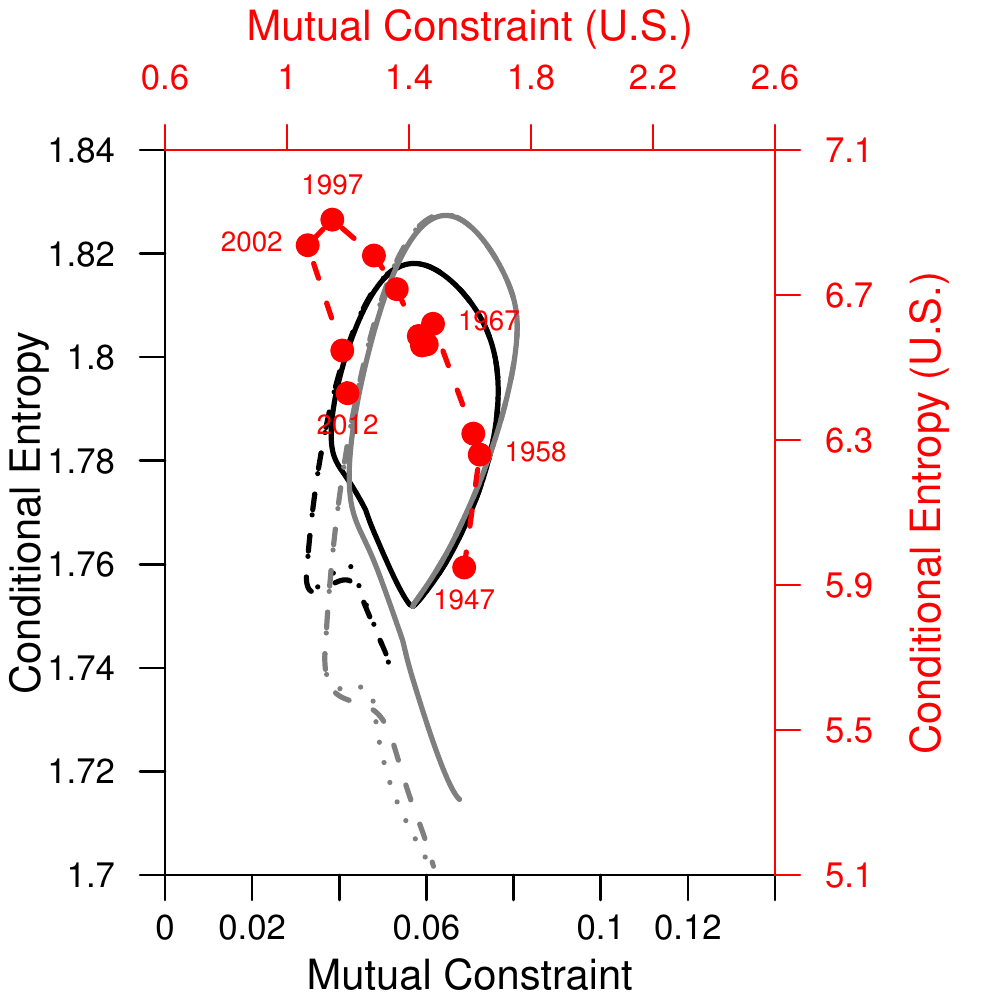}}
\subfloat[Information Entropy]{\includegraphics[width=.40\textwidth]{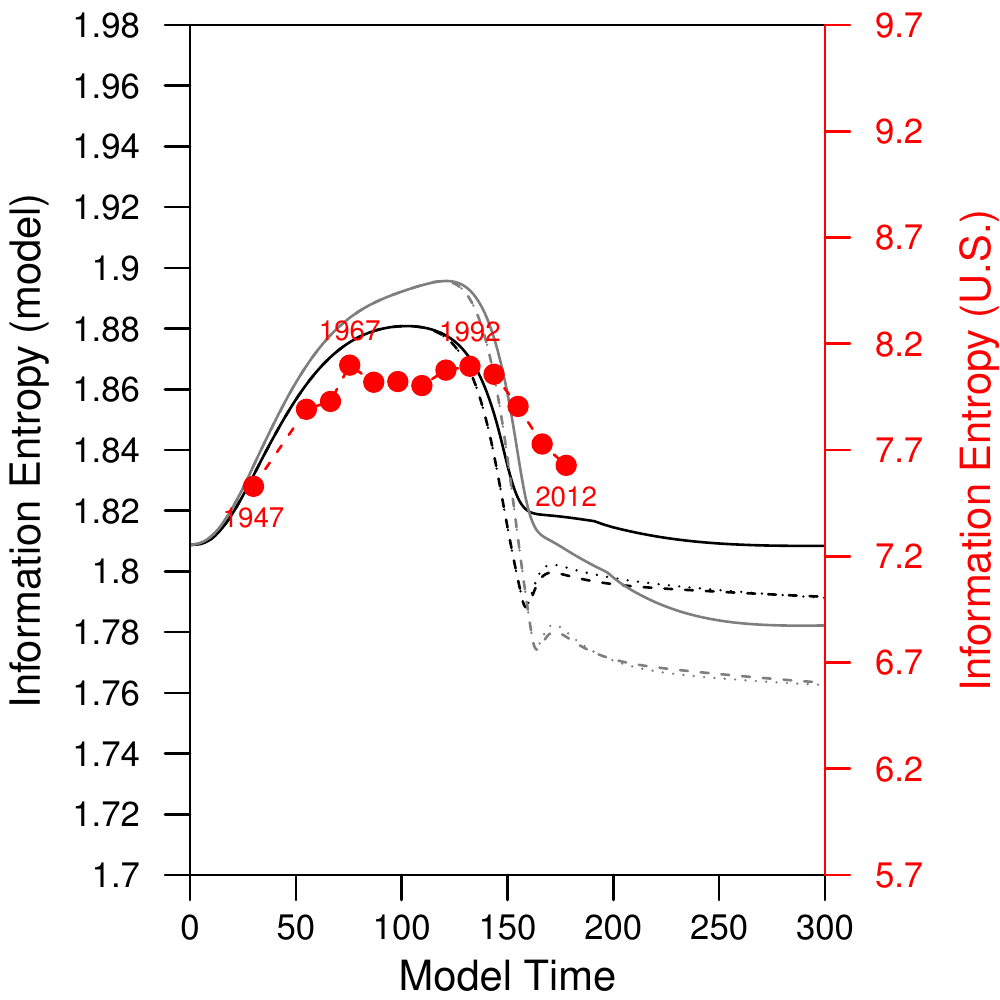}}\\
\subfloat[Conditional Entropy]{\includegraphics[width=.40\textwidth]{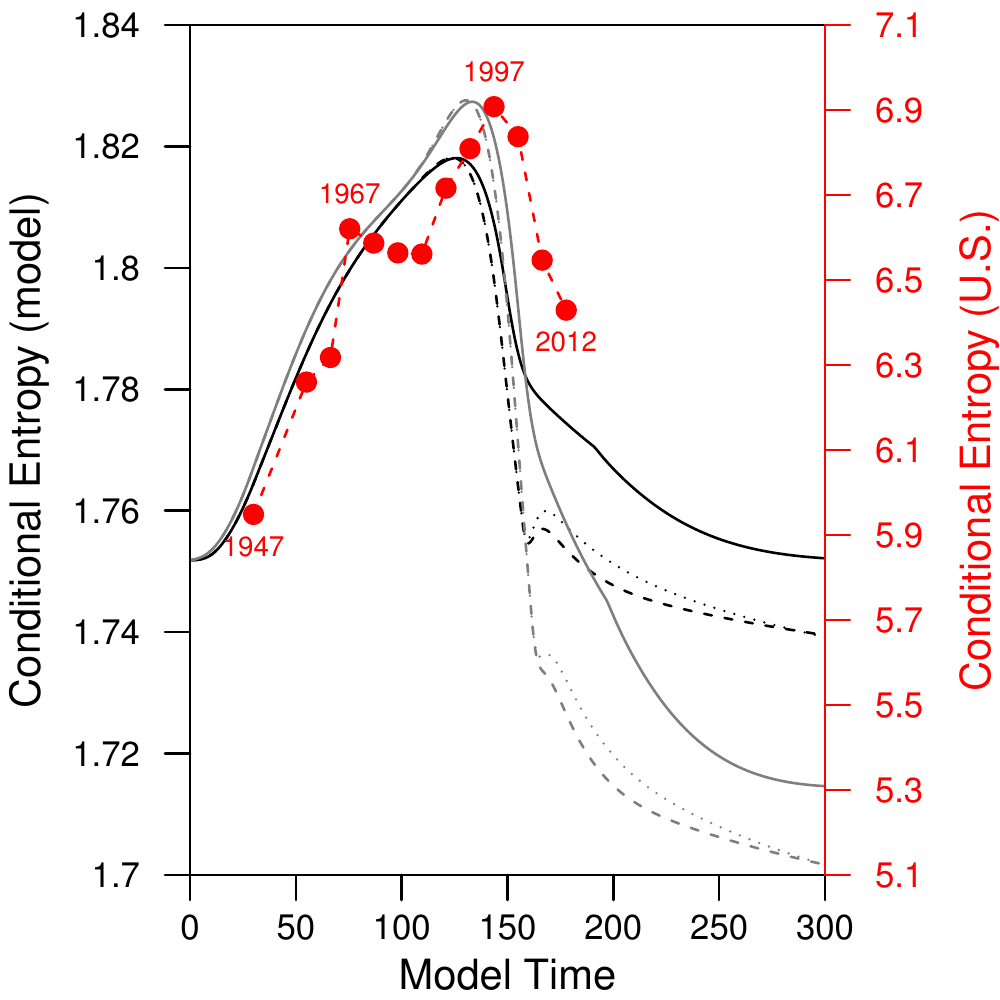}}
\subfloat[Mutual Constraint]{\includegraphics[width=.40\textwidth]{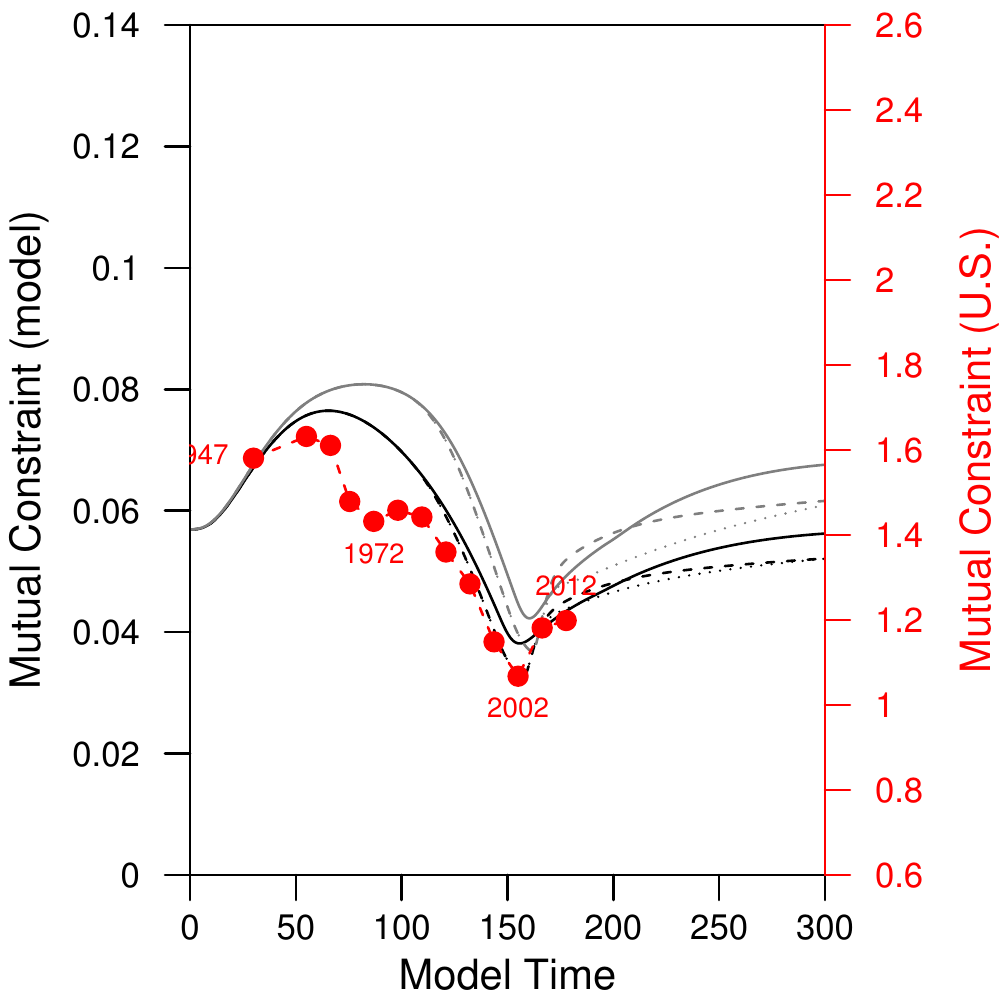}}
\caption{Information theory metrics of the marginal cost scenarios compared to the 37-sector aggregation of the U.S. Use tables from 1947--2012 from King (2016) (red dashed lines, right axis) \cite{King2016}.  Here, information theory metrics are calculated using base 2 logarithm instead of natural logarithm in King (2016).  MC-000 (black solid), MC-010 (black dashed), MC-011 (black dotted), MC-100 (gray solid), MC-110 (gray dashed), MC-111 (gray dotted).  (a) Conditional entropy versus mutual constraint, (b) information entropy vs time, (c) conditional entropy vs time, and (d) mutual constraint vs time.  U.S. information theory metrics are calculated using base 2 logarithm instead of natural logarithm as in \cite{King2016}.}\label{fig:InformationTheory_Comparision_MarginalCost}
\end{center}
\end{figure}

Both the model and the post-World War II U.S. economy show similar temporal patterns in internal structural change as measured via information theory metrics applied to the respective input-output tables.   Even though the HARMONEY model is neither calibrated to the U.S. economy nor explicitly intended to mimic U.S. patterns, we overlay the U.S. economy calculations on the simulation data to enable qualitative comparison.  Figure \ref{fig:InformationTheory_Comparision_MarginalCost} shows information theory calculations for the U.S. from King (2016) and results for six marginal cost pricing simulations, and Supplemental Figure \ref{fig:InformationTheory_Comparision_FullCost} shows results for full cost pricing simulations.  

Considering the information theory phase space in Figure \ref{fig:InformationTheory_Comparision_MarginalCost}(a) both the model and the U.S. data move temporally in a counter-clockwise direction with the initial direction increasing both mutual constraint and conditional entropy.  The counter-clockwise pattern from model simulations is not specific to parameter choices, but comes generally from its structure and growth from initial conditions.  The marginal cost simulations show higher correlation to the U.S. data than the full cost simulations, and we expand our comparison of those results. 

The U.S. information theory metrics from 1947-2012 align well with those of the marginal cost HARMONEY model from $T=30$ to $T=180$.  For both the model and U.S. data \textit{information entropy} rises, remains nearly constant for a significant time, and finally declines.  An interesting similarity between the U.S. data and HARMONEY simulations is that the peaks in per capita energy consumption (U.S. in the 1970s) and resource extraction (model, near $T=100$ per Figure \ref{fig:1p8_A11A00F11F00}(x)) correspond to the approximate times of maximum information entropy that occur when the phase space plots of Figures \ref{fig:InformationTheory_Comparision_MarginalCost}(a) and \ref{fig:InformationTheory_Comparision_FullCost}(a) are moving up and to the left.  Further, as pointed out by Atlan (1974), the rise and fall of information entropy is a natural consequence of a self-organizing system with potential to grow \cite{Atlan1974}.  He discusses system self-organization in the context of being ``induced by the environment,'' via a pattern of rising and then falling information entropy.  Atlan states the further a system starts from maximum information entropy (for the $2 \times 2$ HARMONEY I-O matrix this is $= -log_2(1/4) = 2$), the more the potential for self-organization. Both the full cost (Figure \ref{fig:InformationTheory_Comparision_FullCost}) and marginal cost (Figure \ref{fig:InformationTheory_Comparision_MarginalCost}) results show that increasing fuel efficiency of capital (via decreasing $\eta_g$ and $\eta_e$) enables a higher peak in information entropy as this represents increased capability to use environmental resources.  


\textit{Conditional entropy} generally rises to a maximum point before falling even more quickly (Figure \ref{fig:InformationTheory_Comparision_MarginalCost}(c)). In both the U.S. and model calculations, maximum conditional entropy occurs soon before maximum energy consumption (1997 for the U.S.) and resource extraction (near $T=145$ in the marginal cost results).

\textit{Mutual constraint} remains relatively constant during the early growth period, declines until the time period of maximum energy and resource extraction rates, and then rises after that point with the HARMONEY simulations exhibiting a gradual leveling off toward a steady state (Figure \ref{fig:InformationTheory_Comparision_MarginalCost}(d)).  The important times of comparison are 2002 in the U.S. data and approximately $T=160$ in the simulations when mutual constraint reaches its lowest value.  In the U.S., the year 2002 represents the year with cheapest ``food+energy'' costs relative to GDP \cite{King2016}.  Further, total U.S. primary energy consumption has not appreciably increased since the year 2000, remaining relatively constant since that time (at between 99 and 107 EJ/year, see Table 1.1 of \cite{EIA_MER2021}).  In the same way, near $T=160$, the marginal cost pricing scenarios show both a peak in resource extraction rates (Figures \ref{fig:1p8_A11A00F11F00}(d) and \ref{fig:1p8_A00A00gA00iF00F00gF00i}(d)) and a low point in energy cost measured as the share of GDP associated with the extraction sector ($Y_{\textmd{extract}}/Y_{\textmd{total}}$ in Figures \ref{fig:1p8_A11A00F11F00}(u) and \ref{fig:1p8_A00A00gA00iF00F00gF00i}(u)).  

One important conclusion is that, for both the HARMONEY model and U.S. I-O tables, the turning points in information theoretic metric trends occur at similar critical times of transition from increasing to stagnant per capita or total resource consumption rates.  That is to say, the HARMONEY model effectively captures several important linkages between the definition and structure of the economy in the context of its ability to extract energy and other natural resources from its environment.

\section{Conclusion}
The purpose of this paper was to explore the coupled growth and structural patterns of the HARMONEY model (v1.1) as updated from King (2020) \cite{King2020}.  The differences in the simulation results in this paper versus King (2020) derive from the more robust method in solving for prices and the explicit inclusion of wage bargaining power that augments a short-run Phillips Curve.  Despite the assumption of a single regenerative natural resource (akin to a forest) to support the modeled economy, HARMONEY 1.1 exhibits several important high-level structural, biophysical, and economic patterns that compare well with global and U.S. data.   

The HARMONEY model provides a consistent biophysical and monetary basis for explaining the progression in global and country-level data from an increasing or near constant energy intensity (energy consumption/GDP) to one of decreasing energy intensity.  That is to say, both HARMONEY and global data first show a period of increasing growth rates, when the growth rate of natural resource consumption exceeds or is nearly equal to the growth rate of GDP, followed by a period of decreasing growth rates when the growth rate of resource consumption is lower than that of GDP.  Thus, given this latter condition referred to as a state of relative decoupling, we conclude that it occurs due to a natural progression of self-organized growth, and not necessarily from independent conscious choice by actors within the economy.  

While we show that explicit choices to increase resource consumption efficiency in capital (e.g., machines) do increase the level of relative decoupling, we also show the choice of price formation affects apparent decoupling just as much.  When basing prices on only marginal costs the economy appears more decoupled than if prices are based on full costs that include depreciation and debt interest payments.  Further, marginal cost pricing generates higher debt ratios than full cost pricing, implying higher debt levels might provide only a perception of a more decoupled economy.  Thus, relative decoupling of GDP from resource consumption represents an expected stage of growth, still similarly dependent on resource consumption, rather than a stage during which an economy is less constrained by resource consumption.

When assuming full labor bargaining power for wages, such that  wages increase with inflation, once resource consumption stagnates, profit shares decline to zero and wage share increases.  An explicit reduction in labor bargaining power at peak resource consumption enables some profits to remain.  Thus, the HARMONEY model provides a basis for arguing that because profits decline to zero once resource consumption peaks under a full bargaining power situation, a new pressure emerges to reduce wage bargaining power of labor to ensure some level of profits at the expense of labor.  This reasoning helps explain the wage stagnation and declining wage share experienced in the U.S. since the 1970s.

\section{Competing Interests and Acknowledgments}
The author declares no competing interests or financial conflicts of interest.

\bibliographystyle{unsrt}
\bibliography{C:/Carey/Research/Writings/BibTeX/CareyBibtex_FromJabRef}

\newpage
\renewcommand{\thefigure}{A.\arabic{figure}} 
\setcounter{figure}{1}  

\renewcommand{\thepage}{A.\arabic{page}} 
\setcounter{page}{1}  

\renewcommand{\thesection}{A.\arabic{section}} 
\setcounter{section}{0}  

\renewcommand{\theequation}{A.\arabic{equation}} 
\setcounter{equation}{0}  

\renewcommand{\thetable}{A.\arabic{table}} 
\setcounter{table}{0}  

\newpage
\section{Nomenclature}\label{app:Nomenclature}

\begin{longtable}{llc}
\caption[]{Definition and nomenclature of symbols used in the HARMONEY model equations.}\label{tab:Nomenclature} \\
\hline 	  
Symbol & Description & Units \\
\hline
$a_{e}$	& Labor productivity of resource extraction sector	& resources$/$person$/$time \\	  
$a_{ee}$	& Resources input required per unit of gross nature extraction	& (resource$/$time)$/$(resource$/$time) \\	
$a_{eg}$	& Resources input required per unit of gross goods production	& (resource$/$time)$/$(good$/$time) \\
$a_g$	& Labor productivity of goods sector	& goods$/$person$/$time \\
$a_{ge}$ & Goods input required per unit of gross resource extraction & (good$/$time)$/$(resource$/$time) \\
$a_{gg}$ & Goods input required per unit of gross goods production & (good$/$time)$/$(good$/$time) \\
$c_e$	& Unit cost of production of extraction sector	& money$/$resource\\
$c_g$	& Unit cost of production of goods sector	& money$/$good\\
$C_e$	& Consumption of extraction sector output by households	& money$/$time\\
$C_g$	& Consumption of goods sector output by households	& money$/$time\\
$CU_{e}, CU_g$	& Capacity utilization of extraction and goods capital	& -- \\
$CU_{e,ref}, CU_{g,ref}$	& Reference (target) Capacity utilization of each capital	& -- \\
$D$	& Total private debt (of firms)	& money \\
${D_e, D_g}$ & Debts of extraction and goods sectors    & money \\
$g$	& Physical inventory of goods	& goods \\
$I$	& Total private investment (by firms)	& money$/$time \\
$I_e, I_g$	& Investment by extraction and goods sector (by firms)	& money$/$time \\
$I^g_e, I^g_g$	& Investment by extraction and goods sector (by firms)	& goods$/$time \\
$IC_{e}, IC_g$	& Inventory coverage of extraction and goods capital	& -- \\
$\textmd{INV}_e, \textmd{INV}_g$	& Value of inventory of extraction and goods sector	& money \\
$\Delta\textmd{INV}_e, \Delta\textmd{INV}_g$	& Change in value of inventory of extraction and goods sector	& money$/$time \\
$K$	& Total capital & goods \\
$K_{e}, K_g$	& Capital for resources extraction and goods production & goods \\
$L$	& Total labor	& person \\
$L_{e}, L_g$	& Labor for resources extraction and goods production & person \\
$M$	& Total deposits (of money) in banks	& money \\
$M^h$	& Household deposits (of money)	& money \\
$N$ & Population 	& person \\
$NEPR$ &	Net external power ratio	& (resource$/$time)/(resource$/$time) \\
NPR$_{economy}$ &	Net power ratio	& (resource$/$time)/(resource$/$time) \\
$P_e$ &	Price of extraction output	& money$/$resource \\
$P_g$ &	Price of goods output	& money$/$good \\
$r_D$, $r_M$	& Interest rate	(D: on private debt; M: for household deposits) & 1$/$time \\
$s$ 	& Threshold household consumption of resources per person &  resource$/$person$/$time \\
$S^b$	& Saving of banks	& money$/$time \\
$S^h$	& Saving of households	& money$/$time \\
$V$	& Total value added of the economy 	& money$/$time \\
$V_e$, $V_g$	& Value added of extraction and goods sectors 	& money$/$time \\
$w_H$	& Physical inventory of resources, wealth (per \cite{Motesharrei2014})	& resource \\
$w$	& Wage per person	& money$/$person$/$time \\
$W_e, W_g$	&  Wages for labor in extraction and goods sector ($W = wL$)	& money$/$time \\
\hline 
\pagebreak
\caption[]{(continued)} \\
\hline 	  
Symbol & Description & Units \\
\hline 	
$X_e$	& Gross output of extraction sector (= resources extraction)	& resource$/$time \\
$X_g$	& Gross output of goods sector (= goods production)	& good$/$time \\
$X^b$	& Net worth of banks	& money \\
$X^h$	& Net worth of households	& money \\
$X^f_e, X^f_g$	& Net worth of extraction and goods firms	& money \\
$X^{tot}$	& Net worth of entire economy	& money \\
$y$	& Resources in the environment	& resource \\
$y_{X_g}$	& Resource input required per unit of goods sector output	& goods$/$ (resource$/$time) \\
$Y$	& Total net economic output of the economy 	& money$/$time \\
$Y_e$, $Y_g$	& Net economic output of extraction and goods sectors	& money$/$time \\
$\alpha_N$	& Death rate of population	& 1$/$time \\
$\beta_N$	& Birth rate of population	& 1$/$time \\
$\gamma$	& Regeneration rate of resources 	& 1$/$time \\
$\delta$	& Depreciation rate of capital 	& 1$/$time \\
$\delta_y$	& Technology parameter affecting resources extraction rate	& 1$/$(time $\times$ good) \\
$\eta_e, \eta_g$	& resources input to operate extraction and goods capital 	& resource$/$(time$\times$good) \\
$\kappa_0$	&  Investment function parameter (Equation \ref{eq:investment})	& -- \\
$\kappa_1$	&  Investment function parameter (Equation \ref{eq:investment})	& -- \\
$\lambda_y$	&  Maximum size of natural resources with no depletion 	& resource \\
$\lambda_N$	&  Participation (employment) rate	& -- \\
$\lambda_{N,o}$	&  Wage function parameter (Equation \ref{eq:PhillipsCurve}), equilibrium employment 	& -- \\
$\mu_e$, $\mu_g$ 	&	Cost markup for extraction and goods sector (price Equations \ref{eq:Price_i} \& \ref{eq:dP_over_P})	& -- \\
$\nu_e$	&	Capital to output ratio for extraction output	& goods$/$ (resource$/$time) \\
$\nu_g$	&	Capital to output ratio for goods output	& goods$/$ (goods$/$time) \\
$\Pi_b$	& Bank dividends	& money$/$time \\
$\Pi_e, \Pi_g$	& Net profit of extraction and goods sectors	& money$/$time \\
$\pi_e, \pi_g$	& Profit share of extraction and goods sectors 	& -- \\
$\pi_{r,e}, \pi_{r,g}$	& Profit rate of extraction and goods sectors 	& 1$/$time \\
$\rho_e$	& Minimum per capita household consumption of extraction sector output 	& resources$/$(person$/$time) \\
$\rho_g$	& Minimum per capita household consumption of goods sector output 	& goods$/$(person$/$time) \\
$\tau_{CU,e}, \tau_{CU,g}$ &	Time constant for extraction and goods $CU$ differential (lag) equation & time \\
$\tau_{IC,e}, \tau_{IC,g}$ &	Time constant for extraction and goods $IC$ differential (lag) equation & time  \\
$\tau_{P,e}, \tau_{P,g}$ &	Time constant for extraction and goods price differential (lag) equation & time  \\
$\tau_{V,e}, \tau_{V,g}$ &	Time constant for extraction and goods value added differential (lag) equation & time  \\
$\tau_{Y,e}, \tau_{Y,g}$ &	Time constant for extraction and goods net output differential (lag) equation & time  \\
$\tau_{\Pi,e}, \tau_{\Pi,g}$ &	Time constant for extraction and goods profit differential (lag) equation & time  \\
$\phi_{min}$	& Wage function parameter (Equation \ref{eq:PhillipsCurve}) 	&  1$/$time \\
$\phi_o$	& Wage function parameter (Equation \ref{eq:PhillipsCurve})  	&  1$/$time \\
$\phi_s$	& Wage function parameter (Equation \ref{eq:PhillipsCurve})  	&  1$/$time \\
$\omega$	& Wage share (of value added) 	&  -- \\
\hline 
\end{longtable}

\newpage
\section{Parameter values used in for simulations}
\begin{table}[h]
\caption{Parameter values used in the base run of the HARMONEY model.}\label{tab:Parameters_BaseRun}
\begin{center}
\begin{tabular}{lll}
\hline 	  
Symbol & Description &   Value \\
\hline
$a_{ge}$	& goods input required per unit of gross resources extraction & 0.2 \\
$a_{gg}$	& goods input required per unit of gross goods production & 0.1 \\
$IC_{ref,g}$	& reference (target) inventory coverage for goods sector & 1 \\
$IC_{ref,e}$	& reference (target) inventory coverage for extraction sector & 1 \\
$r_L$	& Interest rate	(L: on loans) & 0.05 \\
$r_M$	& Interest rate	(M: for household deposits) & 0.0 \\
$s$ 	& threshold household consumption of resources per person (Equation \ref{eq:death_rate}) &  0.08 \\
$y_{X_g}$	& resource input required per unit of goods sector output	&  0.1 \\
$\alpha_m$ & minimum death rate of population (Equation \ref{eq:death_rate}) & 0.01 \\
$\alpha_M$ & maximum death rate of population (Equation \ref{eq:death_rate}) & 0.07 \\
$\beta_N$	& Birth rate of population	& 0.03 \\
$\delta_y$	& Resources extraction factor 	& 0.0072 to 0.009 \\
$\gamma$	& Regeneration rate of resources (renewable scenarios) 	& 0.01 \\
$\delta$	& Depreciation rate of capital 	& 0.03 \\
$\eta_e, \eta_g$	& resources input to operate extraction and goods capital 	& 0.16 \\
$\kappa_0$	&  Investment function parameter (Equation \ref{eq:investment})	& 1.0 \\ 
$\kappa_1$	&  Investment function parameter (Equation \ref{eq:investment})	& 1.5 \\
$\lambda_y$	&  Maximum size natural resources stock (renewable scenarios) 	& 100 \\
$\lambda_{N,o}$	&  wage function parameter (Equation \ref{eq:PhillipsCurve}), equilibrium participation rate 	& 0.6 \\
$\nu_g$	&	Capital to output ratio for goods sector	& 1.5 \\
$\rho_e$	& Minimum household consumption of extraction sector output 	& 0.01 \\
$\rho_g$	& Minimum household consumption of goods sector output 	& 0 \\
$\tau_{C,e}$ &	Time constant for extracted resource consumption differential (lag) equation & 0.05 \\
$\tau_{C,g}$ &	Time constant for goods consumption differential (lag) equation & 0.05 \\
$\tau_{CU,e}$ &	Time constant for extraction capacity utilization differential (lag) equation & 0.25 \\
$\tau_{CU,g}$ &	Time constant for goods capacity utilization  differential (lag) equation & 0.25 \\
$\tau_{IC,e}$ &	Time constant for extraction inventory coverage differential (lag) equation & 0.25 \\
$\tau_{IC,g}$ &	Time constant for goods inventory coverage differential (lag) equation & 0.25 \\
$\tau_{P,e}$ &	Time constant for extraction price differential (lag) equation & 1 \\
$\tau_{P,g}$ &	Time constant for goods price differential (lag) equation & 1 \\
$\tau_{V,e}$ &	Time constant for extraction value added differential (lag) equation & 1 \\
$\tau_{V,g}$ &	Time constant for goods net value added differential (lag) equation & 1 \\
$\tau_{Y,e}$ &	Time constant for extraction net output differential (lag) equation & 1 \\
$\tau_{Y,g}$ &	Time constant for goods net output differential (lag) equation & 1 \\
$\tau_{\Pi,e}$ &	Time constant for extraction profit differential (lag) equation & 1 \\
$\tau_{\Pi,g}$ &	Time constant for goods profit differential (lag) equation & 1 \\
$\phi_{min}$	& wage function parameter (Equation \ref{eq:PhillipsCurve}) 	&  -0.05 \\
$\phi_o$	& wage function parameter (Equation \ref{eq:PhillipsCurve})  	&  0 \\
$\phi_s$	& wage function parameter (Equation \ref{eq:PhillipsCurve})  	&  0.05 \\
\hline
\end{tabular}
\end{center}
\end{table}

\clearpage
\section{List of differential equations for model}\label{app:EquationsList}

\begin{table}[ht]
\caption{The differential equations of the merged biophysical and economic model.}\label{tab:EquationsList}
\begin{center}
\begin{tabular}{lll}
\hline
State Variable & Equation &   Equation No. \\
\hline 	  
\multicolumn{3}{l}{\textbf{I. Core equations}}  \\ 
\hline
Available resources & $\dot{y} = \gamma y(\lambda_y - y) - \delta_y y K_e CU_e$ & 
\ref{eq:nature} \\
Population & $\dot{N} = \beta_N N - \alpha_N\left(\frac{C_e}{P_e}\right) N$ & \ref{eq:population} \\ 
Wealth & $\dot{w}_H = (IC_{ref,e} - IC_{percieved,e}) \times$ & \ref{eq:wealth_dot} \\
 & \hspace{20pt}$(C_e/P_e + a_{eg}X_g + a_{ee}X_e)$ &  \\
Goods & $\dot{g} = (IC_{ref,g} - IC_{percieved,g}) \times$  & \ref{eq:goods_dot} \\
 &  \hspace{20pt}$((C_g+I_g+I_e)/P_g + a_{ge}X_e + a_{gg}X_g)$  &  \\
Extraction capital & $\dot{K}_e = I_e/P_g - \delta K_e$ & \ref{eq:Kdot} \\
Goods capital & $\dot{K}_g = I_g/P_g - \delta K_g$ & \ref{eq:Kdot}\\
Debt for extraction sector & $\dot{D}_{e} = I_e - P_g \delta K_e - \Pi_e$ & \ref{eq:Ddot} \\
Debt for goods sector & $\dot{D}_{g} = I_g - P_g \delta K_g - \Pi_g$ & \ref{eq:Ddot} \\
Wage per capita & $\frac{\dot{w}}{w} = \phi(\lambda_N) + w_1 i + w_2 \frac{1}{\lambda_N}\frac{d \lambda_N}{dt}$ & \ref{eq:wages} \\
Price, extraction sector output & $\dot{P}_{e} = \left(\frac{1}{\tau_{P_e}}\right) \left((1 + \mu_e) c_e - P_e \right)$ & \ref{eq:Price_i} (or \ref{eq:dP_over_P}) \\
Price, goods sector output & $\dot{P}_{g} = \left(\frac{1}{\tau_{P_g}}\right) \left((1 + \mu_g) c_g - P_g \right)$ &   \ref{eq:Price_i} (or \ref{eq:dP_over_P})\\

\hline 
\multicolumn{3}{l}{\textbf{II. Lag equations for accessing `past' states needed to solve core equations}}  \\ 
\hline
Capacity utilization for extraction sector  & $\dot{CU}_{e} = \frac{CU_{e,indicated}-CU_{e}}{\tau_{CU,e}}$ & of form of \ref{eq:lag1} \\
Capacity utilization for goods sector  & $\dot{CU}_{g} = \frac{CU_{g,indicated}-CU_{g}}{\tau_{CU,g}}$ & of form of \ref{eq:lag1} \\
Perceived inventory coverage for extraction sector & $\dot{IC}_{e,perceived} = \frac{IC_{e}-IC_{e,perceived}}{\tau_{IC,e}}$ & of form of \ref{eq:lag1} \\
Perceived inventory coverage for goods sector & $\dot{IC}_{g,perceived} = \frac{IC_{g}-IC_{g,perceived}}{\tau_{IC,g}}$  & of form of \ref{eq:lag1} \\
Value added for extraction sector & $\dot{V}_{e,lagged} = \frac{V_{e}-V_{e,lagged}}{\tau_{V,e}}$ & of form of \ref{eq:lag1} \\
Value added for goods sector & $\dot{V}_{g,lagged} = \frac{V_{g}-V_{g,lagged}}{\tau_{V,g}}$ & of form of \ref{eq:lag1} \\
Profit from extraction sector & $\dot{\Pi}_{e,lagged} = \frac{\Pi_{e}-\Pi_{e,lagged}}{\tau_{\Pi,e}}$ & of form of \ref{eq:lag1} \\
Profit from goods sector & $\dot{\Pi}_{g,lagged} = \frac{\Pi_{g}-\Pi_{g,lagged}}{\tau_{\Pi,g}}$ & of form of \ref{eq:lag1} \\
Value of inventory, extraction sector  & $\dot{INV}_{e,lagged} = \frac{INV_{e}-INV_{e,lagged}}{\tau_{P,e}}$ & of form of \ref{eq:lag1} \\
Value of inventory, goods sector  & $\dot{INV}_{g,lagged} = \frac{INV_{g}-INV_{g,lagged}}{\tau_{P,g}}$ & of form of \ref{eq:lag1} \\
Labor of extraction sector  & $\dot{L}_{e,lagged} = \frac{L_{e}-L_{e,lagged}}{\tau_{P,e}}$ & of form of \ref{eq:lag1} \\
Labor of goods sector  & $\dot{L}_{g,lagged} = \frac{L_{g}-L_{g,lagged}}{\tau_{P,g}}$ & of form of \ref{eq:lag1} \\
\hline 
\hline
\end{tabular}
\end{center}
\end{table}

\newpage
\section{Initial conditions for base run of model}\label{app:InitialConditionsList}

\begin{table}[ht]
\caption{Initial values for state equations as used in the base run of our merged biophysical and economic model.}\label{tab:InitialConditions_BaseRun}
\begin{center}
\begin{tabular}{llll}
\hline
\multicolumn{3}{l}{\textbf{I. Initial conditions for core differential equations}}  \\ 
\hline 	  
Symbol & Description & Full Cost & Marginal Cost \\
 &  &  Pricing & Pricing \\
\hline
$D_{e,o}$ & Debt for extraction sector & 0.001 & 0.001 \\
$D_{g,o}$ & Debt for goods sector & 0.001 & 0.001 \\
$g_o$ & Goods  & 1.531 & 0.596 \\
$K_{e,o}$ & Initial extraction capital & 7.980 & 2.883 \\
$K_{g,o}$ & Initial goods capital & 11.116 & 4.276 \\
$w_o$ & Wage per capita & 1.038 & 0.941 \\
$w_{H,o}$ & Wealth & 1.129 & 0.427 \\
$x_o$ & Population & 18.240 & 6.928 \\ 
$y_o$ & Available resources & 95.117 & 98.235 \\
$P_{e,o}$ & Price of extraction sector output & 2.069 & 1.85 \\
$P_{g,o}$ & Price of goods sector output & 2.178 & 1.963 \\
\hline 
\multicolumn{3}{l}{\textbf{II. Initial conditions for (first order) lag differential equations}}  \\ 
\hline
$CU_{e,o}$  & Capacity utilization for extraction sector & 0.85 & 0.85\\
$CU_{g,o}$  & Capacity utilization for goods sector & 0.85 & 0.85 \\
$IC_{e,perceived,o}$ & Perceived inventory coverage for extraction sector & 1 & 1 \\
$IC_{g,perceived,o}$ & Perceived inventory coverage for goods sector & 1 & 1 \\
$V_{e,o}$ & Value added for extraction sector & 4.727 & 1.683 \\
$V_{g,o}$ & Value added for goods sector & 6.427 & 2.366
 \\
$\Pi_{e,o}$ & Profit from extraction sector & 0.0002 & 0.0002 \\
$\Pi_{g,o}$ & Profit from goods sector & 0.0003 & 0.0002 \\
$INV_{e,o}$ & Value of inventory, extraction sector & 2.067 & 0.699 \\
$INV_{g,o}$ & Value of inventory, goods sector & 2.950 & 1.036 \\
$L_{e,o}$ & Labor of extraction sector & 4.645 & 1.733 \\
$L_{g,o}$ & Labor of goods sector & 6.299 & 2.423 \\
\hline 
\end{tabular}
\end{center}
\end{table}

\newpage
\renewcommand{\thefigure}{S.\arabic{figure}} 
\setcounter{figure}{0}  

\renewcommand{\thepage}{SI.\arabic{page}} 
\setcounter{page}{1}  

\renewcommand{\thesection}{SI.\arabic{section}} 
\setcounter{section}{0}  

\renewcommand{\theequation}{S.\arabic{equation}} 
\setcounter{equation}{0}  

\renewcommand{\thetable}{S.\arabic{table}} 
\setcounter{table}{0}  

\clearpage
\section{Input-Output Format of Model}\label{app:IOmatrix}
\begin{table}[h]
\caption{Input-Output matrix with value added and net output.}
\renewcommand{\arraystretch}{1.5} 
\begin{tabular}{llcc|cccc}
\hline
 & & \multirow{3}{*}{Goods} & \multirow{3}{*}{Extraction} & \multicolumn{3}{c|}{Net Output} & \multirow{3}{*}{Gross Output}\\
 & &  &  & \multirow{2}{*}{Consumption} & \multirow{2}{*}{Investment} &  \multicolumn{1}{c|}{Change in Value} & \\
 & &  &  &  &  &  \multicolumn{1}{c|}{of Inventory} & \\
\hline
 & \multicolumn{1}{l|}{Goods} & $P_g a_{gg} X_g$ & $P_g a_{ge} X_e$ & $C_g$ & $P_g(I^g_g + I^g_e)$ & \multicolumn{1}{c|}{$\Delta\textmd{INV}_g$}  & $P_g X_g$ \\
 & \multicolumn{1}{l|}{Extraction} & $P_e a_{eg} X_g$ & $P_e a_{ee} X_e$ & $C_e$ & --  & \multicolumn{1}{c|}{$\Delta\textmd{INV}_e$}  & $P_e X_e$ \\
\hline
\multirow{4}{*}{\rotatebox[origin=c]{90}{Value Added}} & \multicolumn{1}{l|}{Profit} & $\Pi_g$ & $\Pi_e$ &  &  &  & \\
 & \multicolumn{1}{l|}{Wages}  & $w L_g$ & $w L_e$ &  &  &  & \\
 & \multicolumn{1}{l|}{Interest Payments} & $r_L D_g$ & $r_L D_e$ &  &  &  & \\
 & \multicolumn{1}{l|}{Depreciation} & $P_g \delta K_g$ & $P_g \delta K_e$ &  &  &  & \\\cline{1-4}
 & \multicolumn{1}{l|}{Gross Output} & $P_g X_g$ & $P_e X_e$ &  &  &  & \\\cline{1-4}
\end{tabular}\label{tab:IOmatrix}
\end{table}

\section{Stock and Flow Consistency: Balance Sheet, Transactions, and Flow of Funds Table}\label{app:GodleyTable}

The model is stock-flow consistent in both money and physical units of resources and goods.  Table \ref{tab:GodleyTable} shows the balance sheet, transactions, and flow of funds tables which are the same as in King (2020) \cite{King2020}.   The framework is similar to the Bank, Money, World model of Godley and Lavoie (2007) except we allow firms to have profit \cite{Godley2007}. Household deposits, $M^h$, equal total firm debt. The net worth of firms is the value of their capital minus their debt.  We model banks as having zero saving ($S^b = 0$) where bank net interest ($\Pi_b = r_L D - r_M M$) flows to households as bank dividends. Thus, we assume banks have zero net worth ($X^b = 0$).

\clearpage
\begin{landscape} 
\begin{table}[h]
\scriptsize
\caption{\textit{Productive sectors}: Balance sheet, transactions, and flow of funds for \textit{productive sectors} as formulated in the macroeconomic model where each of the productive sectors make individual investment decisions and profits.}\label{tab:GodleyTable_ProdSectors}
\resizebox{1.0\textwidth}{!}{\begin{tabularx}{1.0\textwidth}{lcccc}
\toprule[1pt]\midrule[0.3pt]
& \multicolumn{2}{c}{Extraction Firms} & \multicolumn{2}{c}{Goods Firms}  \\
\hline
\textbf{Balance Sheet} & & & & \\[0.5em]
Capital & \multicolumn{2}{c}{$P_gK_{e}$} & \multicolumn{2}{c}{$P_gK_g$} \\[0.5em]
Deposits & & & & \\[0.5em]
Debt (Loans) & \multicolumn{2}{c}{$-D_{e}$} & \multicolumn{2}{c}{$-D_{g}$} \\[0.5em]
\midrule[0.3pt]
Sum (net worth) & \multicolumn{2}{c}{$X^f_{e}$} & \multicolumn{2}{c}{$X^f_{g}$} \\[0.5em]
& & & & \\
\midrule[0.3pt]\midrule[0.3pt]
\textbf{Transactions} &  Current & Capital & Current & Capital \\[0.5em]
Consumption & $C_{e}$ & & $C_{g}$ & \\[0.5em]
Investment &  & $-P_g I^e_g$ & $P_g I^e_{g} + P_gI^g_g$ & $-P_gI^g_{g}$ \\[0.5em]
Change in value of inventory  & $\Delta INV_e$ & $-\Delta INV_e$ & $\Delta INV_g$ & $-\Delta INV_g$ \\[0.5em]
Intermediate Sales  & $+P_e a_{eg}X_g + P_e a_{ee}X_e$ & & $+P_g a_{ge}X_e + P_g a_{gg}X_g$ & \\[0.5em]
Intermediate Purchases  & $-P_ga_{ge}X_e - P_e a_{ee}X_e$ & & $-P_ea_{eg}X_g - P_ga_{gg}X_g$ & \\[0.5em]
[Value Added] & $[V_{e}]$ & & $[V_{g}]$ & \\[0.5em]
Wages & $-W_{e}$ & & $-W_{g}$ & \\[0.5em]
Depreciation Allowance & $-P_g\delta K_{e}$ & $P_g\delta K_{e}$ & $-P_g\delta K_{g}$ & $P_g\delta K_{g}$ \\[0.5em]
Interest on debt (loans) & $-r_L D_{e}$ & & $-r_LD_{g}$ &  \\[0.5em]
Interest on deposits & & & & \\[0.5em]
Bank Dividends (net interest) & & & & \\[0.3em]
\midrule[0.3pt]
Financial Balances & $\Pi_{e}$ & $P_g(\delta K_{e} - I^e_{g}) - \Delta INV_e$ & $\Pi_{g}$ & $P_g(\delta K_g - I^g_g) - \Delta INV_g$ \\
\midrule[0.3pt]\midrule[0.3pt]
\textbf{Flow of Funds} & & & &  \\[0.5em]
Change in capital stock & & $P_g\dot{K}_e$ & & $P_g\dot{K}_g$ \\[0.5em]
Gross Fixed Capital Formation & & $P_gI^e_{g}$ & & $P_gI^g_{g}$  \\[0.5em]
Change in deposits & & & & \\[0.5em]
Change in debt (loans) & & $-\dot{D}_{e}$ & & $-\dot{D}_{g}$ \\
\midrule[0.3pt]
Column sum & & $\Pi_{e}$ & & $\Pi_{g}$  \\[0.5em]
Change in net worth & & $\dot{X}^f_{e} = \Pi_{e}$ & & $\dot{X}^f_{g} = \Pi_{g}$ \\
\bottomrule[1pt]
\end{tabularx}}
\end{table}
\end{landscape}

\clearpage
\begin{landscape} 
\begin{table}[h]\ContinuedFloat
\scriptsize
\caption{(continued) \textit{Households, Banks, and Sum}: Balance sheet, transactions, and flow of funds tables.}\label{tab:GodleyTable}
\resizebox{1.0\textwidth}{!}{\begin{tabularx}{1.0\textwidth}{lccc}
\toprule[1pt]\midrule[0.3pt]
& Households & Banks & Sum \\
\hline
\textbf{Balance Sheet} & & & \\[0.5em]
Capital & & & $P_g (K_{e}+K_{g})$ \\[0.5em]
Deposits & $M^h$ & $-M$ & \\[0.5em]
Debt (Loans) & & D & \\[0.5em]
\midrule[0.3pt]
Sum (net worth) & $X^h$ & $X^b=0$ & $X^{tot}=P_g (K_{e}+K_{g})$ \\[0.5em]
& & & \\
\midrule[0.3pt]\midrule[0.3pt]
\textbf{Transactions} & & & \\[0.5em]
Consumption & $-C$ & & \\[0.5em]
Investment & & & \\[0.5em]
Change in value of inventory  & & & \\[0.5em]
Intermediate Sales  & & & \\[0.5em]
Intermediate Purchases & & & \\[0.5em]
[Value Added] & & & \\[0.5em]
Wages & $W$ & & \\[0.5em]
Depreciation Allowance & & & \\[0.5em]
Interest on debt (loans) & & & \\[0.5em]
Interest on deposits & $r_M M^h$ & $-r_M M$ & \\[0.5em]
Bank Dividends (net interest) & $\Pi_b$ & $-\Pi_b$ & \\[0.3em]
\midrule[0.3pt]
Financial Balances & $S^h$ & $S^b=0$ & \\
\midrule[0.3pt]\midrule[0.3pt]
\textbf{Flow of Funds} & & & \\[0.5em]
Change in capital stock & & &  \\[0.5em]
Gross Fixed Capital Formation & & &  \\[0.5em]
Change in deposits & $\dot{M}^h$ & $\dot{M}$ & \\[0.5em]
Change in debt (loans) & & $\dot{D}$ & \\
\midrule[0.3pt]
Column sum & $S^h$ & $S^b$ 8&  \\[0.5em]
Change in net worth & $\dot{X}^h=S^h$ & $\dot{X}^b=S^b=0$ &  \\
\bottomrule[1pt]
\end{tabularx}}
\end{table}
\end{landscape}

\section{Additional Equations not in Methods}\label{app:AdditionalEquations}

\subsection{Consumer Price Index (CPI) and GDP deflator}

We calculate inflation as a weighted change in the price of each sector output consumed by households using Equation \ref{eq:inflation}. The consumer price index (CPI) is as in Equation \ref{eq:CPI}.  


\begin{equation}\label{eq:CPI}
CPI = \prod^{T}_{t=1} \left( 1 + i_t \right)
\end{equation}

Assume defining the GDP deflator as nominal GDP divided by real GDP (where $Y_{i,t}$ is the net monetary output of sector $i$ at time $t$).  $P_{i,o}$ is the initial price of output from sector $i$. Real net output, value added, investment, debt, and consumption are calculated by dividing nominal values by the GDP deflator (e.g., $Y_{g,real} = Y_{g,t}/\textmd{GDP deflator}_t$).

\begin{eqnarray}\label{eq:GDP_deflator}
\textmd{GDP deflator}_t & =&  \frac{P_{g,t} \left( \frac{Y_{g,t}}{P_{g,t}} \right) + P_{e,t} \left( \frac{Y_{e,t}}{P_{e,t}} \right)}{P_{g,o} \left( \frac{Y_{g,t}}{P_{g,t}} \right) + P_{e,o} \left( \frac{Y_{e,t}}{P_{e,t}} \right)} \nonumber \\
& = & \frac{Y_{g,t}  + Y_{e,t}}{P_{g,o} \left( \frac{Y_{g,t}}{P_{g,t}} \right) + P_{e,o} \left( \frac{Y_{e,t}}{P_{e,t}} \right)} 
\end{eqnarray}

\subsection{Death Rate}\label{app:DeathRate}
Equation \ref{eq:death_rate} describes the function for death rates, where $s$ is the per capita resource consumption threshold below which death rates rise.  As physical per capita resource consumption, $\frac{C_e}{N P_e}$, declines below the threshold to zero the death rate linearly, the death rate increases from a minimum value of $\alpha_m$ to a maximum ``famine'' death rate of $\alpha_M$ as in \cite{Motesharrei2014}. 

\begin{equation}\label{eq:death_rate}
\alpha_N\left(\frac{C_e}{N P_e}\right) = \alpha_m + \max\left(0,1-\frac{\left(\frac{C_e}{N P_e}\right)}{s}\right)(\alpha_M - \alpha_m)
\end{equation}

\subsection{Inventory Coverage and Capacity Utilization}\label{app:IC_and_CU}

Perceived inventory coverage for each sector:

\begin{eqnarray}
IC_{e,perceived} = \frac{\frac{\textmd{perceived wealth, } w_H}{\textmd{time delay}}}{\textmd{targeted consumption of resources}} \nonumber \\
IC_{e,perceived} = \frac{\frac{w_H}{\tau_{IC,e}}}{C_e/P_e + a_{eg}X_g + a_{ee}X_e} 
\end{eqnarray}\label{eq:ICext_targeted}

\begin{eqnarray}
IC_{g,perceived} = \frac{\frac{\textmd{perceived goods, } g}{\textmd{time delay}}}{\textmd{targeted consumption of goods}} \nonumber \\
IC_{g,perceived} = \frac{\frac{g}{\tau_{IC,g}}}{(C_g + I_e + I_g)/P_g + a_{gg}X_g + a_{ge}X_e} 
\end{eqnarray}\label{eq:ICg_targeted}

Perceived capacity utilization of each sector, $0 \leq CU_{i,perceived} \leq 1$, is a lookup table that is an increasing function of the inverse of its respective inventory coverage \cite{Sterman2000}.   When there is more inventory, capacity utilization decreases, and vice versa.  The reference inventory coverage, $IC_{ref,i}$, is defined in the lookup table for capacity utilization as the amount of inventory present for capacity utilization to be at its reference value, $CU_{ref,i}$. We set $CU_{ref,i}=0.85$ at $IC_{ref,i}=1$. 

\begin{equation}\label{eq:CUgeneric_perceived}
CU_{i,perceived} = f(IC_{i,perceived}^{-1})
\end{equation}

The lookup table for $CU$ of both sectors use input values as \\
$IC_{i,perceived}^{-1} = [0, 0.25, 0.50, 0.75, 1.00, 1.25, 1.50, 1.75, 2.00, 2.25, 1e6]$ and output values as \\
$CU_{i,perceived} = [0, 0.30, 0.55, 0.75, 0.85, 0.90, 0.94, 0.98, 0.99, 1, 1]$.

\subsection{Phillips Curve (Wage Function)}\label{app:WageFunction}

For $\phi(\lambda_N)$ we use Keen's nonlinear exponential curve (Equation \ref{eq:PhillipsCurve}) that allows wages to rise increasingly rapidly at high participation but decrease slowly at low participation rate \cite{Keen2013}. In Equation \ref{eq:PhillipsCurve} $\phi_{min}<0$ is the minimum decline in wages at low participation rate, $\phi_o$ is the change in current wage defined at $\lambda_{N,o}$ (typically set $\phi_o = 0$ at an equilibrium participation rate $\lambda_{N,o}$), and $\phi_s$ defines the exponential rate of increase.  

\begin{equation}\label{eq:PhillipsCurve}
\phi(\lambda_N) = (\phi_o - \phi_{min})e^{\frac{\phi_s}{(\phi_o - \phi_{min})}(\lambda_N - \lambda_{N,o})} + \phi_{min}
\end{equation}

\subsection{Solving for Lagged Variables}\label{app:LaggedStates}

For each sector $i$, the variables modeled using a first order lag are the capacity utilization ($CU_i$), perceived inventory coverage ($IC_{i,perceived}$), price ($P_i$), net output ($Y_i$), profit ($\Pi_i$), value added ($V_i$), and value of inventory ($\textmd{INV}_i$).  Thus, each of $CU_i, IC_{i,perceived}, P_i, Y_i, \Pi_i, V_i$, and $\textmd{INV}_i$ is modeled as an extra model state.  In the model code, these lagged states are used to inform investment and all inputs needed to solve for a new ``current'' price. Additional model calculations use this newly calculated price, including calculations that updated the lagged states themselves. 
The states are updated via Equation \ref{eq:lag1}.  For the inventory coverage of each sector ($IC_i$) it is the perceived inventory coverage ($IC_{i,percieved}$) that is modeled as a lagged state on the left hand side.  

\begin{equation}\label{eq:lag1}
\dot{n}_{lagged} = (n - n_{lagged})/\tau
\end{equation}

\subsection{Net Power Accounting}\label{app:NetPowerAccounting}
Considering the flow of extracted and consumed resources as a flow of ``power'' (energy per time), we summarize the power return ratios (PRR) used in King (2020) \cite{King2020,King2015_part1}. PRRs characterize the power generated by the energy (i.e., resources) sectors relative to the power consumed by the energy sectors themselves. This self-consumption can include both operating inputs and energy embodied in investment.  Researchers have speculated on minimum levels of PRRs ``required'' to sustain society \cite{Halletal2009}, and calculation of PRRs within our model allows internally-consistent investigation of their relation to economic growth, structure, and population.  The net external power ratio (NEPR) is defined in Equation \ref{eq:NEPR} as net resource extraction divided by the extraction sectors' own use of resources.  

NEPR represents what is often termed ``energy return on energy invested'' (EROI) in much of the literature \cite{King2015_part1}.   Since our PRR calculation uses instantaneous resource flow rates in the numerator and denominator, the term \emph{power} return ratio is more appropriate than energy return ratio (ERR) (e.g., energy is power integrated over time).  However, much of the net energy literature uses the terminology EROI and ERR to refer both to ratios of energy and ratios of power.

We calculate the resources embodied in extraction capital, $K_e$, via a resource intensity, $\epsilon$, that measures gross resources extraction required per net physical output of each sector (see \cite{King2014}, \cite{BullardandHerendeen1975}, and \cite{Casler1984} for the methodology for calculating ``energy intensities'').  In Equation \ref{eq:nature_intensity}, $\hat{y}_{\textmd{extract}}$ is a $2 \times 2$ diagonal matrix with non-zero elements \textit{only} for the gross extraction by sectors that extract resources from the environment. For our model $\hat{y}_{\textmd{extract},ee}=\hat{y}_{\textmd{extract},22}=X_{e}$, $\hat{y}_{\textmd{extract},gg}=0$, $\hat{X}$ is a diagonal matrix of the gross physical output of each sector ($X_g$ and $X_e$), $\mathbf{1}$ is the identity matrix, \textbf{A} the technical coefficients matrix, and matrix $E$ is a $2 \times 2$ matrix of resource intensities, $\epsilon_{ij}$.  Further, the first row of $E$ is zero, and the second row contains the resource intensities of $\epsilon_{eg}$ representing the gross resource input per unit of net physical goods output, $\frac{Y_g}{P_g}$, and $\epsilon_{ee}$ representing the gross resource input per unit of net physical resources output, $\frac{Y_e}{P_e}$.  Thus, the embodied resources in extraction capital each time step is equal to $\epsilon_{eg}\frac{I_e}{P_g}$.  Recall that $X_e = \delta_y y K_{e}CU_{e}$ as well as that $\frac{I_e}{P_g}$ has units of goods and represents the physical goods allocated to become new extraction capital. 

\begin{equation}\label{eq:nature_intensity}
\begin{bmatrix}
0 & 0 \\
\epsilon_{eg} & \epsilon_{ee}
\end{bmatrix} = E = \hat{y}_{\textmd{extract}} \hat{X}^{-1} (\mathbf{1} - A)^{-1}  \\
\end{equation}

\begin{eqnarray}\label{eq:NEPR}
NEPR & = & \frac{\textmd{resource extraction} - \textmd{resources required to invest in $K_e$}-\textmd{resources required to operate $K_e$}}{\textmd{resources required to invest in $K_e$}+\textmd{resources required to operate $K_e$}} \nonumber \\
& = & \frac{X_{e} - \epsilon_{eg}\frac{I_e}{P_g} - a_{ee}X_{e}}{\epsilon_{eg}\frac{I_e}{P_g} + a_{ee}X_{e}}
\end{eqnarray}

Per King \textit{et al.} (2015) we can consider an economy level net power ratio (NPR$_\textmd{economy}$) as similar to but distinct from the sector-specific NEPR \cite{King2015_part1}. With only one extraction (``energy'') sector, NPR$_\textmd{economy}$ is calculated using only $\epsilon_{ee}$ in Equation \ref{eq:NPR}.  Economy-wide gross power ratio (GPR$_\textmd{economy}$) equals one plus NPR$_\textmd{economy}$ \cite{King2015_part2}, and it has been referred to as the EROI of the economy \cite{Herendeen2015}.  The upper limit for NPR$_\textmd{economy}$ is defined using $a_{ee}$ as NPR$_\textmd{economy, uppper limit} = \frac{1 - a_{ee}}{a_{ee}}$, where $1 - a_{ee}$ is the fraction of extracted resources left for all other economic activity after operating extraction capital \cite{King2014}.  As such, NPR$_\textmd{economy, uppper limit}$ is largely defined by the resource efficiency of extraction sector capital.

\begin{eqnarray}\label{eq:NPR}
NPR_{\textmd{economy}}  & = & \frac{1}{\frac{\textmd{gross power extracted}}{\textmd{net power output}} - 1} = \frac{1}{\epsilon_{ee} - 1}
\end{eqnarray}

\subsection{Decreasing Resource Consumption to Operate Capital ($\eta_i$) as Function of Cumulative Investment}\label{app:DecreasingEta}
In some simulations we explore the concept an increase in resource consumption efficiency of capital. We do this by decreasing $\eta_i$ as a function of physical capital investment in sector $i$ as in Equation \ref{eq:eta_decrease}.   The equation approximates a logistic decrease in $\eta_i$ as a function of cumulative capital investment, $I_{i,cumulative}$ (Equation \ref{eq:I_cumulative}).  

When simulating, we choose a time ``$T_{critical}$'' at which to start the process of decreasing $\eta_i$ and calculating $I_{i,cumulative}$.   In the paper this critical time is $T = 0.1$, or the first time step in the simulation.  Before this time, we assume $I_{i,cumulative}=0$ such that there is no investment that yet contributes to ``learning by doing'' that would decrease $\eta_i$.  After this time, $I_{i,cumulative}$ is the integral of gross physical capital investment (without depreciation).

The parameter $\eta_{i,adder}$ (Equation \ref{eq:eta_adder}) adjusts adjusts the baseline level.  Parameter $S$ adjusts the steepness of the curve.  $I_{i,mid}$ sets the approximate mid-point, or inflection point, of the logistic curve.  The maximum and minimum values of $\eta_{i}$ are $\eta_{i,max}$ and $\eta_{i,min}$, respectively, where the starting value is by definition $\eta_{i,max}$.

\begin{equation}\label{eq:eta_decrease}
\eta_{i} = \eta_{i,adder} + \frac{\eta_{i,min} - \eta_{i,max}}{1 + e^{S(I_{i,cumulative} - 2I_{i,mid})}} + \eta_{i,max} 
\end{equation}

\begin{equation}\label{eq:eta_adder}
\eta_{i,adder} = \eta_{i,max} - \left( \frac{\eta_{i,min} - \eta_{i,max}}{1 + e^{S(I_{i,cumulative} - 2I_{i,mid})}} + \eta_{i,max} \right)
\end{equation}



%
%


\begin{eqnarray}\label{eq:I_cumulative}
\dot{I}_{i,cumulative} & = & 0, \text{   } (T < T_{critical}) \nonumber \\
\dot{I}_{i,cumulative} & = & \frac{(1 - fP_{I_i}) I_{i}}{P_g},  \text{   } (T \geq T_{critical})
\end{eqnarray}

\clearpage
\section{Additional Simulation Results}\label{app:AdditionalResults}

\subsection{Comparing of results when increasing $\delta_y$ versus increasing $\lambda_y$}\label{app:delta_y_vs_lambda_y}

\begin{figure}[h]
\begin{center}
\subfloat[]{\includegraphics[width=0.33\columnwidth]{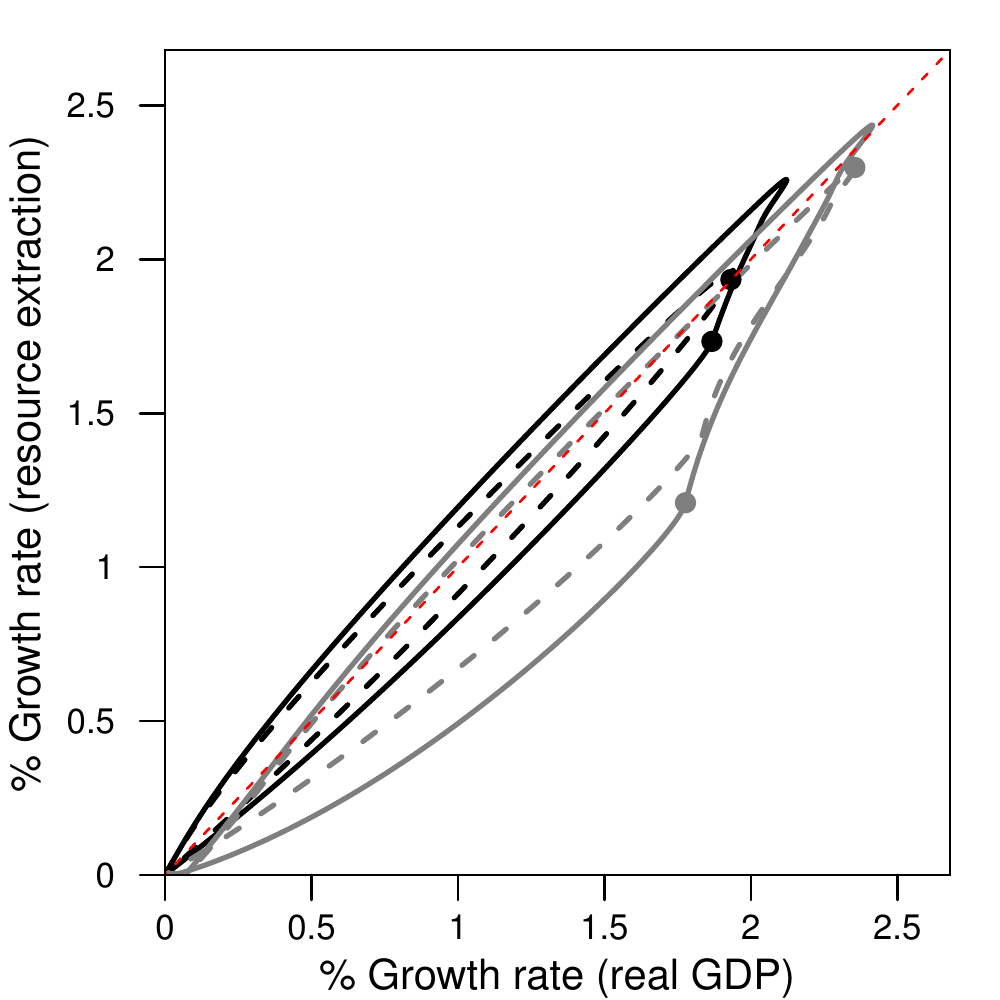}}
\subfloat[]{\includegraphics[width=0.33\columnwidth]{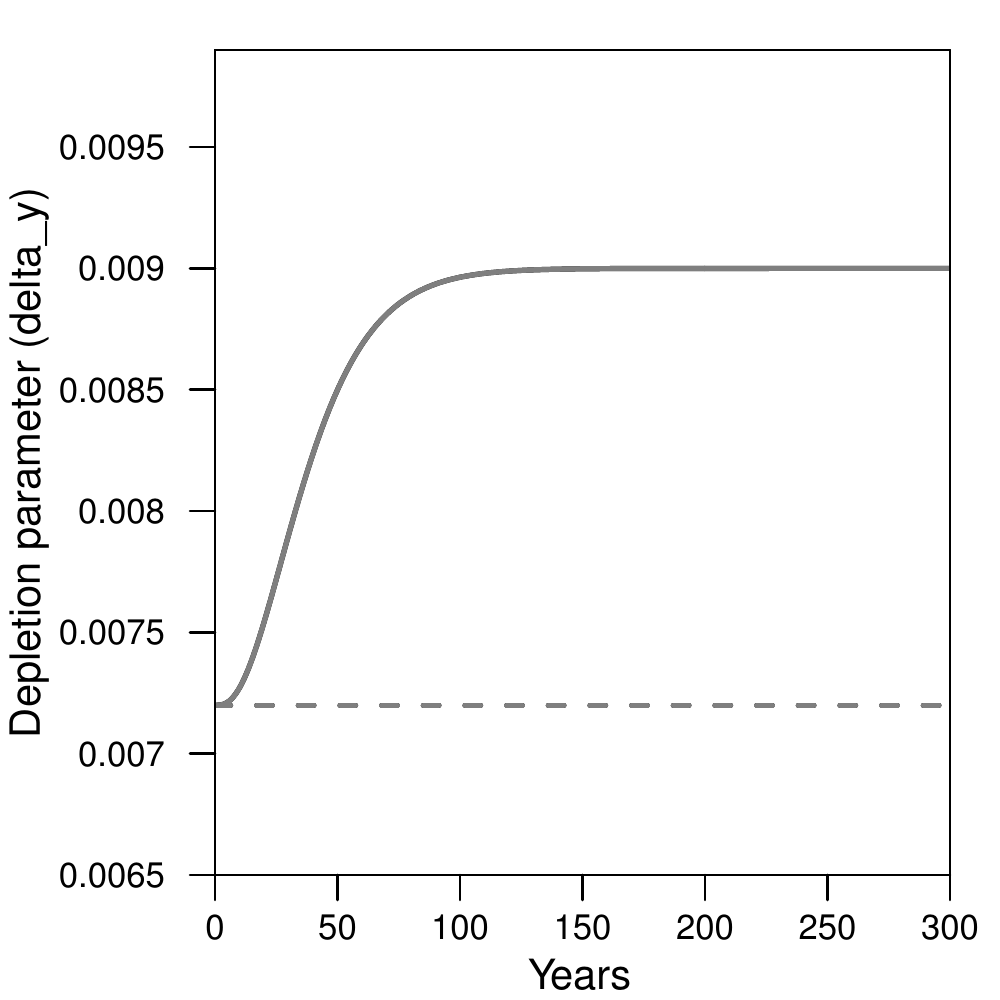}}
\subfloat[]{\includegraphics[width=0.33\columnwidth]{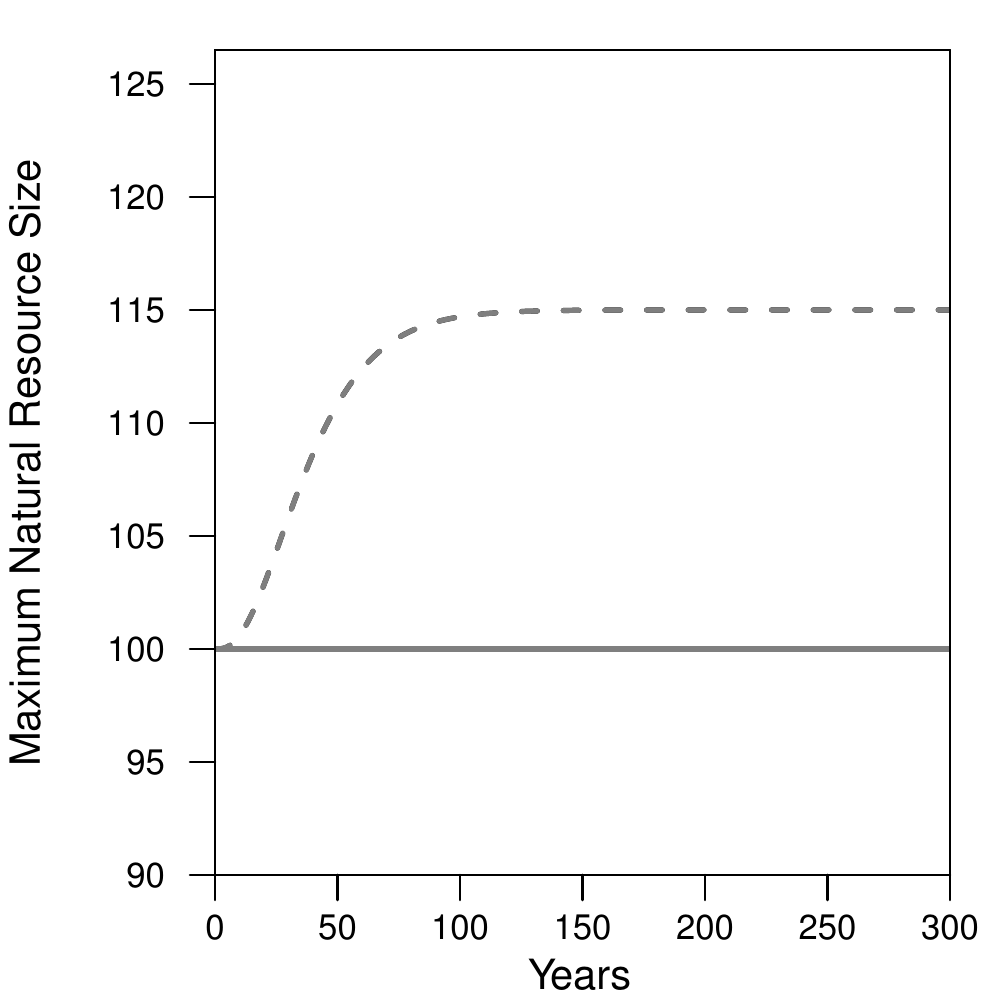}}\\
\caption{(a) HARMONEY simulation results for  FC-000 (black solid), similar to ``FC-000 alternative'' (black dashed), FC-100 (gray solid), and ``FC-100 alternative'' (gray dashed) scenarios. (b)  FC-000 and FC-100 spur growth by increasing $\delta_y$ from 0.0072 to 0.009. (c) The ``alternative'' scenarios spur growth by increasing the maximum resource size $\lambda_{y,max}$ from 100 to 115.}\label{fig:delta_y_vs_lambda_y}
\end{center}
\end{figure}

Figure \ref{fig:delta_y_vs_lambda_y} shows that by only looking at data for growth rates for resource (or energy) consumption and GDP, it is difficult to impossible to explain the difference between growth induced via increased technological capability (increasing $\delta_y$ in scenarios FC-000 and FC-100, Figure \ref{fig:delta_y_vs_lambda_y}(b)) to access a constant maximum resource size versus the ability to access a larger resource at constant technology (increasing $\lambda_{y,max}$ in scenarios ``FC-000 alternative'' and ``FC-100 alternative'', Figure \ref{fig:delta_y_vs_lambda_y}(c)).  The general pattern in the growth rates for resource extraction and GDP follow the same counter-clockwise pattern and change from superlinear (above 1:1 line) to sublinear (below 1:1 line) scaling.  Real world data most assuredly exhibit influence from both drivers of accessing more resources and the ability to access known resources more fully.

\newpage
\subsection{Variations on Full Cost Scenarios with Lower Wage Bargaining and Ponzi}\label{app:ScenariosFulLCost_LowerBargainWithPonzi}

\begin{figure}
\begin{center}
\subfloat[]{\includegraphics[width=0.33\columnwidth]{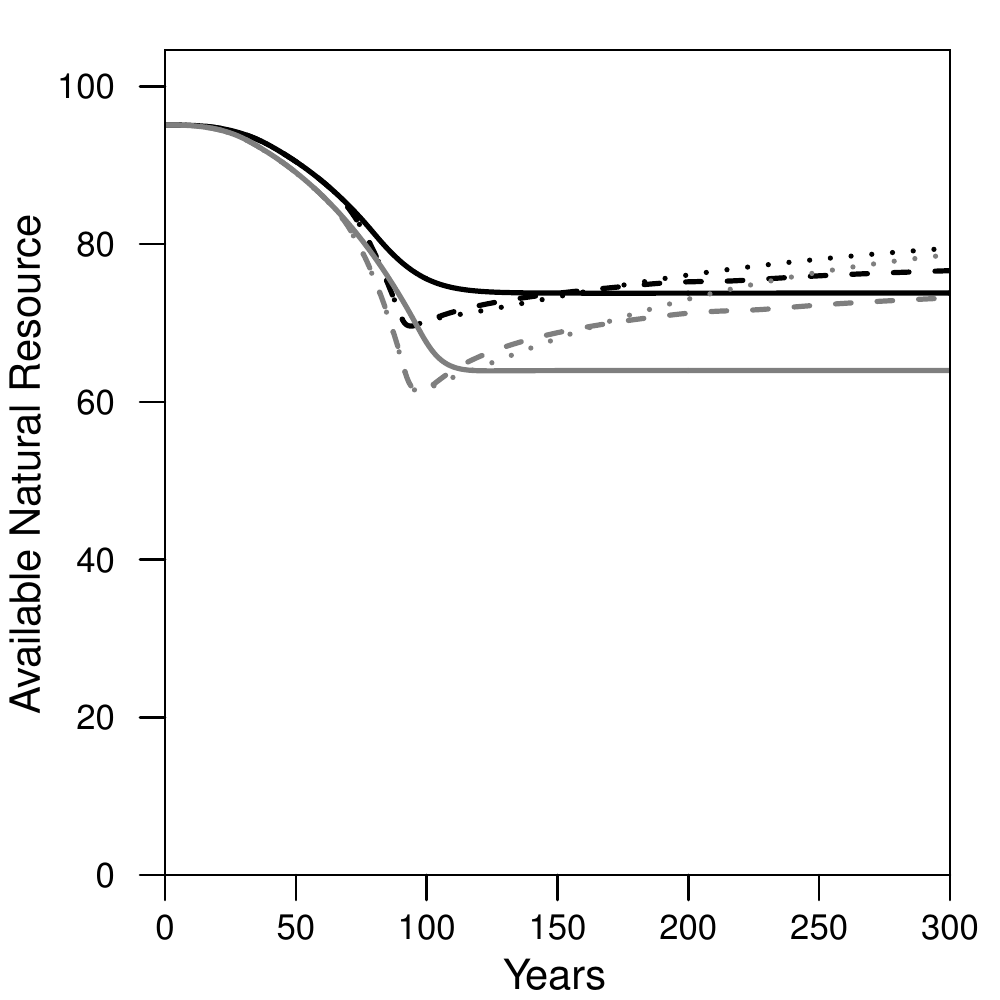}}
\subfloat[]{\includegraphics[width=0.33\columnwidth]{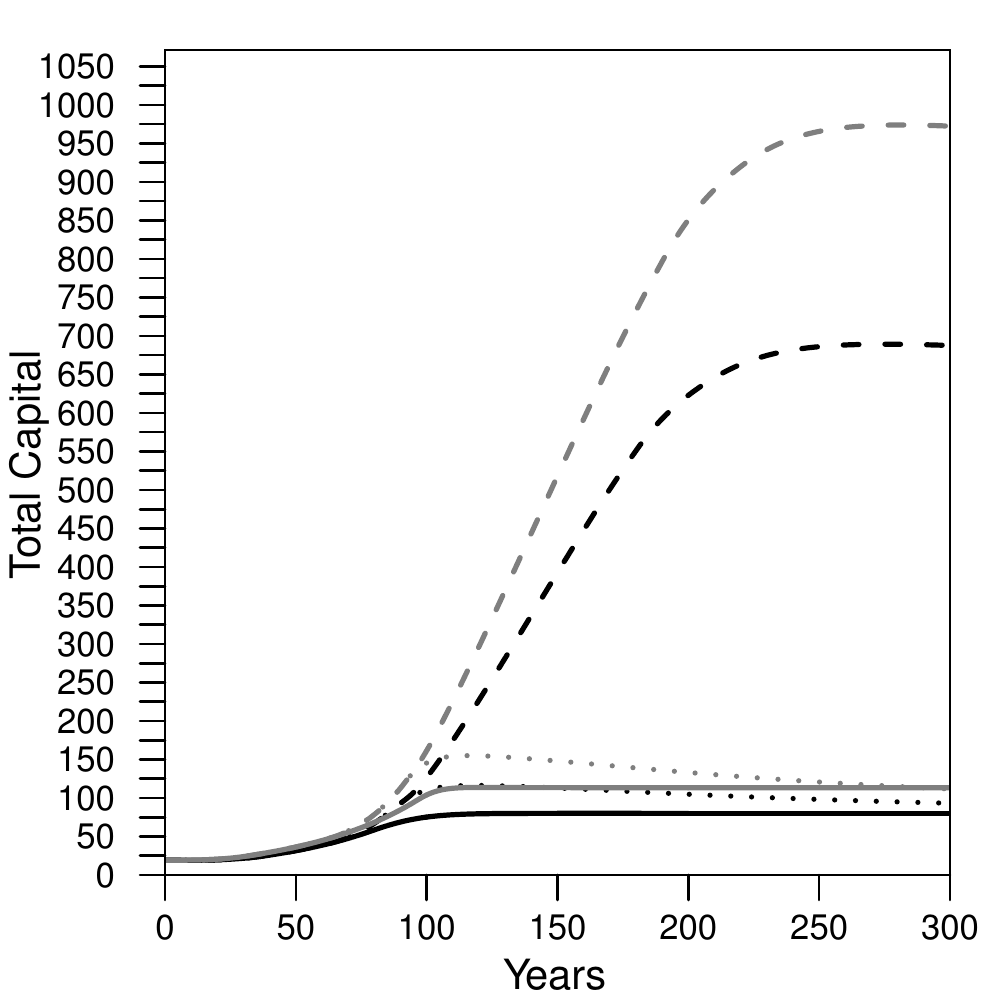}}
\subfloat[]{\includegraphics[width=0.33\columnwidth]{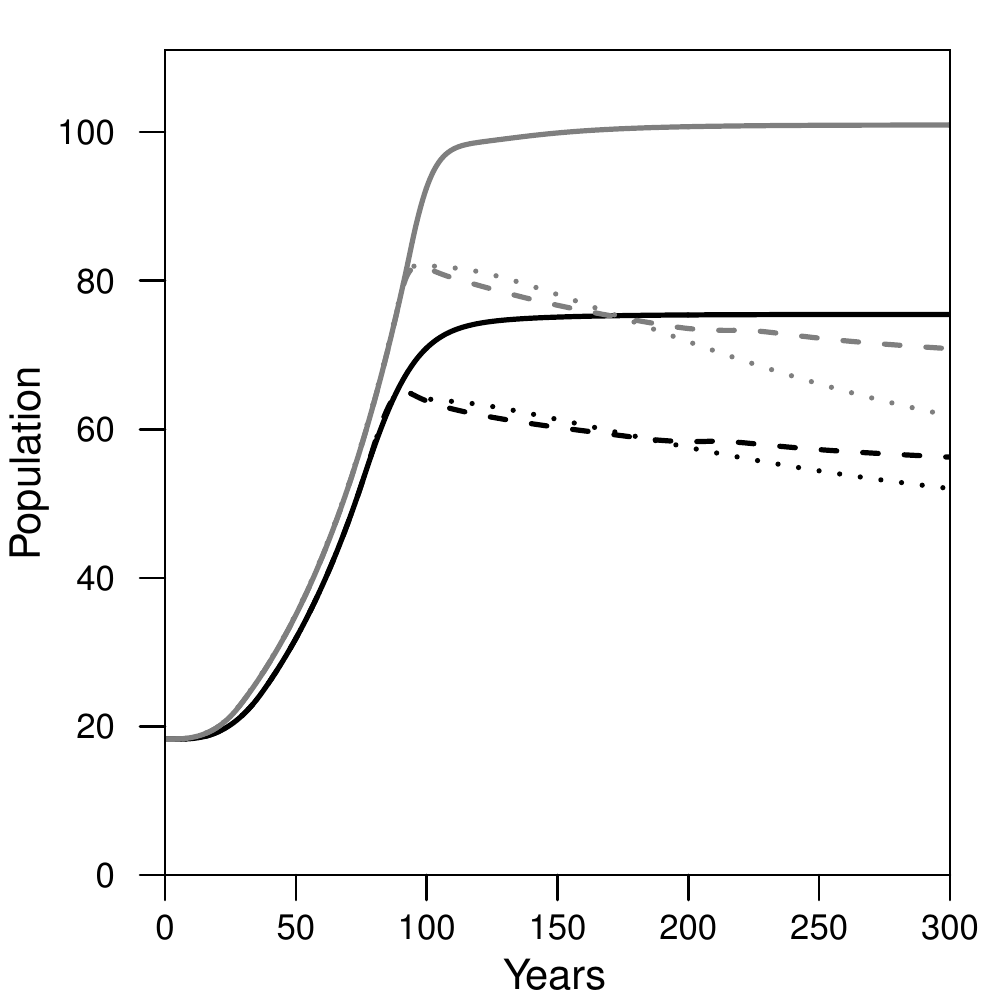}}\\
\subfloat[]{\includegraphics[width=0.33\columnwidth]{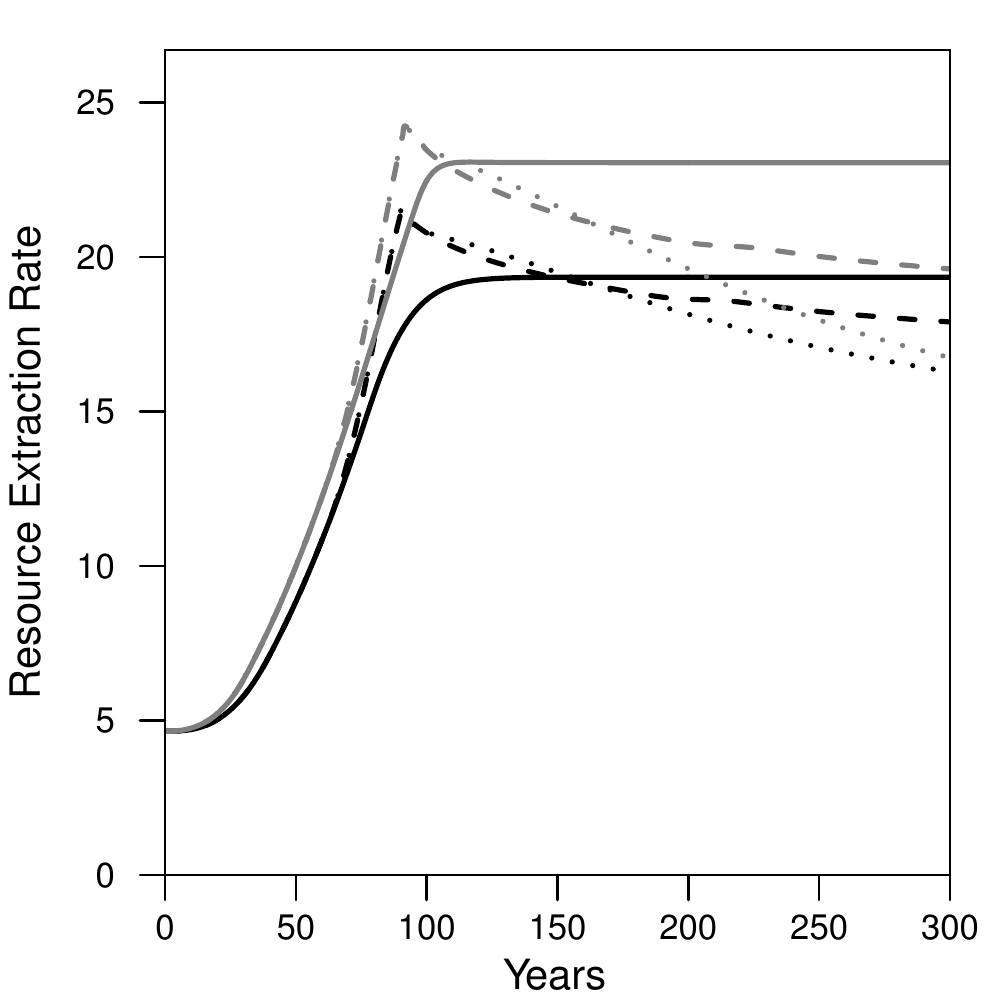}}
\subfloat[]{\includegraphics[width=0.33\columnwidth]{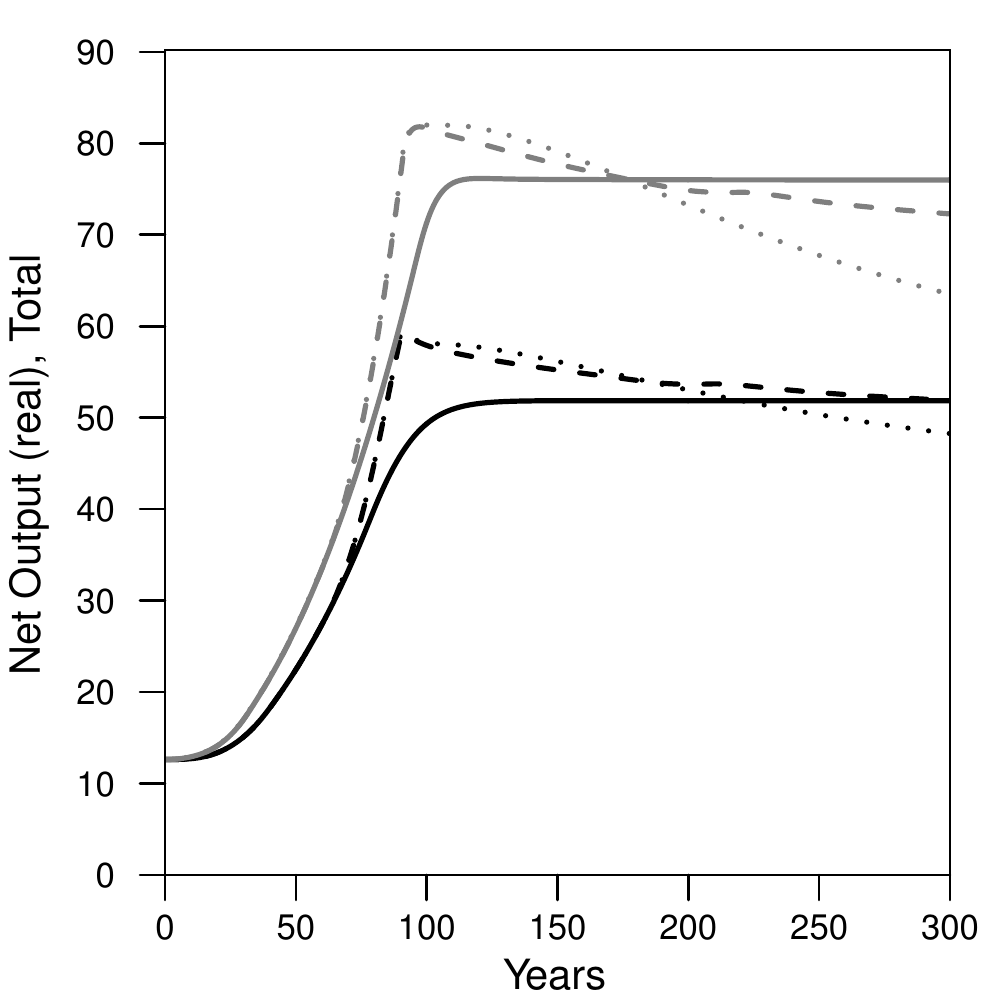}}
\subfloat[]{\includegraphics[width=0.33\columnwidth]{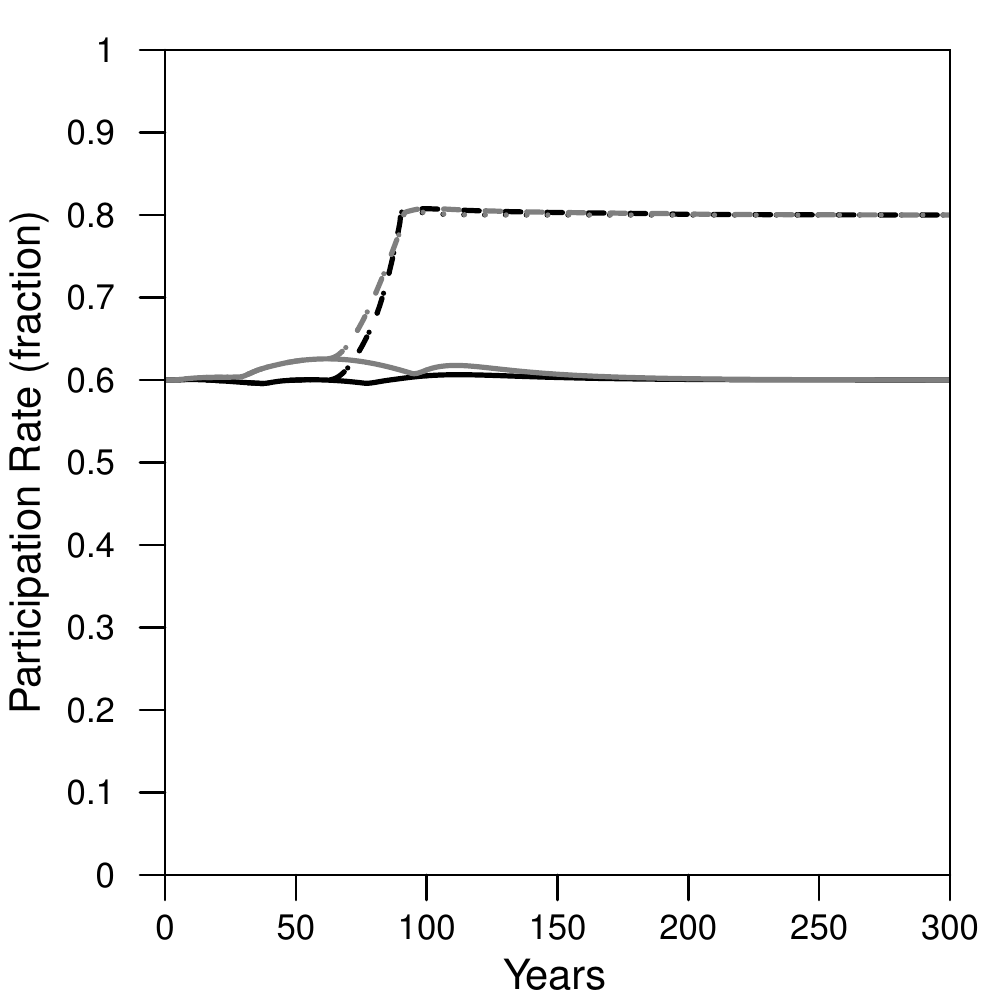}}\\
\subfloat[]{\includegraphics[width=0.33\columnwidth]{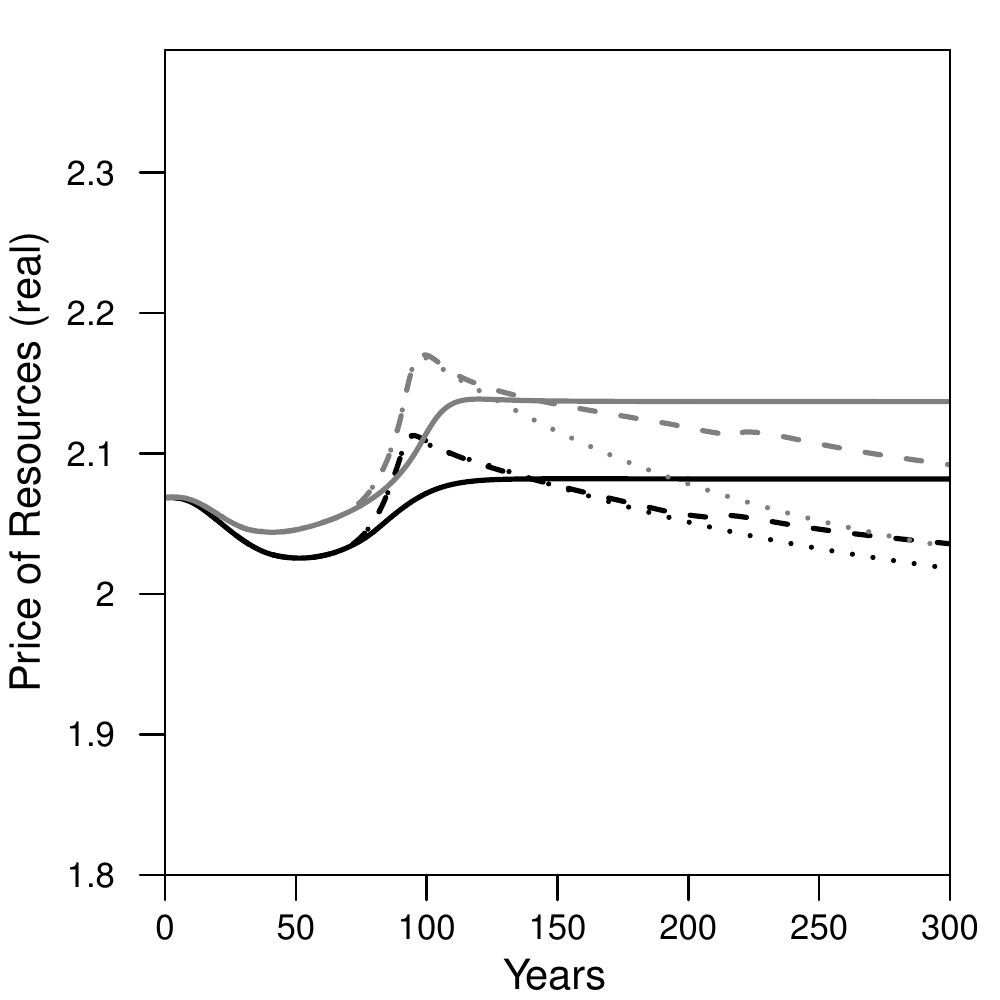}}
\subfloat[]{\includegraphics[width=0.33\columnwidth]{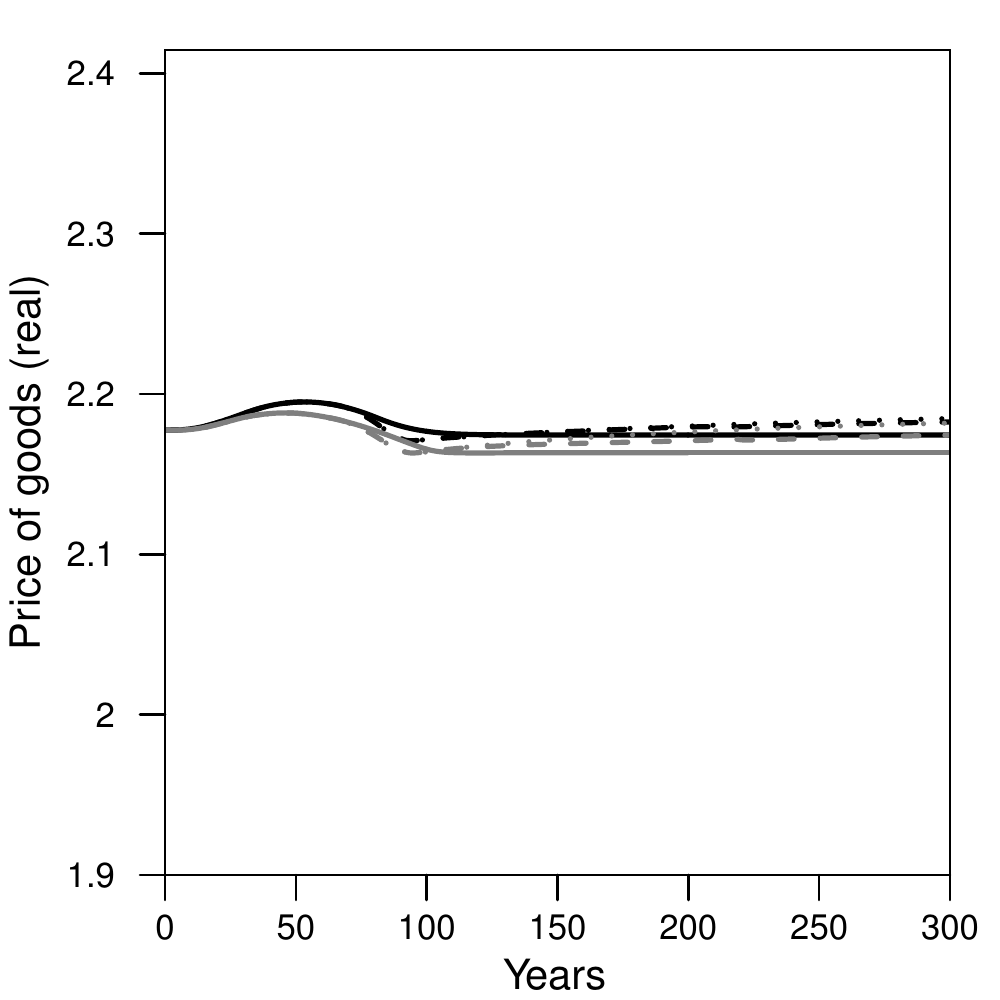}}
\subfloat[]{\includegraphics[width=0.33\columnwidth]{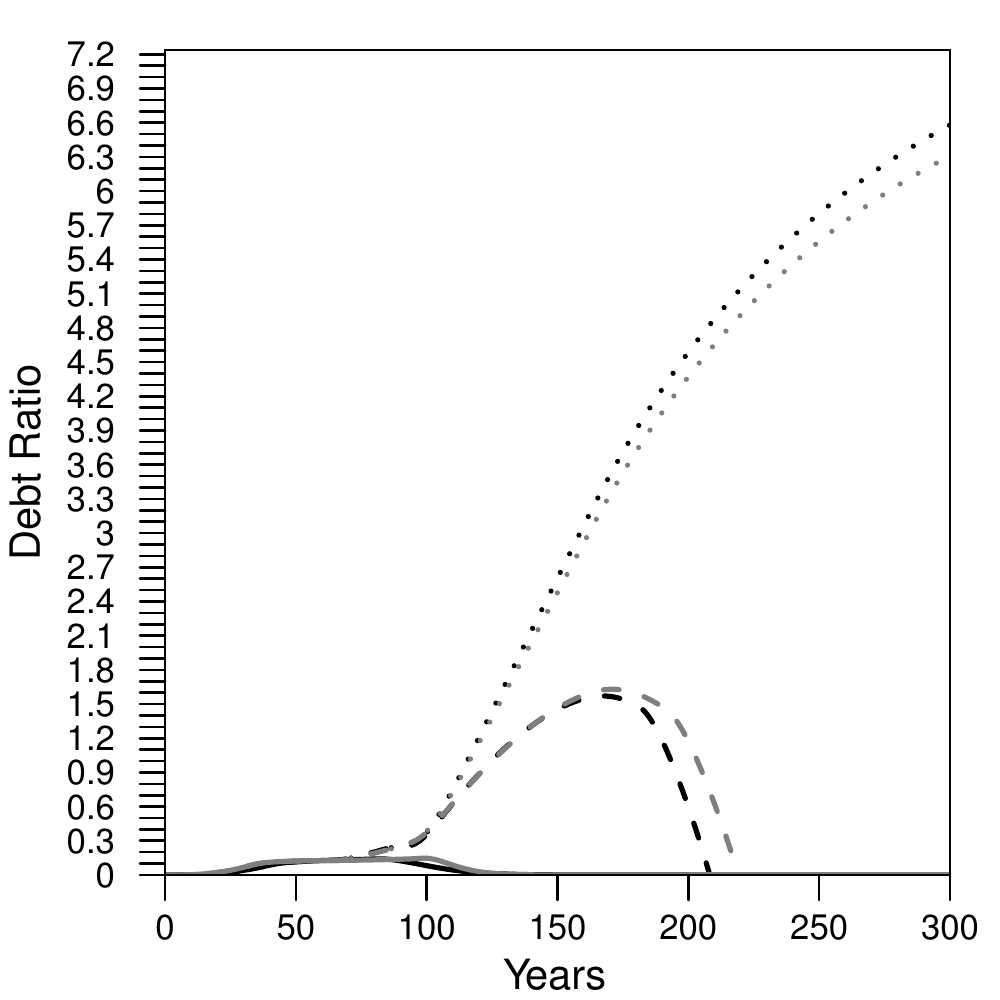}}\\
\caption{Scenarios FC-000, FC-010, FC-011, FC-100, FC-110, FC-111 (black, black dashed, black dotted, gray, gray dashed, gray dotted) results when setting prices via a constant markup ($\mu_i = 0.13$) with $\kappa_0 = 1.0$ and $\kappa_1 = 1.5$. (a) available resources (in the environment), (b)  total capital, (c) population, (d) resource extraction rate, (e) total real net output, (f) participation rate, (g) real price of extracted resources, (h) real price of goods, (i) debt ratio, and (continued) ...}\label{fig:1p8_A11A11gA11iF11F11gF11i}
\end{center}
\end{figure}

\begin{figure}\ContinuedFloat
\begin{center}
\subfloat[]{\includegraphics[width=0.33\columnwidth]{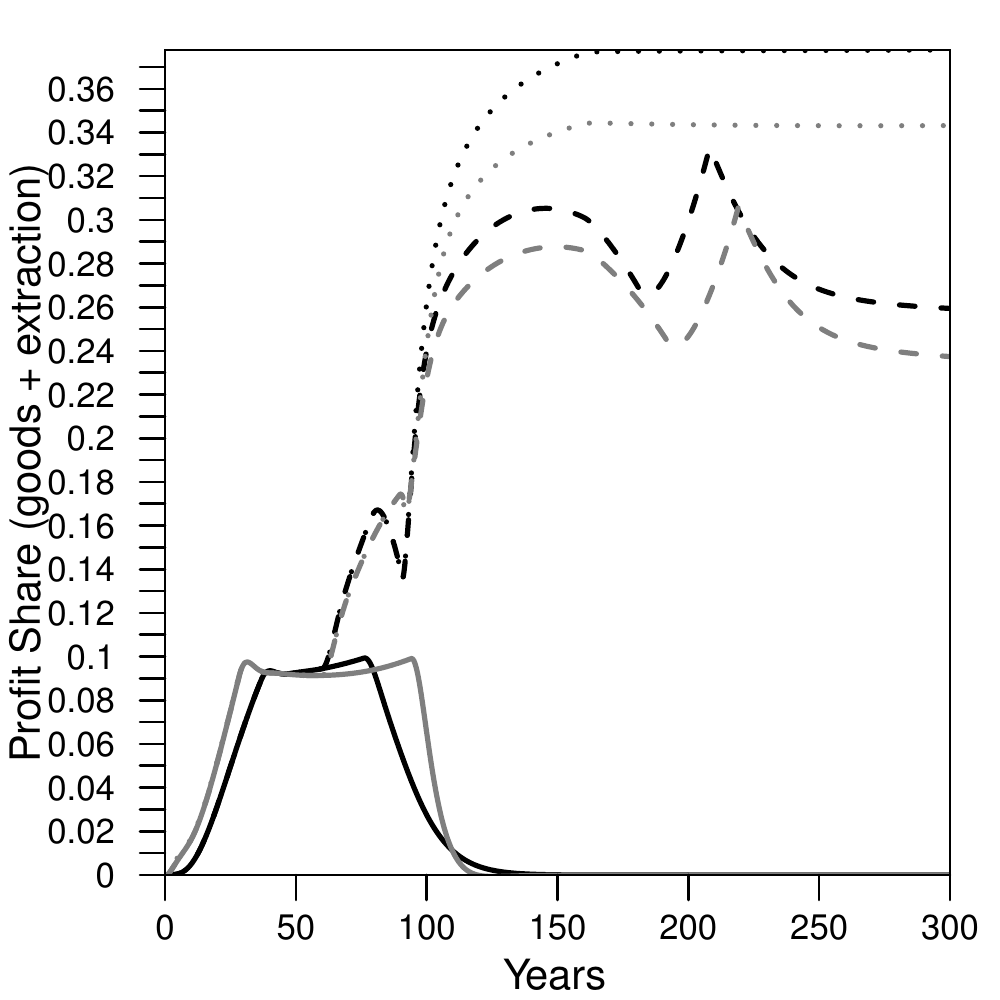}}
\subfloat[]{\includegraphics[width=0.33\columnwidth]{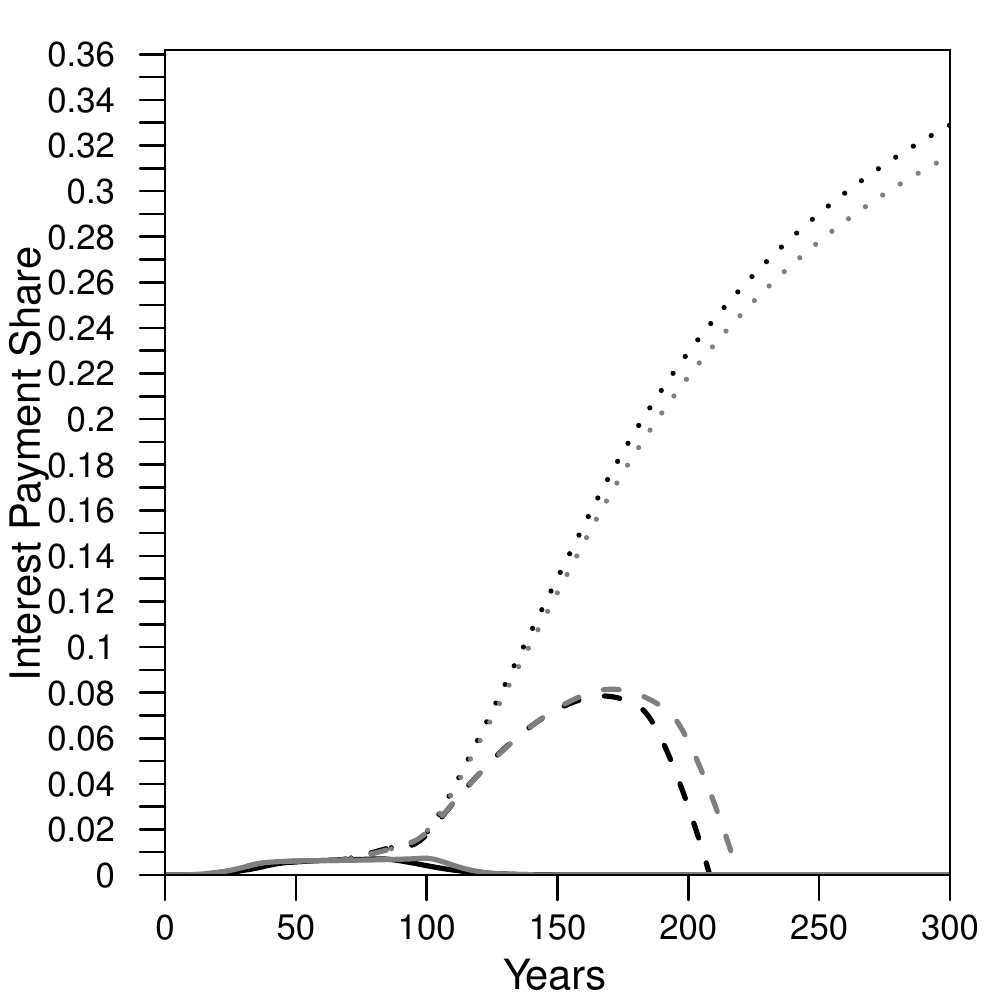}}
\subfloat[]{\includegraphics[width=0.33\columnwidth]{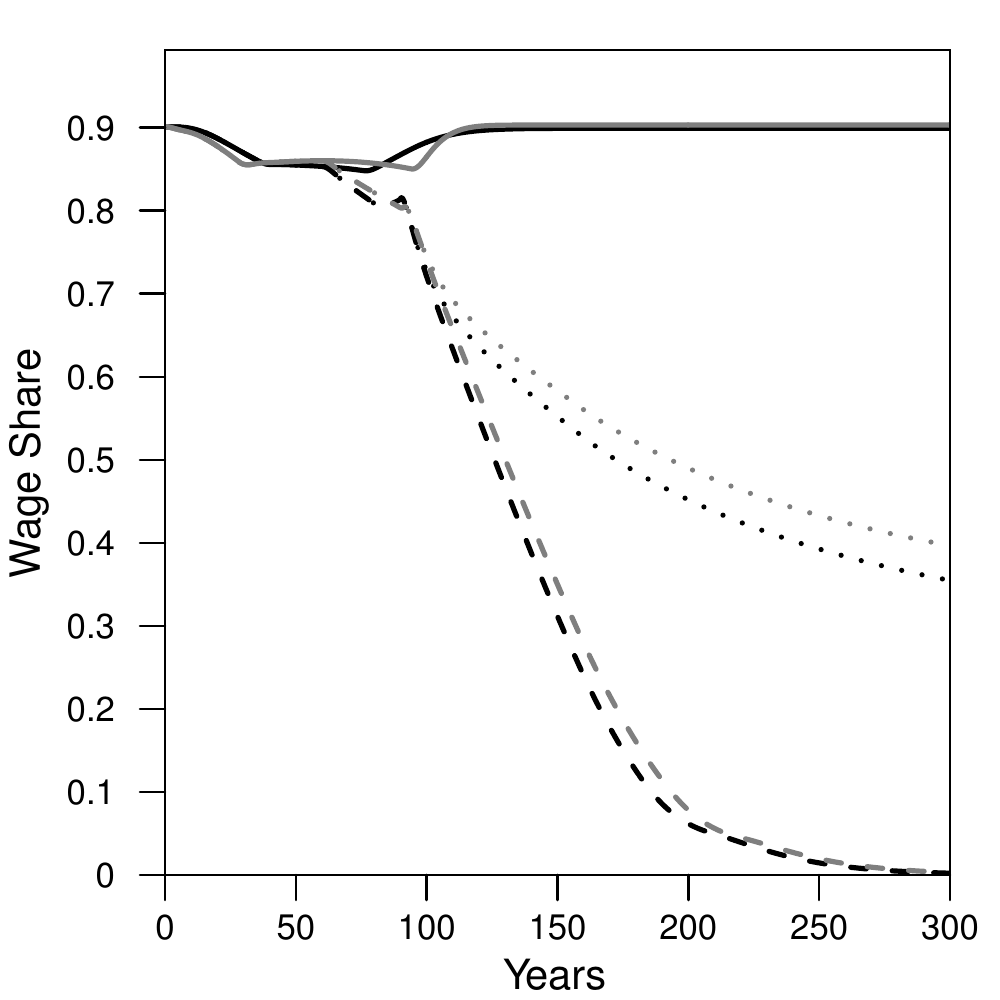}}\\
\subfloat[]{\includegraphics[width=0.33\columnwidth]{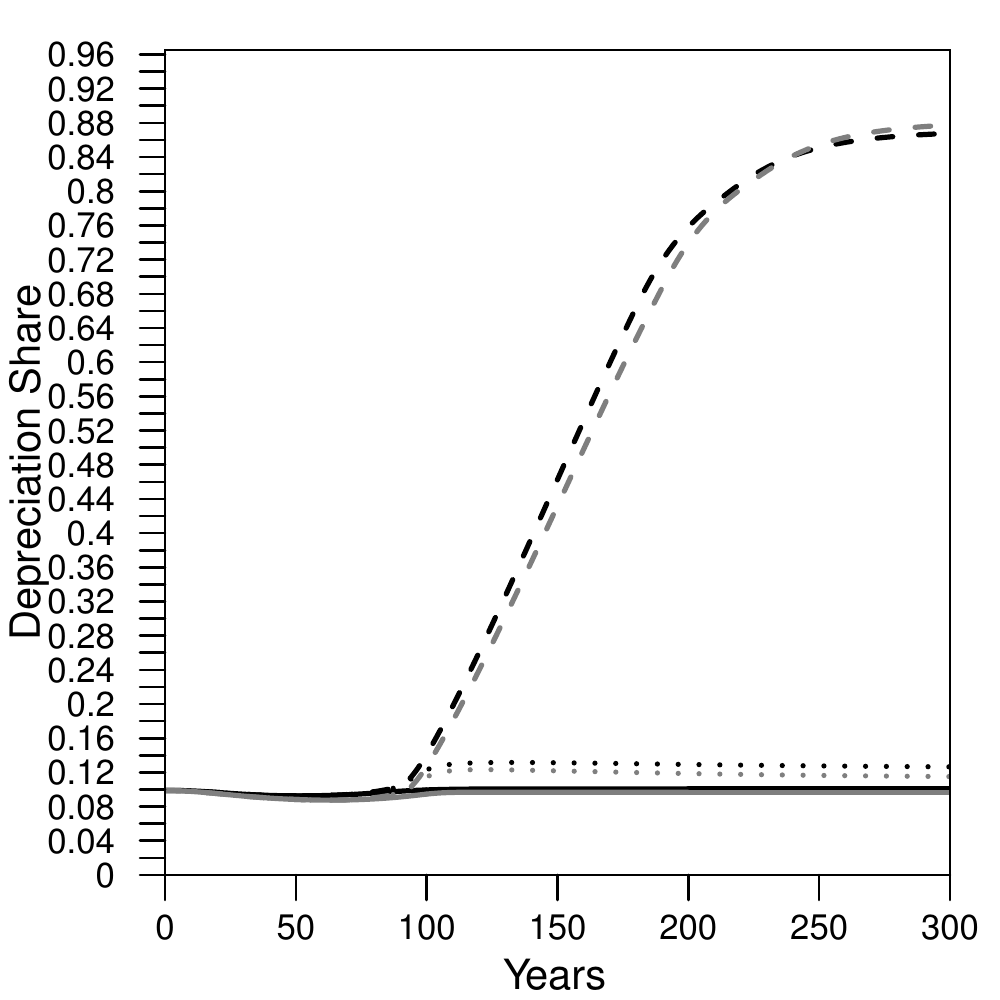}}
\subfloat[]{\includegraphics[width=0.33\columnwidth]{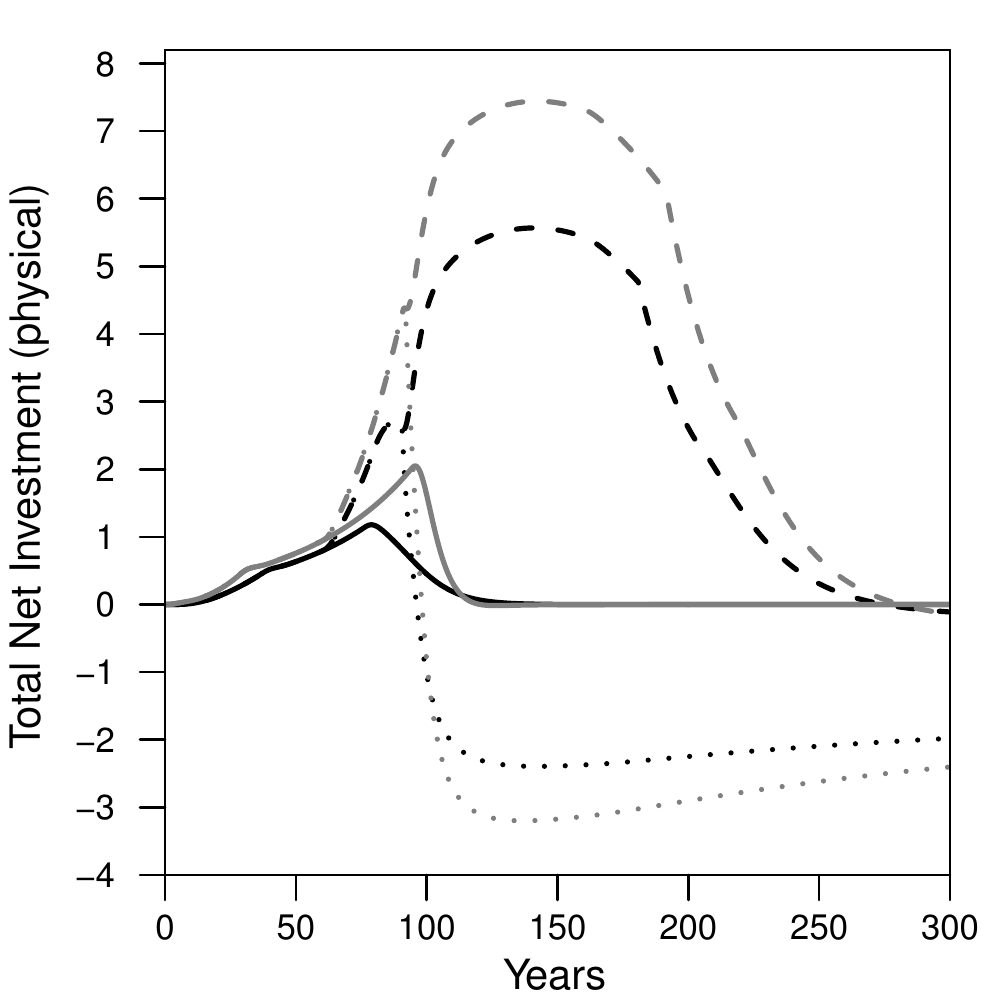}}
\subfloat[]{\includegraphics[width=0.33\columnwidth]{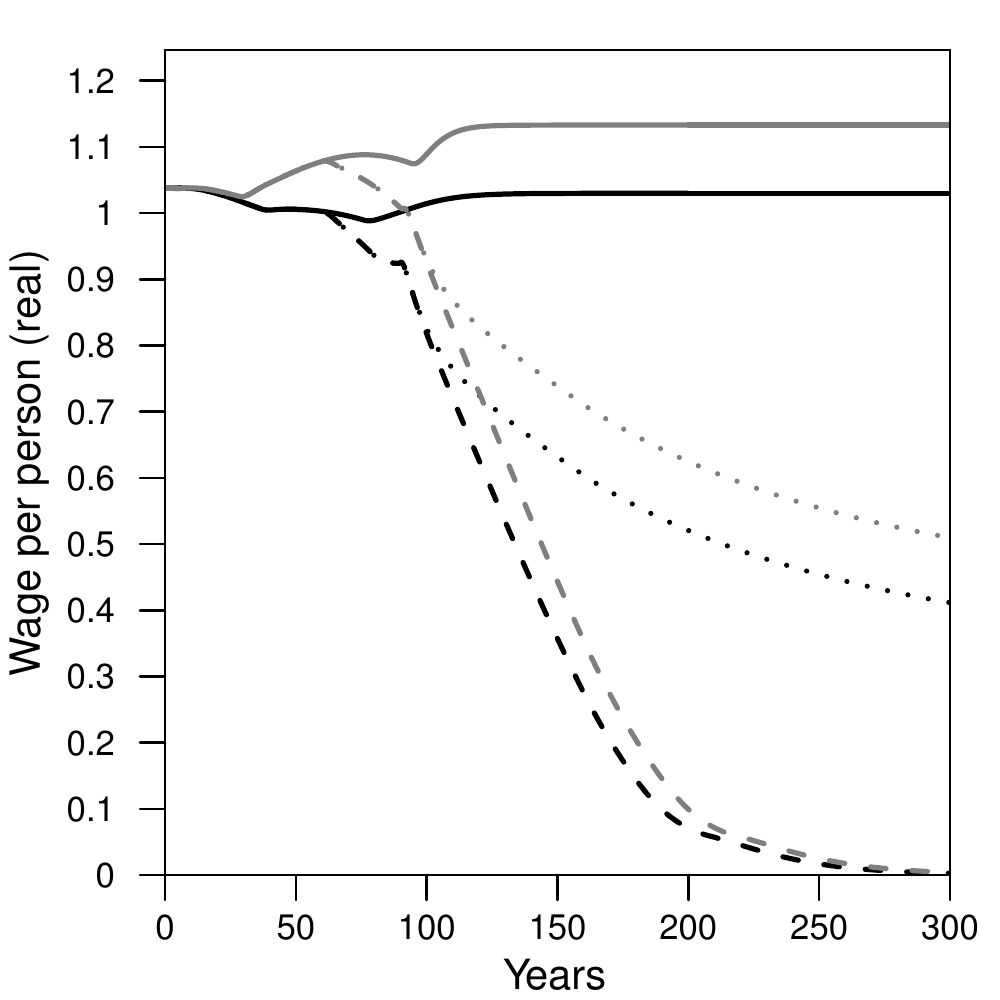}}\\
\subfloat[]{\includegraphics[width=0.33\columnwidth]{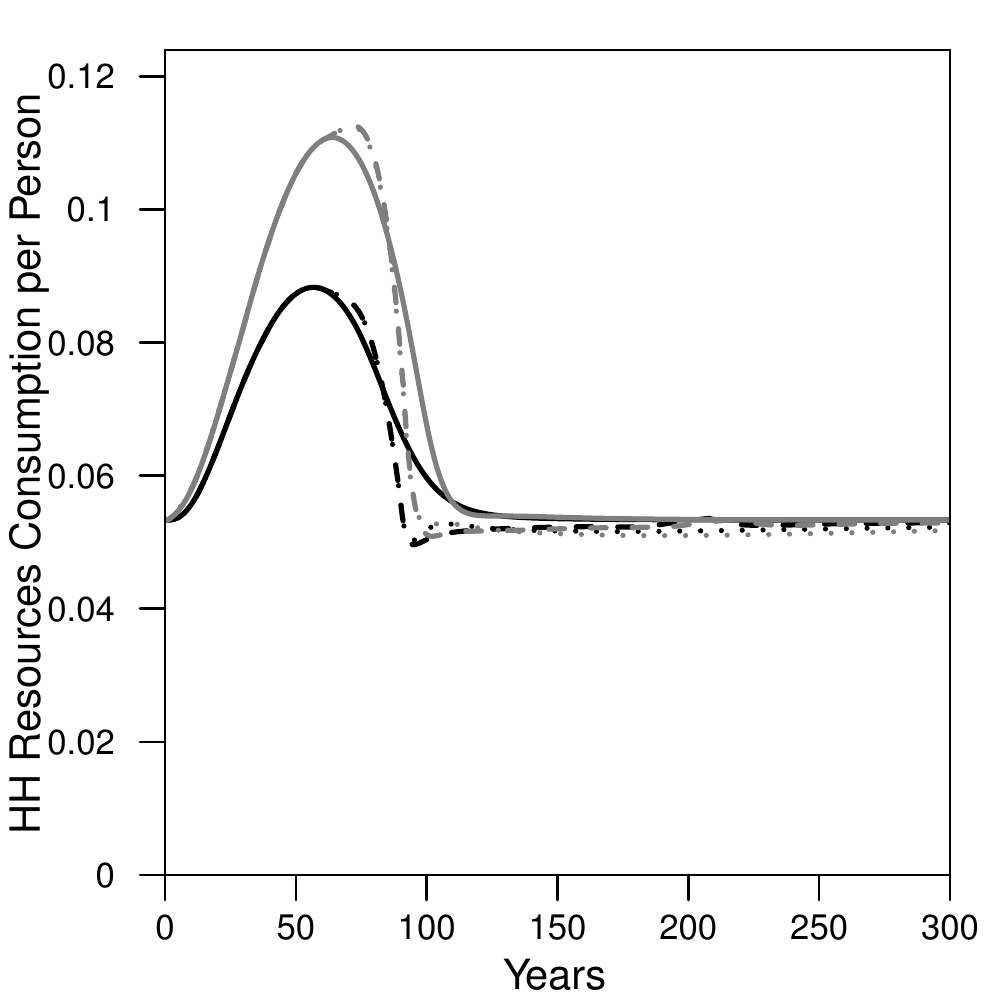}}
\subfloat[]{\includegraphics[width=0.33\columnwidth]{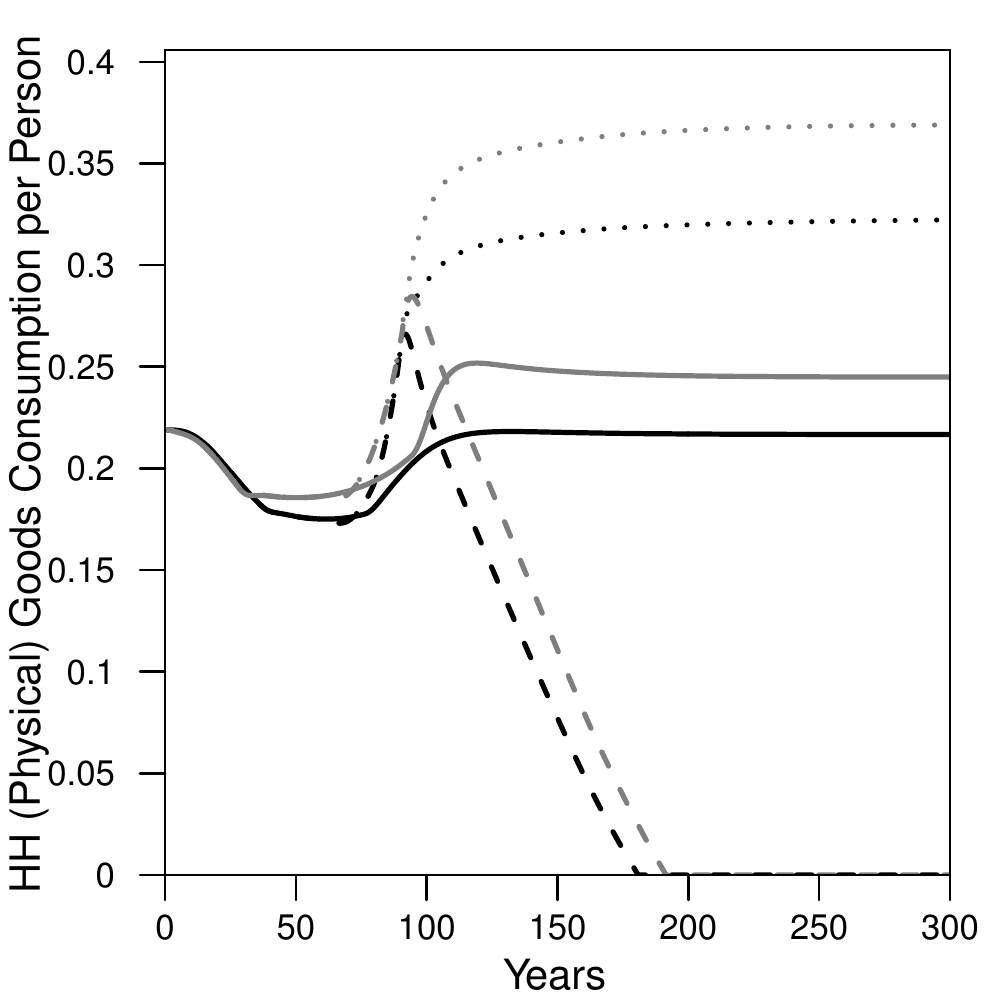}}
\subfloat[]{\includegraphics[width=0.33\columnwidth]{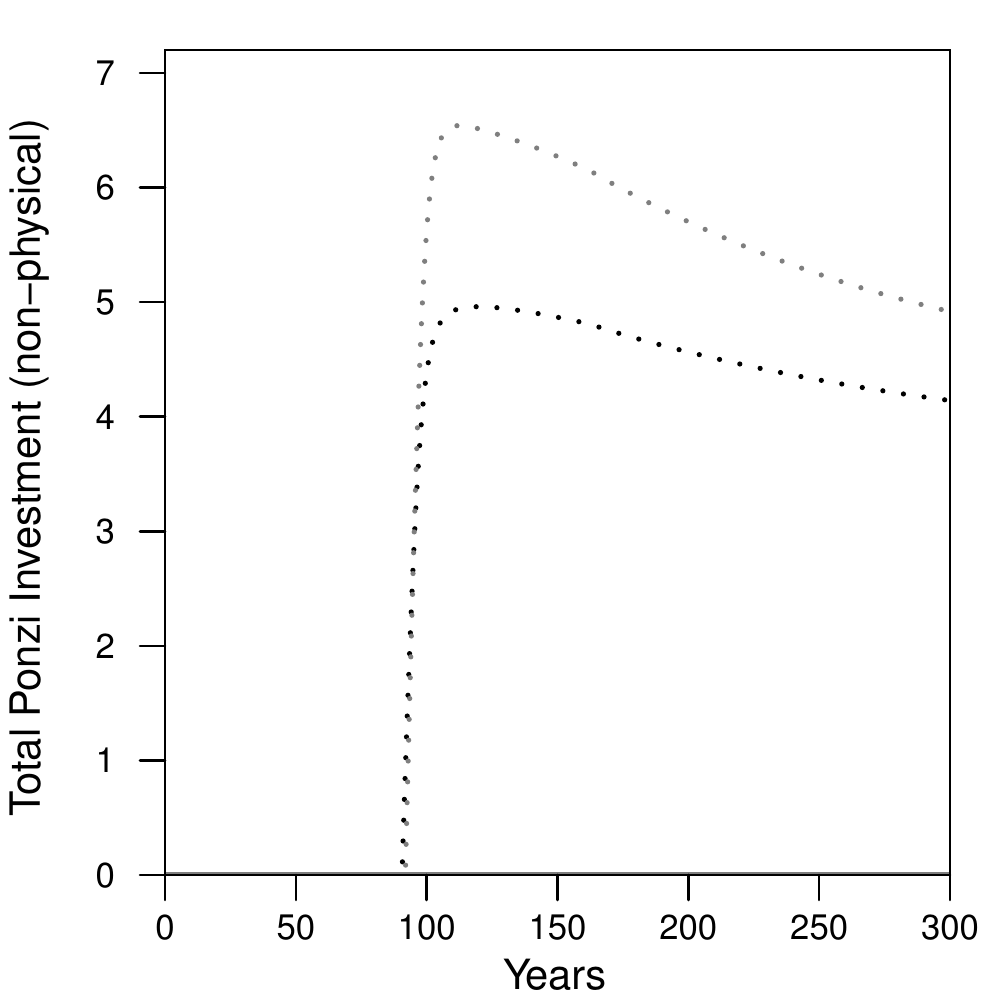}}\\
\caption{(continued) (j) profit share, (k) interest share, (l) wage share, (m) depreciation share, (n) physical net investment in new capital, (o) real wage per person, (p) household consumption of (physical) resources per person, (q) household consumption of (physical) goods per person, (r) total non-physical Ponzi investment, }
\end{center}
\end{figure}

\begin{figure}\ContinuedFloat
\begin{center}
\subfloat[]{\includegraphics[width=0.33\columnwidth]{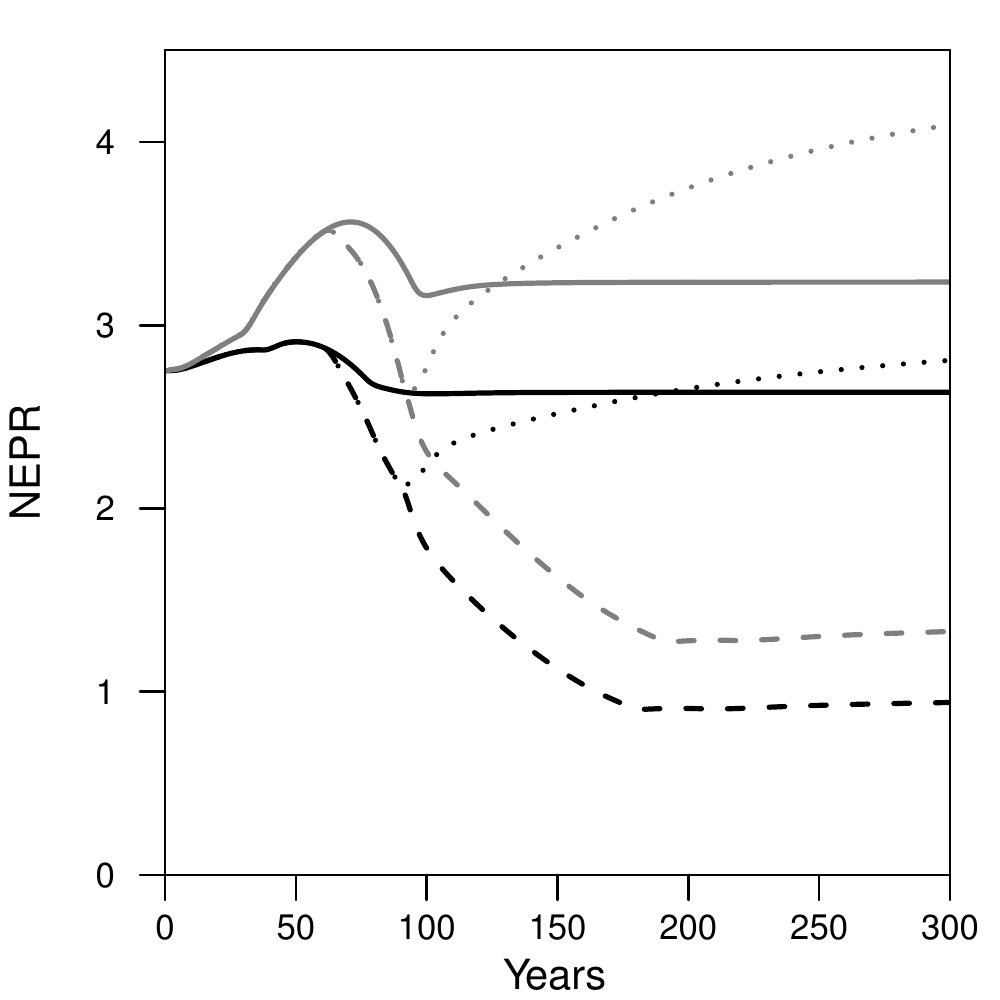}}
\subfloat[]{\includegraphics[width=0.33\columnwidth]{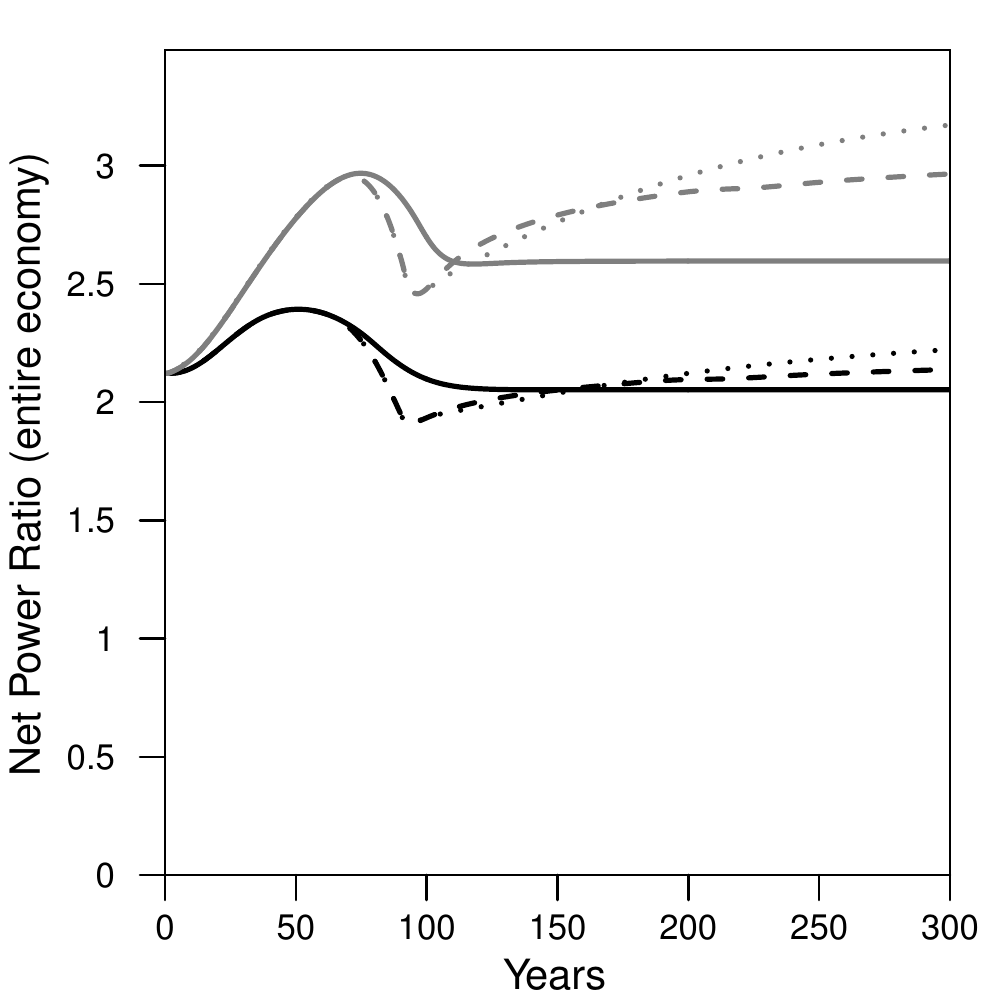}}
\subfloat[]{\includegraphics[width=0.33\columnwidth]{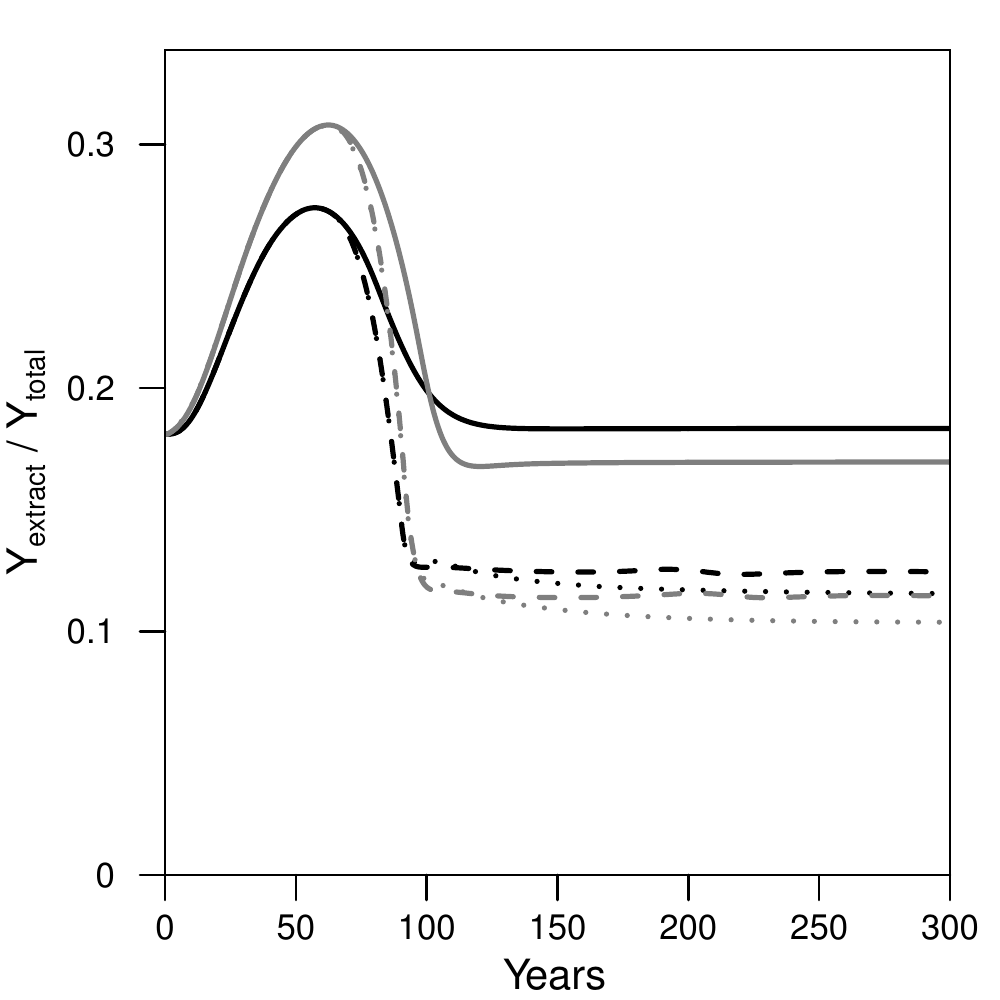}}\\
\subfloat[]{\includegraphics[width=0.33\columnwidth]{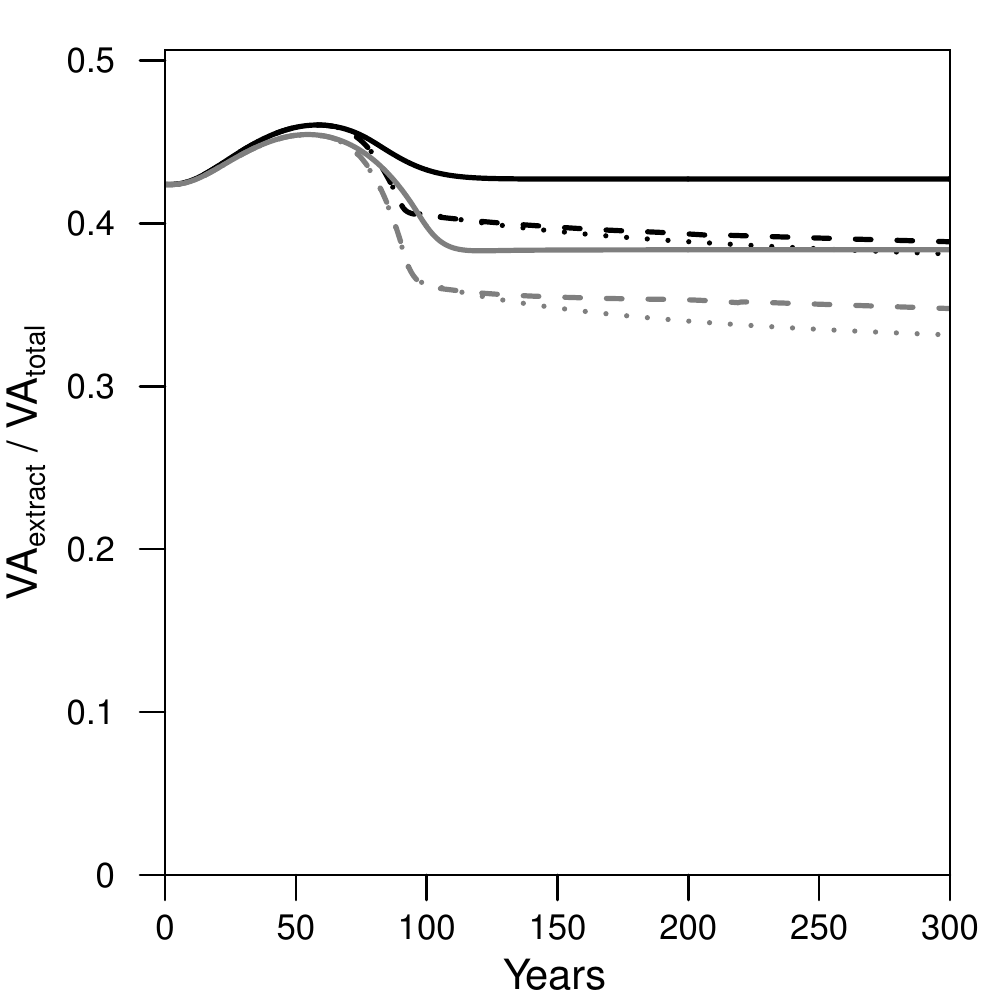}}
\subfloat[]{\includegraphics[width=0.33\columnwidth]{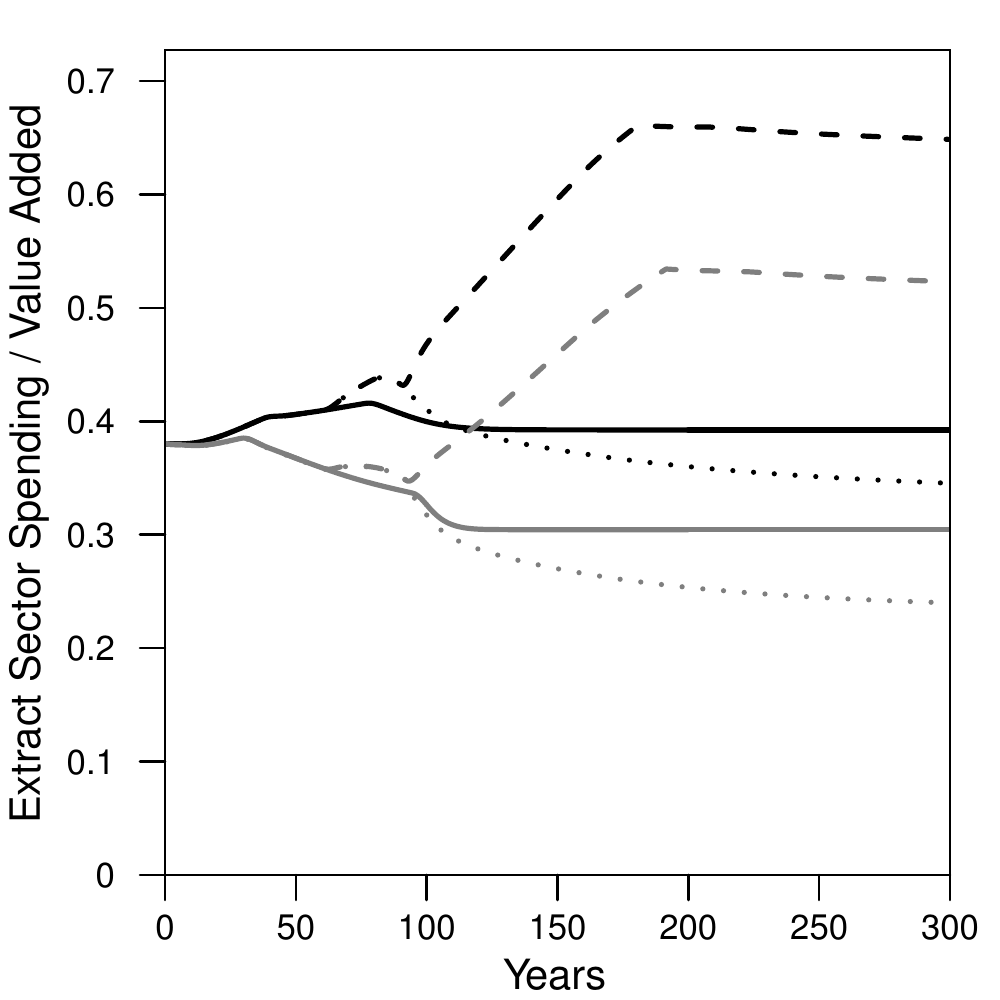}}
\subfloat[]{\includegraphics[width=0.33\columnwidth]{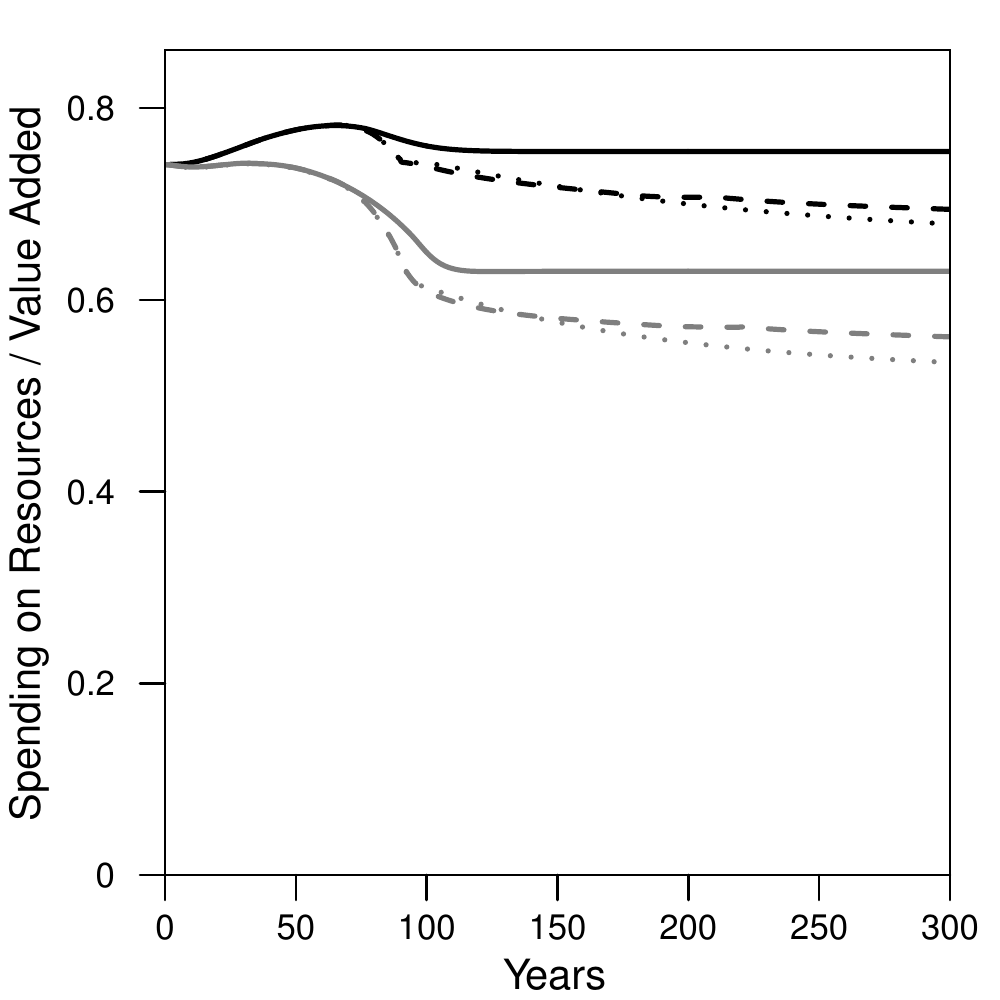}}\\
\subfloat[]{\includegraphics[width=0.33\columnwidth]{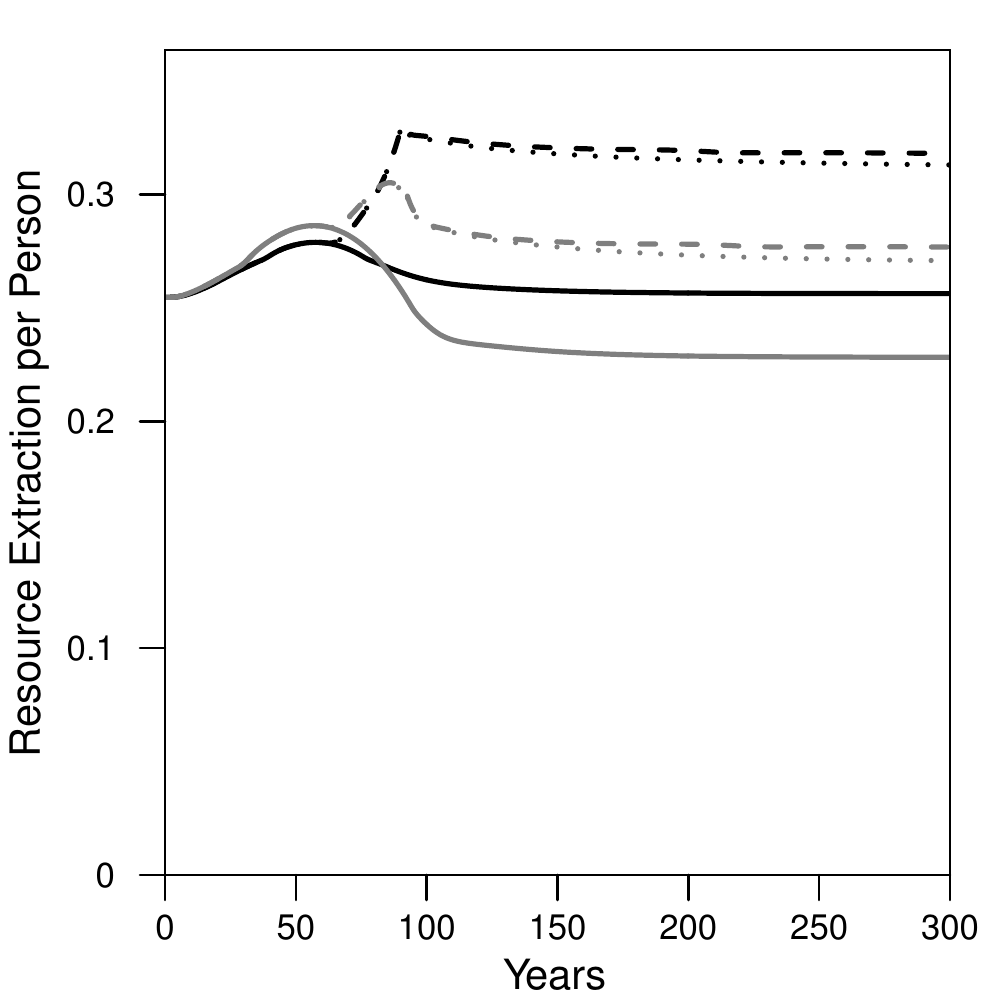}}
\subfloat[]{\includegraphics[width=0.33\columnwidth]{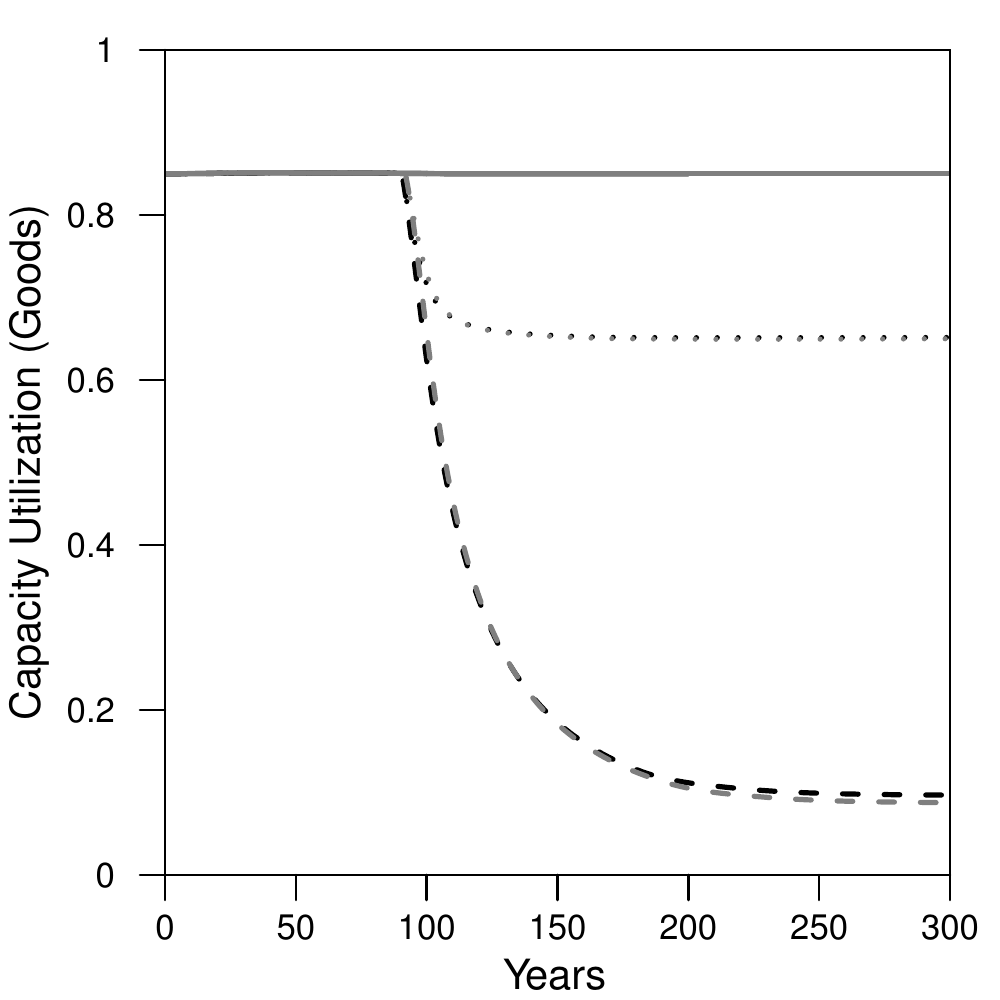}}
\caption{(continued) (s) net external power ratio (extraction sector, NEPR), (t) net power ratio (entire economy, NPR), (u) fraction of net output from extraction sector, (v) fraction of value added in extraction sector, (w) extraction sector spending per total value added (= per total net output), (x) spending on resources per total value added (= per total net output), (y) total resources extraction per person, and (z) capacity utilization of goods capital.} 
\end{center}
\end{figure}

\clearpage
\subsection{Information Theory Calculations for Full Cost Scenarios}\label{app:ScenariosFullCost_InformationTheory}

\clearpage
\begin{figure}[h]
\begin{center}
\subfloat[]{\includegraphics[width=.4\columnwidth]{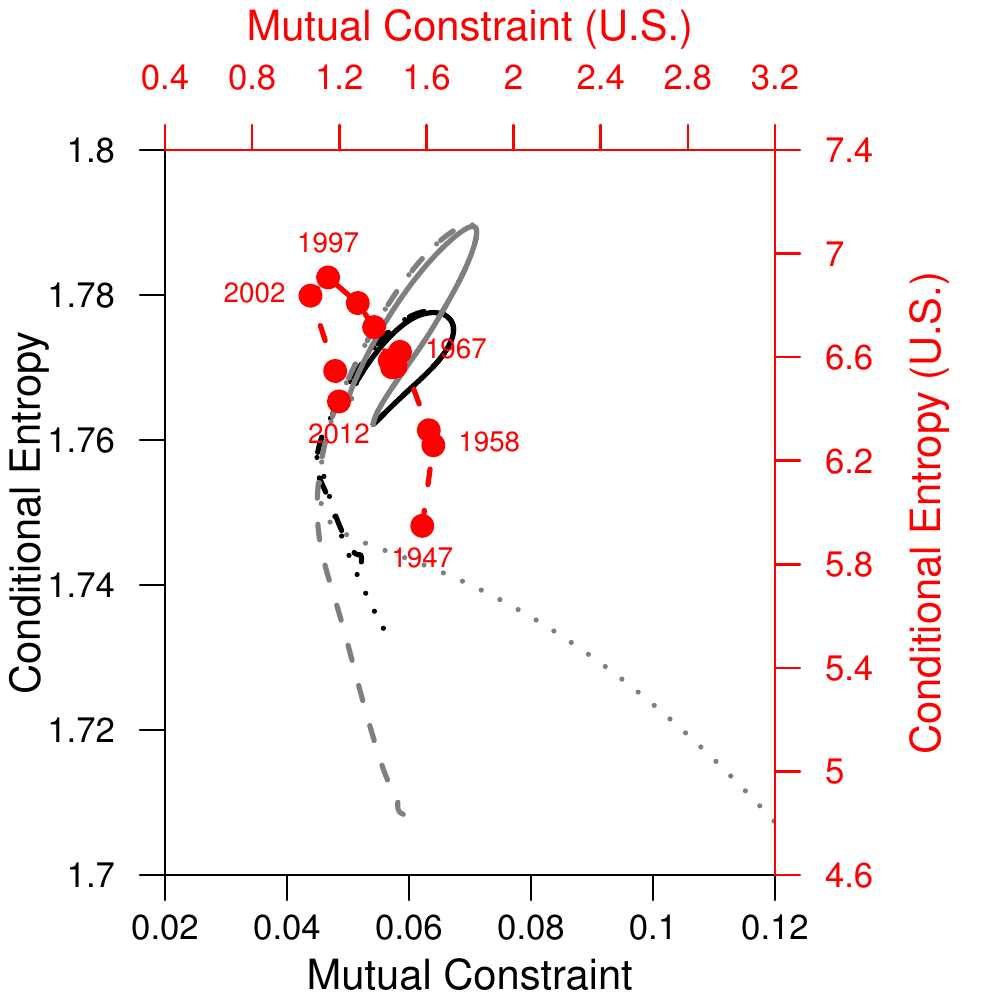}}
\subfloat[Information Entropy]{\includegraphics[width=.4\columnwidth]{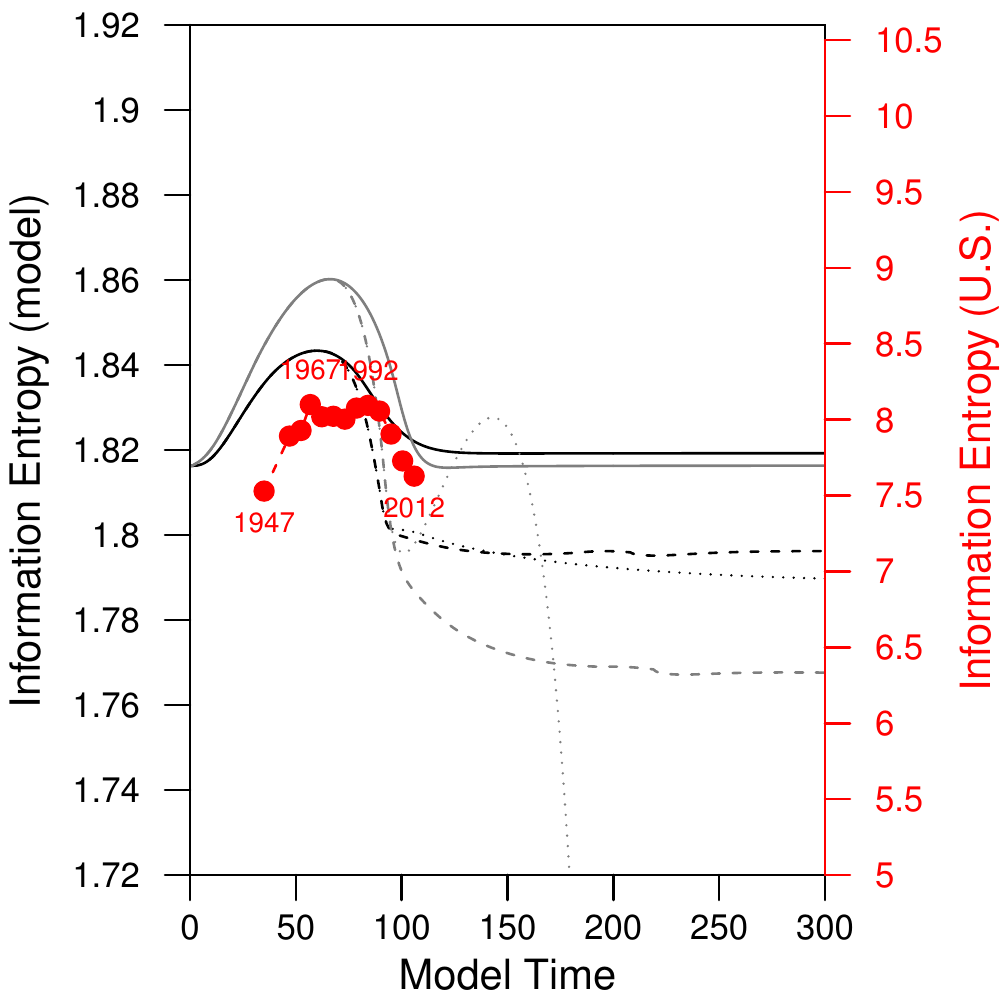}}\\
\subfloat[Conditional Entropy]{\includegraphics[width=.4\columnwidth]{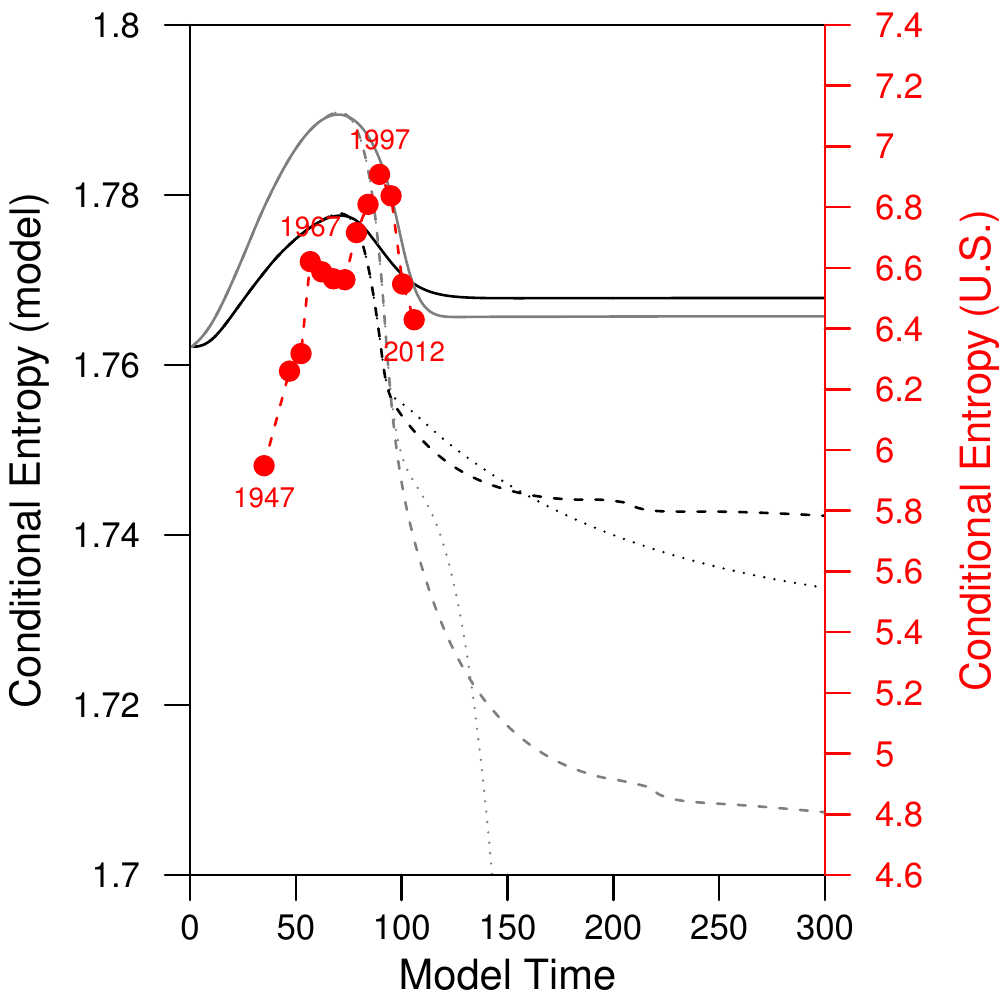}}
\subfloat[Mutual Constraint]{\includegraphics[width=.4\columnwidth]{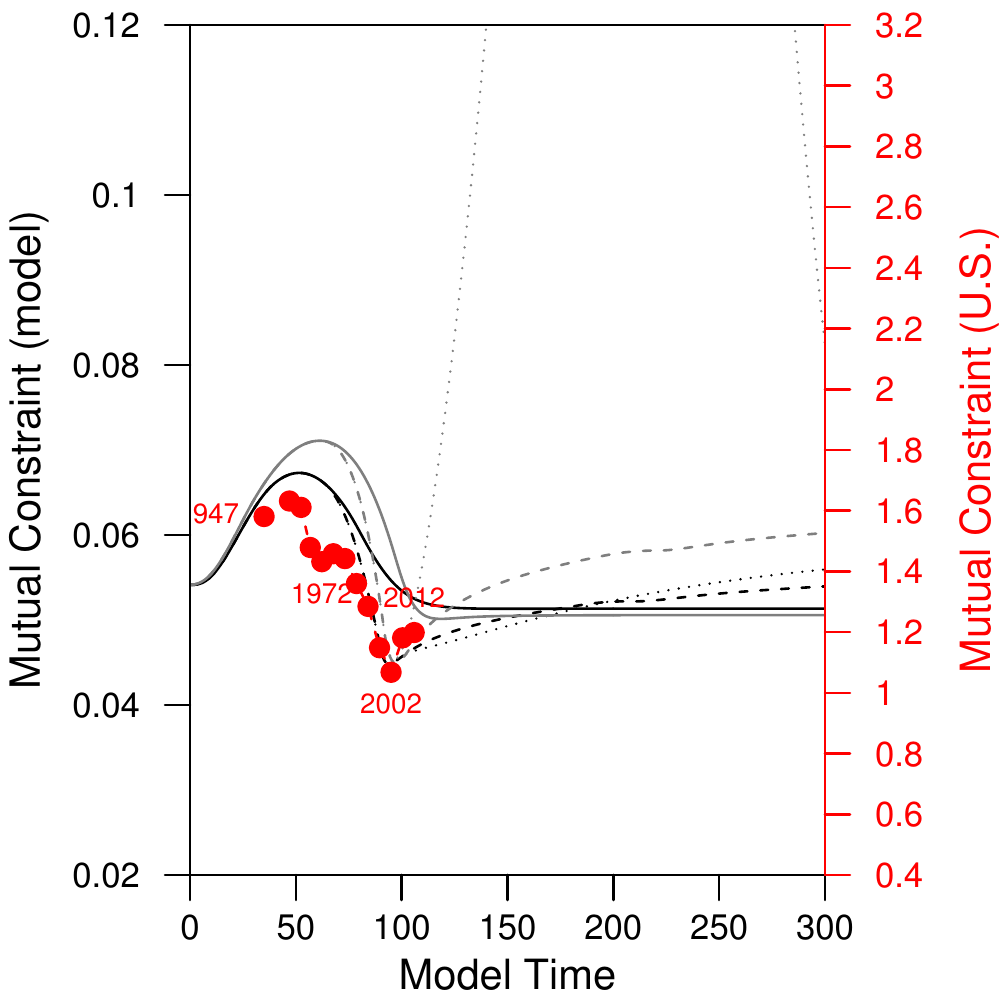}}\\
\caption{Information theory metrics of the full cost scenarios compared to the 37-sector aggregation of the U.S. Use tables from 1947--2012 from King (2016) (red dashed lines, right axis) \cite{King2016}. FC-000 (black solid), FC-010 (black dashed), FC-011 (black dotted), FC-100 (gray solid), FC-110 (gray dashed), FC-111 (gray dotted).  (a) Conditional entropy versus mutual constraint, (b) information entropy vs time, (c) conditional entropy vs time, and (d) mutual constraint vs time.  U.S. information theory metrics are calculated using base 2 logarithm instead of natural logarithm as in \cite{King2016}.}\label{fig:InformationTheory_Comparision_FullCost}
\end{center}
\end{figure}

\clearpage
\subsection{Variations on Marginal Cost Pricing Scenarios with Lower Wage Bargaining and Ponzi}\label{app:ScenariosMarginalCost_LowerBargainWithPonzi}

\begin{figure}
\begin{center}
\subfloat[]{\includegraphics[width=0.33\columnwidth]{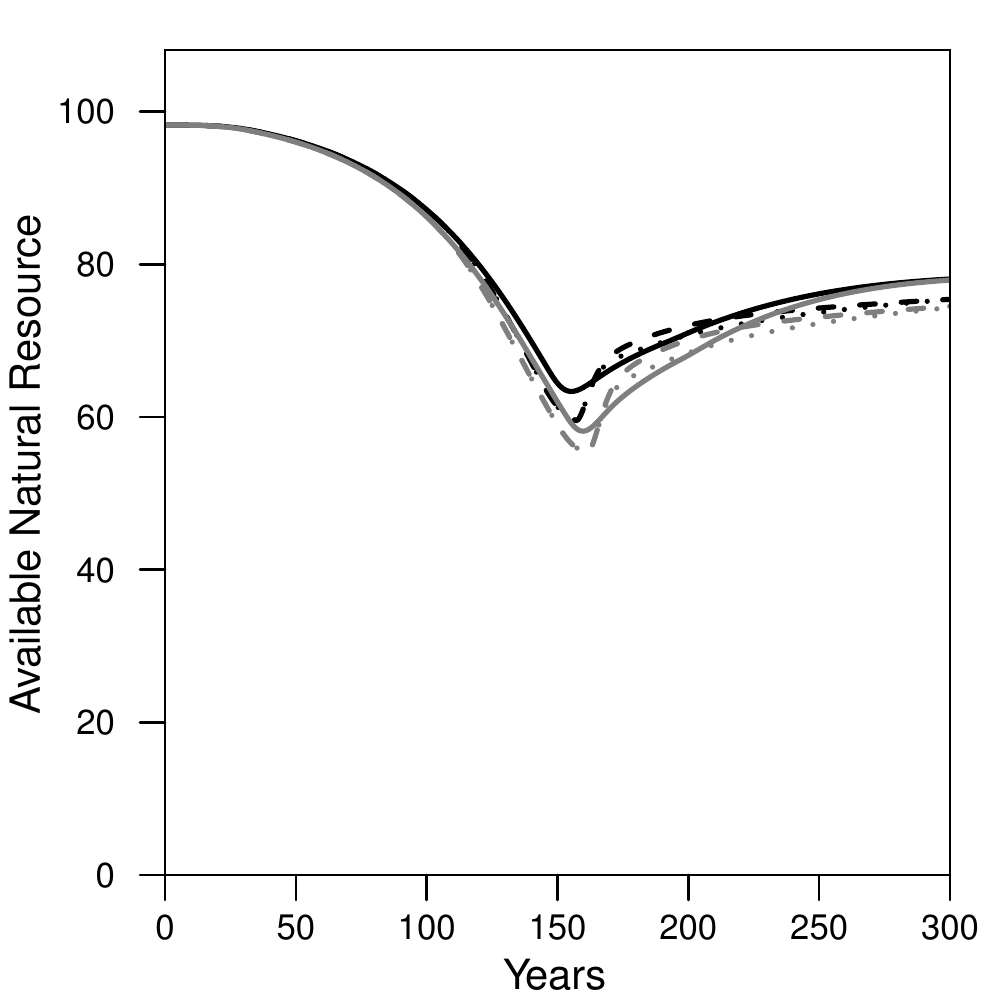}}
\subfloat[]{\includegraphics[width=0.33\columnwidth]{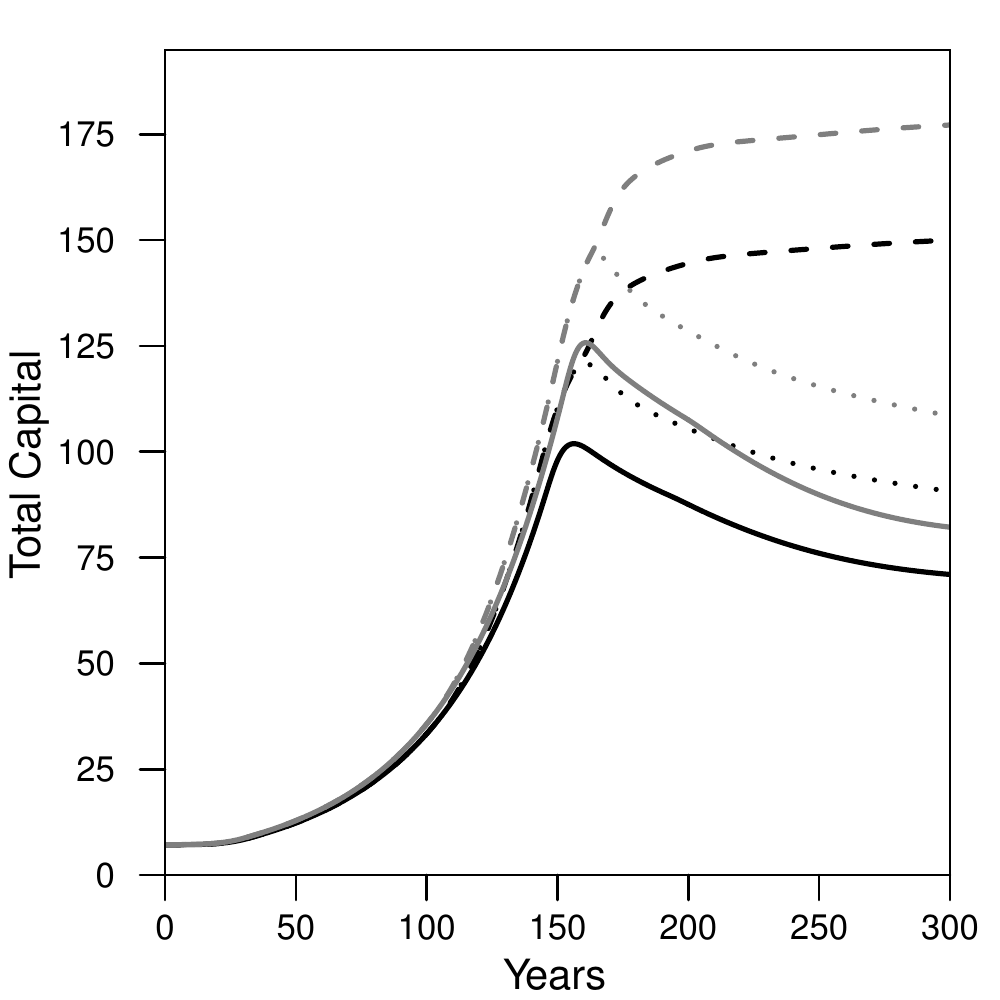}}
\subfloat[]{\includegraphics[width=0.33\columnwidth]{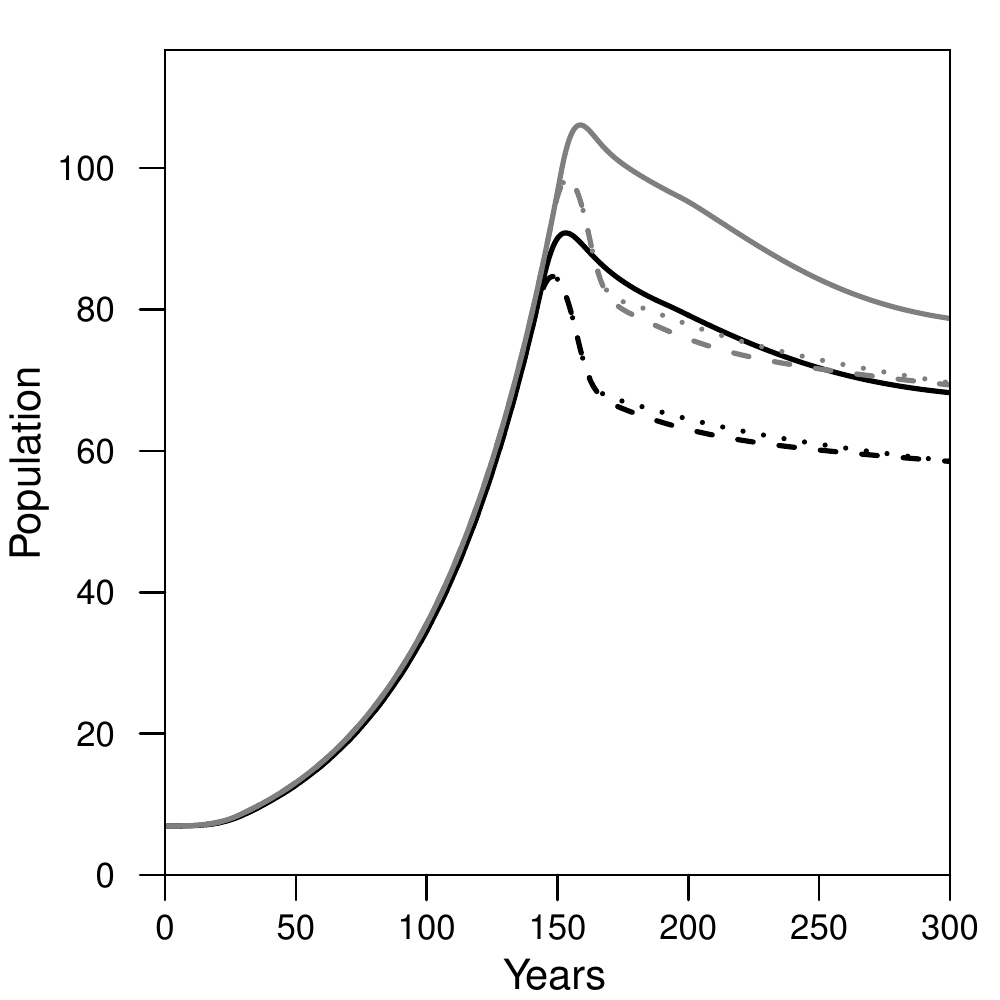}}\\
\subfloat[]{\includegraphics[width=0.33\columnwidth]{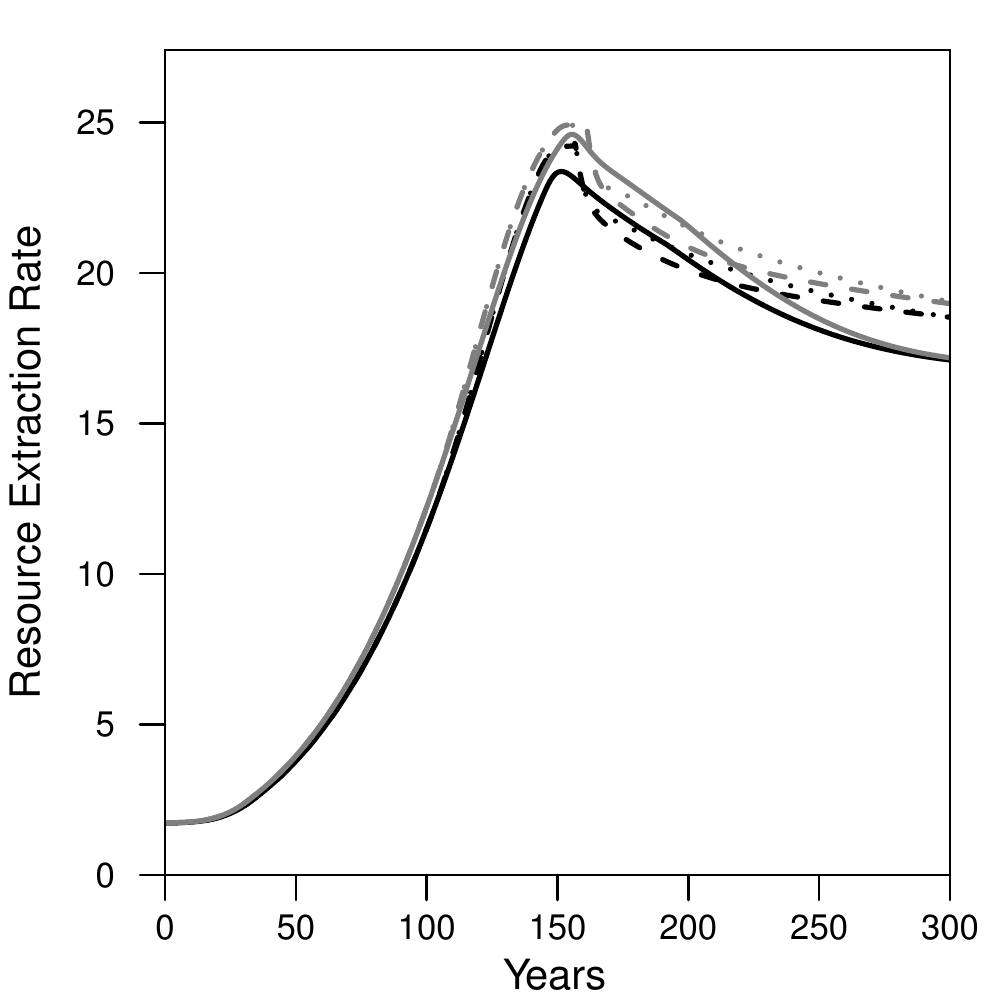}}
\subfloat[]{\includegraphics[width=0.33\columnwidth]{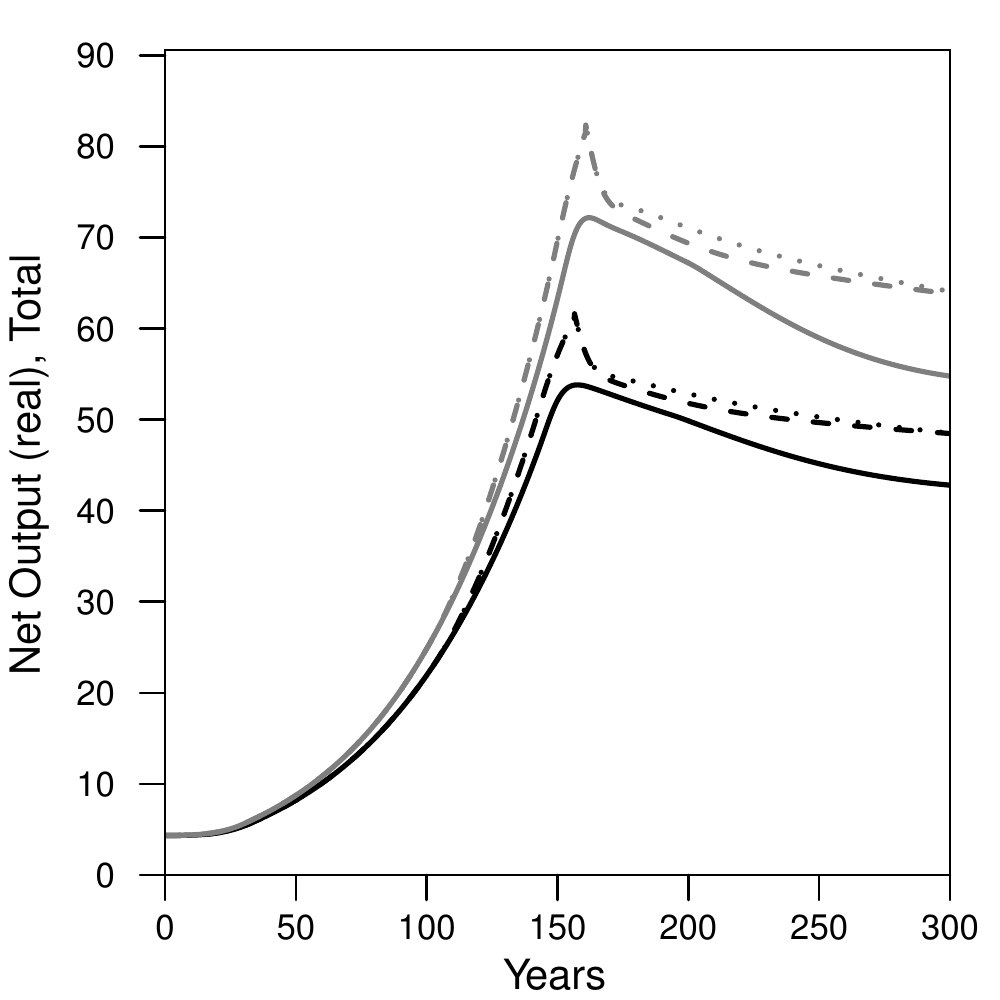}}
\subfloat[]{\includegraphics[width=0.33\columnwidth]{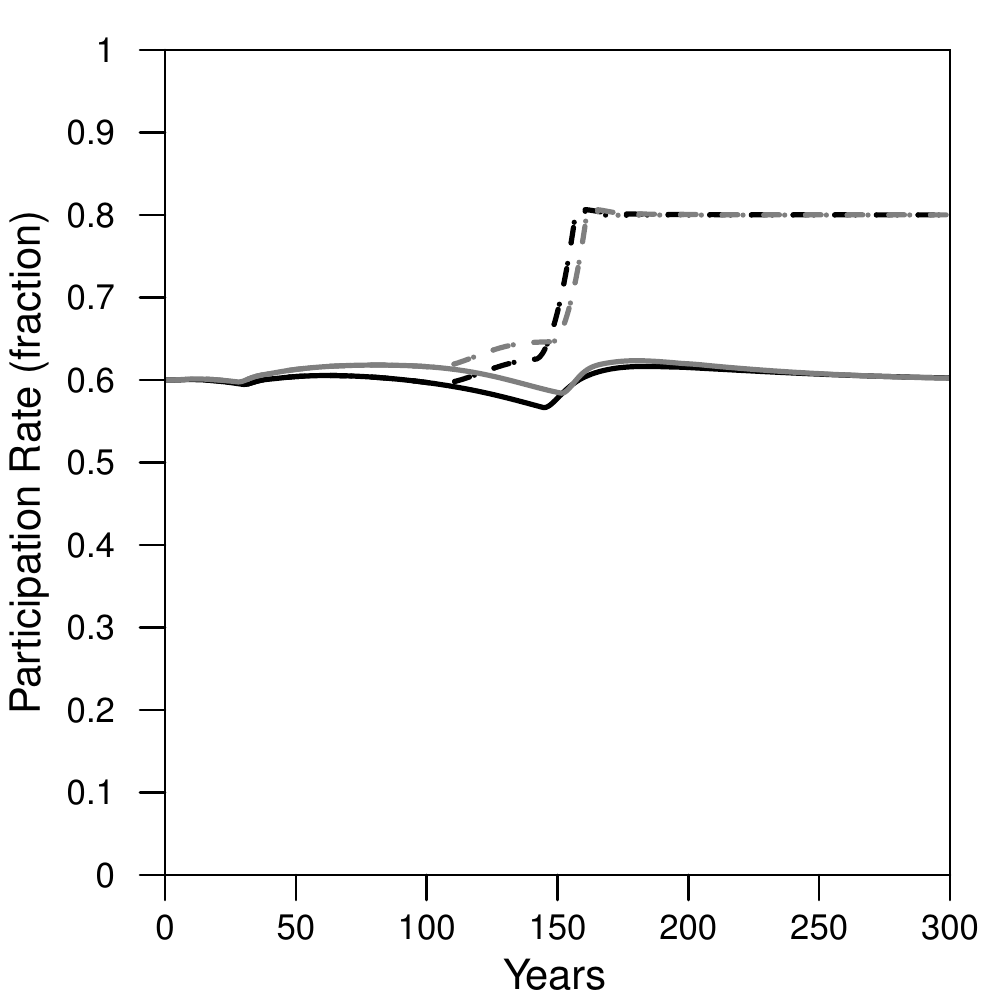}}\\
\subfloat[]{\includegraphics[width=0.33\columnwidth]{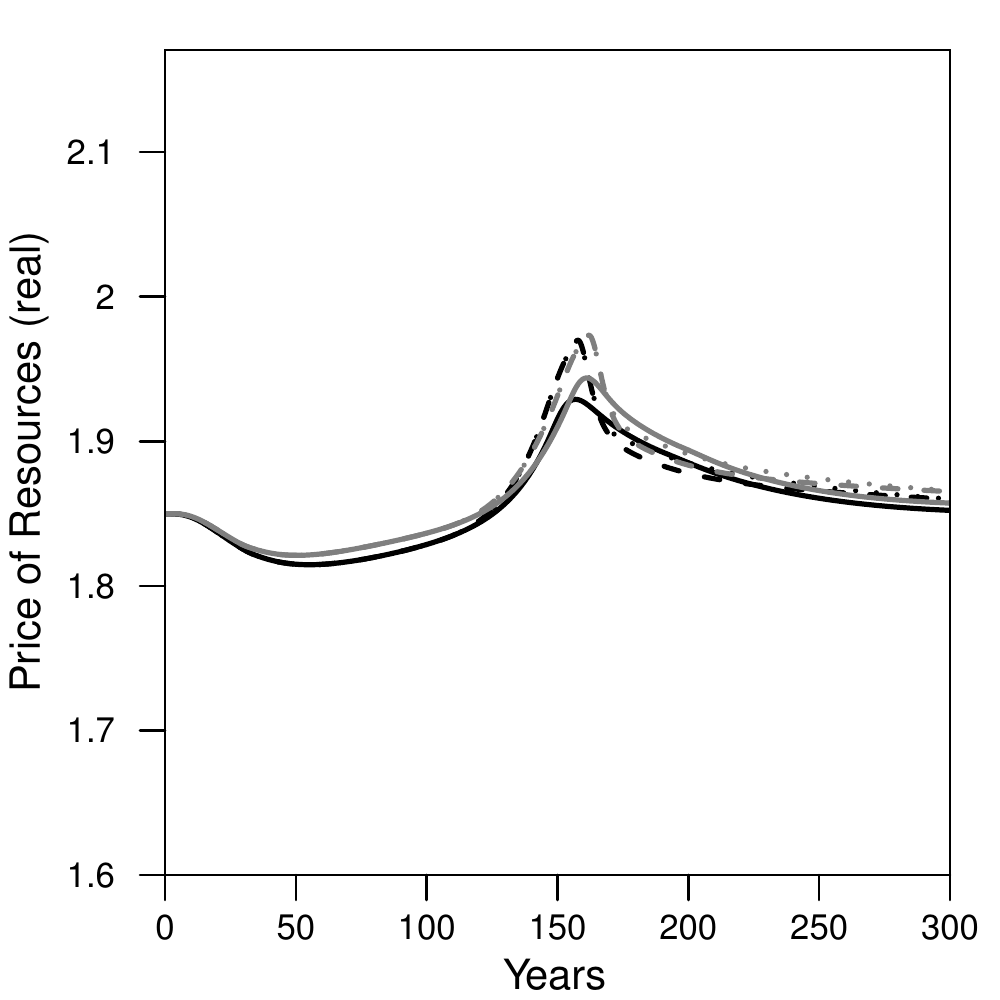}}
\subfloat[]{\includegraphics[width=0.33\columnwidth]{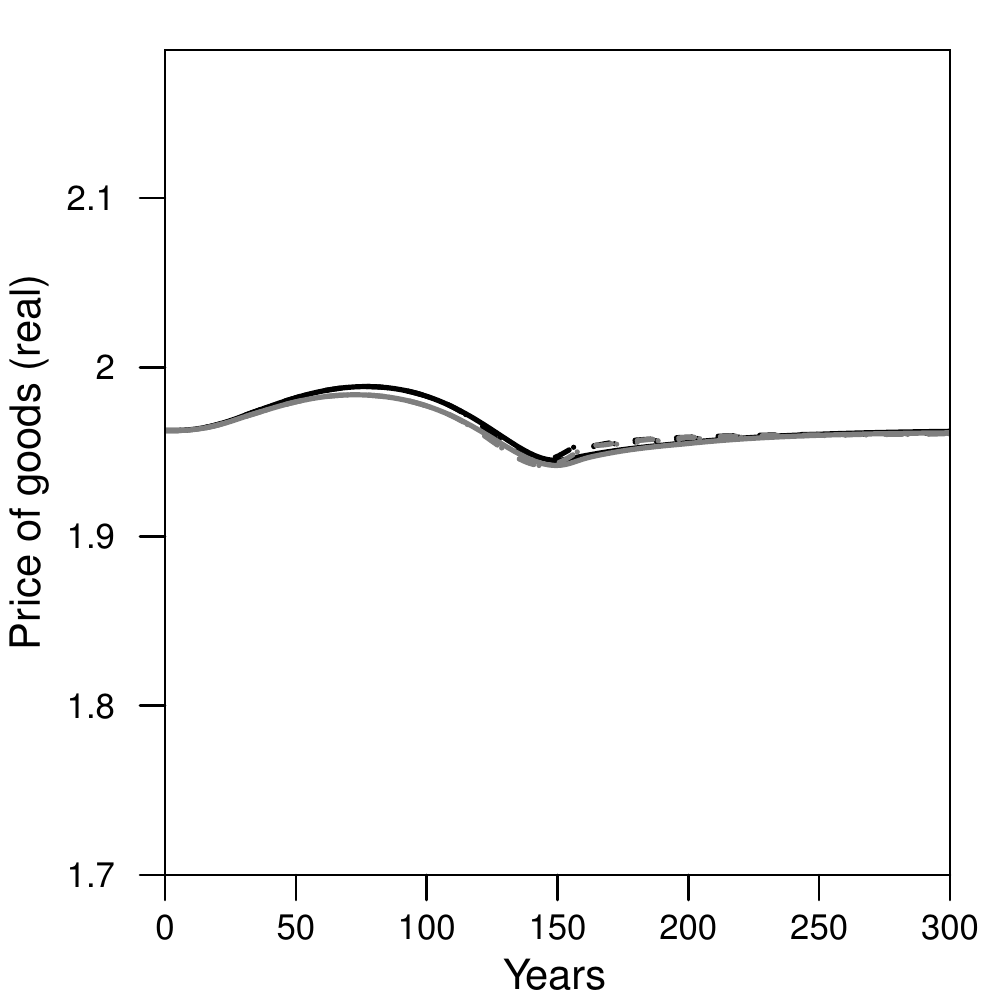}}
\subfloat[]{\includegraphics[width=0.33\columnwidth]{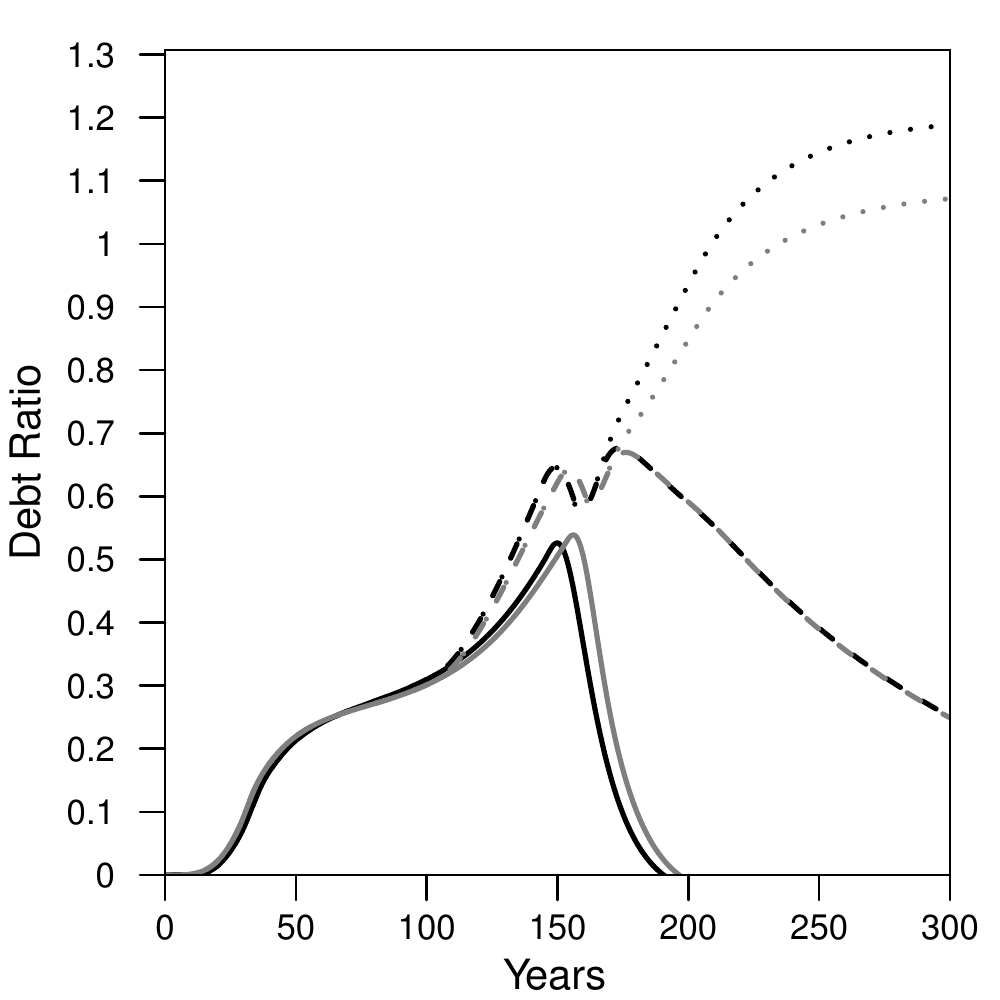}}\\
\caption{Scenarios MC-000, MC-010, MC-011, MC-100, MC-110, MC-111 (black, black dashed, black dotted, gray, gray dashed, gray dotted) results when setting prices via a constant markup ($\mu_i = 0.13$) with $\kappa_0 = 1.0$ and $\kappa_1 = 1.5$. (a) available resources (in the environment), (b)  total capital, (c) population, (d) resource extraction rate, (e) total real net output, (f) participation rate, (g) real price of extracted resources, (h) real price of goods, (i) debt ratio, and (continued) ...}\label{fig:1p8_A00A00gA00iF00F00gF00i}
\end{center}
\end{figure}

\begin{figure}\ContinuedFloat
\begin{center}
\subfloat[]{\includegraphics[width=0.33\columnwidth]{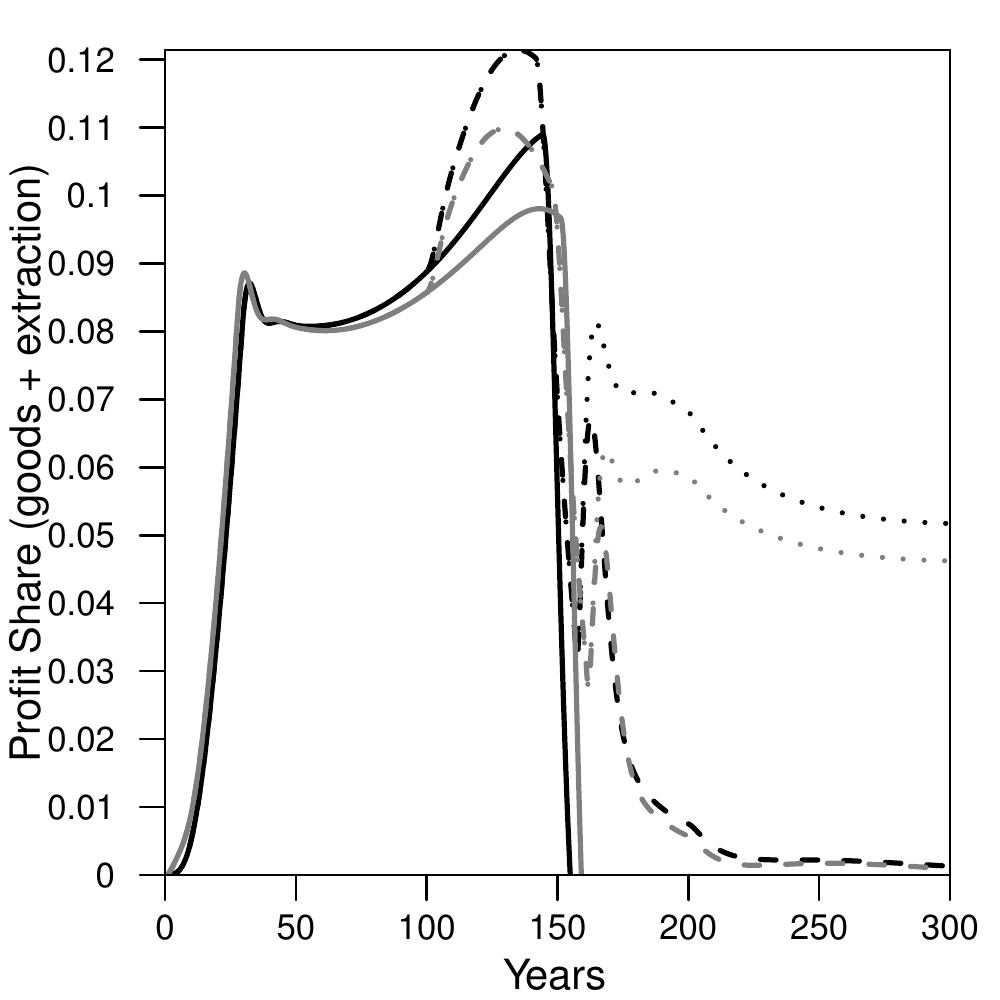}}
\subfloat[]{\includegraphics[width=0.33\columnwidth]{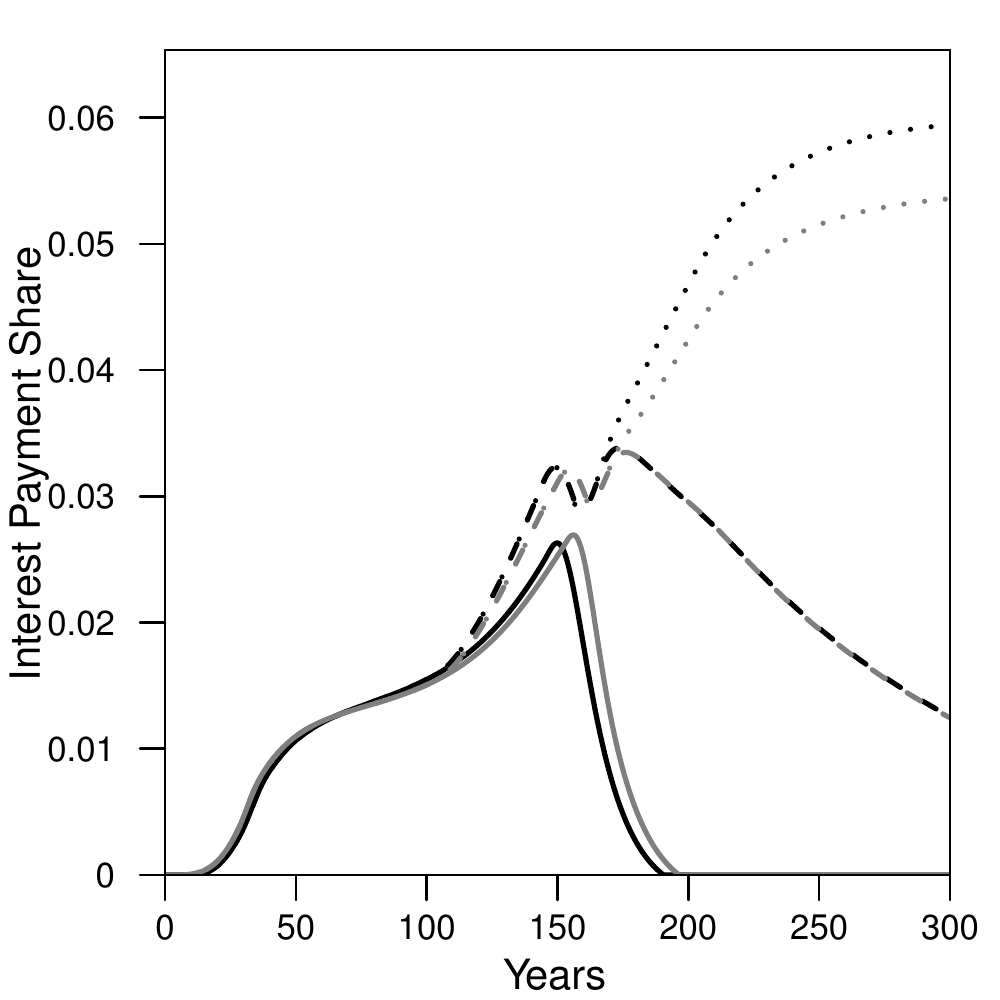}}
\subfloat[]{\includegraphics[width=0.33\columnwidth]{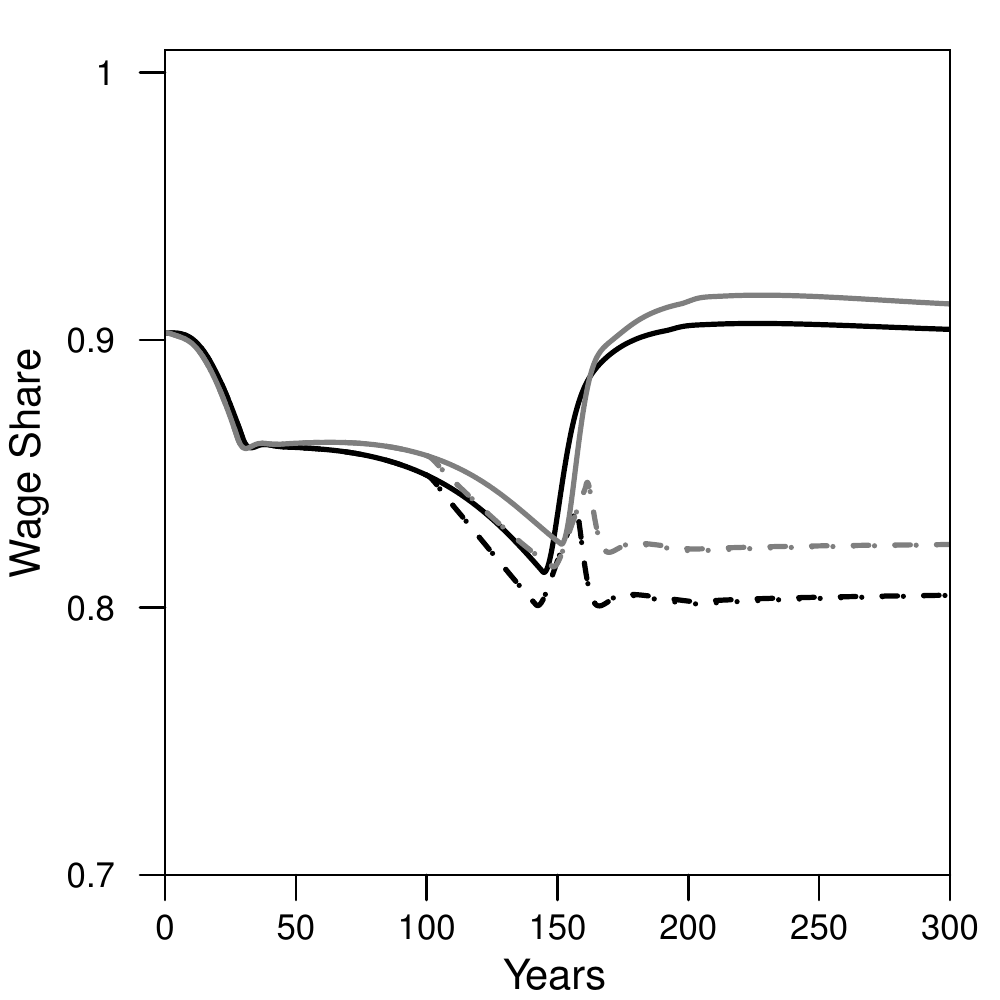}}\\
\subfloat[]{\includegraphics[width=0.33\columnwidth]{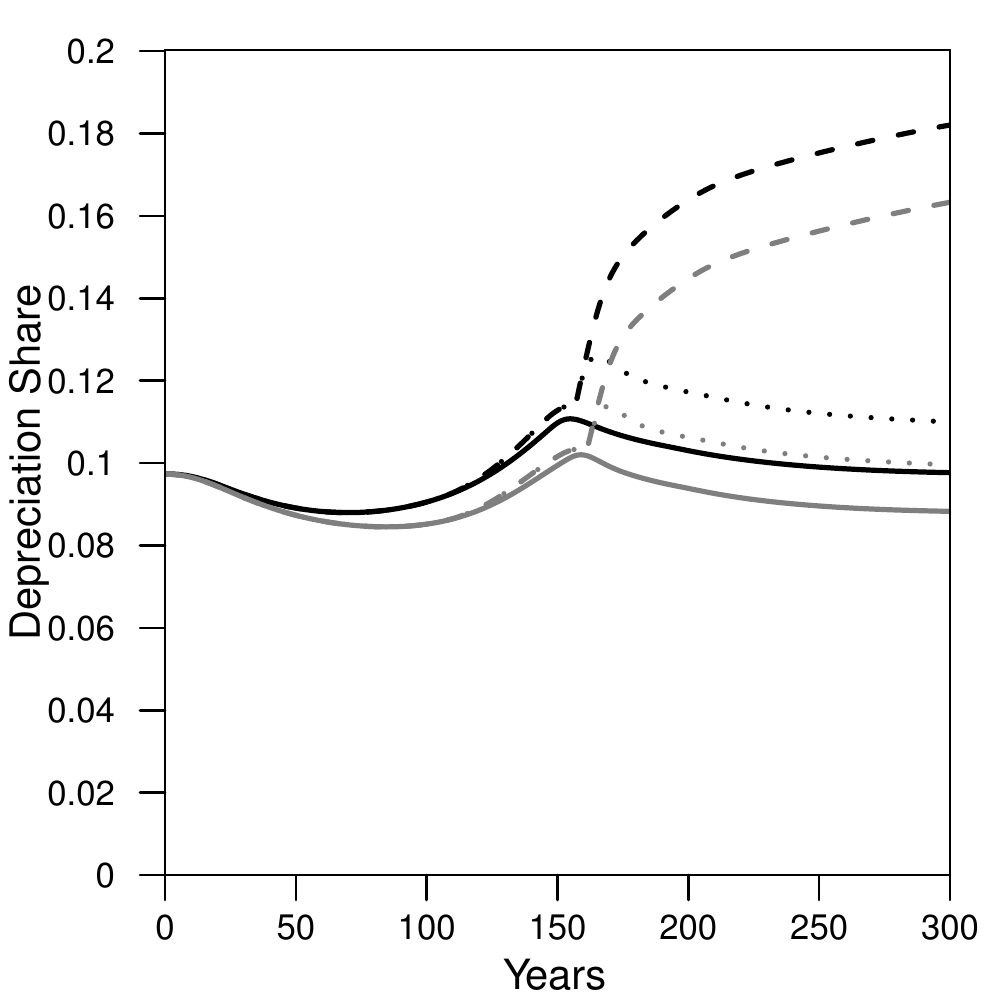}}
\subfloat[]{\includegraphics[width=0.33\columnwidth]{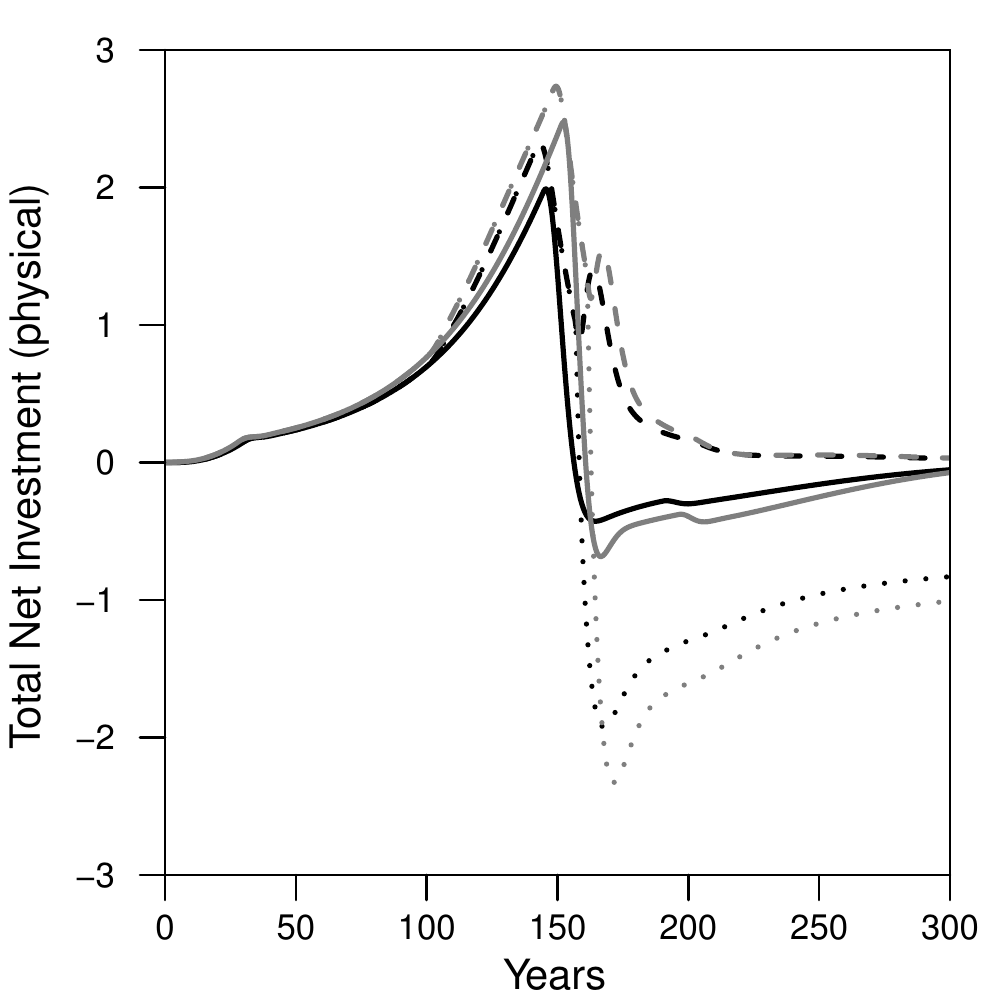}}
\subfloat[]{\includegraphics[width=0.33\columnwidth]{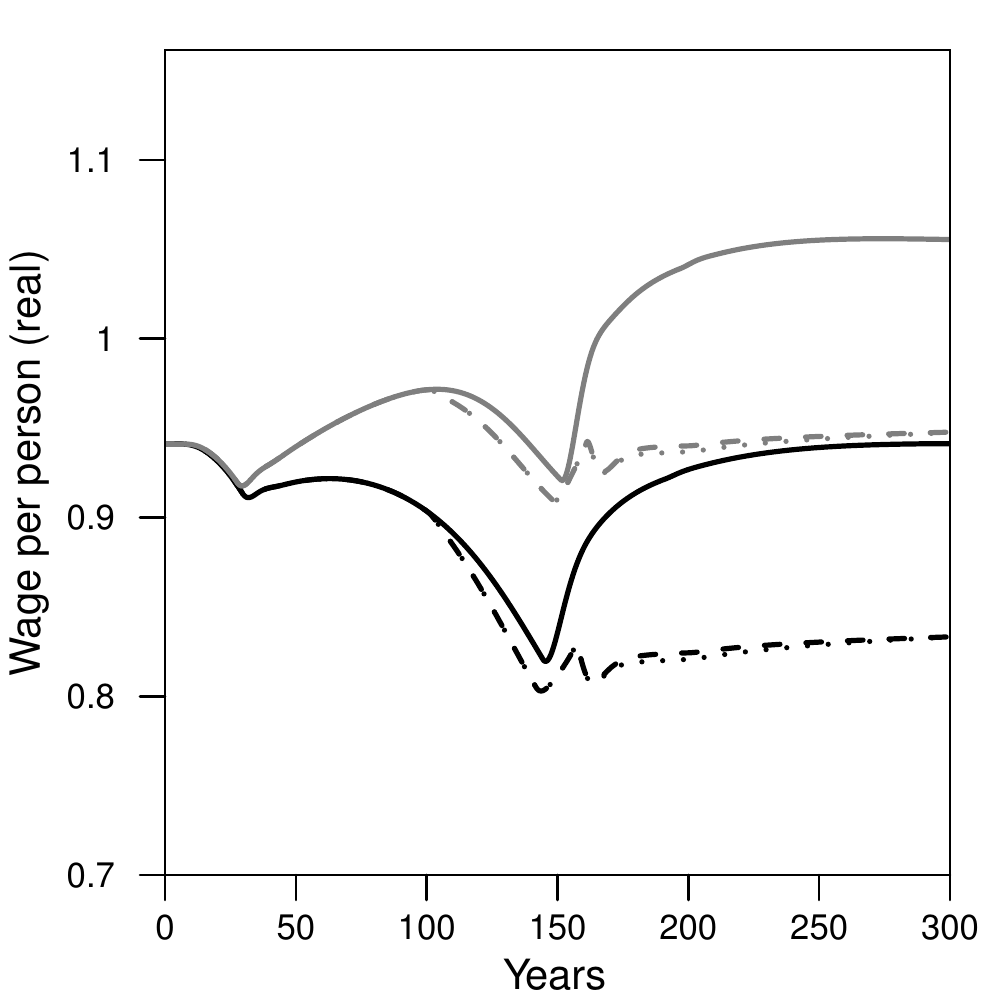}}\\
\subfloat[]{\includegraphics[width=0.33\columnwidth]{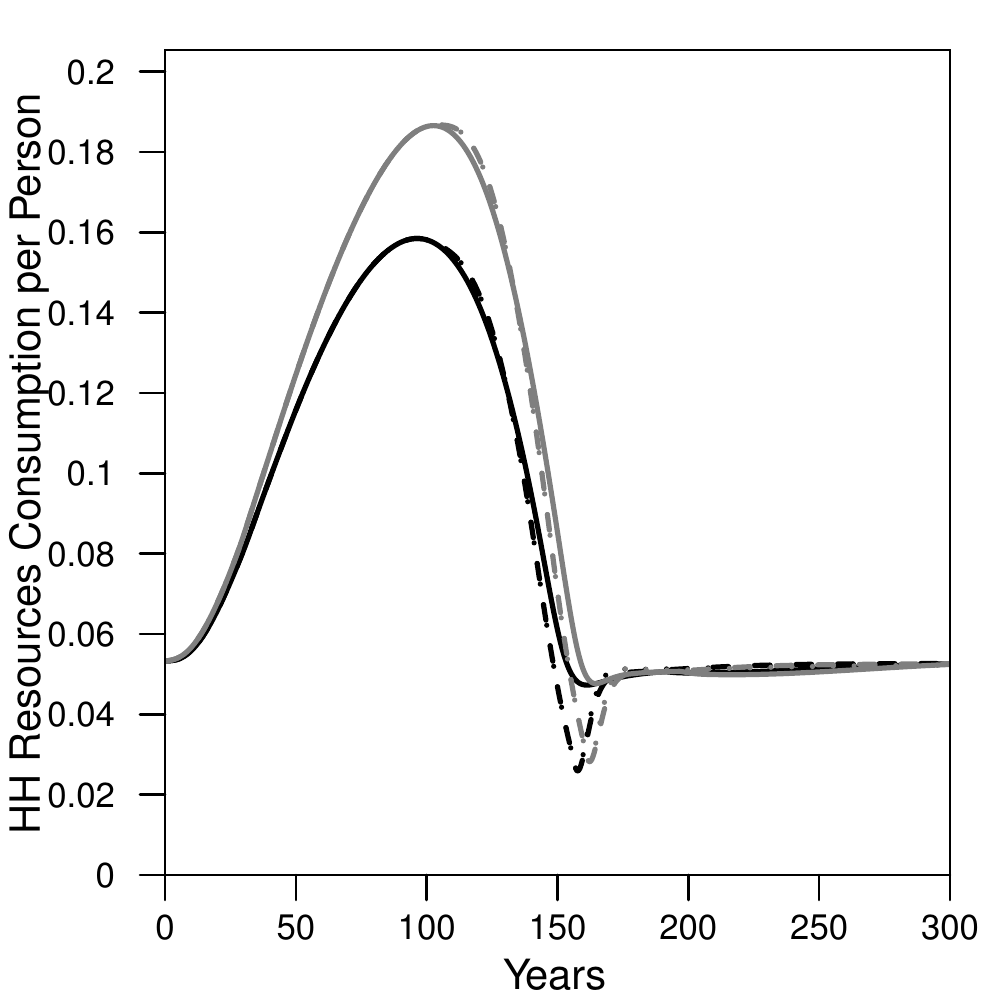}}
\subfloat[]{\includegraphics[width=0.33\columnwidth]{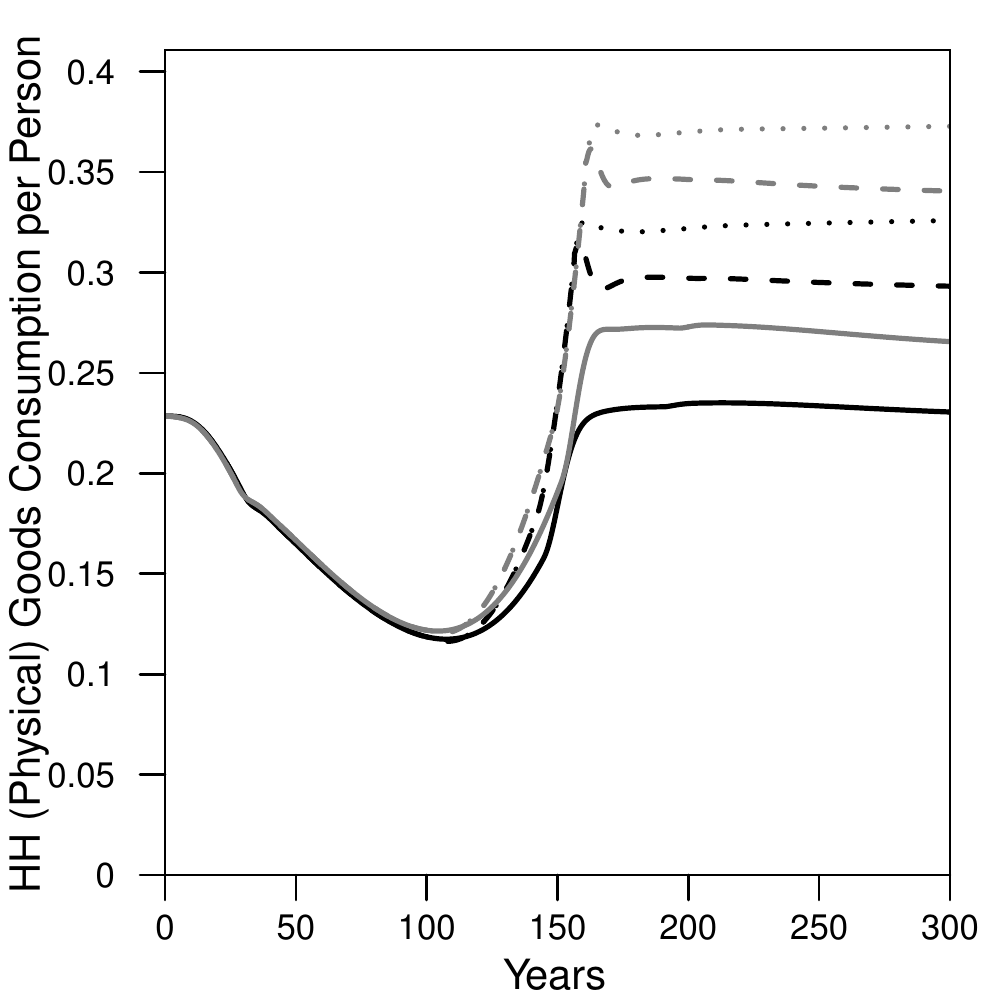}}
\subfloat[]{\includegraphics[width=0.33\columnwidth]{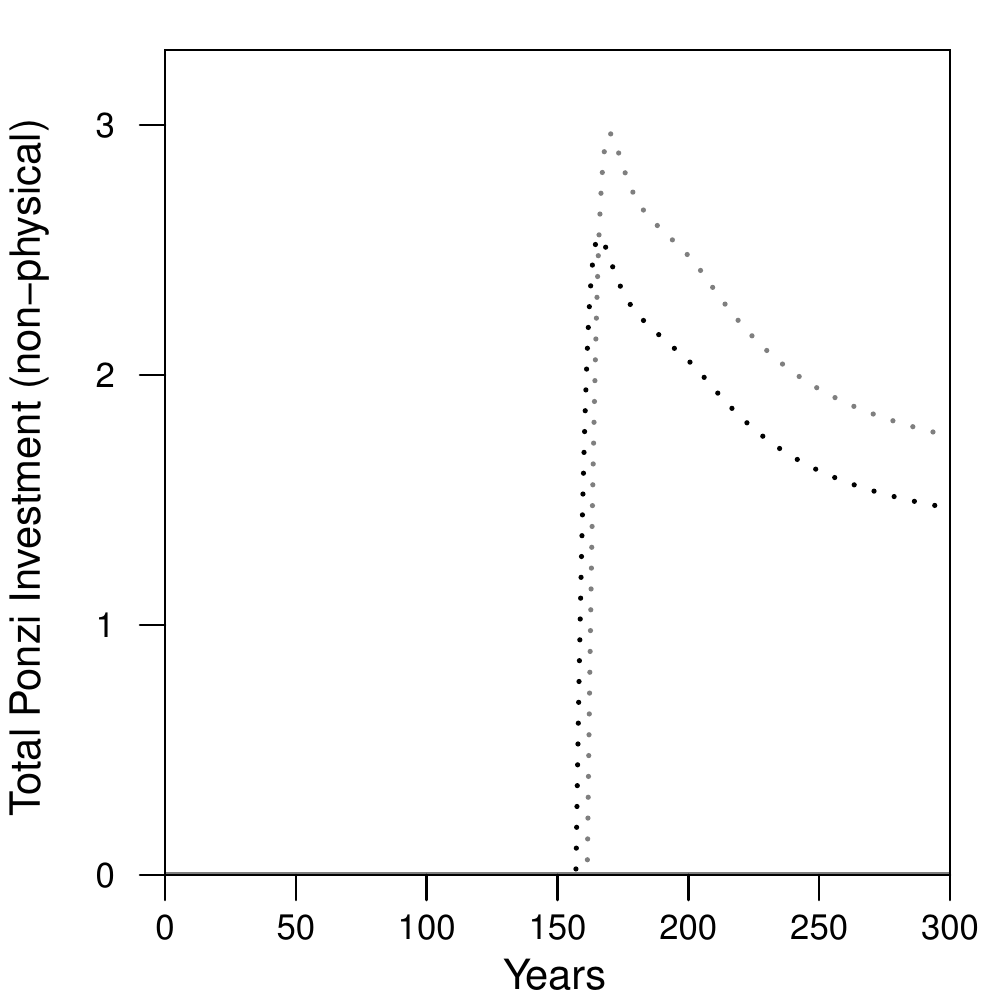}}\\
\caption{(continued) (j) profit share, (k) interest share, (l) wage share, (m) depreciation share, (n) physical net investment in new capital, (o) real wage per person, (p) household consumption of (physical) resources per person, (q) household consumption of (physical) goods per person, (r) total non-physical Ponzi investment, }
\end{center}
\end{figure}

\begin{figure}\ContinuedFloat
\begin{center}
\subfloat[]{\includegraphics[width=0.33\columnwidth]{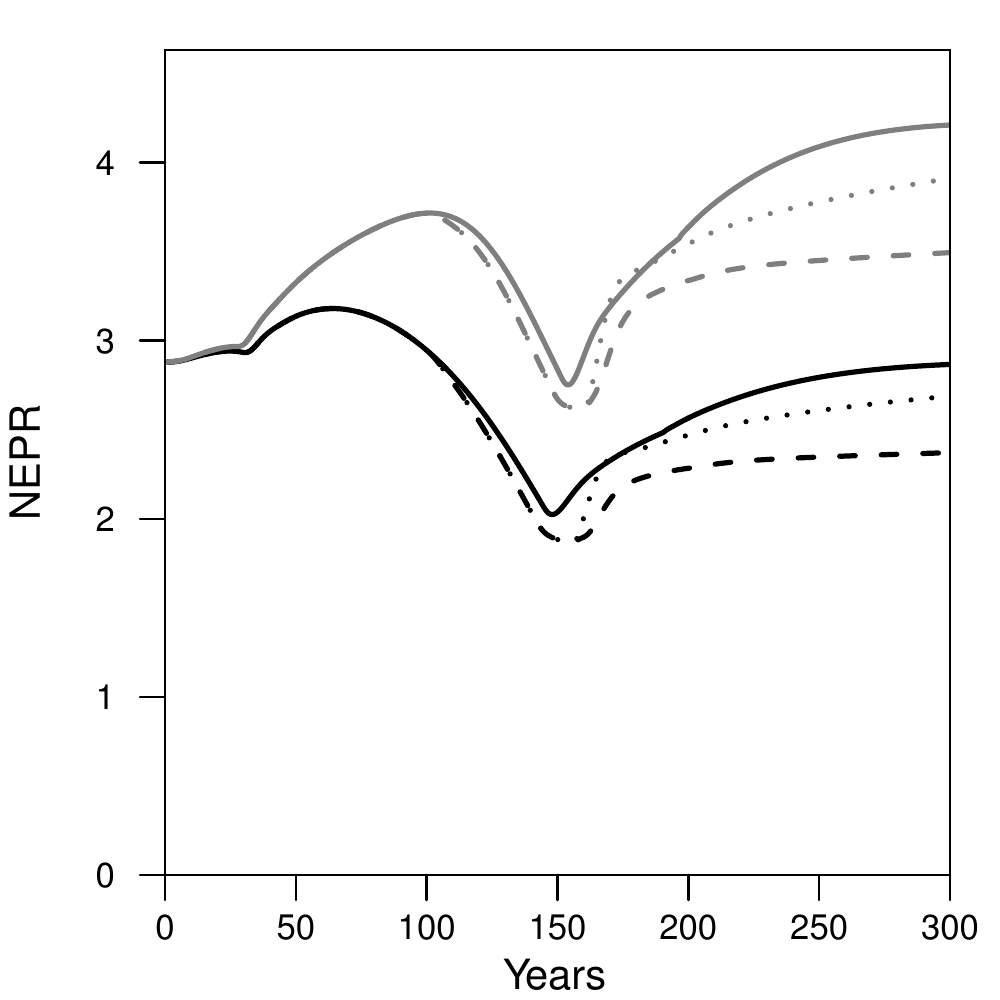}}
\subfloat[]{\includegraphics[width=0.33\columnwidth]{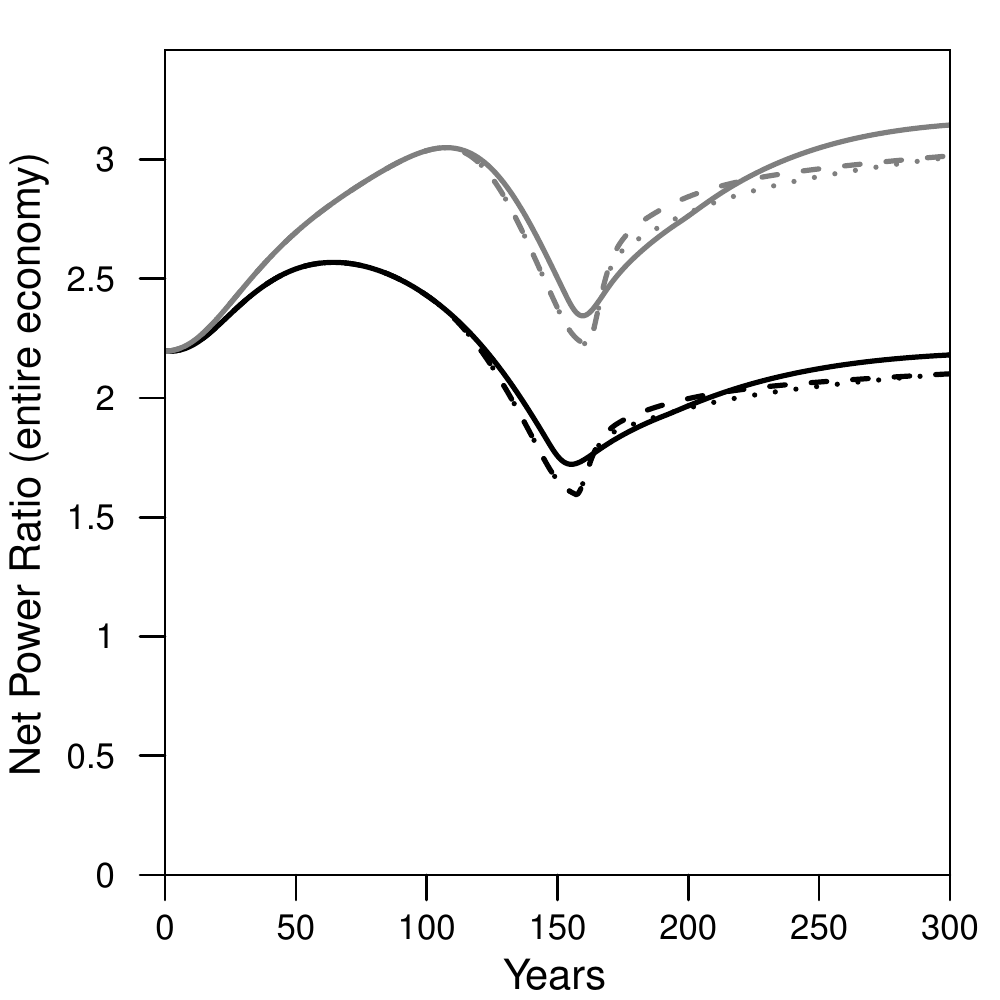}}
\subfloat[]{\includegraphics[width=0.33\columnwidth]{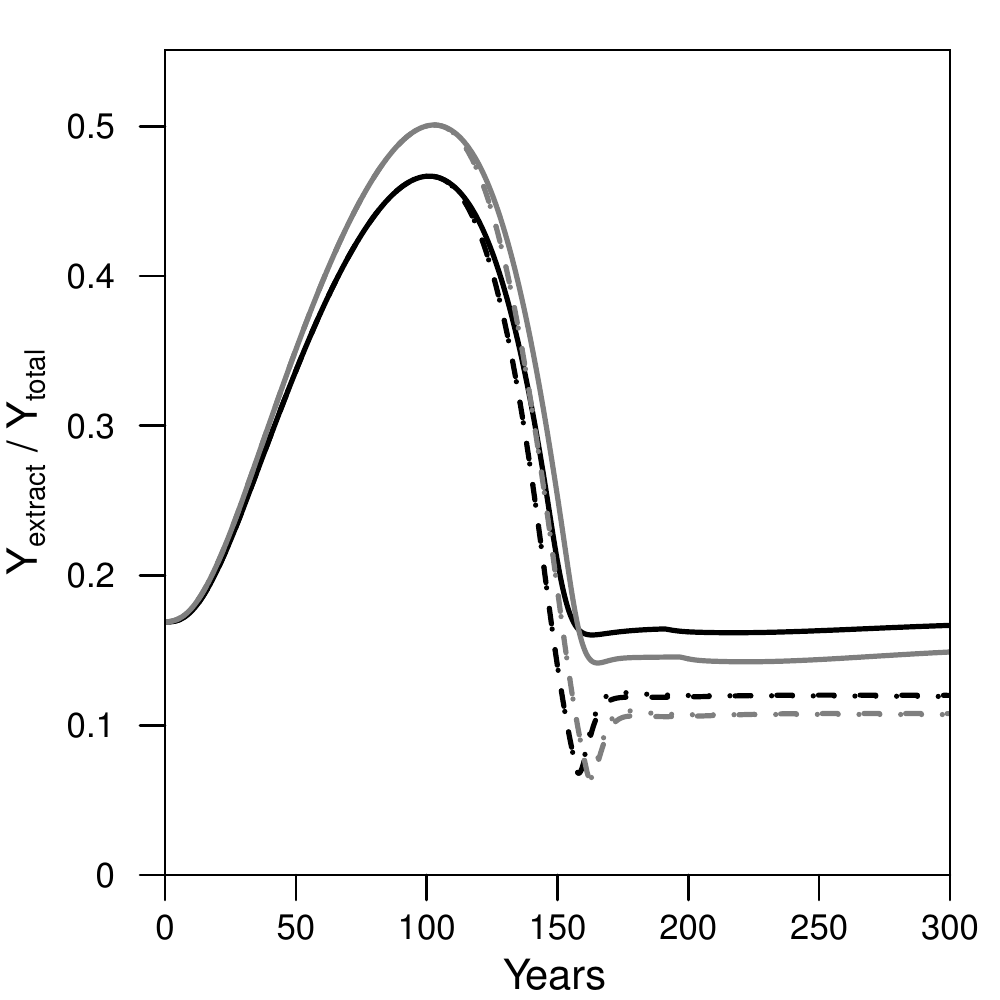}}\\
\subfloat[]{\includegraphics[width=0.33\columnwidth]{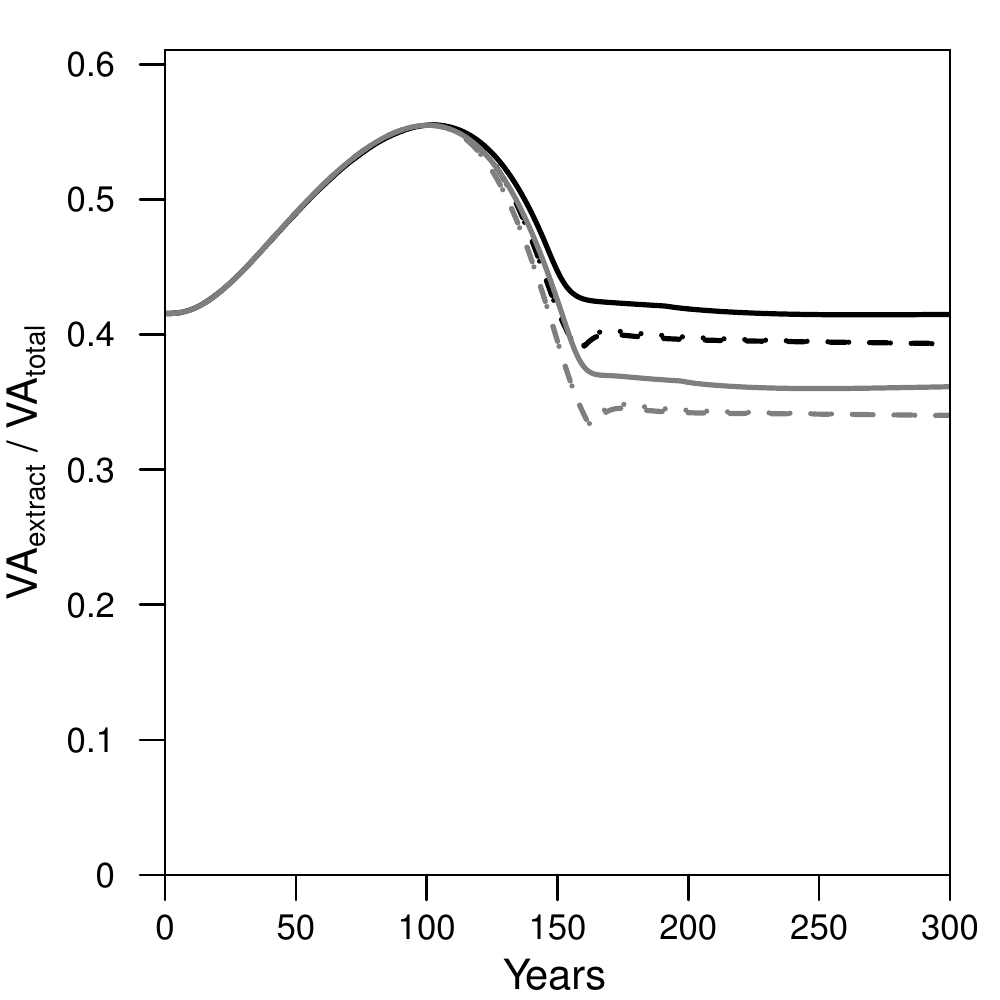}}
\subfloat[]{\includegraphics[width=0.33\columnwidth]{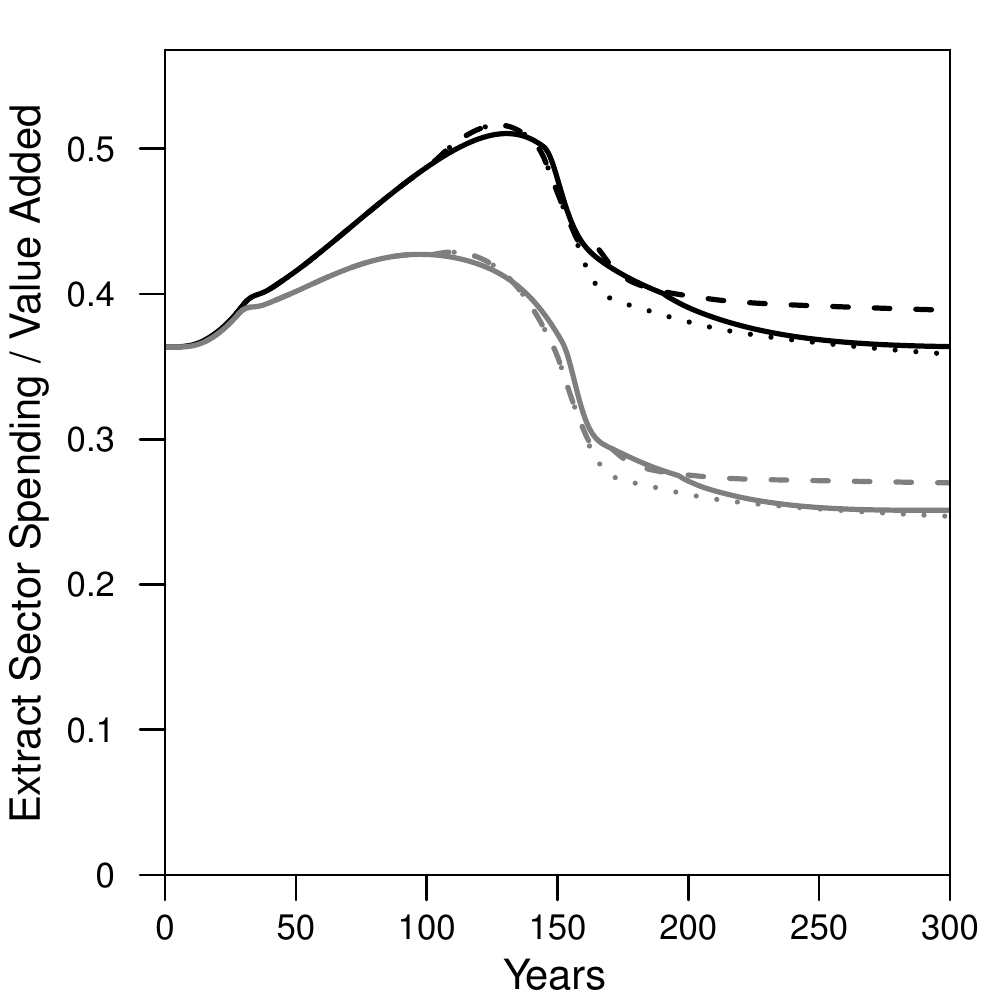}}
\subfloat[]{\includegraphics[width=0.33\columnwidth]{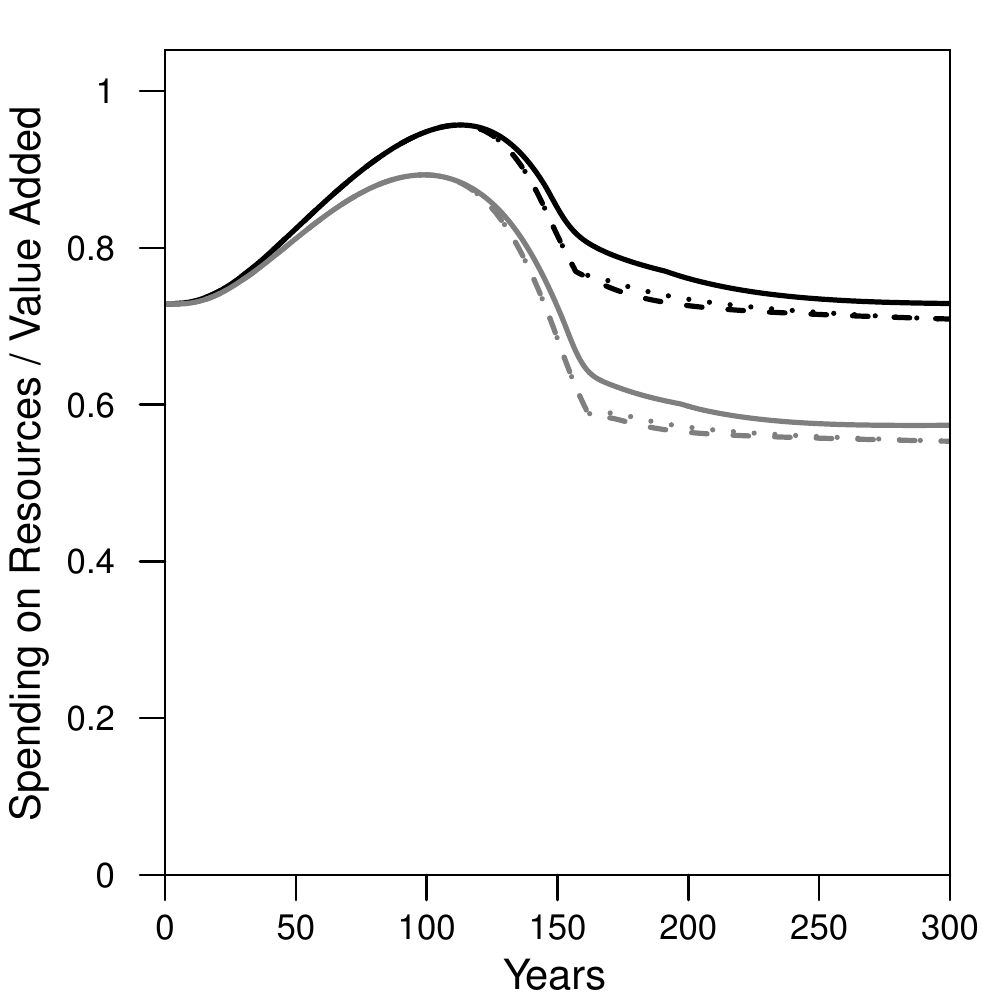}}\\
\subfloat[]{\includegraphics[width=0.33\columnwidth]{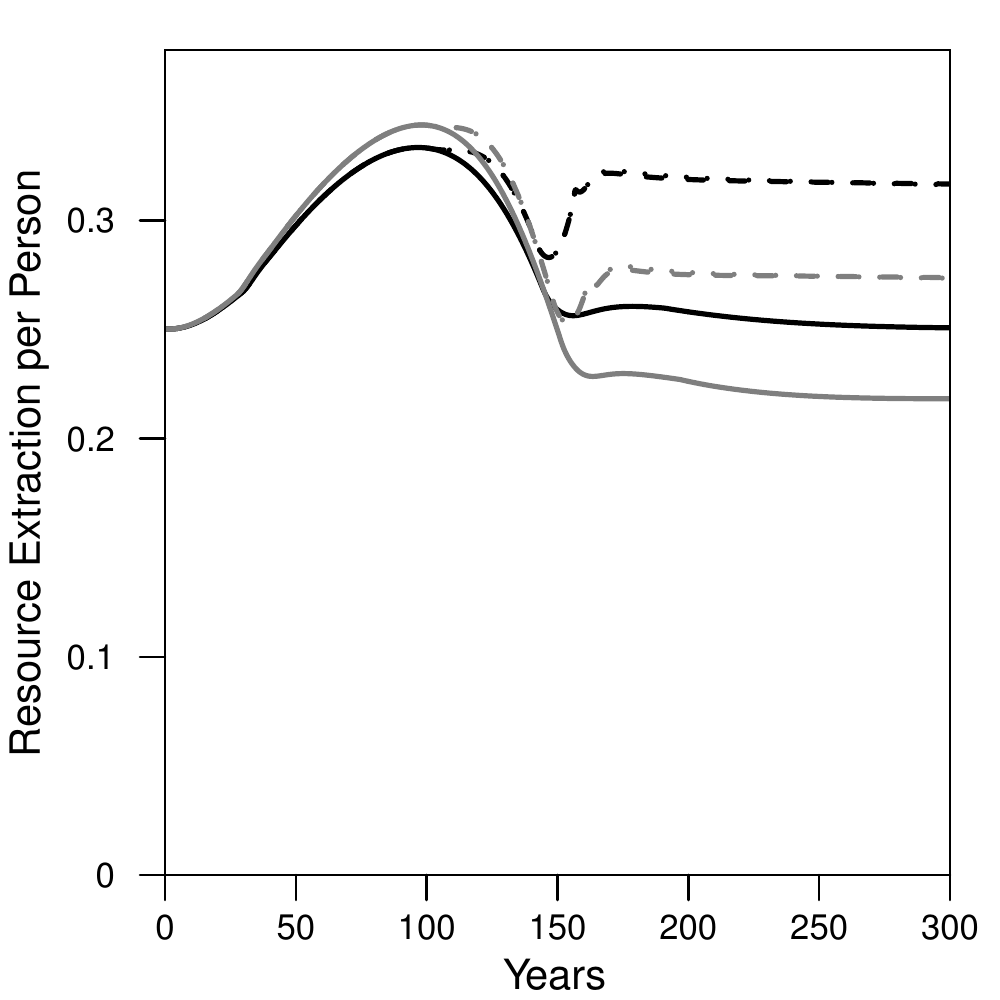}}
\subfloat[]{\includegraphics[width=0.33\columnwidth]{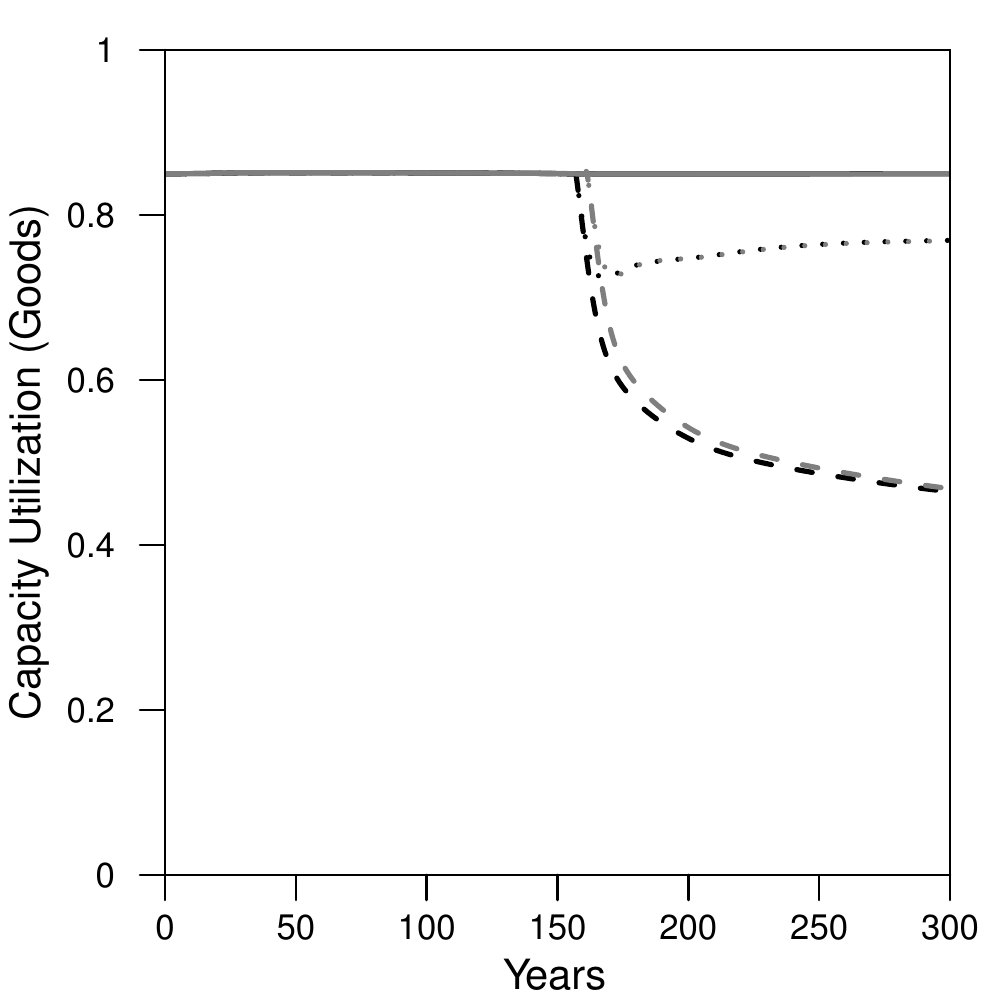}}
\caption{(continued) (s) net external power ratio (extraction sector, NEPR), (t) net power ratio (entire economy, NPR), (u) fraction of net output from extraction sector, (v) fraction of value added in extraction sector, (w) extraction sector spending per total value added (= per total net output), (x) spending on resources per total value added (= per total net output), (y) total resources extraction per person, and (z) capacity utilization of goods capital.} 
\end{center}
\end{figure}

\end{document}